%% file: WebPhD.tex
\documentclass[11pt]{book}

\usepackage{amsfonts}
\usepackage{amscd}
\usepackage{graphicx}
\usepackage{citesort}
\usepackage{mathrsfs}
\usepackage{bbm}
\usepackage{amsmath, amssymb, graphics}
\usepackage{fancyhdr}
\usepackage[Lenny]{fncychap}

\usepackage[OT2,OT1]{fontenc}
\newcommand\cyr{%
\renewcommand\rmdefault{wncyr}%
\renewcommand\sfdefault{wncyss}%
\renewcommand\encodingdefault{OT2}%
\normalfont \selectfont} \DeclareTextFontCommand{\textcyr}{\cyr}
\usepackage[small,bf]{caption}

\def\ad#1{{ #1}}

\newcommand{\fL}{{\mathbb L}}
\newcommand{\fR}{{\mathbb R}}

\newcommand{\fQ}{{\mathbb Q}}

\newcommand{\fC}{{\mathbb C}}

\newcommand{\fHH}{{\mathbb H}}

\newcommand{\comm}[2]{[#1,#2]}
\newcommand{\gen}[1]{\mathfrak{#1}}

\def\[{\begin{equation}}
\def\]{\end{equation}}
\def\<{\begin{eqnarray}}
\def\>{\end{eqnarray}}

\newcommand{\half}{\frac{1}{2}}

\newcommand{\acomm}[2]{\{#1,#2\}}

\newcommand{\nn}{\nonumber}

\newcommand{\alg}[1]{\mathfrak{#1}}
\newcommand{\su}{\alg{su}}

\newcommand{\sls}{\alg{sl}}

\newcommand{\mathsym}[1]{{}}

\newcommand{\stateA}[1]{|#1\rangle^{\rm{I}}}
\newcommand{\stateB}[1]{|#1\rangle^{\rm{II}}}
\newcommand{\stateC}[1]{|#1\rangle^{\rm{III}}}
\newcommand{\costateA}[1]{ ^{\rm{I}}\langle #1}

\def\pare#1{\left\{ #1\right\}}
\def\vev#1{\langle #1\rangle}

\def\ads{{\rm AdS}_5\times {\rm S}^5}
\def\S{{\mathbb S}}

\def\S{{\mathbb S}}

\def\[{\begin{equation}}
\def\]{\end{equation}}
\def\<{\begin{eqnarray}}
\def\>{\end{eqnarray}}

\def\defeq{\ \raisebox{0.2pt}{:}\hspace{-1.2mm}=}
\def\ds{\displaystyle}
\def\ssp{\hspace{0.3mm}}
\def\({\left(}
\def\){\right)}
\renewcommand{\[}{\left[}
\renewcommand{\]}{\right]}

\def\pare#1{\left\{ #1\right\}}
\def\vev#1{\langle #1\rangle}

\renewcommand{\eqref}[1]{$\({\rm \ref{#1}}\)$}

\fancypagestyle{plain}{%
\fancyhead{} 
}

\title{The S-matrix of the $\ads$ superstring}

\author{M de Leeuw}
\date{}

\begin{document}

\thispagestyle{empty}

\begin{center}

\begin{flushright}
\scriptsize{ITP-UU-10-13}\\[-.5ex]
\scriptsize{SPIN-10-11}\\[-.5ex]
\end{flushright}

\vskip 1cm

{{\huge\bf The S-matrix of the $\ads$ superstring}}

\vskip 2cm

Marius de Leeuw\footnote{e-mail: M.deLeeuw@uu.nl}\\
{\it Institute for Theoretical Physics and Spinoza Institute,\\
~~Utrecht University, 3508 TD Utrecht, The Netherlands} \vskip 2cm

{{\bf\large Abstract}}\vskip .5cm

\begin{minipage}[h]{\textwidth}
In this article we review the world-sheet scattering theory of
strings on $\ads$. The asymptotic spectrum of this world-sheet
theory contains both fundamental particles and bound states of the
latter. We explicitly derive the S-matrix that describes
scattering of arbitrary bound states. The key feature that enables
this derivation is the so-called Yangian symmetry which is related
to the centrally extended $\su(2|2)$ superalgebra. Subsequently,
we study the universal algebraic properties of the found S-matrix.
As in many integrable models, the S-matrix plays a key role in the
determination of the energy spectrum. In this context, we employ
the Bethe ansatz approach to compute the large volume energy
spectrum of string bound states.

\end{minipage}

\end{center}

\tableofcontents

\include{Intro2}

\include{IntegrabilityHopf2}

\include{SU22Algebra3}

\include{BoundSmatrix2}

\include{RMatrix2}

\include{UniversalBlocks4}

\include{BetheAnsatz}

\include{Transfer2}

\bibliographystyle{JHEP}

\bibliography{LitRmat}

\end{document}

%% file: Intro2.tex
\chapter{Introduction}

The current understanding of the microscopic world and gravity
originates from the beginning of the twentieth century, when two
revolutionary ideas saw the light of day. In 1900, Max Planck,
with the quantum hypothesis, initiated a field that would become
quantum mechanics and, in 1915, Albert Einstein formulated the
theory of general relativity.

The theory of relativity replaced Newton's theory of gravity by
considering space and time in a conceptually different way. Rather
than having an ambient space in which masses feel gravity and move
around, space and time are unified into an entity called
space-time. Space-time is curved by its matter and energy content.
Curvature is described by a metric which is an object that
measures distances and angles. The metric is a dynamical quantity
and it satisfies Einstein's equations of general relativity.
General relativity successfully describes corrections to the
orbits of planets that Newtonian gravity could not account for. It
also predicts novel effects like gravitational lensing and
gravitational time dilatation. These predictions were indeed
confirmed by experiments.


Where general relativity was more or less a finished theory, it
took quantum mechanics some years to mature into the theory that
is part of the standard physics curriculum of universities today.
Quantum mechanics radically changed the notions of particles and
forces, since it describes nature in a probabilistic way. Outcomes
of measurements can only be given in terms of probabilities and,
moreover, measurements inevitably influence the system under
study.

The quantum mechanical framework to describe systems with an
infinite number of degrees of freedom is known as quantum field
theory. Quantum field theories naturally include the concepts of
particle production and annihilation. The large amount of degrees
of freedom leads to superficial infinities that are caused by
particles being created and annihilated at the same space-time
point. To cope with these infinities, one employs the procedures
of regularization and renormalization to obtain finite and
measurable results for physical quantities.

By now, quantum field theory is one of the cornerstones of
theoretical physics. Quantum field theories are used in the
description of a wide variety of phenomena ranging from the
interactions between fundamental particles to condensed matter
systems. The parameters in a quantum field theory that describe
the interaction strength between particles are called coupling
constants. For example, in Quantum Electrodynamics the coupling
constant between the electron and the photon is given by the
charge of the electron $e$ and it determines the strength of the
electromagentic force on the electron. Quantum field theories with
small coupling constants are called weakly coupled. Weakly coupled
quantum field theories are well-understood. They can be studied in
a perturbative way, based on path integrals and Feynman diagrams.
On the other hand, the knowledge of strongly coupled quantum field
theories (i.e. large coupling constant) is only partial.



A relevant quantum field theory in which strongly coupled
phenomena play a role is Quantum Chromodynamics (QCD). This theory
describes the interactions between quarks and gluons. It is known
to be asymptotically free, meaning that at very small distances
quarks behave like free particles. In other words, in this regime,
QCD is effectively a weakly coupled theory. However, at large
distances, quarks couple strongly, which precludes perturbative
methods to theoretically explain the phenomenon of confinement:
why can hadrons - the bound states of quarks - be observed in
nature, but not the free quarks? Other areas where strong
interactions also are important, include for instance cold atomic
gases and high ${\rm T_c}$-superconductivity.

A class of quantum field theories that have remarkable properties
are the supersymmetric quantum field theories. Supersymmetry is a
symmetry between the bosons and fermions in a theory. Every boson
in the theory has a fermionic partner and vice versa. The theory
is invariant under the interchange of the particles and their
superpartners.

A special supersymmetric theory, that plays an important role in
this work, is $\mathcal{N}=4$ super Yang-Mills theory
($\mathcal{N}=4$ SYM). It is the maximally supersymmetric gauge
theory in four space-time dimensions. The gauge group is $SU(N) $
and the theory has a single coupling constant $g_{YM}$. The rank
$N$ of the gauge group can be seen as a free positive
integer-valued parameter. Introducing the 't Hooft coupling
$\lambda = g_{YM}^2N$, any perturbative Feynman diagrammatic
expansion rearranges itself as \cite{'tHooft:1973jz}
\begin{align}
Z = \sum_{n=0}^{\infty}N^{2-2n}\sum_{k=0}Z_{n,k}\lambda^k,
\end{align}
where the index $n$ can be seen as the genus of the surface on
which the corresponding diagram can be drawn. A particularly
interesting limit to consider is $N\rightarrow\infty$ while
keeping $\lambda$ fixed. In this limit only the diagrams
corresponding to zero genus contribute and for this reason it is
called the planar limit. In the planar limit $\mathcal{N}=4$ SYM
exhibits the remarkable features of a solvable model.

What really sets $\mathcal{N}=4$ SYM apart from a generic quantum
field theory is that it exhibits conformal symmetry at the quantum
level. This means in particular that the theory is invariant under
rescalings. Conformal symmetry puts strong constraints on a
theory. For example, it fixes two-point correlation functions of
scalar fields to be of the form
\begin{align}
\langle\mathcal{O}(x)\mathcal{O}(y)\rangle =
\frac{1}{|x-y|^{2\Delta}}.
\end{align}
The constant $\Delta$ is called the scaling dimension and will
depend non-trivially on the parameters $\lambda,N$. In general it
admits a perturbative expansion
\begin{align}
\Delta = \Delta_0 + \sum_{n=0}\sum_{m=1}
\frac{\lambda^m}{N^{2n}}\Delta_{m,n},
\end{align}
which in the planar limit becomes
\begin{align}
\Delta = \Delta_0 + \sum_{m=1} \lambda^m \Delta_{m}.
\end{align}




Quantum field theories, in the form of the Standard Model,
successfully describe the world of fundamental particles. How to
construct a quantum theory of gravity, however, is currently
unknown. One of the most viable candidates for a quantum theory of
gravity is superstring theory.

The idea of string theory is to consider extended objects, called
strings, rather than point-like particles as fundamental building
blocks. A string can have the topology of a rod (open string) or
of a ring (closed string). Different vibrations of a string
describe different types of particles. One particulary interesting
massless particle is found in the closed string spectrum. It
carries spin two and can be identified with a graviton, a quantum
of the gravitational field. Thus, string theory automatically
incorporates gravity via the quantum mechanical modes of closed
strings. Open strings can end on other extended objects called
D-branes \cite{Polchinski:1995mt}. Massless excitation modes of
such open strings give rise to gauge fields. Hence, open strings
are naturally linked to gauge theories. In this way string theory,
containing open and closed strings, offers a unified framework to
treat gravity and gauge theories.

\begin{figure}[h]
  \centering
  \includegraphics[scale=.5]{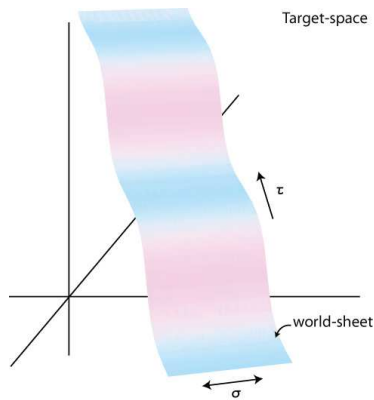}
  \caption{The world-sheet of an open string.}\label{Fig;OpenString}
\end{figure}

Superstring theories describe a string moving in a ten-dimensional
space called the target-space. A propagating string in the
target-space sweeps out a two-dimensional surface which is called
the world-sheet. It can be parameterized by two parameters
$\sigma,\tau$. The $\sigma$-variable is the coordinate
parameterizing the spatial extension along the string whereas
$\tau$ corresponds to the time direction, see figure
\ref{Fig;OpenString}. The vibrations of superstrings can have
bosonic and fermionic degrees of freedom, related to each other by
supersymmetry transformations. Such string theories are said to
have world-sheet supersymmetry. In addition, one can consider a
string which propagates in a target-space that is a superspace (it
has both bosonic and fermionic coordinates). The supersymmetric
structure originating in this way is called target-space
supersymmetry.

There are two coupling constants in string theory, called the
string tension $g$ and string coupling $g_s$. The string tension
$g$ describes the energy per unit length of the string and the
string coupling $g_s$ governs the splitting and joining processes
of strings. When $g_s=0$ there is no string splitting/joining and
the world sheet of a closed string is just a cylinder.

\begin{figure}[h]
  \centering
\parbox[b]{4in}{
  \begin{center}\includegraphics[scale =.9]{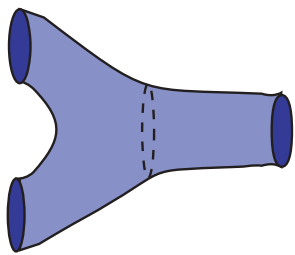}\end{center}
  \caption{Schematic representation of a joining/splitting process of closed strings. Any
such process is weighted with the coupling constant
$g_s$}\label{Fig;StringCoupling}}
\end{figure}

\begin{figure}[h]
  \centering
  \includegraphics{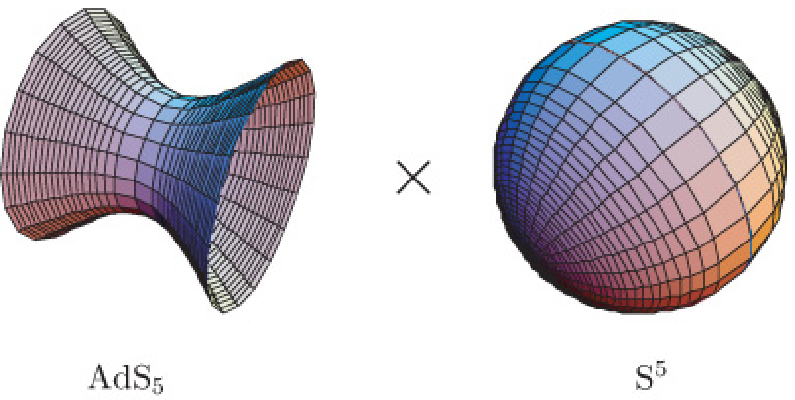}
  \caption{The $\ads$ space.}\label{Fig;AdS}
\end{figure}

An fascinating new insight in the dynamics of strings and strongly
coupled gauge theories came with the advent of the AdS/CFT
correspondence \cite{Maldacena:1997re}. The correspondence states
a duality between superstring theories of {\it closed strings} and
conformal field theories. It assumes that a string theory in an
anti-de Sitter (AdS) target-space is equivalent to a conformal
field theory (CFT) on the conformal boundary of this space. What
is remarkable about this conjecture is that it relates closed
strings, which are inherently related to gravity, to a quantum
field theory that has no gravitational degrees of freedom at all.
In this way it provides a realization of a profound duality
between open and closed strings.
\newpage
We will now continue with describing the conjecture in more detail
by considering the prototype example of the AdS/CFT
correspondence:
\begin{center}
$\mathcal{N}=4$ super Yang-Mills $\Leftrightarrow$ Type IIB $\ads$ superstring.
\end{center}
The $\ads$ superstring is the string theory in a special curved
target space, which is the product of the five-dimensional anti-de
Sitter space ${\rm AdS}_5$ and the five-dimensional sphere ${\rm
S}^5$, see figure \ref{Fig;AdS}. The ${\rm AdS}_5$ space is a
maximally symmetric, five-dimensional space with negative constant
curvature. It can be viewed as a hyperboloid embedded in flat
space, described by the equation $X_0^2+X_5^2-\sum_{i=1}^4X_i^2 =
1$.

According to the AdS/CFT correspondence, the string coupling
constant $g_s$ and the string tension $g$ are related to the
$\mathcal{N}=4$ SYM parameters, $\lambda, N$ as
\begin{align}\label{eqn;CouplingRelation}
&g_s = \lambda/4\pi N, && g = \sqrt{\lambda}/2\pi.
\end{align}
The correspondence also relates gauge invariant operators of
$\mathcal{N}=4$ SYM to string states. The scaling dimensions
$\Delta$ of these operators are mapped to the energies $E$ of
corresponding string states
\begin{align*}
\Delta  =  E.
\end{align*}
In other words, the spectrum of string energies should be equal to
the spectrum of scaling dimensions of $\mathcal{N}=4$ SYM. The
problem of computing both spectra is naturally called the
\emph{spectral problem}.

The AdS/CFT correspondence is a strong-weak duality. This means
that it relates the strongly coupled regime of field theories to
the weakly coupled regime of the corresponding string theory and
vice versa. As such, it enables to probe the strongly coupled
regime of field theories via weakly-interacting strings, giving
access to important strongly coupled phenomena in gauge theories.
This potentially provides a direct way for string theory to have
concrete applications, albeit more as a computational tool rather
than as a fundamental model.

The strong-weak duality, however, is a two-edged sword. It offers
truly exciting possibilities, yet it also presents a challenge in
understanding the precise nature of the relation between strings
and gauge theories. The reason is obvious; if one wishes to
perform computations on both sides of the duality and compare the
results, then generically one of the two sides will be strongly
coupled and hence hard to solve. However, it turns out that both
$\mathcal{N}=4$ SYM and the $\ads$ superstring exhibit a rich
hidden symmetry structure that allows to circumvent this problem.
Namely, in the planar limit both $\mathcal{N}=4$ super Yang-Mills
and the $\ads$ superstring are described by integrable models.
Integrable models are dynamical systems that have an infinite
number of conservation laws, which normally imply the existence of
an {\it exact} solution. One can therefore try to expand and
generalize the methods and tools developed in the theory of
integrable models, like for instance the Bethe ansatz, to explore
and understand this prototype example at the quantitative level.


Thus, through the gauge/string correspondence, integrability
offers the unprecedented possibility to completely solve, at least
in the planar limit, a non-trivial quantum field theory in four
dimensions. Understanding this prototype example would elucidate
underlying physical mechanisms and would provide a solid basis to
move on to other, more interesting, phenomenological models.

Integrability and its implementation for the $\ads$ superstring at
the quantum level is by no means straightforward. The world-sheet
theory turns out to be a two-dimensional non-relativistic quantum
field theory in finite volume. Standard techniques appear to be
insufficient for solving this theory and therefore new methods
have to be invented. One of the most recent developments in this
direction is the generalization of the so-called Thermodynamic
Bethe Ansatz for the $\ads$ mirror model.


In the remainder of this chapter we will first discuss the
emergence of integrability in both $\mathcal{N}=4$ SYM and the
$\ads$ superstring. Subsequently we will elaborate on how one can
use integrability techniques to find a complete solution of the
spectral problem. We close this presentation by giving an outline
of this review and the new results achieved.

\section{Integrability in $\mathcal{N}=4$ Super Yang-Mills Theory}

The constituents of the $\mathcal{N}=4$ SYM theory are: six scalar
fields $\Phi_{i}$, one vector field $A_{\mu}$ and four fermions
$\Psi$. The action is given by:
\begin{eqnarray}\label{ActionSYM4}
S=\frac{1}{g_{YM}^{2}}\int d^{4}x ~ \left\{
\frac{1}{4}(F_{\mu\nu})^{2} + \frac{1}{2}(D_{\mu}\Phi_{i})^{2}
-\frac{1}{4}[\Phi_i,\Phi_j]^2 + {\mathrm{fermions}}\right\}.
\end{eqnarray}
The fields are in the adjoint representation of $SU(N)$, thus
under a local transformation $U(x)\in SU(N)$ they transform as
\begin{align}\label{eqn;GaugeTransSUN}
&\mathcal{X}\rightarrow U\mathcal{X}U^{-1}, && A_{\mu} \rightarrow
U A_{\mu}U^{-1} - i (\partial_{\mu} U) U^{-1},
\end{align}
where $\mathcal{X} = \{\Phi_i,\Psi,F_{\mu\nu}\}$. We are
interested in gauge invariant composite operators. They are formed
by taking the trace over products of various fields $\mathcal{X}$,
for instance
\begin{align}
\mathcal{O}(x) = \mathrm{tr}(\ldots\Phi_i(x)\ldots \Psi^k(x) D_\mu\Phi_j(x)\ldots
F_{\mu\nu}(x)\ldots).
\end{align}
The symmetry algebra of $\mathcal{N}=4$ SYM includes the conformal
algebra $\alg{so}(2,4)\sim \alg{su}(2,2)$ which consists of the
Poincar\'e algebra together with conformal boosts and dilatation.
Adding the supersymmetry transformations extends the conformal
algebra to the superconformal algebra $\alg{psu}(2,2|4)$. This is
the full symmetry algebra of $\mathcal{N}=4$ SYM.

A conformal field theory is characterized by the set of primary
operators $\{\mathcal{O}_i\}$. These operators correspond to
highest weight states, i.e. they are annihilated by conformal
boosts and conformal supercharges of the superconformal algebra.
Primary operators are eigenstates of the dilatation operator
$\mathcal{D}$, which is one of the generators of the conformal
algebra $\alg{psu}(2,2|4)$
\begin{align}
\mathcal{D}\mathcal{O}_n = i \Delta_n \mathcal{O}_n.
\end{align}
In case these operators are scalar fields, one finds that their
two-point function is of the form
\begin{align}
\langle\mathcal{O}_i(x) \mathcal{O}_j(y)\rangle =
\frac{\delta_{ij}}{|x-y|^{2\Delta_i}}.
\end{align}
Thus, the spectrum of scaling dimensions can be determined either
by finding eigenvalues of the dilatation operator or by computing
the corresponding two-point functions.

Concerning the computation of two-point functions, one finds that
generically the operators $\mathcal{O}_i$ loose their tree-level
orthogonality as soon as the leading quantum correction is taken
into account. This can be understood as the appearance of a
non-trivial mixing
\begin{align}
\langle\mathcal{O}_i(x) \mathcal{O}_j(y)\rangle =
\frac{1}{|x-y|^{2\Delta_{classical}}}(\delta_{ij} + \lambda M_{ij}
\ln(|\mu(x-y)|) + \ldots ),
\end{align}
where $\mu$ is a mass scale and $M_{ij}$ is called mixing matrix.
The spectrum of conformal dimensions is then found by
re-diagonalizing the basis of operators. This procedure
effectively introduces a dependence of the scaling dimension on
the 't Hooft coupling $\Delta=\Delta(\lambda,N)$.

In the planar limit, a remarkable simplification happens. The
dilatation operator can be identified with a spin chain
Hamiltonian, while composite gauge invariant operators are then
naturally interpreted as states of this spin chain
\cite{Minahan:2002ve,Beisert:2003tq,Beisert:2003yb,Serban:2004jf,Beisert:2004ry}.

\begin{figure}[h]
  \centering
\parbox[b]{4in}{
  \begin{center}\includegraphics[scale =.5]{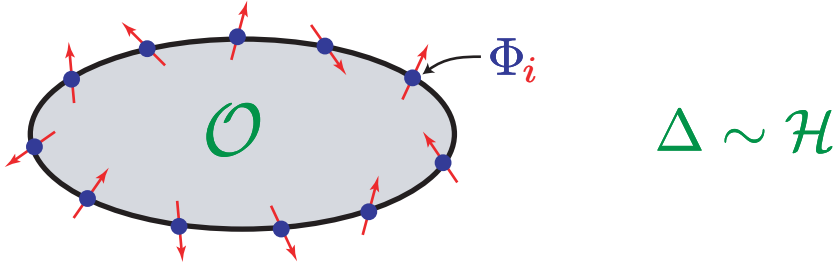}\end{center}
  \caption{Operators correspond to states on a spin chain. The fields correspond to lattice sites.
  The scaling dimensions $\Delta$ are identified with eigenvalues of a Hamiltonian $\mathcal{H}$}\label{Fig;SpinChain}}
\end{figure}

To exemplify this, let us restrict to scalar fields at the
one-loop level. One can make the identification
\begin{align}
\mathrm{tr}(\Phi_{i_1}(x)\ldots \Phi_{i_J}(x)) \longrightarrow
|\Phi_1\rangle\otimes\ldots\otimes |\Phi_J\rangle,
\end{align}
which is a state of a $\alg{so}(6)$ spin chain (the indices of the
scalar fields transform under this algebra) with $J$ sites.
Cyclicity of the trace is then equivalent to the spin chain being
closed.

By explicitly computing one-loop diagrams, one can show that the
dilatation operator acts on neighboring fields only; it is of the
form
\begin{align}
&\mathcal{D}_{1-{\rm loop}} = \sum_{i=1}^J H_{i,i+1}.
\end{align}
This dilatation operator can be recognized as an integrable
Hamiltonian of the $\alg{so}(6)$ spin chain. Finding scaling
dimensions thus reduces to computing the eigenvalues of this
integrable spin chain Hamiltonian (figure \ref{Fig;SpinChain}).
From the gauge theory point of view, the ground state of the
Hamiltonian corresponds to
\begin{align}
&\qquad {\rm tr}(\mathcal{Z}^J), && \mathcal{Z} \equiv \Phi_m +  i
\Phi_n,
\end{align}
for some $n,m=1,\ldots,6$. This operator is known as half-BPS,
which means that it is annihilated by half of the Poincar\'e
supercharges. As a consequence of the superconformal algebra, its
scaling dimension is $\Delta=J$ and it is not affected by quantum
corrections.

\section{String model and integrability}

The action of the $\ads$ superstring string is of the form
\cite{Metsaev:1998it}
\begin{align*}
& S = -\frac{g}{2}\int d\tau d\sigma~g_{\mu\nu}(x)\partial^{\alpha}
x^\mu\partial_{\alpha} x^\nu + {\rm fermions}, && \alpha=\sigma,\tau
\end{align*}
where $g_{ij}$ is the metric of the $\ads$ space. Let $(t,z^i)$ be
the coordinates of ${\rm AdS}_5$ and $(\phi,y^i)$ the coordinates
of ${\rm S}^5$, then it is given by
\begin{align}
&ds^2 = g_{tt}dt^2 + g_{\phi\phi}d\phi^2 + g_{yy}dy^{i}dy^i +
g_{zz}dz^{i}dz^i,
\end{align}
with
\begin{align}
&g_{tt} = \left(\frac{4+z^2}{4-z^2}\right)^2,&& g_{\phi\phi} =
\left(\frac{4-y^2}{4+y^2}\right)^2,&& g_{yy} =
\frac{1}{(1+\frac{y^2}{4})^2},&& g_{zz} =
\frac{1}{(1-\frac{z^2}{4})^2},
\end{align}
where $y^2 = y^iy^i$ and $z^2 = z^i z^i$. It is easily seen that
the metric $g_{ij}$ has isometries corresponding to shifts along
the time direction $t$ of the AdS space and to shifts along the
big circle $\phi$ of the sphere. These correspond to global
symmetries of the string model. The associated Noether charges are
the energy $E$ of the string and its angular momentum $J$
respectively.

It is useful to note that we can write both ${\rm AdS}_5$ and
${\rm S}^5$ as cosets of Lie groups:
\begin{align}
&{\rm AdS}_5 = \frac{{\rm SO(4,2)}}{{\rm SO(4,1)}}, && {\rm S}^5 =
\frac{{\rm SO(6)}}{{\rm SO(5)}}.
\end{align}
More precisely, the supergroup $\mathrm{PSU}(2,2|4)$ contains
$\mathrm{SU}(2,2)\times\mathrm{SU}(4)$ as a bosonic subgroup which
is locally isomorphic to $\mathrm{SO}(4,2)\times\mathrm{SO}(6)$ .
Modding $\mathrm{PSU}(2,2|4)$ out by
$\mathrm{SO}(4,1)\times\mathrm{SO}(5)$ then gives a supersymmetric
space with $\ads$ as bosonic part. This is then a natural
target-space for the $\ads$ superstring and indeed, the
superstring on $\ads$ can be described on the coset
\cite{Metsaev:1998it}
\begin{align*}
\frac{\mathrm{PSU}(2,2|4)}{\mathrm{SO}(4,1)\times\mathrm{SO}(5)},
\end{align*}
see also \cite{Arutyunov:2009ga} for an extensive review. One
advantage of this formulation is that it makes the global
$\alg{psu}(2,2|4)$ symmetry manifest. Notice that
$\alg{psu}(2,2|4)$ is also the symmetry algebra of $\mathcal{N}=4$
SYM.

One can prove that, classically, the string equations of motion
admit a so-called Lax representation \cite{Bena:2003wd}. This
property implies the existence of an infinite number of
conservation laws. In other words, the $\ads$ superstring is
classically integrable. The Lax representation allows one to
explicitly construct the corresponding conserved charges
\cite{Arutyunov:2003rg} and find the solutions to the string
equations of motion \cite{Kazakov:2004qf}.

However, one ultimately is interested in the determination of the
full quantum spectrum. At the quantum level, the $1+1$ dimensional
quantum field theory on the world-sheet also shows signs of being
integrable. The spectrum of spinning strings is compatible with
the assumption of integrability
\cite{Frolov:2002av,Frolov:2003qc,Gubser:2002tv,Frolov:2003tu,Frolov:2003xy,Arutyunov:2003uj,Arutyunov:2003za,Frolov:2004bh,Park:2005ji,Beisert:2003xu,Beisert:2003ea,Beisert:2005mq,Hernandez:2005nf,Beisert:2005bv,Beisert:2005cw,Freyhult:2006vr,Beccaria:2010ry}
and scattering data of world-sheet excitations also seems to
exhibit the properties of integrable theories
\cite{Klose:2006zd,Puletti:2007hq,Klose:2007rz}. From now on we
will assume full quantum integrability of the model and try to
understand its consequences. A necessary check of this assumption
is that all data that has been derived assuming integrability is
in complete agreement with explicit computations.

To find the spectrum of quantum integrable field theories, one can
employ the S-matrix approach, which proved to work well for many
known integrable models. In the context of the AdS/CFT
correspondence this approach was initiated in
\cite{Staudacher:2004tk}. The AdS/CFT S-matrix and its symmetries
constitute one of the main topics of this review. Below we will
discuss a possible route one can undertake to find the string
spectrum with the help of the S-matrix approach.



\section{Large volume spectrum from symmetry}

\begin{figure}[h]
  \centering
\parbox[b]{5.5in}{
\begin{center}\includegraphics[scale =
.47]{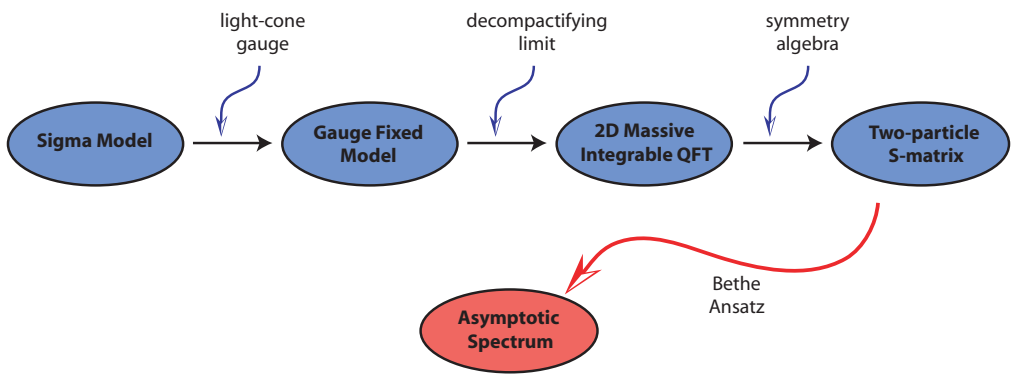}\end{center}
  \caption{Road map for the asymptotic spectrum. For the gauge fixed model
one finds in the infinite volume limit a massive integrable field
theory. This theory has centrally extended $\alg{su}(2|2)$ as
symmetry algebra. The S-matrix can be found by requiring
compatibility with this algebra and is then used to find the
spectrum via the Bethe ansatz.}\label{Fig;Asymptotics}}
\end{figure}

Integrability allows one to find the complete solution of the
`large volume' spectrum. On the gauge theory side this spectrum
describes the scaling dimensions of operators composed of a large
number of fields. On the string theory side, this gives the energy
spectrum when the spatial size of the world-sheet goes to
infinity. The S-matrix will play a crucial role in the derivation
of the spectrum. How one obtains the spectrum is depicted in
figure \ref{Fig;Asymptotics} and in what follows we will elaborate
on the different steps.

\subsection*{Symmetries}

The energy of the string corresponds to the Noether charge
associated with the time direction in the AdS space. It turns out
that it can be related to a world-sheet Hamiltonian. This reduces
the problem of finding the string energy to solving a spectral
problem in a two-dimensional quantum field theory.

To see how this comes about, one works in the Hamiltonian
formalism. We introduce the conjugate momenta $p_i$ to a variable
$x^i$
\begin{align}
p_i = \frac{\delta S}{\delta (\partial_{\tau} x^i)}.
\end{align}
Consider the time direction $t$ in the anti-de Sitter space and
the angle $\phi$ describing one of the big circles of $S^5$. The
associated conserved charges $E$ and $J$ can be written in terms
of the conjugate momenta
\begin{align}
&E \sim \int_{-r}^r d\sigma~ p_t, && J \sim \int_{-r}^r  d\sigma~
p_\phi,
\end{align}
where we take the string world-sheet to be of size
$-r\leq\sigma\leq r$.

To remove the non-physical degrees of freedom one imposes the
so-called the light-cone gauge
\cite{Arutyunov:2004yx,Frolov:2006cc,Arutyunov:2006gs}. To this
end, we define the light-cone coordinates
\begin{align}
&x_-  = \phi-t, && x_+ = t , && p_- = p_\phi + p_t, && p_+ =
p_\phi,
\end{align}
In these coordinates, the conserved charges can be expressed as
follows
\begin{align}
P_- \sim \int_{-r}^r d\sigma p_- = J-E, && P_+ \sim \int_{-r}^r
d\sigma p_+ = J.
\end{align}
The light-cone gauge is now imposed by setting
\begin{align}
&x_+ = \tau, && p_+ = 1,
\end{align}
from which it follows that $r \sim P_+ = J$. In other words, the
size of the world-sheet is proportional to the angular momentum of
the string. To find the complete gauge fixed action, one solves
the Virasoro constraints which give the light-cone momentum in
terms of the transverse coordinates $p_-(x^i,x^{\prime i})$. The
world-sheet Hamiltonian density is given by
\begin{align}
\mathcal{H} = - p_-(x^i,x^{\prime i}).
\end{align}
The world-sheet Hamiltonian is then related to the string energy
and angular momentum via
\begin{align}
H = \int_{-r}^r d\sigma \mathcal{H} = E-J.
\end{align}
Here one sees that the space-time energy $E$ of the string is
directly related to the spectrum of the world-sheet Hamiltonian
$H$ in this gauge. This means that one can find the spectrum of
the superstring (and hence the spectrum of scaling dimensions of
the dual four-dimensional field theory) by solving the spectral
problem of the two-dimensional world-sheet theory.

For closed strings, periodicity of the fields implies that the
total world-sheet momentum $p$ defined by
\begin{align}
p \equiv -\int_{-r}^r d\sigma~ p_i \partial_{\sigma}x^i
\end{align}
has to vanish. This condition is referred to as the level-matching
condition. States that satisfy this condition are called on-shell
and states with non-vanishing momentum are called off-shell. The
level-matching condition can not be solved explicitly for the
fields, but it needs to be imposed on physical states in the
theory. One then proceeds by studying the off-shell theory,
keeping in mind that for physical states the level-matching needs
to be imposed at the end.

The next step is to consider the limit $P_+\rightarrow \infty$
while keeping the string tension $g$ fixed. In this limit, the
world-sheet theory becomes a massive field theory defined on a
plane. Because of this, asymptotic states and the S-matrix are
well-defined. The gauged superstring has eight bosonic fields and
eight fermionic fields.

\begin{figure}[h]
  \centering
  \includegraphics[scale=2]{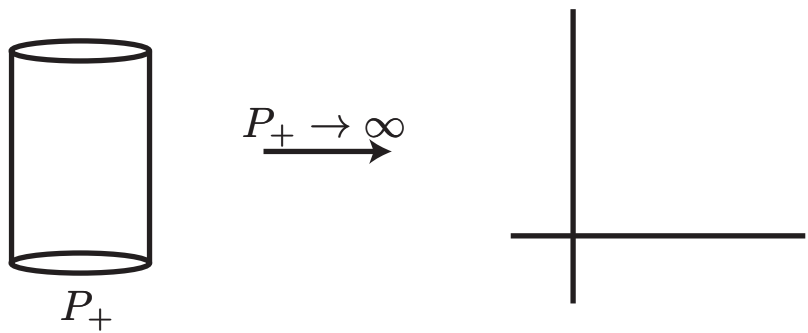}
  \caption{In the infinite volume limit theory is defined on a plane.}
\end{figure}

The gauge-fixing string model still has some residual symmetry
left. It turns out that the $\alg{psu}(2,2|4)$ algebra from the
coset model is reduced to two copies of $\alg{psu}(2|2)$\footnote{
The same algebra also appears on the field theory side
\cite{Beisert:2005tm}.} \cite{Arutyunov:2006ak}. For the off-shell
theory this symmetry algebra gets extended and becomes two copies
of centrally extended $\alg{su}(2|2)$. More precisely, the $16$
degrees of freedom ($8$ bosons and $8$ fermions)
\begin{align}
8_B + 8_F = 16 = 4\times 4.
\end{align}
transform under the tensor product of two fundamental
representations of centrally extended $\alg{su}(2|2)$. This
fundamental representation is four dimensional and is spanned by
two bosonic and two fermionic basis vectors.

The centrally extended $\alg{su}(2|2)$ Lie superalgebra consists
of two copies of $\alg{su}(2)$, whose generators we denote by
$\fL,\fR$ together with two sets of supersymmetry generators
$\fQ,\fQ^{\dag}$. The non-trivial commutation relations are given
by
\begin{align}
&[\mathbb{L}_{a}^{b},\mathbb{J}_{c}] =
\delta_{c}^{b}\mathbb{J}_{a}-\frac{1}{2}\delta_{a}^{b}\mathbb{J}_{c}&
&[\mathbb{R}_{\alpha}^{\beta},\mathbb{J}_{\gamma}] = \delta_{\gamma}^{\beta}\mathbb{J}_{\alpha}-\frac{1}{2}\delta_{\alpha}^{\beta}\mathbb{J}_{\gamma}\nonumber\\
&[\mathbb{L}_{a}^{b},\mathbb{J}^{c}]=-\delta_{a}^{c}\mathbb{J}^{b}+\frac{1}{2}\delta_{a}^{b}\mathbb{J}^{c}&
&[\mathbb{R}_{\alpha}^{\beta},\mathbb{J}^{\gamma}] = -\delta^{\gamma}_{\alpha}\mathbb{J}^{\beta}+\frac{1}{2}\delta_{\alpha}^{\beta}\mathbb{J}^{\gamma}\\
&\{\mathbb{Q}_{\alpha}^{a},\mathbb{Q}_{\beta}^{b}\}=\epsilon_{\alpha\beta}\epsilon^{ab}\mathbb{C}&
&\{\mathbb{Q}^{\dag \alpha}_{a},\mathbb{Q}^{\dag \beta}_{b}\}=\epsilon^{\alpha\beta}\epsilon_{ab}\mathbb{C}^{\dag}\nonumber\\
&\{\mathbb{Q}_{\alpha}^{a},\mathbb{Q}^{\dag\beta}_{b}\} =
\delta_{b}^{a}\mathbb{R}_{\alpha}^{\beta} +
\delta_{\alpha}^{\beta}\mathbb{L}_{b}^{ a}
+\frac{1}{2}\delta_{b}^{a}\delta_{\alpha}^{\beta}\mathbb{H},&&\nonumber
\end{align}
where $\mathbb{J}$ stands for any generator with appropriate index
structure. The central extensions $\fC,\fC^{\dag}$ make the
anti-commutator between the supercharges non-zero. The central
charges are related to the world-sheet momentum $p$ and the string
tension $g$ as
\begin{align}
&\mathbb{C} = \frac{ig}{2}(e^{ip}-1), & &\mathbb{C}^{\dag} =
-\frac{ig}{2}(e^{-ip}-1).
\end{align}
For on-shell configurations these central charges vanish.

\subsection*{Scattering}

The existence of an infinite number of conservation laws greatly
restricts scattering in an integrable theory. More precisely,
scattering processes in integrable quantum field theories exhibit
the following properties
\begin{itemize}
  \item Absence of particle production and annihilation
  \item Conservation of the sets of initial and final momenta
  \item Factorization of multi-particle scattering into a sequence of two-body
  scattering events
  \item The two-body S-matrix satisfies a consistency condition
  which deals with equivalent orderings of scattering of
  multi-particle states called the Yang-Baxter equation.
\end{itemize}
These properties imply that, in integrable models, the two-body
S-matrix is the fundamental building block of the scattering
theory.

\begin{figure}[h]
  \centering
  \includegraphics[scale=.7]{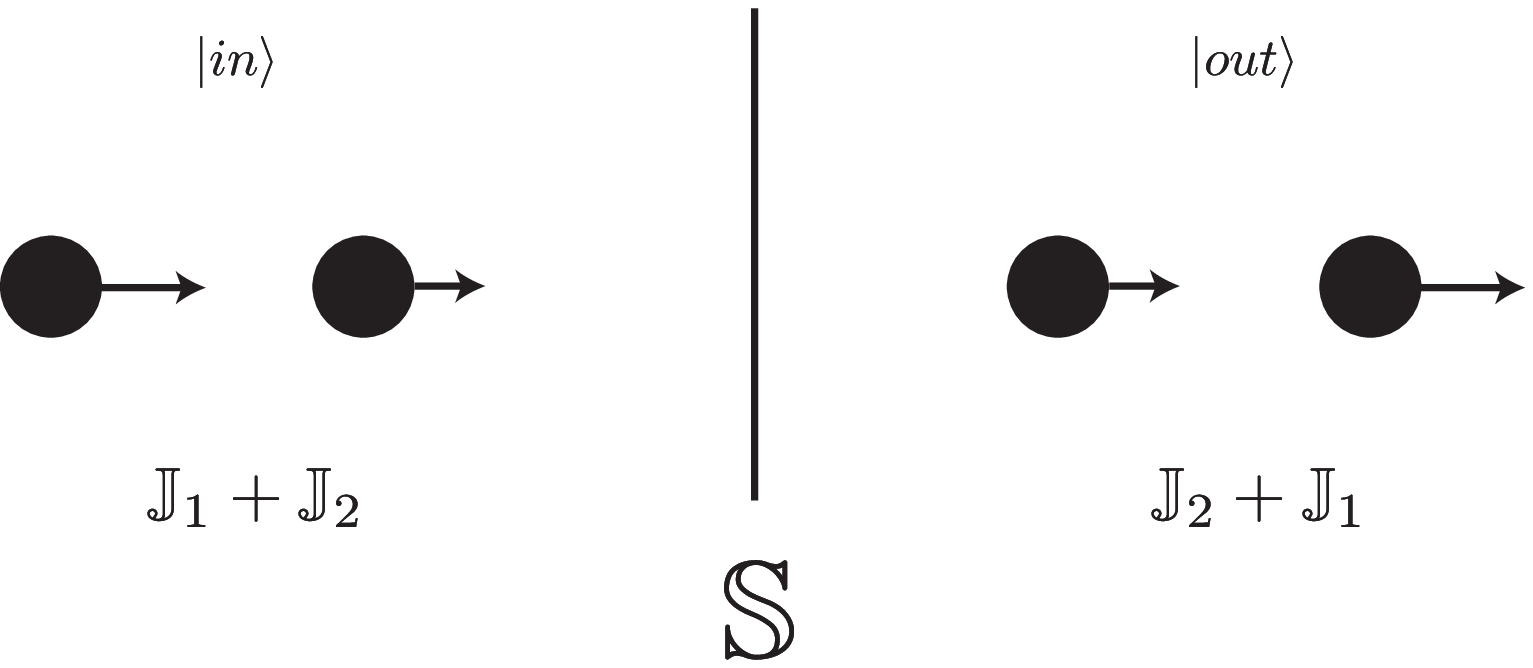}
  \caption{Symmetry commutes with scattering.}\label{Fig;ScatteringSymm}
\end{figure}

Having identified the symmetry properties of the gauge-fixed action, we
should find its implications for scattering processes. The
S-matrix relates $in$-eigenstates to $out$-eigenstates of the
Hamiltonian and it should be compatible with the symmetry of the
underlying model. This means that any such scattering matrix $\S$
should commute with the action of any symmetry generator
$\mathbb{J}$
\begin{align}
\S ~ \mathbb{J} = \mathbb{J} ~ \S.
\end{align}
This is schematically depicted in figure \ref{Fig;ScatteringSymm}.
The action of the symmetry generators on the {\it in}- and {\it
out}-states are encoded in a structure called the coproduct. The
coproduct is an operation in Hopf algebras which naturally encodes
the action of symmetry generators on two-particle states. Since
each of the world-sheet excitations transforms in a sixteen dimensional
representation, the S-matrix will be a $256\times 256$ dimensional
matrix.

The remarkable fact is that requiring the two-body S-matrix for
fundamental excitations to respect centrally extended
$\alg{su}(2|2)$ is enough to fix it up to an overall scalar factor
$S_0$ \cite{Beisert:2005tm,Arutyunov:2006yd}.  For its explicit
form we refer to equation (\ref{eqn;Sfund}), but it can be shown
that it satisfies all physical properties associated with
integrable field theories
\begin{description}
    \item[Unitarity:] $\qquad \, \, \, \, \, \, \, \, \, \, \, \, \S_{12}\S_{21}=\mathbbm{1}$.
    \item[Hermiticity:] $\, \, \, \, \, \, \qquad \S_{12}\S_{12}^{\dag} = \mathbbm{1}$.
    \item[CPT Invariance:] $\, \, \, \, \S_{12}=\S_{12}^t$.
    \item[Yang-Baxter Equation:] $\, \, \, \qquad \S_{12}\S_{13}\S_{23}=\S_{23}\S_{13}\S_{12}$.
\end{description}
Based on the additional requirement of crossing symmetry
\cite{Janik:2006dc}, the overall scalar factor has also been
conjectured \cite{Arutyunov:2004vx,Beisert:2006ib,Beisert:2006ez}
and was found to agree with all known computations so far.

Concluding, the expression for the full S-matrix is conjectured
relying heavily on its symmetry properties. It is an exact quantity in the coupling $g$ and hence interpolates between strong and weak coupling.

\subsection*{Large volume spectrum}

In integrable models one usually can derive the exact large volume
spectrum from the S-matrix by a technique called the Bethe Ansatz.
This technique dates back to 1931, where it was first used to
solve the Heisenberg XXX-spin chain \cite{Bethe:1931hc}. Over the
years this technique has been cast into many different forms and
it has found its way into various physical models.

The Bethe Ansatz offers a method to capture the spectrum of an
integrable Hamiltonian in a set of algebraic equations, called the
Bethe ansatz equations. This is done by making a plane-wave type
ansatz for the eigenstates of the Hamiltonian. This ansatz depends
on a set of momenta which are restricted by imposing periodic
boundary conditions. This is similar, for example, to a free
particle on a circle of circumference $L$ for which periodic
boundary conditions $e^{i p L} = 1$ imply that the momentum is
quantized. The reason periodicity needs to be imposed is that
even though we consider the theory on the plane, it still comes
from a closed string, i.e. a cylindric world-sheet. How to impose
periodicity has a very clear interpretation.

\begin{figure}[h]
  \centering
  \includegraphics{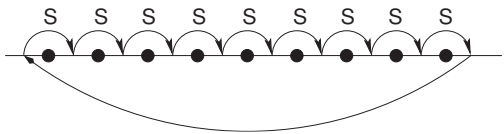}
  \caption{Periodicity is imposed by scattering a particle along the line.}
\end{figure}

\noindent Consider $N$ particles with momenta $p_i$ on a line with
length $L$. If we take one particle and move it along the line, it
will scatter on its way with all the other particles via the
two-body S-matrix. However, when it has travelled a distance $L$,
it returns back at its original position and the state should
remain unchanged, up to a phase factor $e^{ip_1L}$, which is due
to the plane-wave type ansatz. If we denote the S-matrix with $S$,
then the Bethe equations are of the form
\begin{align}
e^{ip_j L} = \prod_{i\neq j} S(p_i,p_j).
\end{align}
This can be seen as a quantization condition on the momenta. The
spectrum is then obtained by first solving the Bethe equations for
the set of momenta $\{p_i\}$, which can then be substituted in the
Hamiltonian. The dispersion relation for the $\ads$ superstring is
known in terms of the momenta $p$ and is given by
\cite{Beisert:2004hm}
\begin{align}\label{eqn;IntroDispersion}
H = \sqrt{1+4 g^2 \sin^2 \frac{p}{2}}.
\end{align}
The total energy of the state is obtained by summing the
contributions of the particles
\begin{align}
H_{tot} = \sum_{i} H(p_i).
\end{align}
The above discussion depends crucially on the two-particle
S-matrix of the theory. Since this matrix is known for the $\ads$
superstring (in the large volume limit) to all orders in the
coupling $g$, the spectrum obtained in this way is also exact.

The situation for the $\ads$ superstring appears to be more
complicated than presented above. The S-matrix has a non-trivial
matrix structure which mixes particles of different types. This
can be taken into account by extending the Bethe ansatz resulting
in a so-called nested structure. The Bethe equations were found in
\cite{Beisert:2005tm,Beisert:2005fw} and further investigated in
\cite{Martins:2007hb,Leeuw:2007uf}. They are given by
\begin{eqnarray}\label{eqn;IntroBAE}
e^{ip_{k}L}&=& \prod_{l=1,l\neq k
}^{K^{\mathrm{I}}}\left[S_{0}(p_{k},p_{l})\frac{x_k^+-x_l^-}{x_k^--x_l^+}\sqrt{\frac{x_l^+x_k^-}{x_l^-x_k^+}}\right]^2\prod_{\alpha=1}^{2}\prod_{l=1}^{K_{(\alpha)}^{\mathrm{II}}}\frac{{x_{k}^{-}-y^{(\alpha)}_{l}}}{x_{k}^{+}-y^{(\alpha)}_{l}}\sqrt{\frac{x^+_k}{x^-_k}}\nonumber \\
1&=&\prod_{l=1}^{K^{\mathrm{I}}}\frac{y^{(\alpha)}_{k}-x^{+}_{l}}{y^{(\alpha)}_{k}-x^{-}_{l}}\sqrt{\frac{x^-_k}{x^+_k}}
\prod_{l=1}^{K_{(\alpha)}^{\mathrm{III}}}\frac{y_{k}^{(\alpha)}+\frac{1}{y_{k}^{(\alpha)}}-w_{l}^{(\alpha)}+\frac{i}{g}}{y_{k}^{(\alpha)}+\frac{1}{y_{k}^{(\alpha)}}-w_{l}^{(\alpha)}-\frac{i}{g}}\\
1&=&\prod_{l=1}^{K_{(\alpha)}^{\mathrm{II}}}\frac{w_{k}^{(\alpha)}-y_{k}^{(\alpha)}-\frac{1}{y_{k}^{(\alpha)}}+\frac{i}{g}}{w_{k}^{(\alpha)}-y_{k}^{(\alpha)}-\frac{1}{y_{k}^{(\alpha)}}-\frac{i}{g}}
\prod_{l\neq
k}^{K_{(\alpha)}^{\mathrm{III}}}\frac{w_{k}^{(\alpha)}-w_{l}^{(\alpha)}-\frac{2i}{g}}{w_{k}^{(\alpha)}-w_{l}^{\alpha}+\frac{2i}{g}}\nonumber,
\end{eqnarray}
where $\alpha=1,2$ reflect the two copies of $\mathfrak{su}(2|2)$
and $S_{0}(p_{k},p_{l})$ is the overall scalar factor of the
S-matrix. The parameters $x^{\pm}$ are related to the coupling $g$
and to the world-sheet momentum via{\footnote {A similar
parameterization also appears in the Hubbard model.} }
\begin{align}
&x^+ + \frac{1}{x^+} - x^- - \frac{1}{x^-} = \frac{2i}{g},&&
\frac{x^+}{x^-} = e^{ip}.
\end{align}
The parameters $y_i,w_i$ are auxiliary variables that one
introduces to deal with the matrix structure of the S-matrix. The
spectrum is then again found by solving this coupled set of
equations and plugging the solution in the dispersion relation
(\ref{eqn;IntroDispersion}).

Because one can only define the scattering theory on the infinite
plane, these Bethe equations only encompass the asymptotic part of
the spectrum. The big remaining challenge is then to compute the
spectrum for finite size world-sheets.

\subsection*{Emergence of bound states}

Before addressing the finite size problem, let us first take a
closer look at the Bethe equations and discuss the emergence of
bound states. Bound states are composite particles which belong to
the physical spectrum and which manifest themselves as poles of a
multi-particle S-matrix, see for example
\cite{Faddeev:1996iy,Dorey:1996gd}. It was found that the
fundamental particles of the $\ads$ superstring model can indeed
form bound states
\cite{Dorey:2006dq,Chen:2006gq,Roiban:2006gs,Beisert:2006qh,Arutyunov:2007tc}.
They transform in symmetric short representations of centrally
extended $\alg{su}(2|2)$ \cite{Beisert:2006qh},
which are discussed in chapter \ref{chap;SU22}.

Consider two particles with complex momenta $p_1=\frac{p}{2}-iq$
and $p_2=\frac{p}{2}+iq$, with $p$ real and $\mathrm{Re}~q>0$. It
is easy to see that in the large volume limit $e^{i p_1 J}$ tends
to $\infty$. When looking at the first line of the Bethe equations
(\ref{eqn;IntroBAE}) we see that such a solution can indeed exist
if the right hand side exhibits a pole
\begin{align}\label{eqn;BoudnStateConditionIntro}
x^-(p_1) - x^+(p_2) = 0,
\end{align}
which corresponds to a pole in the S-matrix. The above relation
implies a non-trivial equation for $p,q$. From
(\ref{eqn;BoudnStateConditionIntro}) one can show that the total
energy becomes
\begin{align}
H = H(p_1) + H(p_2) =\sqrt{2^2+4g^2\sin^2\frac{p}{2}}.
\end{align}
This discussion generalizes to multi-particle bound states. The
latter are composites of $\ell$ fundamental particles whose
momenta are related by the condition
\begin{align}
x^-(p_i) - x^+(p_{i+1}) = 0.
\end{align}
The energy of these bound state particles is given by the
dispersion relation
\begin{align}
H = \sqrt{\ell^2+4g^2\sin^2\frac{p}{2}}.
\end{align}
A bound state that consists of $\ell$ fundamental particles and
transforms in a short representation of centrally extended
$\alg{su}(2|2)$ which is $4\ell$ dimensional. Concluding, the
complete asymptotic spectrum consists of fundamental excitations
and their bound states.

\section{Towards finite size}

To deal with the finite-size spectral problem, two approaches have
been developed in the past  in the context of relativistic models.
One of them is a perturbative approach due to L\"uscher
\cite{Luscher:1985dn} and the other is the Thermodynamic Bethe
Ansatz (TBA) \cite{YangYang,Zamolodchikov:1989cf}. Both approaches
have been recently extended to account for the unconventional
structure of the $\ads$ string model.

\subsection*{L\"uscher's perturbative approach}

\begin{figure}[h]
  \centering
\parbox[b]{5in}{
\begin{center}  \includegraphics{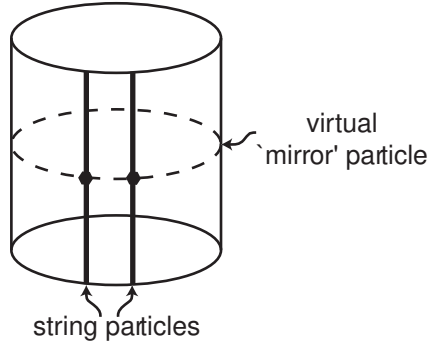}\end{center}
  \caption{Diagrams for finite-size systems. The dashed line depicts a virtual particle moving around the compact direction.}\label{Fig;Luscher}}
\end{figure}

In \cite{Luscher:1985dn} the leading finite-size correction to
energies were computed using diagrammatic methods. This formalism
has been adapted to the $\ads$ superstring
\cite{Ambjorn:2005wa,Janik:2007wt,Heller:2008at,Bajnok:2008bm,Bajnok:2008qj,Bajnok:2009vm}.
The idea is that, in compact spaces, particle energies pick up
corrections coming from virtual particles moving around the
compact direction, see figure \ref{Fig;Luscher}. Where the
particle meets the virtual particle from the loop they scatter via
the S-matrix. The virtual particles that run in the loop can be
both fundamental or bound state.



The first successful application of this procedure was the
computation of the four-loop scaling dimension of the Konishi
operator in $\mathcal{N}=4$ SYM. The Konishi operator is of the form
\begin{align}
K = {\rm tr} (D\mathcal{Z}D\mathcal{Z})-{\rm tr}
(\mathcal{Z}D^2\mathcal{Z}),
\end{align}
where $D= D_1 + i D_2$. One can compute its scaling dimension
directly in field theory
\cite{Fiamberti:2007rj,Velizhanin:2008jd,Fiamberti:2010ra}. It
gives
\begin{align}
\Delta_{K} = 4 + 3g^2 - 3 g^4 + \frac{21}{4} g^6 +
\left[-\frac{39}{4} + \frac{9\zeta(3)}{4} -
\frac{45\zeta(5)}{8}\right]g^8.
\end{align}
To derive this result one has to take into account around 200
supergraphs or equivalently 130000 Feynman diagrams. Needless to
say, these are complicated and demanding computations. However,
this result can also be derived from string theory in a very
simple and elegant way, as we will explain below.


First, one has to identify the string state to which the Konishi
operator corresponds. On the string theory side, $K$ corresponds
to the state built up out of two world-sheet excitations
corresponding to the light-cone derivatives $D$, with momenta
$p_1,p_2$. The level-matching condition implies that $p=p_1=-p_2$.
The angular momentum of the string is found to be $J=2$. From the
Bethe equations (\ref{eqn;IntroBAE}) one can solve perturbatively
for the momentum $p$ to find
\begin{align}
p = \frac{2\pi}{3} - \frac{\sqrt{3}}{4}g^2 +
\frac{9\sqrt{3}}{32}g^4 + \ldots.
\end{align}
From this momentum, one can then compute the energy via the
dispersion relation (\ref{eqn;IntroDispersion})
\begin{align}
E_{BAE} = J+H  = 4 + \sqrt{1+4g^2\sin^2\frac{p_1}{2}}+
\sqrt{1+4g^2\sin^2\frac{p_2}{2}}.
\end{align}
The field theory computation is done in perturbation theory at
weak coupling, so in order to compare with this, one expands the
energy $E_{BAE}$ around $g=0$ and finds
\begin{align} E_{BAE} = 4 +
3g^2 - 3 g^4 + \frac{21}{4} g^6 + \left[-\frac{705}{64}
+\frac{9\zeta(3)}{8}\right]g^8.
\end{align}
We see a disagreement in the $g^8$ term. However, it turns out
that the L\"uscher correction exactly contributes to this term.
Indeed, when the L\"uscher correction is taken into account, one
{\it does} find perfect agreement \cite{Bajnok:2008bm}
\begin{align}
E_{BAE + L} = 4 + 3g^2 - 3 g^4 + \frac{21}{4} g^6 +
\left[\frac{39}{4} + \frac{9\zeta(3)}{4} -
\frac{45\zeta(5)}{8}\right]g^8.
\end{align}
Summarizing, by using the Bethe ansatz for the string model
supplemented by the leading L\"uscher correction, one finds
beautiful agreement with the result of a highly non-trivial
quantum field theory computation.

\begin{figure}[h]
  \centering
\parbox[b]{5.5in}{
\begin{center}\includegraphics{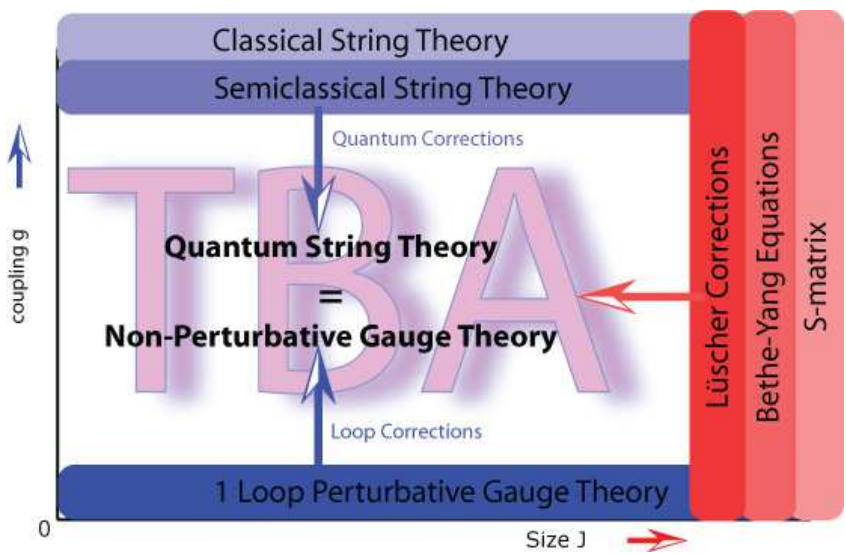}\end{center}
  \caption{Overview of the known parts of the spectrum. The coupling constant $g$ runs
  along the vertical axis and the size of the system on the horizontal. In the weak coupling regime perturbative gauge theory computations are
  possible and in the strong coupling regime one can perform perturbative string computations. For infinite sizes one can compute the
  asymptotic spectrum via Bethe ansatz techniques. The TBA should provide a complete covering of the diagram. }\label{Fig;ParameterPlane}}
\end{figure}

Although generating these nice results, L\"uscher's approach is
perturbative in nature and, therefore, has its limitations.  The
problem of establishing the exact spectrum is further addressed by
the TBA approach, to which L\"uscher's technique can be seen as a
certain approximation.  The regions of the $(g,J)$-parameter plane
where the various techniques to study the string/gauge theory
spectrum are applicable are schematically depicted in figure
\ref{Fig;ParameterPlane}. The TBA should cover the entire diagram.

\subsection*{Thermodynamic Bethe Ansatz}

The TBA approach in the AdS/CFT spectral problem was first advocated in \cite{Ambjorn:2005wa} where it was used to explain wrapping effects in gauge theory. We will follow \cite{Arutyunov:2007tc} where the first results towards an explicit construction of the corresponding TBA approach were obtained. 

\begin{figure}[h]
  \centering
\parbox[b]{4.5in}{
\begin{center}  \includegraphics{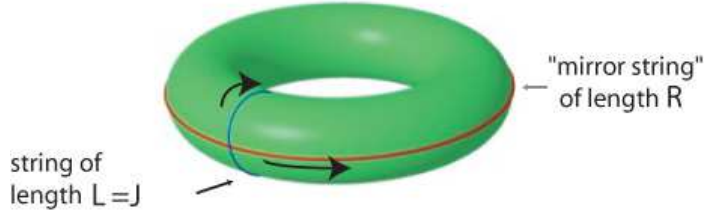}\end{center}
  \caption{String model and mirror model both come from the same torus.
  They are related via a double Wick rotation.}\label{Fig;TBATorus}}
\end{figure}

Consider the world-sheet of a closed string which is parameterized
by variables $\sigma$ and $\tau$. In the thermodynamic Bethe
ansatz (TBA) approach one considers a
closed string of size $L$ which wraps a `time'-loop of size $R$.
In this way one makes the time variable $\tau$ periodic. The
surface that is formed in this way is a torus; the product of two
circles of circumferences $L,R$, see figure \ref{Fig;TBATorus}.

By performing a Wick rotation $\tau\rightarrow i\tau$ one obtains
an Euclidean theory associated with this torus. In this Euclidean
theory there is no distinction between the coordinates and because
of this, one can associate two different `Minkowski' models to
this theory, namely one can apply an inverse Wick rotation to both
variables.

Applying the inverse Wick rotation to the $\tau$ variable
inevitably returns us back to the string model we started out
with. Applying it to the $\sigma$ variable, however, gives a new
model, called `mirror' model. This mirror model is then related to
the original string model via a double Wick rotation
\begin{align}
&\tilde{\sigma} =  -i\tau, && \tilde{\tau} = i\sigma.
\end{align}
Notice that the roles of position and time are interchanged.
Consequently, the mirror model also has a different Hamiltonian
$\tilde{H}$, which is now defined with respect to $\tilde{\tau}$.

It turns out that the mirror theory is different from the
world-sheet theory of the $\ads$ superstring. For example, the
dispersion relation is now given by
\begin{align}
\tilde{H} = 2 \mathrm{arcsinh} \frac{\sqrt{1+\tilde{p}^2}}{2g},
\end{align}
where $\tilde{p}$ is the momentum of the mirror particles, cf.
(\ref{eqn;IntroDispersion}). Nevertheless, one can show that the
partition functions $Z(R,L),\tilde{Z}(L,R)$ of both models are
equal. They are given by
\begin{align}
&Z(R,L) \equiv \sum_n \langle\psi_n|e^{-HR}|\psi_n\rangle = \sum_n
e^{-E_n R}\\
&\tilde{Z}(L,R) \equiv \sum_n
\langle\tilde{\psi}_n|e^{-\tilde{H}L}|\tilde{\psi}_n\rangle.
\end{align}
In the limit $R\rightarrow\infty$ one obtains
\begin{align}
&\log Z(R,L) \stackrel{R\rightarrow\infty}{\longrightarrow} -R
E(L), && \log \tilde{Z}(R,L)
\stackrel{R\rightarrow\infty}{\longrightarrow} -RL f(L),
\end{align}
where $E(L)$ is the ground state energy and $f(L)$ is the free
energy per unit length of the mirror model at temperature $1/L$.
Since both partition functions agree, one finds
\begin{align}
E(L) = Lf(L).
\end{align}
We see that by sending $R\rightarrow\infty$, the ground state
energy of the $\ads$ superstring in finite volume $L$ is described
by the free energy of the mirror theory in infinite volume but at
finite temperature $1/L$. This is the basic idea of the TBA; the
spectrum of the original theory can be computed through
thermodynamic quantities in the mirror model.

The importance of these observations is
that, for the mirror model, one can still use all the large volume
techniques that have been described earlier to compute the exact
spectrum. In fact, the mirror model in the infinite volume limit
also exhibits centrally extended $\alg{su}(2|2)$ symmetry. This
means that the same symmetry arguments are applicable to the large
volume spectrum for the mirror model. Finally, one has to work at
finite temperature and identify all states that contribute in the
thermodynamic limit \cite{Arutyunov:2009zu}.

In the mirror theory, the S-matrix again contains poles that
correspond to bound states \cite{Arutyunov:2007tc}. In other words, the complete
asymptotic spectrum is composed of fundamental particles and their
bound states. More precisely, this means that bound states will
contribute in the thermodynamic limit. Because of this, their
scattering data and spectrum needs to be understood.

The thermodynamic limit is implemented by sending
$R\rightarrow\infty$ while keeping $N_i/R$ fixed, where $N_i$ is
the number of particles of a certain species $i$. From the Bethe
equations that describe the large volume spectrum, one can then
derive the TBA equations which constitute an infinite number of
coupled integral equations. These equations are supposed to
describe the finite-size spectrum of the original string theory
\cite{Gromov:2009bc,Arutyunov:2009ur,Gromov:2009bc,Bombardelli:2009ns}. The TBA
approach perfectly accommodates L\"uscher's correction, which
emerges from the large $J$ asymptotic solutions of the TBA
equations for excited states. The future challenge is to obtain a
detailed understanding of the TBA solutions. The first results in
this direction are very promising
\cite{Gromov:2009tv,Gromov:2009zb,Frolov:2009in,Arutyunov:2009ax,Arutyunov:2009ux,Cavaglia:2010nm,Frolov:2010wt}. The agreement
between the TBA approach and L\"uscher's approach has even been
extended to the five-loop level
\cite{Bajnok:2009vm,Arutyunov:2010gb,Balog:2010xa,Balog:2010vf}.

\section{Bound States and Yangian}

Bound states and their scattering data play an important role in both
the TBA and in L\"uscher's perturbative approach. The S-matrix
for fundamental representations was fixed by the requirement that
it respects the $\alg{su}(2|2)$ symmetry. One might wonder whether
similar symmetry arguments can also be applied to bound states.

Bound states of fundamental particles transform in higher
dimensional representations of centrally extended $\alg{su}(2|2)$.
Because of this, it turns out that invariance under the extended
$\alg{su}(2|2)$ algebra is no longer enough to fix the matrix
structure for these higher dimensional representations
\cite{Beisert:2007ds,Arutyunov:2008zt}. In addition, one has to invoke the
Yang-Baxter equation to completely fix the S-matrix. However, it was shown
that the fundamental S-matrix is actually invariant under a bigger
symmetry group; the Yangian of $\alg{su}(2|2)$
\cite{Beisert:2007ds}.

The Yangian of a Lie algebra can be seen as an infinite
dimensional deformation of the associated loop algebra. Consider
an algebra with structure constants $f^{AB}_C$
\begin{align*}
[\mathbb{J}^A,\mathbb{J}^B] = f^{AB}_C \mathbb{J}^{C}
\end{align*}
and introduce a new set of generators $\hat{\mathbb{J}}^A$ that
satisfy the relation
\begin{align*}
[\mathbb{J}^A,\hat{\mathbb{J}}^B] = f^{AB}_C \hat{\mathbb{J}}^{C}.
\end{align*}
The Yangian of the algebra is now spanned by the generators
$\mathbb{J}^A,\hat{\mathbb{J}}^A$ and commutators thereof,
generating an infinite dimensional algebra. It turns out that
bound state S-matrices are fixed by requiring invariance under the
{\it Yangian} of centrally extended $\alg{su}(2|2)$ rather than
only the algebra itself \cite{deLeeuw:2008dp}. In fact, one can
construct any bound state S-matrix by using Yangian symmetry\footnote{
Similarly, (Yangian) symmetry is also crucial in finding boundary S-matrices for open strings, see e.g. \cite{Ahn:2010xa,MacKay:2010zb}.} \cite{Arutyunov:2009mi}. 




\section{Different models}

The main focus here is on the $\ads$ superstring and its partner
$\mathcal{N}=4$ super Yang-Mills. One might wonder how useful this
is for physical applications since $\mathcal{N}=4$ SYM is a rather
special quantum field theory. It is conformal and highly
supersymmetric, properties that are not shared with real-world
theories like QCD. Nevertheless, one can smoothly deform the
$\ads$ space-time \cite{Lunin:2005jy} to obtain a deformed
$\mathcal{N}=1$ SYM theory. Even though the number of
supersymmetries is reduced, one still finds integrable structures
\cite{Frolov:2005dj,Frolov:2005ty,Beisert:2005if}. One might hope
to apply the TBA approach to this case as well. There even exists
a more general class of deformations \cite{Alday:2005ww} which is
expected to be dual to a non-supersymmetric gauge theory. A
thorough understanding of the prime example which is central in
this work can then, via these related models, be used to extend
our understanding of the AdS/CFT correspondence to more realistic
models.\newpage

A different example of a pair of dual theories is
\cite{Aharony:2008ug}
\begin{center}
$\mathcal{N}= 6$ Chern-Simons Theory $\leftrightarrow$ IIA strings
on ${\rm AdS}_4\times \mathbb{CP}^3 $.
\end{center}
It also admits a formulation in terms of a coset model
\cite{Arutyunov:2008if,Stefanski:2008ik} and one can show that the
model is classically integrable. The asymptotic symmetry is again
centrally extended $\alg{su}(2|2)$ and, because of this, it has
many features in common with the $\ads$ superstring and its
partner $\mathcal{N}=4$ SYM. Also for this model the asymptotic
Bethe equations have been proposed \cite{Gromov:2008qe} and the
set of TBA equations have recently been conjectured
\cite{Bombardelli:2009xz,Gromov:2009at}. Even though there are
quite some similarities between this instance of the AdS/CFT
correspondence and the prototype example, there are also
differences. Elucidating those can provide a partial guide to what
new features could be expected on the way towards understanding
more realistic models.

The importance of a complete understanding of the aforementioned
theories goes beyond giving evidence for the AdS/CFT
correspondence. On the one hand, the integrable structures can
allow for a determination of the complete spectrum of
$\mathcal{N}=4$ SYM. This would then be the first exact solution
of a non-trivial four-dimensional quantum field theory. On the
other hand, a comprehensive description of the corresponding
string model could prove invaluable in the future. A full solution
might serve as a benchmark against which future computational
methods in string theory can be checked and refined.

\section{Outline}

In this chapter we outlined a very promising road that hopefully
leads to a full determination of the spectrum of the $\ads$
superstring and through the AdS/CFT correspondence to the spectrum
of scaling dimensions of $\mathcal{N}=4$ SYM. The material in this
review particulary focusses on two aspects that were encountered
along this way, namely scattering data of bound states and the
Bethe ansatz.

First, the S-matrix describing the scattering of bound states with
arbitrary bound state numbers is explicitly constructed by using
Yangian symmetry, see equations
(\ref{eqn;SCase1}),(\ref{eqn;SCase2}),(\ref{eqn;SCase3}). We study
its classical limit and compare it against two different proposals
in the literature \cite{Moriyama:2007jt,Beisert:2007ty}, finding
agreement only with the latter. We also examine some of its
mathematical features. More precisely we find blocks in this
S-matrix that exhibit universal algebraic properties.

The next topic is the computation of the large volume spectrum of
bound state configurations. The fact that we explicitly know the
S-matrix allows us to apply the Bethe ansatz and impose periodic
boundary conditions. This results in a set of Bethe equations that
describe the large volume bound state spectrum
(\ref{eqn;FullBAE}). We have also found an alternative way to
derive these equations by using the underlying Yangian symmetry of
the S-matrix. These Bethe equations play a crucial role in the
derivation of the TBA equations. A second approach that we have
examined is that of the algebraic Bethe ansatz. We present the
explicit transfer matrices (\ref{eqn;FullEignvalue}), from which
one, once again, can derive the Bethe equations. The algebraic
Bethe ansatz is also a useful framework for possible future
investigations like form factors.

This remainder of this review is organized as follows. In chapter
2, we will first give a brief introduction to integrable models
and Hopf algebras. In this chapter we also introduce the notion of
a Yangian.

After this we will discuss in chapter \ref{chap;SU22} the algebra
that plays a key role in the whole discussion: centrally extended
$\alg{su}(2|2)$. We will give its defining relations and discuss
in detail the representations describing both fundamental
particles and their bound states. Subsequently, we introduce the
Hopf algebra structure that is used to derive the S-matrix. We
finish the chapter with a discussion of the Yangian of
$\alg{su}(2|2)$, which is important for the determination of the
bound state scattering data.

In chapter \ref{chap;BoundSmat} the bound state S-matrix will be
derived by making use of Yangian symmetry. Using the two
$\alg{su}(2)$ subalgebras contained in $\alg{su}(2|2)$ one can
split states into three different types. We first solve for one
specific type of states (equation (\ref{eqn;SCase1})) and then,
via the different supersymmetry generators, extend the solution to
the whole space, (equations (\ref{eqn;SCase2}) and
(\ref{eqn;SCase3})). This results in an explicit formula for the
bound state scattering data that agrees with all S-matrices that
were previously found.

Since the underlying algebra determines the entire scattering
data, one might wonder whether it can be written purely in
algebraic terms. Such an algebraic object corresponding to the
S-matrix is called a universal R-matrix in the framework of Hopf
algebras. This is the subject of chapters \ref{chap;Rmat} and
\ref{chap;UnivBlocks}. In the first chapter the classical limit of
the bound state S-matrix is studied and one finds agreement with a
universal expression that has been proposed in the literature. In
the subsequent chapter we study certain blocks in the S-matrix
that exhibit universality at the full quantum level.

After this more mathematically oriented part we move on to the
determination of the spectrum. We will first do this by applying
the coordinate Bethe ansatz. After a discussion of this technique
applied to the non-linear Schr\"odinger model we will introduce
nesting to deal with particles with different colors and
generalize the discussion to the $\ads$ superstring. This Bethe
ansatz procedure can then be reformulated by making use of the
Yangian symmetry, which allows for a derivation of the Bethe
equations describing the asymptotic bound state spectrum
\cite{deLeeuw:2008ye}.

Alternatively one can use the algebraic Bethe ansatz, which is
done in chapter \ref{chap;Transfer}. In this approach one derives
the eigenvalues of the transfer matrix, that play a crucial role
in the TBA. We will derive an explicit expression for the
eigenvalues of the transfer matrix for generic bound state
representations (\ref{eqn;FullEignvalue}). Some of these
eigenvalues were already conjectured in the literature
\cite{Beisert:2006qh} via a fusion procedure. We work this
procedure out explicitly and compare it to our results, finding
perfect agreement.

%% file: IntegrabilityHopf2.tex
\chapter{Integrable Models and Hopf Algebras}\label{chap;IntegrabilityAndHopf}

The notion of integrability was already briefly touched upon in
the introduction. In this chapter we will expand this discussion.
We will particulary focus on the relation between integrability
and Hopf algebras. The mathematical language of Hopf algebras is
useful in describing symmetries of integrable field theories and
we will use it frequently.

We will first briefly discuss the notion of integrability in
classical mechanics and in the theory of partial differential
equations. We then continue with an overview of scattering
processes in integrable field theories and introduce the notion of
Hopf algebras and Yangians. After this we will show in an example
how Yangians can arise in an integrable theory.

\section{Classical Integrable Systems}

Examples of classical integrable systems can be encountered {\it
e.g.} when solving problems of Newtonian mechanics. Most of such
problems, like Kepler's one, are well-known. However {\it exact}
solutions, especially when dealing with multiple degrees of
freedom, are rather rare. In the 19th century Liouville derived a
theorem in which a big class of exactly solvable models was
identified, the so-called integrable models. For an extensive
treatment on this topic we refer to \cite{Babelon}.

\subsection*{Finite-Dimensional Integrable Models}

Consider a system with Hamiltonian $\mathcal{H}$, coordinates
$q_i$ and conjugate momenta $p_i$. Its equations of motion are
written as
\begin{align}
&\dot{q}_i = \frac{\partial\mathcal{H}}{\partial p_i}, &&
\dot{p}_i = -\frac{\partial\mathcal{H}}{\partial q_i}.
\end{align}
The time evolution of any quantity $F(p,q)$ is given by
\begin{align}
\dot{F} = \{\mathcal{H},F\},
\end{align}
where $\{,\}$ is the Poisson bracket. Suppose that the phase space
is $2N$ dimensional. The system is called {\it integrable} if
there are $N$ (functionally) independent conserved quantities
$F_i$ (of which the Hamiltonian is one), which mutually Poisson
commute
\begin{align}
&\{F_i,F_j\} = 0, & &\{F_i,\mathcal{H}\} = 0,\qquad \mathrm{for\ all\ } i,j.
\end{align}
Liouville's theorem states that an integrable system can be solved
by quadratures, i.e. by solving a finite number of algebraic
equations and computing a finite number of integrals. In this
sense integrable models are exactly solvable.

Examples of integrable systems include the harmonic oscillator
(the conserved quantity is the Hamiltonian $F=\mathcal{H}$) and
Kepler's problem (conserved quantities are the Hamiltonian, the
total angular momentum and the z-component of the angular
momentum: $F_1=\mathcal{H},F_2=J_3,F_3=J^2$).

A different way to formulate integrability is in terms of a
so-called Lax pair. Suppose there are matrices $L(p,q),M(p,q)$
such that one can write the equations of motion in the following
way
\begin{align}\label{eqn;LaxPair}
\partial_0 L -[L,M] = 0.
\end{align}
(From now on we denote the time derivative as $\partial_0$ rather
then using a dot). If this is the case, then one can
straightforwardly see that the quantities
\begin{eqnarray}
I_k = {\rm{tr}} L^k
\end{eqnarray}
are conserved. The property that these quantities Poisson commute
it related to an object called classical $r$-matrix. Assuming one
can prove that these quantities are independent and Poisson
commute with one another, we see that such a system is integrable.

The harmonic oscillator admits a Lax-pair. Define the matrices
\begin{align}
&L = \begin{pmatrix}
p & \omega q \\
\omega q & -p
\end{pmatrix},
&& M = \begin{pmatrix}
0 & -\frac{\omega}{2} \\
\frac{\omega}{2} & 0
\end{pmatrix}.\
\end{align}
Then one can readily check that these indeed encode the equations
of motion in the form (\ref{eqn;LaxPair}). The Hamiltonian is
given by $\mathcal{H} = \frac{1}{4}{\rm{tr}} L^2$.

Since conserved charges correspond to symmetries via Noether's
theorem, one finds that integrable models have enough symmetries
to be solved by quadratures.


\subsection*{Integrable Partial Differential Equations}

There is a related notion of integrability for two-dimensional
(non-linear) partial differential equations (PDE). These equations
admit a reformulation analogous to the Lax pair one discussed
above.

Let $\Psi(t,x,z)$ be a rank $n$ vector, and consider the
overdetermined set of equations
\begin{align}
&\partial_0 \Psi(t,x,z) = L_0(t,x,z)\Psi(t,x,z), &&
\partial_1\Psi(t,x,z) = L_1(t,x,z)\Psi(t,x,z),
\end{align}
where $\partial_1= \partial_x$ and $L_i(t,x,z)$ are $n\times n$
matrices. By considering the double derivative
$\partial_0\partial_1 \Psi(t,x,z)=\partial_1\partial_0
\Psi(t,x,z)$ one sees that the matrices $L_i(t,x,z)$ need to
satisfy the consistency condition
\begin{align}
\partial_0 L_1 - \partial_1 L_0 + [L_1,L_0] =0.
\end{align}
The above equation can be seen as the flatness condition for a
two-dimensional (non-abelian) connection. This connection is
called the Lax connection and it is a generalization of the notion
of a Lax pair (\ref{eqn;LaxPair}).

If a PDE can be written as the flatness condition for a Lax
connection, then it is called integrable. For example, the
sine-Gordon equation
\begin{align}
\phi_{tt} - \phi_{xx} + \frac{m^2}{\beta}\sin (\beta\phi)=0,
\end{align}
can be written in this way via the following Lax connection:
\begin{align}
&L_1 = \frac{\beta \phi_t}{4i}
\begin{pmatrix}
1 & 0 \\
0 & -1 \end{pmatrix}
+ \frac{k_0
\sin\frac{\beta\phi}{2}}{i}
\begin{pmatrix}
0 & 1 \\
1 & 0
\end{pmatrix}
+ \frac{k_1
\cos\frac{\beta\phi}{2}}{i}
\begin{pmatrix}
0 & -i \\
i & 0
\end{pmatrix}\\
&L_0 = \frac{\beta \phi_x}{4i}
\begin{pmatrix}
1 & 0 \\
0 & -1
\end{pmatrix}
+ \frac{k_1 \sin\frac{\beta\phi}{2}}{i}
\begin{pmatrix}
0 & 1 \\
1 & 0
\end{pmatrix}
+ \frac{k_0 \cos\frac{\beta\phi}{2}}{i}
\begin{pmatrix}
0 & -i \\
i & 0
\end{pmatrix}
\end{align}
with
\begin{align}
&k_0 = \frac{m}{4}\left(z+\frac{1}{z}\right), && k_1 =
\frac{m}{4}\left(z-\frac{1}{z}\right).
\end{align}
From a Lax connection one can construct an infinite tower of
conserved charges. In this case this can be achieved by defining a
monodromy matrix $T(z)$ as the path ordered exponential of the Lax
connection. Expanding this quantity in the parameter $z$ then
generates the conserved quantities \cite{Babelon}.

\section{Integrable 2d Relativistic Field Theories}\label{sec;ZFalgebra}

The notion of integrability can be extended to two dimensional
quantum field theories. Also in these theories, integrability
corresponds to the system having an enhanced symmetry resulting in
an infinite set of conservation laws. This is translated into the
fact that scattering processes in these theories have very special
features. The scattering processes have the following properties
\cite{Zamolodchikov:1978xm,Delius:1995tc,Dorey:1996gd}

\begin{description}
\item[Absence of Particle Production] There is no particle production in these systems.
\begin{figure}[h]
\centering
\includegraphics[scale=.5]{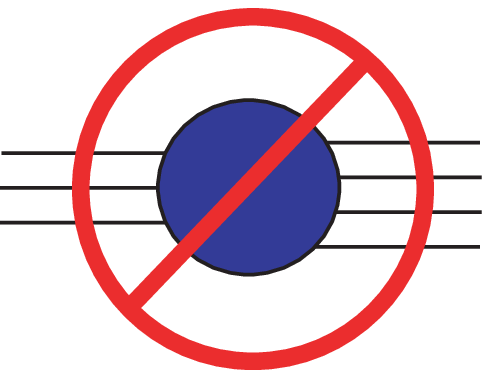}
\end{figure}
\item[Momentum Conservation] The sets of initial and final momenta are the same
\begin{eqnarray}
\{p_i\}_{in} = \{p_i\}_{out}.
\end{eqnarray}
\item[Factorizability] Any scattering process reduces to a chain of two-body interactions
\begin{eqnarray}
(n \rightarrow n)~~ =~~ \prod~ (2\rightarrow 2).
\end{eqnarray}
\item[Unitarity]\
\begin{figure}[h]
\centering
\includegraphics{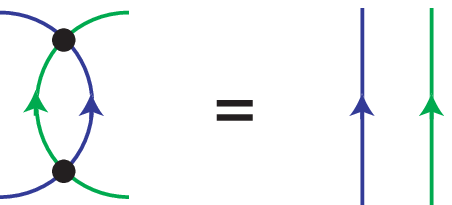}
\end{figure}
\begin{eqnarray}
\S_{12}(p_1,p_2)\S_{21}(p_2,p_1) = \mathbbm{1}.
\end{eqnarray}
\item[Crossing] Scattering is symmetric under particle to anti-particle transformations
\begin{figure}[h]
\centering
\includegraphics{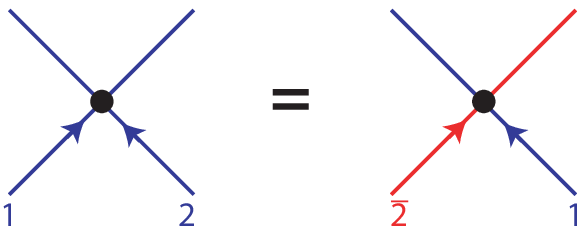}
\end{figure}
\begin{eqnarray}\label{eqn;EasyCrossing}
\S_{12}(p_1,p_2) = \S_{\bar{2}1}(p_1,\tilde{p_2})
\end{eqnarray}
\item[Yang-Baxter Equation] The two body S-matrix satisfies a consistency condition which
can be derived by considering the factorization of three particle scattering.
\begin{figure}[h]
\centering
\includegraphics{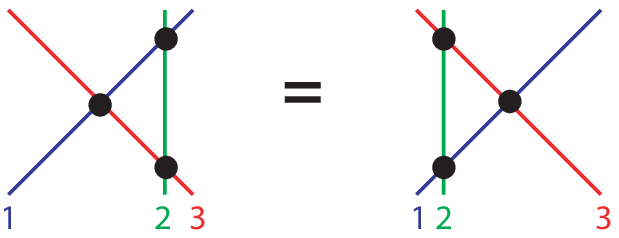}
\end{figure}
\begin{eqnarray}\label{eqn;EasyYBE}
\S_{12}\S_{13}\S_{23} = \S_{23}\S_{13}\S_{12}.
\end{eqnarray}
\end{description}
One can often take the above set of conditions as a definition of
an integrable (quantum) field theory. Examples of such a theories
are for instance the Sine-Gordon model \cite{Zamolodchikov:1977py}
and the Sinh-Gordon model \cite{Dorey:1996gd}.

From the factorization property we see that all scattering
information in such theories is encoded in the two-body S-matrix.
This means that computing this scattering process is the key to
solving these systems.


\section{Hopf Algebras}\label{sec;HopfAlgebra}

A convenient mathematical framework to deal with symmetries of
integrable models is the language of Hopf algebras. Hopf algebras
carry with them the structure of a coproduct and an antipode. From
the physical point of view the coproduct describes the action of
symmetry generators on multi-particle states and the antipode
corresponds to a particle to anti-particle transformation.

We will review basic facts and definitions of Hopf algebras. We
will be mostly interested in those associated to Lie
(super)algebras. A very special family of Hopf algebras are the
Yangians. These are infinite dimensional algebras associated to
Lie algebras. They appear as symmetry algebras in various
integrable models and are important in our understanding of the
spectrum of the $\ads$ superstring.

There is a vast literature on this subject and for a more detailed
analysis of Yangians and Hopf algebras we refer to
\cite{twi,stuko,MacKay:2004tc,Molev,Etingof,Chari,Drin,Dsecond}.

\subsection*{Definitions}

An associative algebra $A$ with unit is a vector space over
$\mathbb{C}$ (the notion of algebra is more general, but will
restrict to complex vector spaces) that is equipped with a
multiplication $\mu$
\begin{align}
\mu:A\otimes A\rightarrow A,\qquad a_1\otimes a_2\rightarrow \mu(a_1\otimes a_2) \equiv a_1 a_2.
\end{align}
and a unit under multiplication $\eta:\mathbb{C}\rightarrow A$,
such that
\begin{align}
&\mu(\eta(\lambda)\otimes a) = \lambda a = \mu(a \otimes
\eta(\lambda)), && \lambda\in\mathbb{C},a\in A.
\end{align}
The multiplication is bilinear and associativity is formulated as
\begin{align}
a(bc) = (ab)c, \qquad a,b,c\in A.
\end{align}
An obvious example of a such an algebra is the vector space of
$n\times n$ complex matrices with the standard matrix
multiplication and $\eta(\lambda) = \lambda\mathbbm{1}$. Let
$\mathcal{P}:A\otimes A\rightarrow A\otimes A$ be the (graded)
permutation operator, then we say that the algebra is commutative
if $\mu\circ\mathcal{P} = \mu$.

A coalgebra $A$ is an object which has a structure `dual' to that
of an algebra; it has a comultiplication $\Delta$ and a co-unit
$\epsilon$
\begin{align}
&\Delta: A\rightarrow A\otimes A, && \epsilon: A \rightarrow \mathbb{C}.
\end{align}
Similarly these maps have to be bilinear and the coproduct needs
to satisfy coassociativity
\begin{align}
(\Delta\otimes\mathbbm{1})\Delta = (\mathbbm{1}\otimes\Delta)\Delta.
\end{align}
A coalgebra is called cocommutative if $\mathcal{P}\Delta =
\Delta$. We call $\Delta^{op}= \mathcal{P}\Delta$ the opposite
coproduct.

An algebra $A$ is called a bialgebra if it also is endowed with
the structure of a coalgebra such that $\Delta$ and $\epsilon$ are
algebra homomorphisms, i.e. they respect the multiplicative
structure
\begin{align}
&\Delta(a_1 a_2) = \Delta(a_1)\cdot \Delta(a_2), && \epsilon(a_1 a_2)= \epsilon(a_1)\epsilon(a_2),\\
&\epsilon(\mathbbm{1})=1, & & \Delta\mathbbm{1} = \mathbbm{1}\otimes \mathbbm{1}.
\end{align}
Finally a Hopf algebra $H$ is a bialgebra which is equipped with
an anti-homomorphism called the antipode ${\cal S}: A\rightarrow
A$, which satisfies
\begin{align}\label{eqn;DefAntipode}
&\mathcal{S}(a_1a_2) =\mathcal{S}(a_2) \mathcal{S}(a_1),&&
\mu({\cal S}\otimes \mathbbm{1})\circ\Delta = \epsilon = \mu(
\mathbbm{1}\otimes {\cal S})\circ\Delta,
\end{align}
where in the graded case one has to pick up relevant signs due to
the graded structure. The first property is what defines
$\mathcal{S}$ to be an anti-homomorphism. An example of a Hopf
algebra is the universal enveloping algebra of a Lie algebra. This
algebra is automatically equipped with a multiplication, which in
case of a matrix representation is just matrix multiplication. One
can equip it with a Hopf algebra structure by specifying
\begin{align}\label{eqn;UnivCoprod}
&\Delta(\mathbb{J}^A) = \mathbb{J}^A\otimes \mathbbm{1} +
\mathbbm{1} \otimes \mathbb{J}^A, && \epsilon(\mathbb{J}^A) = 0,&&
{\cal S}(\mathbb{J}^A) = -\mathbb{J}^A,
\end{align}
and one can use the homomorphism properties to extend the above
maps to products of elements. It is easy to check that the maps
defined in this way respect multiplication (and hence also the Lie
bracket). This Hopf algebra is cocommutative, but generically not
commutative as an algebra.

The coproduct offers a natural way to extend a representation of a
Lie algebra $\alg{g}$ on a vector space $V$ to a tensor product
representation on $V\otimes V$. One can use
$(\Delta\otimes\mathbbm{1})\Delta$ to define it on a triple tensor
product and so on. By coassociativity one gets the same structure
from using $(\mathbbm{1}\otimes\Delta)\Delta$.

The above definition of the coproduct for the universal enveloping
algebra (\ref{eqn;UnivCoprod}) is natural in the language of
symmetries in physics. For example, consider two particles
$|m_1\rangle,|m_2\rangle$ in quantum mechanics with z-component of
the angular momenta $m_1,m_2$. One expects to find that in their
tensor product state $|m_1\rangle\otimes |m_2\rangle$ their
charges $m_1$ and $m_2$ add. The operator $\hat{S}_z$ is just an
element of $\alg{su}(2)$, in other words one has
\begin{align}
\hat{S}_z(|m_1\rangle\otimes |m_2\rangle)\equiv\hat{S}_z|m_1,m_2\rangle = (m_1+m_2)|m_1,m_2\rangle = \Delta S_z|m_1,m_2\rangle,
\end{align}
by the above definition of the coproduct (\ref{eqn;UnivCoprod}).
This indeed indicates that the coproduct encodes the natural
action of symmetry generators on multi-particle states.

\subsection*{Quasitriangular Hopf algebras}

To any Hopf algebra $(H,\epsilon,\mathcal{S},\Delta)$ one can
associate another Hopf algebra
$(H,\epsilon,\mathcal{S},\Delta^{op})$ by equipping it with the
opposite coproduct $\Delta^{op}$ rather than $\Delta$. This Hopf
algebra is called the opposite Hopf algebra $H^{op}$.

Generically there need not be a relation between the two different
Hopf algebra structures. However, there is a class of Hopf
algebras called quasi-cocommutative Hopf algebras, where there is
an invertible element $\S\in H\otimes H$ such that
\begin{align}\label{eqn;SymmProp}
&\Delta^{op}\mathbb{J} ~ \S = \S ~ \Delta\mathbb{J}, && \mathbb{J}\in H.
\end{align}
The element $\S$ is called the R-matrix in mathematics. In physics
this object corresponds to the S-matrix. The motivation for this
is as follows. In physics the S-matrix relates in-states to
out-states of the Hamiltonian.

\begin{figure}[h]
\centering
\includegraphics[scale=.7]{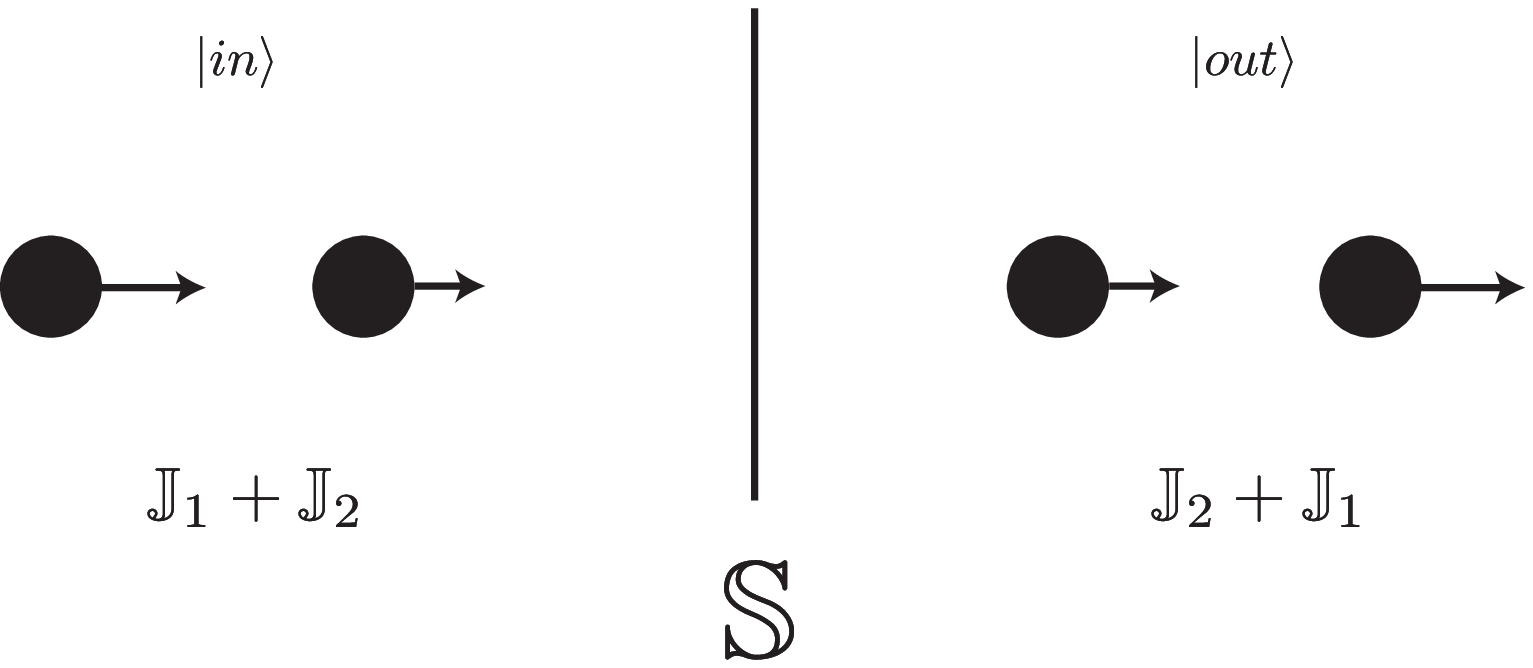}
\caption{Symmetry commutes with scattering.}\label{Fig;ScatteringSymm2}
\end{figure}

For definiteness, consider an elastic scattering process of two to
two particles. The in-state can then be seen as two particles with
the fastest one to the left and the slowest one on the right,
figure \ref{Fig;ScatteringSymm2}. After they scatter, the
particles have changed places and the fastest one is now on the
right. If the theory possesses some symmetry algebra, then the
S-matrix should be compatible with this algebra. This now
translates into (\ref{eqn;SymmProp}). In the context of integrable
models, this can be made more precise by using the so-called
Faddeev-Zamolodchikov algebra
\cite{Zamolodchikov:1978xm,Faddeev:1980zy}.


Suppose now that the R-matrix satisfies the extra conditions
\begin{align}
&(\Delta\otimes\mathbbm{1})(\S) = \S_{13}\S_{23}, && (\mathbbm{1}\otimes\Delta)(\S) = \S_{13}\S_{12},
\end{align}
then the Hopf algebra is called quasitriangular. The R-matrix of a
quasitriangular Hopf algebra satisfies the following properties
\begin{align}
&\S_{12}\S_{13}\S_{23} = \S_{23}\S_{13}\S_{12}, && (\mathcal{S}\otimes\mathbbm{1})\S = (\mathbbm{1}\otimes \mathcal{S}^{-1})\S = \S^{-1}.
\end{align}
The first relation is easily recognized as the Yang-Baxter
equation (\ref{eqn;EasyYBE}) and the second corresponds to the
crossing equation (\ref{eqn;EasyCrossing}). These equations were
important properties of scattering processes in integrable field
theories and they arise naturally in the context of Hopf algebras.

In the above we assumed that the R-matrix was an element of
$H\otimes H$, but the abstract form of this element might be hard
to find. However, in a specific representation, the intertwining
R-matrix can usually be computed explicitly. In what follows we
will refer to the R-matrix seen as an abstract element in
$H\otimes H$ as the universal R-matrix. The intertwining operator
in an explicit representation will be referred to as the S-matrix.

\section{Yangians}\label{sec;DefYang}

A class of Hopf algebras that play an important role in integrable
systems are the so-called Yangians. The Yangians are a family of
infinite dimensional algebras that are associated to Lie algebras.
They are constructed by introducing an additional set of
generators to the Lie algebra ones. In this section we collect
some basic facts about Yangians. For more details see e.g
\cite{MacKay:2004tc,Molev,Drin,Dsecond}.

\subsection*{Definition}

Consider the universal enveloping algebra $U(\alg{g})$ of a
(simple) Lie algebra $\alg{g}$, with structure constants
$f^{AB}_{C}$
\begin{align}
[\mathbb{J}^{A},\mathbb{J}^{B} ] = f^{AB}_{C}\mathbb{J}^{C}.
\end{align}
The Yangian $Y(\mathfrak{g})$ of this Lie algebra is the algebra
generated by the generators $\mathbb{J}$ of $\alg{g}$ and a new
set of generators $\hat{\mathbb{J}}$, subject to
\begin{align}\label{eqn;DefYangGenerators}
[\mathbb{J}^{A},\hat{\mathbb{J}}^{B} ] = f^{AB}_{C}\hat{\mathbb{J}}^{C}.
\end{align}
Higher level generators can then be obtained by commuting two
level one generators and so on. The above commutation relations
should obey the Jacobi and Serre relations (these are for algebras
other than $\alg{su}(2)$)
\begin{align}
&\left[\mathbb{J}^{[A} ,[\mathbb{J}^B,\mathbb{J}^{C]}]\right]=0\nonumber\\
&\left[\mathbb{J}^{[A} ,[\mathbb{J}^B,\hat{\mathbb{J}}^{C]}]\right]=0\nonumber\\
&\left[\hat{\mathbb{J}}^{[A} ,[\hat{\mathbb{J}}^B,\mathbb{J}^{C]}]\right]=
\frac{1}{4} f^{AG}_{D}f^{BH}_{E}f^{CK}_{F}f_{GHK}\mathbb{J}^{(D}\mathbb{J}^{E}\mathbb{J}^{F)},\nonumber
\end{align}
where $(),[]$ in the indices stand for total symmetrization and
anti-symmetrization, respectively. The indices of the structure
constants are lowered with the Cartan-Killing matrix.\smallskip

The Yangian can be given the structure of a Hopf algebra by
specifying the following coproduct
\begin{align}\label{eqn;StandardYangHopf}
&\Delta \mathbb{J}^A = \mathbb{J}^A\otimes\mathbbm{1} + \mathbbm{1}\otimes\mathbb{J}^A & &
\Delta \hat{\mathbb{J}}^A = \hat{\mathbb{J}}^A\otimes\mathbbm{1} + \mathbbm{1}\otimes\hat{\mathbb{J}}^A
+ \frac{1}{2}f_{BC}^{A}\mathbb{J}^B\otimes\mathbb{J}^C.
\end{align}
The coproducts of the higher level generators are obtained by
using the fact that $\Delta$ should respect commutators. Then
antipode is given by
\begin{align}\label{eqn;YangAntipode}
&{\cal S}(\mathbb{J}^A)=-\mathbb{J}^A&
& {\cal S}(\hat{\mathbb{J}}^A)=-\hat{\mathbb{J}}^A + \frac{1}{4}f^{A}_{BC}f^{BC}_{D}\mathbb{J}^D,
\end{align}
and the counit is
\begin{align}
&\epsilon(\mathbb{J}^A)=\epsilon(\hat{\mathbb{J}}^A)=0&
& \epsilon(\mathbbm{1})=1.
\end{align}
This realization of the Yangian is called Drinfeld's first
realization.

\subsection*{Drinfeld's second realization}

There is also a second realization of a Yangian. This realization
turns out to be particulary useful when checking the Serre
relations of a representation explicitly or in constructing the
universal R-matrix \cite{Khoroshkin:1994uk,Cai:q-alg9709038}.

The second realization of the Yangian \cite{Dsecond} is given in
terms of Chevalley-Serre type generators $\kappa_{i,m},
\xi^\pm_{i,m}$, $i=1,\dots, \text{rank} \alg{g}$, $m=0,1,2,\dots$
satisfying relations
\begin{align}\label{def:drinf2General}
&[\kappa_{i,m},\kappa_{j,n}]=0,\quad [\kappa_{i,0},\xi^+_{j,m}]=a_{ij} \,\xi^+_{j,m},\nonumber\\
&[\kappa_{i,0},\xi^-_{j,m}]=- a_{ij} \,\xi^-_{j,m},\quad \comm{\xi^+_{j,m}}{\xi^-_{j,n}}=\delta_{i,j}\, \kappa_{j,n+m},\nonumber\\
&[\kappa_{i,m+1},\xi^+_{j,n}]-[\kappa_{i,m},\xi^+_{j,n+1}] = \frac{1}{2} a_{ij} \{\kappa_{i,m},\xi^+_{j,n}\},\nonumber\\
&[\kappa_{i,m+1},\xi^-_{j,n}]-[\kappa_{i,m},\xi^-_{j,n+1}] = - \frac{1}{2} a_{ij} \{\kappa_{i,m},\xi^-_{j,n}\},\nonumber\\
&\comm{\xi^\pm_{i,m+1}}{\xi^\pm_{j,n}}-\comm{\xi^\pm_{i,m}}{\xi^\pm_{j,n+1}} = \pm\frac{1}{2} a_{ij} \acomm{\xi^\pm_{i,m}}{\xi^\pm_{j,n}},\nn\\
&i\neq j,\, \, \, \, n_{ij}=1+|a_{ij}|,\, \, \, \, \, Sym_{\{k\}} [\xi^\pm_{i,k_1},[\xi^\pm_{i,k_2},\dots [\xi^\pm_{i,k_{n_{ij}}}, \xi^\pm_{j,l}]\dots]]=0.
\end{align}
In these formulas, $a_{ij}$ is the (symmetric) Cartan matrix. The
index $m$ in the generators is referred to as the level.

Drinfeld \cite{Dsecond} gave an explicit isomorphism between the
two realizations as follows. Let us define a Chevalley-Serre basis
for $\alg{g}$ as composed of Cartan generators $\gen{H}_i$, and
positive (negative) simple roots $\gen{E}^+_i$ ($\gen{E}^-_i$,
respectively). Then one has
\begin{align}\label{def:isom}
&\kappa_{i,0}=\gen{H}_i,\quad \xi^+_{i,0}=\gen{E}^+_i,\quad \xi^-_{i,0}=\gen{E}^-_i,\nonumber\\
&\kappa_{i,1}=\hat{\gen{H}}_i-v_i,\quad \xi^+_{i,1}=\hat{\gen{E}}^+_i-w_i,\quad \xi^-_{i,1}=\hat{\gen{E}}^-_i-z_i,
\end{align}
where $v_i,w_i,z_i$ are certain quadratic combinations of
level-zero generators that we will not list here explicitly. From
the level-zero and level-one generators, one can then recursively
construct all higher-level generators by repeated use of the
relations (\ref{def:drinf2General}).

We will employ both the first and second realization of the
Yangian in the other chapters.

\subsection*{Evaluation representation}

An important representation of the Yangian $Y(\alg{g})$ is the
evaluation representation. Let us work in Drinfeld's first
realization. Consider a representation $V$ of $\alg{g}$ and
introduce a parameter $u$. Then $V(u)$ can host a so-called
evaluation representation of $Y(\alg{g})$ by setting
\begin{align}\label{eqn;EvalRep}
&\hat{\mathbb{J}}^A |v\rangle = u\, \mathbb{J}^A |v\rangle, &
&|v\rangle\in V.
\end{align}
Of course in order for this to be a valid representation, one
needs to check that the Serre relations are satisfied.

\subsection*{Double Yangian}

The Yangian $Y(\alg{g})$ of a Lie (super)algebra is not
quasitriangular, i.e. there is no element $\S\in Y(\alg{g})\otimes
Y(\alg{g})$ that intertwines the coproduct with the opposite
coproduct. However, there exists a Hopf algebra called the double
Yangian $DY(\alg{g})$ which is defined by relations
(\ref{def:drinf2General}) but now one takes the level
$m\in\mathbb{Z}$ \cite{Dsecond,Khoroshkin:1994uk}. On evaluation
representations this means that also negative powers of the
evaluation parameter $u$ are considered.

This enhanced algebra is quasitriangular for simple Lie algebras
\cite{Khoroshkin:1994uk}. For the Yangian of Lie superalgebras
this is not known, although it has been found in specific cases see
for instance
\cite{Cai:q-alg9709038,Spill:2008yr,stuko,Frappat7334}. The
Yangian $Y(\alg{g})$ can be identified with a subalgebra of
$DY(\alg{g})$ by restricting to elements with positive level. This
means that in any representation, the universal R-matrix of
$DY(\alg{g})$ also gives the S-matrix that relates the coproduct
and opposite coproduct for $Y(\alg{g})$.

\section{Integrability and Yangians}

We have already seen that the R-matrix of a quasitriangular Hopf
algebra is closely related to the S-matrix of integrable systems.
Yangians also appear naturally in the context of integrable
models. They appear in a variety of systems like the Hubbard model
\cite{Korepin}, the XXX spin chain \cite{Bernard:1992ya} and also,
as we will see later, in the $\ads$ superstring
\cite{Beisert:2007ds}. In this section we will treat an example of
how a Yangian can arise in integrable models \cite{MacKay:2004tc}.

Consider a two-dimensional field theory with Noether currents
taking values in some Lie-algebra $\alg{g}$
\begin{eqnarray}
J_{\mu} = J_{\mu;a}(t,x)t^a, \qquad t\in\alg{g}.
\end{eqnarray}
These currents are conserved on-shell
\begin{eqnarray}
\partial^{\mu}J_{\mu} = 0,
\end{eqnarray}
and hence they define conserved charges
\begin{eqnarray}
Q_a = \int dx J_{0;a}.
\end{eqnarray}
Assume that the currents satisfy the flatness condition
\begin{eqnarray}\label{eqn;PrincChrModCurrents}
\partial_0 J_1-\partial_1 J_0 +[J_0,J_1]=0.
\end{eqnarray}
Both conservation of the current and the above condition
(\ref{eqn;PrincChrModCurrents}) are equivalent to the flatness of
the following Lax connection
\begin{align}
L_\mu(t,x,z) = \frac{1}{1-z^2}(J_{\mu}(t,x) + z \epsilon_\mu^\nu J_\nu(t,x)).
\end{align}
In this sense this model is integrable. An example of such a model
is the principal chiral model see e.g.
\cite{Sochen:1995dm,Arutyunov:2009ga}. Due to integrability one
expects to find more conserved charges than just the ones
corresponding to the Lie algebra. Because of the flatness
condition it turns out that one can define an additional non-local
current
\begin{eqnarray}
\hat{J}_{\mu;a}(t,x) = \epsilon_{\mu\nu}J^{\nu}_a(t,x)-\frac{i}{2}f_{abc}J_{\mu;b}(t,x)\int_{-\infty}^x dy J_{0;c}(t,x).
\end{eqnarray}
One can then show that the corresponding charge
\begin{eqnarray}
\hat{Q}_{a} =\int dx \hat{J}_{0;a}(t,x)
\end{eqnarray}
is conserved by using conservation of the currect $J_\mu$,
integration by parts and finally the flatness condition.

\begin{figure}[h]
\centering
\includegraphics{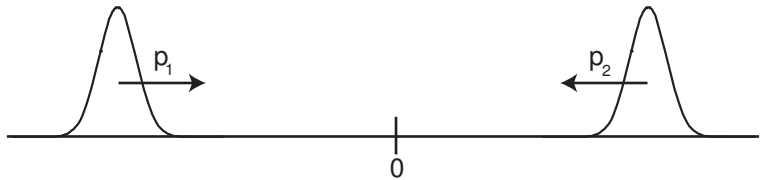}
\caption{A schematic state of a pair of well-separated particles.}\label{Fig;2ParticleState}
\end{figure}

The conserved charges $\hat{Q}_{a}$ form an algebra which can be
studied by computing their Poisson brackets. We will now study the
Hopf algebra structure of this model. Assume that there are
particle-like solutions of the equations of motion that can be
taken to be well separated, see figure \ref{Fig;2ParticleState}.

Let us first compute the charges forming the Lie algebra $\alg{g}$
on such a profile:
\begin{eqnarray}
\left.Q_a\right|_{\rm{profile}} = \int_{-\infty}^0 dx~ J_{0;a}(t,x) + \int^{\infty}_0 dx ~J_{0;a}(t,x).
\end{eqnarray}
This can be though of as a semi-classical analog of the coproduct
of $Q$, and we recognize
\begin{align}
\Delta Q_a &= \left.Q_a\right|_{\rm{particle\ } 1} + \left.Q_a\right|_{\rm{particle\ } 2}\nonumber\\
&= Q_a\otimes\mathbbm{1} + \mathbbm{1}\otimes Q_a.
\end{align}
This exactly agrees with the coproduct for the enveloping algebra
that was discussed earlier (\ref{eqn;UnivCoprod}).

For the non-local charges $\hat{Q}_a$ the discussion becomes more
involved. When evaluating this charge on the same profile we can
split the integral
\begin{align}
\Delta\hat{Q}_a =& \int_{-\infty}^0 dx J^1_a (x) + \int^{\infty}_0
dx J^1_a(x)- \left.\frac{if_{abc}}{2}\right\{\int_{-\infty}^0
dx J_{0;b}(x)\int_{-\infty}^x dy J_{0;c}(y) + \nonumber\\
& \left.+ \int_0^{\infty} dx J_{0;b}(x)\int_0^x dy
J_{0;c}(y)+\int_0^{\infty} dx J_{0;b}(x) \int_{-\infty}^0 dy
J_{0;c}(y)\right\}.
\end{align}
The first terms clearly give $\hat{Q}_a|_{\rm{particle\ } 1} +
\hat{Q}_a|_{\rm{particle\ } 2}$ and the last piece is given in
terms of the charges $Q_a$. We can write this as
\begin{eqnarray}
\Delta\hat{Q}_a = \hat{Q}_a\otimes\mathbbm{1} + \mathbbm{1}\otimes
\hat{Q}_a -\frac{i}{2}f_{abc}Q_b\otimes Q_c.
\end{eqnarray}
Here one recognizes the coproduct of a Yangian generator
(\ref{eqn;StandardYangHopf}).

Since integrability is closely related to symmetries, it is
important to know the underlying symmetry algebra of an integrable
model. For the $\ads$ superstring in the decompactifying limit
this algebra consists of two copies of centrally-extended
$\alg{su}(2|2)$ \cite{Arutyunov:2006ak}. Since the classical model
allows a Lax reformulation \cite{Bena:2003wd} one expects, in view
of the above example, Yangian symmetry to be present. This indeed
turns out to be the case. In the next chapter we will study
centrally-extended $\alg{su}(2|2)$ and its Yangian in detail.

%% file: SU22Algebra3.tex
\chapter{Centrally extended $\mathfrak{su}(2|2)$}\label{chap;SU22}

The algebra which plays a key role in the developments we will
present is the centrally extended $\mathfrak{su}(2|2)$ Lie
superalgebra, which we will denote with $\alg{h}$ throughout the
rest of this work. This chapter will be devoted to the discussion
of the basic properties of this algebra. The main focus will be on
symmetric representations and the underlying Hopf algebra
structure. $\alg{h}$ is the symmetry algebra of the light-cone
Hamiltonian of the $\ads$ superstring \cite{Arutyunov:2006ak} and
it also appears as the symmetry algebra of the spin chain
describing single-trace operators in $\mathcal{N}=4$ super
Yang-Mills \cite{Beisert:2005tm}.

We will first discuss the definition of this algebra and its
automorphisms. After this we will describe its representations,
their tensor products and the (twisted) Hopf algebra structure.
The last part of this chapter will be dealing with the Yangian of
$\alg{h}$ and its corresponding Hopf algebra structure. This
mathematical framework forms the basis for subsequent chapters.

In this chapter we will amply utilize the language of Hopf
algebras and Yangians, cf. chapter
\ref{chap;IntegrabilityAndHopf}.

\section{Defining Relations}

$\alg{h}$ consists of two sets of bosonic
generators $\mathbb{R}^{\alpha}_{\beta},\mathbb{L}^{a}_b$ that
constitute two copies of the $\alg{su}(2)$ algebra. We will use
the convention that roman letters denote bosonic indices and take
values $a,b,\ldots=1,2$ and greek letters are used for fermionic
indices and take values $\alpha,\beta,\ldots=3,4$. There are
supersymmetry generators
$\mathbb{Q}^{a}_{\alpha},\mathbb{Q}^{\dag\alpha}_a$ and central
elements $\mathbb{H},\mathbb{C},\mathbb{C}^{\dag}$. The
non-trivial commutation relations between the generators are given
by
\begin{align}\label{eqn;DefRelsSU22}
&[\mathbb{L}_{a}^{b},\mathbb{J}_{c}] =
\delta_{c}^{b}\mathbb{J}_{a}-\frac{1}{2}\delta_{a}^{b}\mathbb{J}_{c}&
&[\mathbb{R}_{\alpha}^{\beta},\mathbb{J}_{\gamma}] = \delta_{\gamma}^{\beta}\mathbb{J}_{\alpha}-\frac{1}{2}\delta_{\alpha}^{\beta}\mathbb{J}_{\gamma}\nonumber\\
&[\mathbb{L}_{a}^{b},\mathbb{J}^{c}]=-\delta_{a}^{c}\mathbb{J}^{b}+\frac{1}{2}\delta_{a}^{b}\mathbb{J}^{c}&
&[\mathbb{R}_{\alpha}^{\beta},\mathbb{J}^{\gamma}] = -\delta^{\gamma}_{\alpha}\mathbb{J}^{\beta}+\frac{1}{2}\delta_{\alpha}^{\beta}\mathbb{J}^{\gamma}\\
&\{\mathbb{Q}_{\alpha}^{a},\mathbb{Q}_{\beta}^{b}\}=\epsilon_{\alpha\beta}\epsilon^{ab}\mathbb{C}&
&\{\mathbb{Q}^{\dag \alpha}_{a},\mathbb{Q}^{\dag \beta}_{b}\}=\epsilon^{\alpha\beta}\epsilon_{ab}\mathbb{C}^{\dag}\nonumber\\
&\{\mathbb{Q}_{\alpha}^{a},\mathbb{Q}^{\dag\beta}_{b}\} =
\delta_{b}^{a}\mathbb{R}_{\alpha}^{\beta} +
\delta_{\alpha}^{\beta}\mathbb{L}_{b}^{ a}
+\frac{1}{2}\delta_{b}^{a}\delta_{\alpha}^{\beta}\mathbb{H}.&&\nonumber
\end{align}
The first two lines show how the indices of an arbitrary generator $\mathbb{J}$
with the appropriate indices transform.

\section{Fundamental Representation}

The fundamental representation of $\alg{h}$ is four-dimensional.
It is realized on a graded vector space with two bosonic basis
vectors $|e_a\rangle$, $a=1,2$ and two fermionic basis vectors
$|e_\alpha\rangle$, $\alpha=1,2$.

\subsection{Matrix Realization}

The two copies of $\alg{su}(2)$ act canonically on both the
bosonic and fermionic subspace
\begin{align}
\mathbb{L}^a_b|e_c\rangle &=
\delta^a_c|e_b\rangle-\frac{1}{2}\delta^a_b|e_c\rangle &
\mathbb{L}^a_b|e_{\gamma}\rangle &=0\\
\mathbb{R}^{\alpha}_{\beta}|e_c\rangle &= 0
&\mathbb{R}^{\alpha}_{\beta}|e_{\gamma}\rangle &=
\delta^{\alpha}_{\gamma}|e_{\beta}\rangle-\frac{1}{2}\delta^{\alpha}_{\beta}|e_{\gamma}\rangle.
\end{align}
The supercharges act as follows
\begin{align}
\mathbb{Q}^a_{\beta}|e_c\rangle &= a\delta^a_c|e_{\beta}\rangle
&\mathbb{Q}^a_{\beta}|e_{\gamma}\rangle &= b\epsilon_{\beta\gamma}\epsilon^{ab}|e_b\rangle\\
\mathbb{Q}^{\dag\alpha}_{b}|e_c\rangle &=
c\epsilon^{\alpha\gamma}\epsilon_{bc}|e_{\gamma}\rangle
&\mathbb{Q}^{\dag\alpha}_{b}|e_{\gamma}\rangle &=
d\delta^{\alpha}_{\gamma}|e_{b}\rangle.
\end{align}
It is easily seen that all the defining commutation
relations are satisfied, provided the parameters satisfy $ad-bc=1$.

The values of the central charges are found by commuting the
supercharges. They are all proportional to the identity matrix
\begin{align}
\mathbb{H} &= H\mathbbm{1}, & \mathbb{C} = C\mathbbm{1}, & &
\mathbb{C}^{\dag} = C^{\dag}\mathbbm{1}.
\end{align}
and have eigenvalues
\begin{align}
H &= ad + bc, & C = ab, & & C^\dag = cd.
\end{align}
Because of the constraint $ad-bc=1$, the central charges satisfy
the `shortening' condition
\begin{align}\label{eqn;ShorteningSU22}
\mathbb{H}^2 - 4 \mathbb{C}\mathbb{C}^{\dag}=\mathbbm{1}.
\end{align}
Since the central charges are related the representation is
atypical; such a representation is called short
\cite{Beisert:2006qh}. In order for this representation to be
unitary one needs
\begin{align}
&a = d^*, && b=c^*.
\end{align}
In unitary representations we find that $\mathbb{C}$ is the
Hermitian conjugate of $\mathbb{C}^{\dag}$, which justifies our notation, and that the
Hamiltonian $\mathbb{H}$ is Hermitian and its eigenvalues are
real.

For completeness, we explicitly list the matrix
representations of a choice of simple roots of $\alg{h}$:
\begin{align}
&\mathbb{L}^2_1=\left(
\begin{smallmatrix}
 0 & 1 & 0 & 0 \\
 0 & 0 & 0 & 0 \\
 0 & 0 & 0 & 0 \\
 0 & 0 & 0 & 0
\end{smallmatrix}\right),& &
\mathbb{L}^1_2=\left(
\begin{smallmatrix}
 0 & 0 & 0 & 0 \\
 1 & 0 & 0 & 0 \\
 0 & 0 & 0 & 0 \\
 0 & 0 & 0 & 0
\end{smallmatrix}\right),\nonumber\\
&\mathbb{R}^4_3=\left(
\begin{smallmatrix}
 0 & 0 & 0 & 0 \\
 0 & 0 & 0 & 0 \\
 0 & 0 & 0 & 1 \\
 0 & 0 & 0 & 0
\end{smallmatrix}\right),& &
\mathbb{R}^3_4=\left(
\begin{smallmatrix}
 0 & 0 & 0 & 0 \\
 0 & 0 & 0 & 0 \\
 0 & 0 & 0 & 0 \\
 0 & 0 & 1 & 0
\end{smallmatrix}\right),\\
&\mathbb{Q}^1_3=\left(
\begin{smallmatrix}
 0 & 0 & 0 & 0 \\
 0 & 0 & 0 & b \\
 a & 0 & 0 & 0 \\
 0 & 0 & 0 & 0
\end{smallmatrix}\right),& &
\mathbb{Q}^{\dag 4}_2=\left(
\begin{smallmatrix}
 0 & 0 & 0 & 0 \\
 0 & 0 & 0 & d \\
 c & 0 & 0 & 0 \\
 0 & 0 & 0 & 0
\end{smallmatrix}\right).\nonumber
\end{align}
The other elements are easily obtained by making use of the
defining commutation relations.

\subsection{Parameterizations}\label{sect;Parameterization}

From the study of the light-cone gauged superstring on $\ads$ we
know that the central charges are related to the world-sheet
momentum $p$ and the string tension $g$
\begin{align}
&\mathbb{C} = \frac{ig}{2}(e^{ip}-1), & &\mathbb{C}^{\dag} =
-\frac{ig}{2}(e^{-ip}-1).
\end{align}
From this one can express the parameters $a,b,c,d$ that describe
the fundamental representation in terms of $p,g$. Introduce
parameters $x^{\pm}$ that are related to $p,g$ by
\begin{align}\label{eqn;ParamtersPandG}
x^{+} + \frac{1}{x^{+}}-x^{-}-\frac{1}{x^{-}}&=\frac{2i}{g}, &
\frac{x^{+}}{x^{-}} &= e^{ip}.
\end{align}
Then we can write
\begin{align}\label{eqn;ABCDparameters}
  a &= \sqrt{\frac{g}{2}}\eta, &  b &= \sqrt{\frac{g}{2}}
\frac{i}{\eta}\left(\frac{x^{+}}{x^{-}}-1\right), \nonumber\\
  c &= -\sqrt{\frac{g}{2}}\frac{\eta}{x^{+}}, &
d
&=\sqrt{\frac{g}{2}}\frac{x^{+}}{i\eta}\left(1-\frac{x^{-}}{x^{+}}\right).
\end{align}
The parameter $\eta$ corresponds to a rescaling of the bosonic
basis vectors relative to the fermionic ones. However, upon
insisting on unitarity of the representation it is easily seen that
we are led to
\begin{align}\label{eqn;DefEta}
\eta = e^{\frac{ip}{4}}\sqrt{i x^- - ix^+}.
\end{align}
The factor $e^{\frac{ip}{4}}$ is not a consequence of unitarity,
but this particular choice will turn out to be convenient
later on.

The central charges take the form
\begin{align}\label{eqn;centralcharges}
C &= \frac{ g}{2i}\left(1-\frac{x^+}{x^-}\right) =
\frac{g}{2i}\left(1-e^{ip}\right), & & C^\dag = \frac{i
g}{2}\left(1-\frac{x^-}{x^+}\right)= \frac{i
g}{2}\left(1-e^{-ip}\right),\nonumber\\
H &= i g\left(x^- - x^+ +\frac{i}{g} \right).
\end{align}
From the shortening condition (\ref{eqn;ShorteningSU22}) one sees
that the Hamiltonian satisfies a lattice type dispersion relation
\begin{align}\label{eqn;HviaMomentum}
H^2 &= 1 + 4 g^2\sin^2(p/2).
\end{align}
It is also worthwhile mentioning that instead of $p$, one can
define a rapidity variable $z$ living on an elliptic curve
(torus). Equation (\ref{eqn;HviaMomentum}) can be uniformized on
the torus, which prevents the appearance of branch cuts that arise
when taking square roots in order to solve for one of the
variables. In terms of Jacobi elliptic functions one finds
\begin{align}
&p = 2 \mathrm{am} z,& & \sin\frac{p}{2} = 2 \mathrm{sn}(z,k),& H=
2 \mathrm{dn}(z,k),
\end{align}
where $k  = -4g^2$. Written in terms of this rapidity variable the
parameters $x^{\pm}$ become
\begin{align}
&x^{\pm} = \frac{1+\mathrm{dn} z}{2g}\left(\frac{\mathrm{cn} z}{\mathrm{sn} z}\pm i\right).
\end{align}
The periods of the torus $\omega_1,\omega_2$ are defined by
complete elliptic integrals of the first kind
\begin{align}
&2\omega_1 = 4K(k)& & 2\omega_2 = 4iK(1-k)-4K(k),
\end{align}
where
\begin{align}
K(k) = \frac{\pi}{2}\sum_{n=0}^{\infty}\left[\frac{(2n-1)!!}{2n !!}\right]^2k^{2n}.
\end{align}

\section{The Outer Automorphism and $\alg{gl}(2|2)$}\label{sec;OuterAndGL22}

The algebra $\alg{h}$ admits a useful family of outer automorphisms
\cite{Beisert:2006qh,Arutyunov:2008zt} that form an
$SL(2)$ group. It is defined by
\begin{eqnarray}\label{eqn;OuterAut}
\begin{pmatrix}
\mathbb{Q}^a_{\alpha}\\
\epsilon^{ab}\epsilon_{\alpha\beta}\mathbb{Q}^{\dag\beta}_{b}
\end{pmatrix}
\rightarrow
\begin{pmatrix}
u_1 & u_2 \\
v_1 & v_2
\end{pmatrix}
\begin{pmatrix}
\mathbb{Q}^a_{\alpha}\\
\epsilon^{ab}\epsilon_{\alpha\beta}\mathbb{Q}^{\dag\beta}_{b}
\end{pmatrix}.
\end{eqnarray}
The parameters $u_1,u_2,v_1,v_2$ satisfy
\begin{eqnarray}
u_1 v_2-u_2 v_1 =1.
\end{eqnarray}
This condition precisely defines an $SL(2)$ transformation.
Under this automorphism the central charges transform as
\begin{eqnarray}
\begin{pmatrix}
\mathbb{H} \\
\mathbb{C} \\
\mathbb{C}^{\dag}
\end{pmatrix}
\rightarrow
\begin{pmatrix}
(u_1v_1+u_2v_2)\mathbb{H} +2u_1v_2\mathbb{C} +2u_2v_1\mathbb{C}^{\dag} \\
u_1^2\mathbb{C} + u_2^2 \mathbb{C}^{\dag} + u_1u_2\mathbb{H}\\
v_1^2\mathbb{C} + v_2^2 \mathbb{C}^{\dag} + v_1v_2\mathbb{H}
\end{pmatrix}.
\end{eqnarray}
It can be checked that the shortening condition
(\ref{eqn;ShorteningSU22}) is invariant under this transformation.

A very useful application of this automorphism is that one can
transform $\alg{h}$ into a normal
$\alg{su}(2|2)$ algebra and vice versa. Explicitly, the
parameters that transform the central charges $\fC,\fC^\dag$ to zero are given by
\begin{align}
&u_1=\frac{\mathbb{H}-\sqrt{\mathbb{H}^2-4\mathbb{C}\mathbb{C}^{\dag}}}{2
\mathbb{C}}, & & u_2 = -1, \\
& v_1 = \frac{1}{2}
\left(1+\frac{\mathbb{H}}{\sqrt{\mathbb{H}^2-4\mathbb{C}\mathbb{C}^{\dag}}}\right),
&& v_2 =  -\frac{\mathbb{C}}{\sqrt{\mathbb{H}^2 - 4
\mathbb{C}\mathbb{C}^\dag}}\nonumber.
\end{align}
Notice the appearance of the shortening condition
(\ref{eqn;ShorteningSU22}). A special case of this automorphism
occurs when $u_2 = 0 = v_1 $ and $u_1 = e^{i\phi}$ for some real
phase $\phi$.
\begin{align}\label{eqn;U1automorphism}
&\mathbb{Q}^{a}_{\alpha} \rightarrow e^{i\phi} \mathbb{Q}^{a}_{\alpha} & & \mathbb{Q}^{\dag \alpha}_{a} \rightarrow e^{-i\phi} \mathbb{Q}^{\dag \alpha}_{a} &
&\mathbb{C} \rightarrow e^{2i\phi} \mathbb{C} & & \mathbb{C}^{\dag} \rightarrow e^{-2i\phi} \mathbb{C}^{\dag}.
\end{align}

\subsection*{Relation to $\alg{gl}(2|2)$}

By the SL$(2)$ automorphism one can now transform any
representation of the bigger Lie superalgebra $\alg{gl}(2|2)$ into
a representation of $\alg{h}$. This is convenient, since in the
paper \cite{Zhang:2004qx} all finite-dimensional irreducible
representations of $\alg{gl} (2|2)$ are explicitly constructed in
an oscillator basis. Generators of $\alg{gl}(2|2)$ are denoted by
$E_{ij}$, with commutation relations
\begin{align}
[E_{ij},E_{kl}]=\delta_{jk} E_{il} - (-)^{(d[i]+d[j])(d[k]+d[l])}
\delta_{il} E_{kj}.
\end{align}
Indices $i,j,k,l$ run from $1$ to $4$, and the fermionic grading
is assigned as $d[1]=d[2]=0$, $d[3]=d[4]=1$. The quadratic Casimir
of this algebra is $C_2 = \sum_{i,j=1}^4 (-)^{d[j]} E_{ij}E_{ji}$.
One finds that the finite dimensional irreps are labelled by two
half-integers $j_1,j_2 = 0,\frac{1}{2},...$, and two complex
numbers $q$ and $y$. These numbers correspond to the values taken
by appropriate generators on the highest weight state
$|\omega\rangle$ of the representation, defined by the following
conditions:
\begin{align} \label{zghw} &&H_1 |\omega\rangle = (E_{11}-E_{22})
|\omega\rangle = 2 j_1 |\omega\rangle, \qquad
H_2 |\omega\rangle = (E_{33}-E_{44}) |\omega\rangle = 2 j_2 |\omega\rangle,\nonumber\\
&&I |\omega\rangle = \sum_{i=1}^4 E_{ii} |\omega\rangle = 2 q
|\omega\rangle, \, \, \, N |\omega\rangle = \sum_{i=1}^4 (-)^{[i]}
E_{ii} |\omega\rangle = 2 y |\omega\rangle, \, \, \, \,
E_{i<j}|\omega\rangle =0.
\end{align}
The generator $N$ never appears on the right hand side of the
commutation relations, therefore it is defined up to the addition
of a central element $\beta I$, with $\beta$ a
constant\footnote{We drop the term $\beta I$ since it will not
affect our discussion.}. This also means that we can consistently
mod out the generator $N$, and obtain $\sls (2|2)$ as a subalgebra
of the original $\alg{gl} (2|2)$ algebra\footnote{Further modding
out of the center $I$ produces the simple Lie superalgebra
$\alg{psl} (2|2)$. The representation theory of $\alg{psl} (2|2)$
has been completely classified in \cite{Gotz:2005ka}.}. In order
to construct representations of the centrally-extended $\su (2|2)$
Lie superalgebra, we then first mod out $N$, and subsequently perform
an $\sls (2)$ rotation by means of the outer automorphism
(\ref{eqn;OuterAut}).

The way the outer automorphism is implemented is by mapping the
$\alg{gl} (2|2)$ non-diagonal generators into new generators as
follows:
\begin{align}
&\fL^b_a = E_{a b} \, \, \, \forall \, \, a \neq b, \qquad
\fR^\beta_\alpha = E_{\alpha \beta} \, \, \,
\forall \, \, \alpha \neq \beta,\nonumber\\
&\fQ^a_\alpha = a \, E_{\alpha a} + b \, \epsilon_{\alpha \beta} \epsilon^{a b} E_{b \beta},\nonumber\\
&\fQ^{\dag\alpha}_a = c \, \epsilon_{a b} \epsilon^{\alpha \beta}
E_{\beta b} + d \, E_{a \alpha},
\end{align}
subject to the constraint
\begin{align} ad - bc = 1.
\end{align}
The diagonal elements are automatically obtained by commuting
positive and negative roots. In particular, one obtains the
following values of the central charges:
\begin{align}
\fHH = (a d + b c) \, I, \qquad \fC = a b \, I, \qquad \fC^\dagger
= c d \, I.
\end{align}
Note that in \cite{Zhang:2004qx} $I$ is just proportional to the
identity operator $I = 2 q \mathbbm{1}$. Moreover, it turns out
that the generator $N$ will play an important role later on. Let us discuss its properties here. We define the following operator
\begin{align}\label{eqn;defB}
\mathbb{B} = \frac{1}{2}\frac{1}{a d + bc }N,
\end{align}
which satisfies the following commutation relations
\begin{align}\label{eqn;commB}
&[\mathbb{B},\mathbb{Q}^{a}_{\alpha} ] =- \mathbb{Q}^{a}_{\alpha}
+ 2 \mathbb{C}\mathbb{H}^{-1}\epsilon_{\alpha\beta}\epsilon^{ab}
\mathbb{Q}_{b}^{\dag\alpha}\nonumber\\
&[\mathbb{B},\mathbb{Q}^{\dag\alpha}_{a} ] =
\mathbb{Q}_{a}^{\dag\alpha} - 2\mathbb{C}^{\dagger}\mathbb{H}^{-1}
\epsilon^{\alpha\beta}\epsilon_{ab}
\mathbb{Q}^{b}_{\beta}\\
&[\mathbb{B},\mathbb{L}^{a}_{b} ]=\
[\mathbb{B},\mathbb{R}^{\alpha}_{\beta} ]=\ [\mathbb{B},\mathbb{H}
]=0\nonumber.
\end{align}
Notice that since central charges are actually proportional to $I$
we have replaced terms like $\frac{ab}{ad+bc}$ by $\fC\fHH^{-1}$.
Furthermore, by defining the quadratic operator
\begin{align}
\mathcal{T}=
\mathbb{R}^{\alpha}_{\beta}\mathbb{R}^{\beta}_{\alpha}-
\mathbb{L}^{a}_{b}\mathbb{L}^{b}_{a}+
\mathbb{Q}_{a}^{\dag\alpha}\mathbb{Q}_{\alpha}^{a}-
\mathbb{Q}_{\alpha}^{a}\mathbb{Q}_{a}^{\dag\alpha},
\end{align}
it follows that $\mathbb{B}$ and $\mathcal{T}$ can be used to construct a generalized
Casimir operator $C_2$
\begin{eqnarray}\label{eqn;Casimir}
C_2 = \mathbb{B}\mathbb{H} - \mathcal{T}.
\end{eqnarray}
Generalized here means that this operator can be shown to be
central upon using the non-linear commutation relations
(\ref{eqn;commB}). Hence if one constructs a representation of
$\alg{h}$ by using the $\sls (2)$ rotation procedure we have
described before, one can supply it with an extra generator
$\mathbb{B}$. This generator would be the missing element to
complete the algebra to $\alg{gl}(2|2)$, if it were not for its
non-linear commutation relations.

\section{Symmetric Short Representations}\label{sec;boundstaterep}

An important class of representations are the symmetric short
representations. It turns out that they describe bound states
of $\ads$ world sheet excitations. These representations are
indexed by a positive integer $\ell$ and the corresponding
representation is $4\ell$ dimensional. This $4\ell$ dimensional
atypical symmetric representation is realized on a graded vector
space with basis $|e_{a_{1}\ldots a_{\ell}}\rangle,
|e_{a_{1}\ldots a_{\ell-1}\alpha}\rangle$ and $|e_{a_{1}\ldots
a_{\ell-2}\alpha\beta}\rangle$, where $a_{i}$ are bosonic indices
and $\alpha,\beta$ are fermionic indices. Each of the basis
vectors is totally symmetric in the bosonic indices and
anti-symmetric in the fermionic indices
\cite{Arutyunov:2008zt,Beisert:2006qh,Dorey:2006dq,Chen:2006gq,Roiban:2006gs}.
We will refer to these representations as bound state representations.

The most convenient way to describe these representations is by
the so-called superspace formalism, introduced in
\cite{Arutyunov:2008zt}. In this formalism the basis vectors
correspond to monomials and the algebra generators are
differential operators. The big advantage of this formalism is that
it allows one to treat all bound states at once instead of
dealing with matrices of arbitrary (big) size $4\ell$.


Consider the vector space of analytic functions of two bosonic
variables $w_{1,2}$ and two fermionic variables $\theta_{3,4}$.
Since we are dealing with analytic functions we can expand any
such function $\Phi(w,\theta)$:
\begin{align}
\Phi(w,\theta) =&\ \sum_{\ell=0}^{\infty}\Phi_{\ell}(w,\theta),\nonumber\\
\Phi_{\ell} =&\ \phi^{a_{1}\ldots a_{\ell}}w_{a_{1}}\ldots
w_{a_{\ell}} +\phi^{a_{1}\ldots a_{\ell-1}\alpha}w_{a_{1}}\ldots
w_{a_{\ell-1}}\theta_{\alpha}+\nonumber\\
&\ +\phi^{a_{1}\ldots a_{\ell-2}\alpha\beta}w_{a_{1}}\ldots
w_{a_{\ell-2}}\theta_{\alpha}\theta_{\beta}.
\end{align}
In terms of the above analytic functions, the basis vectors of the
totally symmetric representation can clearly be identified as
$|e_{a_{1}\ldots a_{\ell}}\rangle \leftrightarrow w_{a_{1}}\ldots
w_{a_{\ell}}$,$|e_{a_{1}\ldots a_{\ell-1}\alpha}\rangle$     $
\leftrightarrow w_{a_{1}}\ldots w_{a_{\ell-1}}\theta_{\alpha}$ and
$|e_{a_{1}\ldots a_{\ell-1}\alpha\beta}\rangle \leftrightarrow
w_{a_{1}}\ldots w_{a_{\ell-2}}\theta_{\alpha}\theta_{\beta}$,
respectively. In other words, we find the atypical totally
symmetric representation of dimension $4\ell$ when we restrict to
terms $\Phi_{\ell}$, i.e. monomials of degree $\ell$.\smallskip

In this representation the algebra generators can be written in
differential operator form as
\begin{align}\label{eqn;AlgDiff}
\mathbb{L}_{a}^{\ b} &= w_{a}\frac{\partial}{\partial w_{b}}-\frac{1}{2}\delta_{a}^{b}w_{c}\frac{\partial}{\partial w_{c}}, & \mathbb{R}_{\alpha}^{\ \beta} &= \theta_{\alpha}\frac{\partial}{\partial \theta_{\beta}}-\frac{1}{2}\delta_{\alpha}^{\beta}\theta_{\gamma}\frac{\partial}{\partial \theta_{\gamma}}, \\
\mathbb{Q}_{\alpha}^{\ a} &= a
\theta_{\alpha}\frac{\partial}{\partial
w_{a}}+b\epsilon^{ab}\epsilon_{\alpha\beta}
w_{b}\frac{\partial}{\partial \theta_{\beta}}, &
\mathbb{Q}_{a}^{\dag \alpha} &= d w_{a}\frac{\partial}{\partial
\theta_{\alpha}}+c\epsilon_{ab}\epsilon^{\alpha\beta}
\theta_{\beta}\frac{\partial}{\partial w_{b}},
\end{align}
and the central charges are
\begin{align}
 \mathbb{C} &= ab \left(w_{a}\frac{\partial}{\partial w_{a}}+\theta_{\alpha}\frac{\partial}{\partial
 \theta_{\alpha}}\right),& \mathbb{C}^{\dag} &= cd \left(w_{a}\frac{\partial}{\partial w_{a}}+\theta_{\alpha}\frac{\partial}{\partial
 \theta_{\alpha}}\right),\\
 \mathbb{H} &= (ad +bc)\left(w_{a}\frac{\partial}{\partial w_{a}}+\theta_{\alpha}\frac{\partial}{\partial
 \theta_{\alpha}}\right).
\end{align}
To form a representation, the parameters $a,b,c,d$ again must
satisfy the condition $ad-bc=1$. The central charges become $\ell$
dependent:
\begin{eqnarray}
H = \ell(ad+bc),\qquad C =\ell ab , \qquad C^{\dag} =\ell cd.
\end{eqnarray}
The parameters $a,b,c,d$ can be expressed in terms of the bound state
momentum $p$ and the coupling $g$:
\begin{eqnarray}
\begin{array}{lll}
  a = \sqrt{\frac{g}{2\ell}}\eta, & \quad & b = \sqrt{\frac{g}{2\ell}}
\frac{i}{\eta}\left(\frac{x^{+}}{x^{-}}-1\right), \\
  c = -\sqrt{\frac{g}{2\ell}}\frac{\eta}{x^{+}}, & \quad &
d=\sqrt{\frac{g}{2\ell}}\frac{x^{+}}{i\eta}\left(1-\frac{x^{-}}{x^{+}}\right),
\end{array}
\end{eqnarray}
where the parameters $x^{\pm}$ satisfy
\begin{eqnarray}
x^{+} +
\frac{1}{x^{+}}-x^{-}-\frac{1}{x^{-}}=\frac{2i\ell}{g},\qquad
\frac{x^{+}}{x^{-}} = e^{ip}
\end{eqnarray}
and the parameter $\eta$ is given by
\begin{eqnarray}\label{eqn;ScatteringBasis}
\eta = \eta(p),\qquad \eta(p)=
e^{i\frac{p}{4}}\sqrt{ix^{-}-ix^{+}}.
\end{eqnarray}
The fundamental representation is obtained by taking $\ell=1$.

Tensor product representations are easily obtained by multiplying
these superfields. Note that due to the fermionic nature of the
$\theta$ variables we are automatically dealing with graded tensor
products. For example the tensor product of two fundamental
representations is described by monomials
\begin{align}
\{w_a v_b, w_a \vartheta_{\beta},  \theta_{\alpha}v_b, \theta_{\alpha}\vartheta_{\beta}\},
\end{align}
where the variables $w,\theta$ and $v,\vartheta$ describe the
first and the second fundamental representation respectively.

These bound state representations can be obtained starting from a
$\alg{gl}(2|2)$ representation by identifying $E_{ab} =
w_a\partial_{w_b}, \ldots$ and proceeding as described in section
\ref{sec;OuterAndGL22}. This means that one can find an analog of
(\ref{eqn;defB}) in these representations
\begin{align}\label{eqn;defBoperator}
\mathbb{B} = \frac{1}{2(a d+ b
c)}\left(w_a\frac{\partial}{\partial w_a}
-\theta_\alpha\frac{\partial}{\partial \theta_\alpha}\right).
\end{align}
The quadratic Casimir takes values
\begin{align}
C_2  = \ell(\ell-1)\mathbbm{1}
\end{align}
and is indeed central.

\section{Hopf Algebra Structure}

In order to have a consistent Hopf algebra structure for $\alg{h}$
and for its Yangian which we introduce in the next section, one
needs to consider a modified coproduct structure. To this end we
introduce an additional central generator $\mathbb{U}$, which is
closely related to the central charges. Let us equip the symmetry
algebra with the following deformed Hopf-algebra (opposite)
coproduct \cite{Gomez:2006va,Plefka:2006ze}
\begin{align}\label{eqn;BraidCoproduct}
&\Delta (\mathbb{J}) = \mathbb{J} \otimes \mathbb{U}^{[[\mathbb{J}]]} +
\mathbbm{1} \otimes \mathbb{J}, & &\Delta^{op} (\mathbb{J}) = \mathbb{J} \otimes \mathbbm{1} +
\mathbb{U}^{[[\mathbb{J}]]} \otimes \mathbb{J}, \nonumber\\
&\Delta(\mathbb{U})=\mathbb{U}\otimes \mathbb{U}, & &\Delta^{op}(\mathbb{U})=\mathbb{U}\otimes \mathbb{U},
\end{align}
where $\mathbb{J}$ is any generator of $\alg{h}$,
$[[\mathbb{J}]]=0$ for the bosonic $\su (2) \oplus \su (2)$
generators and for the energy generator $\mathbb{H}$,
$[[\mathbb{J}]]=1$ (resp., $-1$) for the $\mathbb{Q}$ (resp.,
$\mathbb{Q}^{\dag}$) supercharges, and $[[\mathbb{J}]]=2$ (resp.,
$-2$) for the central charge $\mathbb{C}$ (resp.
$\mathbb{C}^\dagger$). The fact that $[[\mathbb{C}]]=2$, even
though it is central like the Hamiltonian, is a consequence of
$\Delta$ respecting the Lie bracket. The value of $\mathbb{U}$ is
determined by the consistency requirement that the coproduct is
cocommutative on the center. Since the S-matrix should commute
with the center, one finds that this is a necessary condition for
the existence of an S-matrix (\ref{eqn;SymmProp}). This produces
the algebraic condition
\begin{align}\label{eqn;CviaBraiding}
\mathbb{U}^2 \, = \, \kappa
\, \mathbb{C} \, + \, \mathbbmss{1}
\end{align}
for some representation-independent constant $\kappa$. With our
choice of parameterization of the central charge for short
representations (\ref{eqn;centralcharges}) it follows that $\kappa
= \frac{2}{i g}$, and we obtain the relation
\begin{align}
\mathbb{U} \, = \,
\sqrt{\frac{x^+}{x^-}} \, \mathbbmss{1} \, =\, e^{i \frac{p}{2}}
\, \mathbbmss{1}.
\end{align}
We call $\mathbb{U}$ a braiding factor. In order to have a
complete realization of the Hopf algebra, one needs to specify the
antipode map $\cal{S}$
\cite{Janik:2006dc,Plefka:2006ze,Arutyunov:2008zt} and the co-unit
\begin{align}
&\epsilon(\mathbbm{1})=1, & & \epsilon(\mathbb{J})=0, &&
\epsilon(\mathbbm{U})=1.
\end{align}
From (\ref{eqn;DefAntipode}) one deduces that the inclusion of
$\mathbb{U}$ also alters the antipode which becomes
\begin{align}
\mathcal{S}(\mathbb{J}) = -\mathbb{U}^{-[[\mathbb{J}]]}\mathbb{J}.
\end{align}
The antipode map corresponds to particle to anti-particle
transformations, and has an alternative description in terms of a
charge conjugation matrix $\cal{C}$ as we will now explain. The $z$-torus offers a
convenient way to describe anti-particles. One finds that
\begin{align}
&H(z+\omega_2) = -H(z) & & p(z+\omega_2) = - p(z).
\end{align}
This is similar to the crossing transformation in relativistic
models. On the level of the parameters $x^{\pm}$ this crossing
transformation is
\begin{align}
x^{\pm}\rightarrow \frac{1}{x^{\pm}}.
\end{align}
Consider the map
\begin{align}
\mathbb{J} \rightarrow -\mathbb{J}^{st},
\end{align}
which preserves the $\alg{su}(2|2)$ commutation relations. Letting
this map act on an irrep of centrally $\alg{su}(2|2)$ with central
elements $\mathbb{H},\mathbb{C}$ clearly gives a different irrep
of centrally extended $\alg{su}(2|2)$ with central elements
$-\mathbb{H},-\mathbb{C}$. Under the crossing relation the central
charge $\mathbb{C}$ transforms as
\begin{align}
\mathbb{C}(z+\omega_2) = -e^{-ip}\mathbb{C}(z).
\end{align}
The phase $e^{-ip}$ can be absorbed by the $U(1)$-automorphism
(\ref{eqn;U1automorphism}), i.e. we choose the phase in this
automorphism to be $e^{\frac{ip}{2}} =\mathbb{U}$. We see that for
short representations acting with the antipode on the algebra
generators produces the same set of central charges as the above
described anti-particle transformation. This indicates that there
should be a similarity transformation by which we can relate the
two
\begin{align}
{\cal{S}} (\mathbb{J}) = -\mathbb{U}^{-[[\mathbb{J}]]} \,
\mathbb{J} \, = \, {\cal{C}} \, \overline{\mathbb{J}}^{st} \,
{\cal{C}}^{-1},
\end{align}
where $\overline{\mathbb{J}} = \mathbb{J}(z+\omega_2)$. This
transformation matrix is called the charge conjugation matrix
$\cal{C}$. One finds for the fundamental representation it is
given by
\begin{align}
{\cal C} =
\left(\begin{smallmatrix}
 0 & -i & 0 & 0 \\
 i & 0 & 0 & 0 \\
 0 & 0 & 0 & 1 \\
 0 & 0 & -1 & 0
\end{smallmatrix}\right).
\end{align}
For generic bound state representations in the operator language,
the conjugation operator is
\begin{align}
{\cal C} = -i\epsilon^{ab} w_a \partial_{w_b} + \epsilon^{\alpha\beta}\theta_\alpha \partial_{\theta_\beta}.
\end{align}
On the uniformizing torus, applying the particle to anti-particle
transformation four times gives the identity.


It is readily checked that the structure we just introduced
satisfies all the defining relations of Hopf algebras. In
particular, the braided (opposite) coproduct respects the Lie
bracket
\begin{align}
\Delta [\mathbb{J}^{A},\mathbb{J}^{B}] = [\Delta\mathbb{J}^{A},
\Delta\mathbb{J}^{B}].
\end{align}
Here we can see the convenience of our particular choice of $\eta$
(\ref{eqn;DefEta}). It turns out that for this choice
\begin{align}
&(\Delta \mathbb{L}^a_b)^t = \Delta^{op} \mathbb{L}^b_a & & (\Delta \mathbb{R}^{\alpha}_{\beta})^t = \Delta^{op} \mathbb{L}^{\beta}_{\alpha}\\
&(\Delta \mathbb{Q}^{a}_{\alpha})^t \sim \Delta^{op}
\mathbb{Q}^{\dag\alpha }_a & & (\Delta
\mathbb{Q}^{\dag\alpha}_a)^t \sim \Delta^{op} \mathbb{Q}^{\dag
a}_{\alpha}.
\end{align}
This means that the R-matrix will be symmetric.


Finally, we would like to discuss a more technical rewriting of
the braided structure. It is worthwhile to notice that the
braiding factors appearing in the coproducts can be absorbed
explicitly in the parameters $a,b,c,d$. Explicitly, the parameters
for the tensor products of two bound state representations
appearing in the coproduct (\ref{eqn;BraidCoproduct}) are given
by:
\begin{eqnarray}\label{eqn;LabelWithBrainding1}
\begin{array}{lll}
  a_{1} = \sqrt{\frac{g}{2\ell_1}}\eta_{1}, & ~ & b_{1} =
 -i e^{i p_{2}}\sqrt{\frac{g}{2\ell_1}}~
\frac{1}{\eta_{1}}\left(\frac{x_{1}^{+}}{x_{1}^{-}}-1\right), \\
  c_{1} = -e^{-i p_{2}}\sqrt{\frac{g}{2\ell_1}}\frac{\eta_{1}}{ x_{1}^{+}}, & ~ &
d_{1}=i\sqrt{\frac{g}{2\ell_1}}\frac{x_{1}^{+}}{\eta_{1}}\left(\frac{x_{1}^{-}}{x_{1}^{+}}-1\right),\\
\eta_{1} =
e^{i\frac{p_{1}}{4}}e^{i\frac{p_{2}}{2}}\sqrt{ix^{-}_{1}-ix^{+}_{1}},&~&~\\
~ &~& ~ \\
  a_{2} = \sqrt{\frac{g}{2\ell_2}}\eta_{2}, & ~ & b_{2} = -i\sqrt{\frac{g}{2\ell_2}}
\frac{1}{\eta_{2}}\left(\frac{x_{2}^{+}}{x_{2}^{-}}-1\right), \\
  c_{2} = -\sqrt{\frac{g}{2\ell_2}}\frac{\eta_{2}}{x_{2}^{+}}, & ~ &
d_{2}=i\sqrt{\frac{g}{2\ell_2}}\frac{x_{2}^{+}}{i\eta_{2}}\left(\frac{x_{2}^{-}}{x_{2}^{+}}-1\right),\\
\eta_{2} = e^{i\frac{p_{2}}{4}}\sqrt{ix^{-}_{2}-ix^{+}_{2}},&~&~
\end{array}
\end{eqnarray}
where the indices $1,2$ refer to first and second space
respectively. One sees now that the effect of the braiding factor
$\mathbb{U}$ causes the parameters of the first space to depend on
the momentum $p_2$ of the second particle.

Accordingly, the labels used in $\Delta^{op}$ are given by (we
supply them with indices $3,4$ to make notational distinction
between opposite and normal coproduct):
\begin{eqnarray}
\begin{array}{lll}\label{eqn;LabelWithBrainding2}
  a_{3} = \sqrt{\frac{g}{2\ell_1}}\eta^{op}_{1}, & ~ & b_{3} = -i\sqrt{\frac{g}{2\ell_1}}
\frac{1}{\eta^{op}_{1}}\left(\frac{x_{1}^{+}}{x_{1}^{-}}-1\right), \\
  c_{3} = -\sqrt{\frac{g}{2\ell_1}}\frac{\eta^{op}_{1}}{x_{1}^{+}}, & ~ &
d_{3}=i\sqrt{\frac{g}{2\ell_1}}\frac{x_{1}^{+}}{i\eta^{op}_{1}}\left(\frac{x_{1}^{-}}{x_{1}^{+}}-1\right),\\
\eta^{op}_{1} =
e^{i\frac{p_{1}}{4}}\sqrt{ix^{-}_{1}-ix^{+}_{1}},&~&~\\
 ~ &~& ~ \\
  a_{4} = \sqrt{\frac{g}{2\ell_2}}\eta^{op}_{2}, & ~ &
b_{4} =
 -i e^{i p_{1}}\sqrt{\frac{g}{2\ell_2}}~
\frac{1}{\eta^{op}_{2}}\left(\frac{x_{2}^{+}}{x_{2}^{-}}-1\right), \\
  c_{4} = -e^{-i p_{1}}\sqrt{\frac{g}{2\ell_2}}\frac{\eta^{op}_{2}}{ x_{2}^{+}}, & ~ &
d_{4}=i\sqrt{\frac{g}{2\ell_2}}\frac{x_{2}^{+}}{\eta^{op}_{2}}\left(\frac{x_{2}^{-}}{x_{2}^{+}}-1\right),\\
\eta^{op}_{2} =
e^{i\frac{p_{2}}{4}}e^{i\frac{p_{1}}{2}}\sqrt{ix^{-}_{2}-ix^{+}_{2}}.&~&~
\end{array}
\end{eqnarray}
When using this parameters these parameters the coproduct looks
standard again
\begin{align}
&\Delta\mathbb{J}^A = \mathbb{J}^A(a_1,b_1,c_1,d_1)\otimes\mathbbm{1} + \mathbbm{1}\otimes\mathbb{J}^A(a_2,b_2,c_2,d_2)\\
&\Delta^{op}\mathbb{J}^A = \mathbb{J}^A(a_3,b_3,c_3,d_3)\otimes\mathbbm{1} + \mathbbm{1}\otimes\mathbb{J}^A(a_4,b_4,c_4,d_4).
\end{align}
The non-trivial braiding factors are all hidden in the parameters
of the four representations involved. The reason for this
technical excursion is that it will make future computations and
results more transparent.

\section{The Yangian of centrally extended $\alg{su}(2|2)$}

Next we discuss the Yangian of $\alg{h}$. One can check that the
Cartan matrix of $\alg{h}$ is not invertible. Actually one finds
that this algebra also does not allow for a non-zero bilinear
(Killing) form. This indicates that one cannot straightforwardly
apply formula (\ref{eqn;StandardYangHopf}) since we have no means
of lowering the Lie algebra indices with the Killing form.
Nevertheless, it turns out that $\alg{su}(2|2)$ admits a Yangian.
The Yangian structure is somewhat unconventional since in
evaluation representations (\ref{eqn;EvalRep})
quasi-cocommutativity implies a relation between the evaluation
parameter $u$ and the parameters of the representation as we will
explain later on.


\subsection{First realization}

Let us introduce an additional set of generators
$\hat{\mathbb{J}}^A$ that satisfy
\begin{align}
[\mathbb{J}^{A},\hat{\mathbb{J}}^{B} ]  =
f^{AB}_{C}\hat{\mathbb{J}}^{C},
\end{align}
where $f^{AB}_{C}$ are the structure constants of $\alg{h}$. The
algebra generated by these generators, together with the
generators $\mathbb{J}^A$ of $\alg{h}$ is called the Yangian
$Y(\alg{h})$ of $\alg{h}$, see section \ref{sec;DefYang}.

The absence of a non-zero Killing form prevents one from using
(\ref{eqn;StandardYangHopf}) to find the Hopf algebra structure of
$Y(\alg{h})$. In order to be able to still derive coproducts of
the Yangian type (\ref{eqn;StandardYangHopf}) one can apply a
number of techniques, for example one can make use of a limiting
procedure on the exceptional algebra $D(2,1;\varepsilon)$
\cite{Matsumoto:2008ww}. This is done in Appendix A of this
chapter. A different approach was followed in
\cite{Beisert:2007ds} where automorphisms were used. Of course,
these procedures are not direct computations of the coproduct
structure and one has to check afterwards whether the found
coproduct satisfies all the defining relations. Let us list the
explicit formulae for the coproducts here
\begin{align}\label{eqn;YangianCoprod}
\Delta(\hat{\mathbb{L}}^{a}_{b}) &=
\hat{\mathbb{L}}^{a}_{b}\otimes\mathbbm{1} +
\mathbbm{1}\otimes\hat{\mathbb{L}}^{a}_{b} +
\left.\frac{1}{2}\right[\mathbb{L}^{c}_{b}\otimes\mathbb{L}^{a}_{c}-
\mathbb{L}^{a}_{c}\otimes\mathbb{L}^{c}_{b}+\\
&\ \left.-
\mathbb{Q}^{\dag\gamma}_{b}\otimes\mathbb{U}^{-1}\mathbb{Q}^{a}_{\gamma}-\mathbb{Q}^{a}_{\gamma}\otimes\mathbb{U}\mathbb{Q}^{\dag
\gamma}_{b}+\frac{\delta^{a}_{b}}{2}(\mathbb{Q}^{\dag\gamma}_{c}\otimes\mathbb{U}^{-1}\mathbb{Q}^{c}_{\gamma}+
\mathbb{Q}^{c}_{\gamma}\otimes\mathbb{U}\mathbb{Q}^{\dag\gamma}_{c})~\right]\nonumber\\
\Delta(\hat{\mathbb{R}}^{\alpha}_{\beta}) &=
\hat{\mathbb{R}}^{\alpha}_{\beta}\otimes\mathbbm{1} +\mathbbm{1}\otimes\hat{\mathbb{R}}^{\alpha}_{\beta}
+ \left.\frac{1}{2}\right[ -\mathbb{R}^{\gamma}_{\beta}\otimes\mathbb{R}^{\alpha}_{\gamma} + \mathbb{R}^{\alpha}_{\gamma}\otimes
\mathbb{R}^{\gamma}_{\beta} + \\
&\ \left.+\mathbb{Q}^{\dag\alpha}_{c}\otimes\mathbb{U}^{-1}\mathbb{Q}^{c}_{\beta}+\mathbb{Q}^{c}_{\beta}\otimes
\mathbb{U}\mathbb{Q}^{\dag\alpha}_{c} - \frac{\delta^{\alpha}_{\beta}}{2}(\mathbb{Q}^{\dag\gamma}_{c}\otimes\mathbb{U}^{-1}\mathbb{Q}^{c}_{\gamma} +  \mathbb{Q}^{c}_{\gamma}\otimes\mathbb{U}\mathbb{Q}^{\dag\gamma}_{c})\right]\nonumber\\
\Delta(\hat{\mathbb{Q}}^{a}_{\alpha}) &= \hat{\mathbb{Q}}^{a}_{\alpha}\otimes\mathbbm{U} + \mathbbm{1}\otimes\hat{\mathbb{Q}}^{a}_{\alpha} +
\left.\frac{1}{2}\right[
-\mathbb{R}^{\gamma}_{\alpha}\otimes\mathbb{Q}^{a}_{\gamma} +  \mathbb{Q}^{a}_{\gamma}\otimes\mathbb{U}\mathbb{R}^{\gamma}_{\alpha}+\\
&\ -\mathbb{L}^{a}_{1;c}\otimes\mathbb{Q}^{c}_{\alpha} + \mathbb{Q}^{c}_{\alpha}\otimes\mathbb{U}\mathbb{L}^{a}_{c} -\frac{1}{2}\mathbb{H}_{1}\otimes\mathbb{Q}^{a}_{\alpha} + \frac{1}{2}\mathbb{Q}^{a}_{\alpha}\otimes\mathbb{U}\mathbb{H} + \nonumber\\
&\ +\left.\epsilon_{\alpha\gamma}\epsilon^{ac}\mathbb{C}\otimes\mathbb{U}^2\mathbb{Q}^{\dag\gamma}_{c}-\epsilon_{\alpha\gamma}\epsilon^{ac}
\mathbb{Q}^{\dag\gamma}_{c}\otimes\mathbb{U}^{-1}\mathbb{C}\right],\nonumber\\
\Delta(\hat{\mathbb{Q}}^{\dag\alpha}_{a}) &= \hat{\mathbb{Q}}^{\dag\alpha}_{a}\otimes\mathbb{U}^{-1} + \mathbbm{1}\otimes\hat{\mathbb{Q}}^{\dag\alpha}_{a} +
\left.\frac{1}{2}\right[ \mathbb{L}^{c}_{a}\otimes\mathbb{Q}^{\dag\alpha}_{a}-
\mathbb{Q}^{\dag\alpha}_{c}\otimes\mathbb{U}^{-1}\mathbb{L}^{c}_{a}+\\
&\ +\mathbb{R}^{\alpha}_{\gamma}\otimes\mathbb{Q}^{\dag\gamma}_{a} - \mathbb{Q}^{\dag\gamma}_{a}\otimes\mathbb{U}^{-1}\mathbb{R}^{\alpha}_{\gamma}+ \frac{1}{2}\mathbb{H}\otimes\mathbb{Q}^{\dag\alpha}_{a}-\frac{1}{2}\mathbb{Q}^{\dag\alpha}_{a}\otimes\mathbb{U}^{-1}\mathbb{H}+\nonumber\\
&\ -\epsilon_{ac}\epsilon^{\alpha\gamma}\mathbb{C}^{\dag}\otimes\mathbb{U}^{-2}\mathbb{Q}^{c}_{\gamma}+ \epsilon_{ac}\epsilon^{\alpha\gamma}\mathbb{Q}^{c}_{\gamma}\otimes\mathbb{U}\mathbb{C}^{\dag}\mbox{\Huge]}.\nonumber
\end{align}
and central charges
\begin{align}
\Delta(\hat{\mathbb{H}}) &= \hat{\mathbb{H}}\otimes\mathbbm{1} + \mathbbm{1}\otimes\hat{\mathbb{H}} + \mathbb{C}\otimes\mathbb{U}^{2}\mathbb{C}^{\dag} - \mathbb{C}^{\dag}\otimes\mathbb{U}^{-2}\mathbb{C},\\
\Delta(\hat{\mathbb{C}}) &= \hat{\mathbb{C}}\otimes\mathbb{U}^2 + \mathbbm{1}\otimes\hat{\mathbb{C}} +
\frac{1}{2}\left[ \mathbb{H}\otimes\mathbb{C} - \mathbb{C}\otimes\mathbb{U}^2\mathbb{H}\right],\\
\Delta(\hat{\mathbb{C}}^{\dag}) &=
\hat{\mathbb{C}}^{\dag}\otimes\mathbb{U}^{-2} + \mathbbm{1}\otimes\hat{\mathbb{C}}^{\dag} - \frac{1}{2}\left[
\mathbb{H}\otimes\mathbb{C}^{\dag} - \mathbb{C}^{\dag}\otimes\mathbb{U}^{-2}\mathbb{H}\right].
\end{align}
It is indeed readily seen that the above introduced coproduct
respects the commutator structure of the Yangian. Finally, to
complete the Hopf algebra structure, we give the antipode and the
counit
\begin{align}
&{\cal S}(\hat{\mathbb{J}}^A) = -\mathbb{U}^{-[[A]]}\hat{\mathbb{J}}^A,& & \epsilon(\hat{\mathbb{J}}^A) = 0.
\end{align}
The reason the antipode does not have the extra (structure
constant dependent) term as in (\ref{eqn;YangAntipode}) is because
of the vanishing of the Killing form.


In order for a evaluation type representation (\ref{eqn;EvalRep})
to be quasi-cocommutative we need to have that $\Delta
\hat{\mathbb{C}} = \Delta^{op} \hat{\mathbb{C}}$ because this element is central. Recall that the
central charge is related to the braiding factor via
(\ref{eqn;CviaBraiding}). From this we derive
\begin{align}
0&=\Delta\hat{\mathbb{C}}-\Delta^{op}\hat{\mathbb{C}} \nonumber\\
&=\frac{ig}{2}
\left[u_1(\mathbb{U}^2-1)-\frac{\mathbb{H}(\mathbb{U}^2+1)}{2}\right]\otimes(\mathbb{U}^2-1)\nonumber\\
&\qquad\qquad + (\mathbb{U}^2-1)\otimes\frac{ig}{2}
\left[u_2(\mathbb{U}^2-1)-\frac{\mathbb{H}(\mathbb{U}^2+1)}{2}\right].
\end{align}
This implies
\begin{align}\label{eqn;defU}
u &= \frac{\mathbb{H}}{2} \frac{\mathbb{U}^2+1}{\mathbb{U}^2-1}.
\end{align}
Notice that the non-triviality of the braiding factor
$\mathbb{U}\neq\mathbbm{1}$ is important here. Strictly speaking
one can also add a representation independent constant to $u$.
However, it will be clear from the explicit derivations in later
chapters that this constant does not enter the discussion and for
that reason we set it to 0. In our explicit parameterization in
terms of $x^{\pm}$ (\ref{eqn;ParamtersPandG}) the evaluation
parameter $u$ becomes
\begin{align}\label{eqn;defUparameter}
u = \frac{g}{4i}(x^+ + x^- )\left(1+\frac{1}{x^+ x^-}\right).
\end{align}
This feature is rather unusual since in most models the evaluation
parameter $u$ is unrelated to the representation of the Lie
algebra.

In each of the representations of $\alg{su}(2|2)$ discussed above
one can consider the evaluation representation of the Yangian.
However, explicitly checking the Serre identities proves to be
difficult and becomes more transparent in Drinfeld's second
realization.

\subsection{Second realization}

Let us continue with the discussion of Drinfeld's second
realization \cite{Dsecond}. As already clear from the previous
discussion, also in this case the peculiar features of $\alg{h}$
make the analysis more complicated than for standard Lie
superalgebras.

Let us first indicate the Chevalley-Serre generators
\begin{align}
&\alg{E}^+_1 = \mathbb{Q}^{\dag4}_2, & &\alg{E}^-_1 = \mathbb{Q}^{2}_4, & &\alg{H}_1 = -\mathbb{L}^1_1-\mathbb{R}^3_3+\frac{1}{2}\mathbb{H},\\
&\alg{E}^+_2 = i\mathbb{Q}^{1}_4, & &\alg{E}^-_2 = i\mathbb{Q}^{\dag4}_1, & &\alg{H}_2 = -\mathbb{L}^1_1+\mathbb{R}^3_3-\frac{1}{2}\mathbb{H},\\
&\alg{E}^+_3 = i\mathbb{Q}^{2}_3, & &\alg{E}^-_3 =
i\mathbb{Q}^{\dag3}_2, & &\alg{H}_3 =
\mathbb{L}^1_1-\mathbb{R}^3_3-\frac{1}{2}\mathbb{H}.
\end{align}
They satisfy the defining relations
\begin{align}
&[\mathfrak{H}_i,\mathfrak{H}_j] = 0, &&
[\mathfrak{H}_i,\mathfrak{E}^{\pm}_j] = \pm a_{ij}
\alg{E}^{\pm}_j, && \{\alg{E}^+_i,\alg{E}^-_j\} =
\delta_{ij}\alg{H}_i,
\end{align}
and the Serre relations
\begin{align}\label{eqn;SerreDinfeldII}
&\mathrm{ad}(\alg{E}^{\pm}_{1})^2(\alg{E}^{\pm}_{2}) =
\mathrm{ad}(\alg{E}^{\pm}_{2})^2(\alg{E}^{\pm}_{1}) = 0, &&
\{\alg{E}^{\pm}_2,\alg{E}^{\pm}_3\} = \mathrm{central},
\end{align}
with Cartan matrix
\begin{align}
a_{ij} = \begin{pmatrix}
0 & 1 & -1\\
1 & 0 & 0\\
-1 & 0 & 0
\end{pmatrix}.
\end{align}
Notice that the Cartan matrix is degenerate.

Drinfeld's second realization \cite{Spill:2008tp} is now expressed
in terms of Cartan generators $\kappa_{i,m}$ and fermionic simple
roots $\xi^\pm_{i,m}$, $i=1,2,3$, $m=0,1,2,\dots$, subject to the
following relations:
\begin{align}
\label{relazionizero}
&[\kappa_{i,m},\kappa_{j,n}]=0,\quad [\kappa_{i,0},\xi^+_{j,m}]=a_{ij} \,\xi^+_{j,m},\nonumber\\
&[\kappa_{i,0},\xi^-_{j,m}]=- a_{ij} \,\xi^-_{j,m},\quad \{\xi^+_{i,m},\xi^-_{j,n}\}=\delta_{i,j}\, \kappa_{j,n+m},\nonumber\\
&[\kappa_{i,m+1},\xi^+_{j,n}]-[\kappa_{i,m},\xi^+_{j,n+1}] = \frac{1}{2} a_{ij} \{\kappa_{i,m},\xi^+_{j,n}\},\nonumber\\
&[\kappa_{i,m+1},\xi^-_{j,n}]-[\kappa_{i,m},\xi^-_{j,n+1}] = - \frac{1}{2} a_{ij} \{\kappa_{i,m},\xi^-_{j,n}\},\nonumber\\
&\{\xi^+_{i,m+1},\xi^+_{j,n}\}-\{\xi^+_{i,m},\xi^+_{j,n+1}\} = \frac{1}{2} a_{ij} [\xi^+_{i,m},\xi^+_{j,n}],\nonumber\\
&\{\xi^-_{i,m+1},\xi^-_{j,n}\}-\{\xi^-_{i,m},\xi^-_{j,n+1}\} = - \frac{1}{2} a_{ij} [\xi^-_{i,m},\xi^-_{j,n}],
\end{align}
\begin{eqnarray}
&&i\neq j, \, \, \, \, \, n_{ij}=1+|a_{ij}|,\, \, \, \, \, Sym_{\{k\}} [\xi^+_{i,k_1},[\xi^+_{i,k_2},\dots \{\xi^+_{i,k_{n_{ij}}}, \xi^+_{j,l}\}\dots\}\}=0,\nonumber\\
&&i\neq j, \, \, \, \, \, n_{ij}=1+|a_{ij}|,\, \, \, \, \, Sym_{\{k\}} [\xi^-_{i,k_1},[\xi^-_{i,k_2},\dots \{\xi^-_{i,k_{n_{ij}}}, \xi^-_{j,l}\}\dots\}\}=0,\nonumber\\
&&\text{except for} \, \, \, \, \, \, \, \, \,
\{\xi^+_{2,n},\xi^+_{3,m}\} = \fC_{n+m}, \qquad
\{\xi^-_{2,n},\xi^-_{3,m}\} = \fC^\dag_{n+m}.
\end{eqnarray}
In the last relations $\fC_{n+m}$ and $\fC^{\dag}_{n+m}$ are
central elements. The exact relation in terms of Drinfeld I
operators will be discussed below. The last equation differs from
the standard relations due to the central charges and is
reminiscent of last Serre relation in the Chevalley-Serre basis of
the underlying algebra (\ref{eqn;SerreDinfeldII}). We call the
index $m$ of the generators in this realization the level.

We now construct the isomorphism (Drinfeld's map) \cite{Dsecond}
between the first and the second realization as follows
\begin{align}
\label{mappa1}
&\kappa_{i,0}=\alg{H}_i,\quad \xi^+_{i,0}=\alg{E}^+_i,\quad \xi^-_{i,0}=\alg{E}^-_i,\nonumber\\
&\kappa_{i,1}=\hat{\alg{H}}_i-v_i,\quad
\xi^+_{i,1}=\hat{\alg{E}}^+_i-w_i,\quad
\xi^-_{i,1}=\hat{\alg{E}}^-_i-z_i,
\end{align}
the special elements are given by
\begin{align}
v_1 &= - \frac{1}{2} \kappa_{1,0}^2 + \frac{1}{4} \mathbb{R}^4_3 \mathbb{R}^3_4 +  \frac{1}{4} \mathbb{R}^3_4 \mathbb{R}^4_3 +
\frac{3}{4} \mathbb{L}^2_1 \mathbb{L}^1_2 - \frac{1}{4} \mathbb{L}^1_2 \mathbb{L}^2_1 +  \nonumber\\
&\ \ \ - \frac{1}{4} \mathbb{Q}^2_3 \mathbb{Q}^{\dag3}_2 -
\frac{1}{4} \mathbb{Q}^1_4 \mathbb{Q}^{\dag4}_1  - \frac{3}{4}
\mathbb{Q}^{\dag4}_1 \mathbb{Q}^1_4
+ \frac{1}{4} \mathbb{Q}^{\dag3}_2 \mathbb{Q}^2_3 + \half \mathbb{C} \mathbb{C}^\dag ,\nonumber\\
v_2 &= - \frac{1}{2}\left[ \kappa_{2,0}^2 +2 \mathbb{R}^4_3 \fR^3_4 - \fR^3_4 \fR^4_3 - \fL^1_2 \fL^2_1 -2 \fQ^1_3 \fQ^{\dag3}_1 - \fQ^{\dag3}_1 \fQ^1_3  + \fQ^{\dag4}_2 \fQ^2_4 + \fC \fC^\dag\right] ,\nonumber\\
v_3 &= - \frac{1}{2}\left[ \kappa_{3,0}^2 - \fR^4_3 \fR^3_4  + \fL^2_1 \fL^1_2 - \fQ^{\dag3}_1 \fQ^1_3  - \fQ^{\dag4}_2 \fQ^2_4 + \fC \fC^\dag\right] ,\nonumber \\
w_1 &= - \frac{1}{4}\left[\xi^+_{1,0} \kappa_{1,0} + \kappa_{1,0} \xi^+_{1,0} - 3\fQ^{\dag4}_1 \fL^1_2 + \fL^1_2 \fQ^{\dag4}_1 - \fQ^{\dag3}_2 \fR^4_3 - \fR^4_3 \fQ^{\dag3}_2 - 2\fQ^1_3 \fC^\dag\right],\nonumber\\
w_2 &= \frac{i}{4}\left[i\xi^+_{2,0} \kappa_{2,0} + i\kappa_{2,0} \xi^+_{2,0} + 3\fQ^1_3 \fR^3_4 - \fL^1_2 \fQ^2_4 - \fQ^2_4 \fL^1_2 - \fR^3_4 \fQ^1_3 -2 \fQ^{\dag3}_2 \fC\right],\nonumber\\
w_3 &= \frac{i}{4}\left[i\xi^+_{3,0} \kappa_{3,0} + i\kappa_{3,0} \xi^+_{3,0} - \fQ^1_3 \fL^2_1 + 3 \fL^2_1 \fQ^1_3 - \fQ^2_4 \fR^4_3 - \fR^4_3 \fQ^2_4 -2 \fQ^{\dag4}_1 \fC\right],\nonumber\\
z_1 &= -\frac{1}{4}\left[\xi^-_{1,0} \kappa_{1,0} + \kappa_{1,0} \xi^-_{1,0} + \fQ^1_4 \fL^2_1 -3 \fL^2_1 \fQ^1_4 - \fQ^2_3 \fR^3_4 - \fR^3_4 \fQ^2_3 -2 \fQ^{\dag3}_1 \fC\right],\nonumber\\
z_2 &= \frac{i}{4}\left[i\xi^-_{2,0} \kappa_{2,0} + i\kappa_{2,0} \xi^-_{2,0} - \fQ^{\dag3}_1 \fR^4_3 + 3 \fR^4_3 \fQ^{\dag3}_1 - \fQ^{\dag4}_2 \fL^2_1 - \fL^2_1 \fQ^{\dag4}_2 -2 \fQ^2_3 \fC^\dag\right],\nonumber\\
z_3 &= \frac{i}{4}\left[i\xi^-_{3,0} \kappa_{3,0} +
i\kappa_{3,0}\xi^-_{3,0} - \fQ^{\dag4}_2 \fR^3_4 - \fR^3_4
\mathbb{Q}^{\dag4}_2 +3 \mathbb{Q}^{\dag3}_1 \fL^1_2 - \fL^1_2
\mathbb{Q}^{\dag3}_1 -2 \fQ^1_4 \fC^\dag\right].\nonumber
\end{align}
One can check that the above identifications indeed define the
second realization of the Yangian (\ref{relazionizero}). We can
now also make the relation between $\hat{\fC},\hat{\fC}^{\dag}$ and
$\fC_{1},\fC_{1}^{\dag}$ precise. After explicitly computing the
corresponding anti-commutator of supercharges we find
\begin{align}
\fC_1 = \hat{\fC} -\frac{\mathbb{H}+1}{2}\fC.
\end{align}
Or, in the evaluation representation (of the first realization)
\begin{align}
\fC_1 = \left(u -\frac{\mathbb{H}+1}{2}\right)\fC.
\end{align}
Alternatively, if we assume an evaluation representation for the
second realization with parameters $\omega_i$
\begin{align}
&\kappa_{i,n} = \omega_i^n \kappa_{i,0}, & &\xi^+_{i,n} =
\omega_i^n \xi^+_{i,0}, & &\xi^-_{i,n} = \omega_i^n \xi^-_{i,0},
\end{align}
then one has to explicitly work through the defining relations and
see whether this ansatz solves it. For example, this works for the
fundamental representation where one finds
\begin{align}
&\omega_1  = u, &&  \omega_2 = \omega_3 = u-\frac{1}{2}\mathbb{H}.
\end{align}
The fact that the parameters $\omega_{2,3}$ are shifted has to do
with the modified Serre relations.


\section{Long Representations}\label{sec;LongReps}

The representations that describe physical particles for the
$\ads$ superstring are the short (atypical) symmetric
representations. However, there is a also a big class of
$\alg{su}(2|2)$ representations that is more general. These are
the typical (or long) representations. A very convenient way to
construct such representations is to make use of the $\alg{sl}(2)$
outer automorphism (\ref{eqn;OuterAut}) to construct them out of
long representations of $\alg{gl}(2|2)$ \cite{Zhang:2004qx}.

We can identify the values of the labels which will produce the
representations we are particularly interested in.
First of all, the fundamental $4$-dimensional short representation
\cite{Beisert:2005tm} corresponds to $j_1=\frac{1}{2},j_2=0$ (or,
equivalently, $j_1=0,j_2=\frac{1}{2}$) and $q=\frac{1}{2}$
($q=-\frac{1}{2}$). More generally, the bound state (symmetric
short) representations
\cite{Dorey:2006dq,Chen:2006gp,Chen:2006gq,Roiban:2006gs,Beisert:2006qh,Arutyunov:2008zt}
are given by $j_2=0,q=j_1$, with $j_1 =\frac{1}{2},1,...$ and
bound state number $\ell \equiv s = 2 j_1$. In addition, there are
the antisymmetric short representations given by $j_1=0,q=1+j_2$,
with $j_2 =\frac{1}{2},1,...$ and the bound state number $M\equiv
a =2j_2$. Both symmetric and antisymmetric representations have
dimension $4\ell$. We see that symmetric and antisymmetric
representations are associated with the different shortening
conditions $\pm q=j_1-j_2$ and $\pm q=1+j_1+j_2$.

Second, we consider the simplest long representation of dimension
16. In terms of the $\alg{gl} (2|2)$ labels introduced above, this
is the $16$-dimensional long representation characterized by
$j_1=j_2=0$, and arbitrary $q$. It is instructive to see how it
branches under the $\su (2) \oplus \su (2)$ algebra. We denote as
$[l_1,l_2]$ the subset of states which furnish a representation of
$\su (2) \oplus \su (2)$ with angular momentum $l_1$ w.r.t the
first $\su (2)$, and $l_2$ w.r.t the second $\su (2)$,
respectively. The branching rule is
\begin{align} \label{split}
(2,2) \, \rightarrow \, 2 \times [0,0] \oplus 2 \times
[\frac{1}{2},\frac{1}{2}] \oplus [1,0] \oplus [0,1].
\end{align}
One can straightforwardly verify that the total dimension adds up
to $16$, since $[l_1,l_2]$ has dimension $(2 l_1 + 1)(2 l_2 + 1)$.

We have explicitly constructed the oscillator representation by
using the formulas of \cite{Zhang:2004qx}, and derived from it the
$16 \times 16$ matrix realization of the algebra generators. We
have done this before acting with the outer automorphism, in such
a way that the subsequent $\sls (2)$ rotation provides an explicit
matrix representation of centrally-extended $\su (2|2)$. We report
this explicit realization in appendix B. In particular, from the
explicit matrix realization one obtains the following values of
the central charges:
\begin{align}
\fHH = 2 q \, (a d + b c) \, \mathbbm{1}, \qquad \fC = 2 q \, a b
\, \mathbbm{1}, \qquad \fC^\dagger = 2 q \, c d \, \mathbbmss{1},
\end{align}
($\mathbbm{1}$ is the $16$-dimensional identity matrix),
satisfying the condition
\begin{align} \label{conditio}
\frac{\fHH^2}{4} - \fC \fC^\dagger = q^2 \,
\mathbbmss{1}.
\end{align}
When $q^2 =1$, this becomes a shortening condition. In fact, for
$q =1$ the $16$-dimensional representation becomes reducible but
indecomposable. Formula (\ref{conditio}) above,
however, tells us that we can conveniently think of $q$ as a {\it
generalized} bound state number, since for short representations
$2 q$ would be replaced by the bound state number $\ell$ in the
analogous formula for the central charges. This is particularly
useful, since it allows us to parameterize the labels $a,b,c,d$ in
terms of the familiar bound state variables $x^\pm$, just
replacing the bound state number $\ell$ by $2 q$. The explicit
parameterization is given by
\begin{align}
\label{param}
a &= \sqrt{\frac{g}{4q}}\eta,  &b &=-\sqrt{\frac{g}{4q}}\frac{i}{\eta}\left(1-\frac{x^+}{x^-}\right) ,\nonumber\\
c &=-\sqrt{\frac{g}{4q}} \frac{\eta}{x^+}  , &d&= \sqrt{\frac{g}{4q}}\frac{x^+}{i\eta}\left(1-\frac{x^-}{x^+}\right),
\end{align}
where
\begin{align}
\label{eta} \eta = e^{\frac{ip}{4}}\sqrt{i(x^- - x^+)}
\end{align}
and
\begin{align}
x^+ + \frac{1}{x^+} - x^- - \frac{1}{x^-} \, = \, \frac{4 i q}{g}.
\end{align}
As in the case of short representations, there exist a
uniformizing torus with variable $z$ and periods depending on $q$
\cite{Janik:2006dc} . The choice (\ref{eta}) for $\eta$ is
historically preferred in the string theory analysis
\cite{Arutyunov:2006yd,Arutyunov:2007tc,Arutyunov:2008zt,Arutyunov:2009mi},
and will again ensure a symmetric S-matrix.

\addcontentsline{toc}{section}{Appendix A: Exceptional Lie
algebra}

\section*{Appendix A: Exceptional Lie algebra}

$\alg{h}$ can be obtained via a limiting procedure from the
exceptional Lie algebra $D(2,1;\varepsilon)$
\cite{Beisert:2005tm,Matsumoto:2008ww}. The advantage is that
$D(2,1;\varepsilon)$ has a non-degenerate Killing form which
allows for a standard derivation of the Yangian coproducts. In
this appendix we will give the definitions of
$D(2,1;\varepsilon)$, compute its inner product and use it to
derive the coproduct of the first Yangian generators. For more
details on this exceptional algebra we refer to
\cite{Heckenberger:2007ry,VanDerJeugt:1985hq}.

The algebra $D(2,1;\varepsilon)$ consists of three copies of
$\alg{su}(2)$
\begin{align}\label{eqn;CommRelD1}
&[\mathbb{L}^a_b,\mathbb{L}^c_d] = \delta^{c}_{b}\mathbb{L}^a_d - \delta^a_d\mathbb{L}^c_b, &
&[\mathbb{R}^\alpha_\beta,\mathbb{R}^\gamma_\delta] = \delta^{\gamma}_{\beta}\mathbb{R}^\alpha_\delta - \delta^\alpha_\delta\mathbb{R}^\gamma_\beta, \nonumber\\
&[\mathbb{C}^\alg{a}_\alg{b},\mathbb{C}^\alg{c}_\alg{d}] = \delta^{\alg{c}}_{\alg{b}}\mathbb{C}^\alg{a}_\alg{d} - \delta^\alg{a}_\alg{d}\mathbb{C}^\alg{c}_\alg{b},
\end{align}
and eight supersymmetry generators $\mathbb{F}^{a\alpha\alg{a}}$
that transform in the fundamental representation of each
$\alg{su}(2)$
\begin{align}\label{eqn;CommRelD2}
&[\mathbb{L}^a_b,\mathbb{F}^{c\gamma\alg{c}}] = \delta^c_b \mathbb{F}^{a\gamma\alg{c}}-\frac{1}{2}\delta^a_b\mathbb{F}^{c\gamma\alg{c}}, &
&[\mathbb{R}^\alpha_\beta,\mathbb{F}^{c\gamma\alg{c}}] = \delta^\gamma_\beta \mathbb{F}^{c\alpha\alg{c}}-\frac{1}{2}\delta^\alpha_\beta\mathbb{F}^{c\gamma\alg{c}}, \nonumber\\
&[\mathbb{C}^\alg{a}_\alg{b},\mathbb{F}^{c\gamma\alg{c}}] = \delta^\alg{c}_\alg{b} \mathbb{F}^{c\gamma\alg{a}}-\frac{1}{2}\delta^\alg{a}_\alg{b}\mathbb{F}^{c\gamma\alg{c}}.
\end{align}
Finally the anti-commutator between the fermionic generators is
\begin{align}\label{eqn;CommRelD3}
\{\mathbb{F}^{a\alpha\alg{a}},\mathbb{F}^{b\beta\alg{b}}\} = \sigma_1\epsilon^{ak}\epsilon^{\alpha\beta}\epsilon^{\alg{a}\alg{b}}\mathbb{L}^{a}_{k}+
\sigma_2\epsilon^{\beta\kappa}\epsilon^{ab}\epsilon^{\alg{a}\alg{b}}\mathbb{R}^{\alpha}_{\kappa}+
\sigma_3\epsilon^{ab}\epsilon^{\alpha\beta}\epsilon^{\alg{a}\alg{k}}\mathbb{C}^{\alg{a}}_{\alg{k}},
\end{align}
with $\sigma_1+\sigma_2+\sigma_3 = 0$. The algebra is invariant
under rescaling of the supersymmetry generators and hence the only
independent parameter in the algebra is $\varepsilon =
-\sigma_3/\sigma_1$. We make the dependence on $\varepsilon$
explicit by setting
\begin{align}
&\sigma_1 = -1, && \sigma_2 = 1-\varepsilon, && \sigma_3 = \varepsilon.
\end{align}
To obtain $\alg{h}$ one needs to identify
\begin{align}
&(\mathbb{C})^{\alg{a}}_{\alg{b}} =
\frac{1}{\varepsilon}
\begin{pmatrix}
\frac{\mathbb{H}}{2} & \mathbb{C}^{\dag}\\
-\mathbb{C} & -\frac{\mathbb{H}}{2}
\end{pmatrix},
& & (\mathbb{F}^{a\alpha})^{\alg{a}} =
\begin{pmatrix}
\epsilon^{ak} \mathbb{Q}^{\dag\alpha}_k \\
\epsilon^{\alpha\kappa}\mathbb{Q}^a_{\kappa}
\end{pmatrix}.
\end{align}
The above identifications have to be understood in the sense that
the $D(2,1;\varepsilon)$ generators reduce to the $\alg{su}(2|2)$
ones in the limit $\varepsilon\rightarrow0$, e.g. $\mathbb{C}^1_1
= \frac{\mathbb{H}}{2\varepsilon} + \mathcal{O}(1)$. It is now
easily seen that in the limit $\varepsilon\rightarrow0$ the
commutation relations
(\ref{eqn;CommRelD1}),(\ref{eqn;CommRelD2}),(\ref{eqn;CommRelD3})
reduce to (\ref{eqn;DefRelsSU22}), e.g. we see that
\begin{align}
\{\mathbb{Q}^{\dag3}_2,\mathbb{Q}^2_3\}=
-\{\mathbb{F}^{111},\mathbb{F}^{222}\} = -\mathbb{L}^1_1 +
(1-\varepsilon)\mathbb{R}^4_4 + \frac{1}{2}\mathbb{H}.
\end{align}
It is also readily checked that the elements
$\mathbb{C}^{\alg{a}}_{\alg{b}}$ become indeed central.

Normally in superalgebras one lowers indices by making use of the
Killing form $K^{AB}$. The Killing form for superalgebras is
defined as
\begin{align}
K^{AB} = \mathrm{str}(\mathrm{ad}(\mathbb{J}^{A})\mathrm{ad}(\mathbb{J}^{B})).
\end{align}
Computing this from the commutation relations is straightforward
and we find
\begin{align}
K^{AB} = (-1)^{d(D)}f^{AC}_Df^{BD}_C = 0,
\end{align}
where $d(A)=0,1$ for bosonic and fermionic generators
respectively. Nevertheless, $D(2,1;\varepsilon)$ admits an
invariant, non-degenerate inner product. An inner product
$\tilde{K}(\mathbb{J}^A,\mathbb{J}^B) \equiv \tilde{K}^{AB}$ on a
Lie superalgebra needs to satisfy \cite{Frappat:1996pb}
\begin{align}
\tilde{K}(\mathbb{J}^A,\mathbb{J}^B) &= 0\ \mathrm{if}\ d(A)\neq d(B)\\
\tilde{K}(\mathbb{J}^A,\mathbb{J}^B) &= (-1)^{d(A)d(B)}\tilde{K}(\mathbb{J}^B,\mathbb{J}^A)\\
\tilde{K}(\mathbb{J}^A,[\mathbb{J}^B,\mathbb{J}^C\}) &=
\tilde{K}([\mathbb{J}^A,\mathbb{J}^B\},\mathbb{J}^C).
\end{align}
In terms of components and structure constants, the last equation
becomes $\tilde{K}^{AD}f^{BC}_D = \tilde{K}^{DC}f^{AB}_D$ for all
$A,B,C$. Solving it gives an unique solution (up to an overall
scalar) for $\tilde{K}^{AB}$. Let us enumerate the algebra
generators as
\begin{align}
&J[1]= \mathbb{L}^1_1 & &J[2]= \mathbb{L}^2_1 & &J[3]= \mathbb{L}^1_2 \\
&J[4]= \mathbb{R}^1_1 & &J[5]= \mathbb{R}^2_1 & &J[6]= \mathbb{R}^1_2 \\
&J[7]= \mathbb{C}^1_1 & &J[8]= \mathbb{C}^2_1 & &J[9]= \mathbb{C}^1_2
\end{align}
and supersymmetry generators
\begin{align}
&J[10]= \mathbb{F}^{111} & & J[11]= \mathbb{F}^{112} & & J[12]= \mathbb{F}^{121} & & J[13]= \mathbb{F}^{211} \\
&J[14]= \mathbb{F}^{122} & & J[15]= \mathbb{F}^{212} & & J[16]= \mathbb{F}^{221} & & J[17]= \mathbb{F}^{222}.
\end{align}
Then the elements of the inner product can be conveniently encoded
in a matrix and are given by
\begin{align}
\tilde{K}^{AB}=\left(\begin{smallmatrix}
 \frac{1}{\sigma_1} & 0 & 0 & 0 & 0 & 0 & 0 & 0 & 0 & 0 & 0 & 0 & 0 & 0 & 0 & 0 & 0 \\
 0 & 0 & \frac{2}{\sigma_1} & 0 & 0 & 0 & 0 & 0 & 0 & 0 & 0 & 0 & 0 & 0 & 0 & 0 & 0 \\
 0 & \frac{2}{\sigma_1} & 0 & 0 & 0 & 0 & 0 & 0 & 0 & 0 & 0 & 0 & 0 & 0 & 0 & 0 & 0 \\
 0 & 0 & 0 & \frac{1}{\sigma_1} & 0 & 0 & 0 & 0 & 0 & 0 & 0 & 0 & 0 & 0 & 0 & 0 & 0 \\
 0 & 0 & 0 & 0 & 0 & \frac{2}{\sigma_1} & 0 & 0 & 0 & 0 & 0 & 0 & 0 & 0 & 0 & 0 & 0 \\
 0 & 0 & 0 & 0 & \frac{2}{\sigma_2} & 0 & 0 & 0 & 0 & 0 & 0 & 0 & 0 & 0 & 0 & 0 & 0 \\
 0 & 0 & 0 & 0 & 0 & 0 & \frac{1}{\sigma_3} & 0 & 0 & 0 & 0 & 0 & 0 & 0 & 0 & 0 & 0 \\
 0 & 0 & 0 & 0 & 0 & 0 & 0 & 0 & \frac{2}{\sigma_3} & 0 & 0 & 0 & 0 & 0 & 0 & 0 & 0 \\
 0 & 0 & 0 & 0 & 0 & 0 & 0 & \frac{2}{\sigma_3} & 0 & 0 & 0 & 0 & 0 & 0 & 0 & 0 & 0 \\
 0 & 0 & 0 & 0 & 0 & 0 & 0 & 0 & 0 & 0 & 0 & 0 & 0 & 0 & 0 & 0 & -2 \\
 0 & 0 & 0 & 0 & 0 & 0 & 0 & 0 & 0 & 0 & 0 & 0 & 0 & 0 & 0 & 2 & 0 \\
 0 & 0 & 0 & 0 & 0 & 0 & 0 & 0 & 0 & 0 & 0 & 0 & 0 & 0 & 2 & 0 & 0 \\
 0 & 0 & 0 & 0 & 0 & 0 & 0 & 0 & 0 & 0 & 0 & 0 & 0 & 2 & 0 & 0 & 0 \\
 0 & 0 & 0 & 0 & 0 & 0 & 0 & 0 & 0 & 0 & 0 & 0 & -2 & 0 & 0 & 0 & 0 \\
 0 & 0 & 0 & 0 & 0 & 0 & 0 & 0 & 0 & 0 & 0 & -2 & 0 & 0 & 0 & 0 & 0 \\
 0 & 0 & 0 & 0 & 0 & 0 & 0 & 0 & 0 & 0 & -2 & 0 & 0 & 0 & 0 & 0 & 0 \\
 0 & 0 & 0 & 0 & 0 & 0 & 0 & 0 & 0 & 2 & 0 & 0 & 0 & 0 & 0 & 0 & 0
\end{smallmatrix}\right).
\end{align}
In order to compute the coproducts of the Yangian generators we
need to lower indices. This is done by the contravariant form
$\tilde{K}_{AB}$. In matrix form it becomes
\begin{align}
\tilde{K}_{AB}=\left(\begin{smallmatrix}
 \sigma_1  & 0 & 0 & 0 & 0 & 0 & 0 & 0 & 0 & 0 & 0 & 0 & 0 & 0 & 0 & 0 & 0 \\
 0 & 0 & \frac{\sigma_1}{2} & 0 & 0 & 0 & 0 & 0 & 0 & 0 & 0 & 0 & 0 & 0 & 0 & 0 & 0 \\
 0 & \frac{\sigma_1}{2} & 0 & 0 & 0 & 0 & 0 & 0 & 0 & 0 & 0 & 0 & 0 & 0 & 0 & 0 & 0 \\
 0 & 0 & 0 & \sigma_2  & 0 & 0 & 0 & 0 & 0 & 0 & 0 & 0 & 0 & 0 & 0 & 0 & 0 \\
 0 & 0 & 0 & 0 & 0 & \frac{\sigma_2}{2} & 0 & 0 & 0 & 0 & 0 & 0 & 0 & 0 & 0 & 0 & 0 \\
 0 & 0 & 0 & 0 & \frac{\sigma_2}{2} & 0 & 0 & 0 & 0 & 0 & 0 & 0 & 0 & 0 & 0 & 0 & 0 \\
 0 & 0 & 0 & 0 & 0 & 0 & \sigma_3 & 0 & 0 & 0 & 0 & 0 & 0 & 0 & 0 & 0 & 0 \\
 0 & 0 & 0 & 0 & 0 & 0 & 0 & 0 & \frac{\sigma_3}{2} & 0 & 0 & 0 & 0 & 0 & 0 & 0 & 0 \\
 0 & 0 & 0 & 0 & 0 & 0 & 0 & \frac{\sigma_3}{2} & 0 & 0 & 0 & 0 & 0 & 0 & 0 & 0 & 0 \\
 0 & 0 & 0 & 0 & 0 & 0 & 0 & 0 & 0 & 0 & 0 & 0 & 0 & 0 & 0 & 0 & \frac{1}{2} \\
 0 & 0 & 0 & 0 & 0 & 0 & 0 & 0 & 0 & 0 & 0 & 0 & 0 & 0 & 0 & -\frac{1}{2} & 0 \\
 0 & 0 & 0 & 0 & 0 & 0 & 0 & 0 & 0 & 0 & 0 & 0 & 0 & 0 & -\frac{1}{2} & 0 & 0 \\
 0 & 0 & 0 & 0 & 0 & 0 & 0 & 0 & 0 & 0 & 0 & 0 & 0 & -\frac{1}{2} & 0 & 0 & 0 \\
 0 & 0 & 0 & 0 & 0 & 0 & 0 & 0 & 0 & 0 & 0 & 0 & \frac{1}{2} & 0 & 0 & 0 & 0 \\
 0 & 0 & 0 & 0 & 0 & 0 & 0 & 0 & 0 & 0 & 0 & \frac{1}{2} & 0 & 0 & 0 & 0 & 0 \\
 0 & 0 & 0 & 0 & 0 & 0 & 0 & 0 & 0 & 0 & \frac{1}{2} & 0 & 0 & 0 & 0 & 0 & 0 \\
 0 & 0 & 0 & 0 & 0 & 0 & 0 & 0 & 0 & -\frac{1}{2} & 0 & 0 & 0 & 0 & 0 & 0 & 0
\end{smallmatrix}\right).
\end{align}
This form defines the two-site Casimir
\begin{align}
T_{12} = \tilde{K}_{AB}\mathbb{J}^A\otimes\mathbb{J}^B.
\end{align}
The operator $T_{12}$ can for example be used to compute the
classical $r$-matrix of this algebra. Computing the coproduct of
the Yangian generator is now straightforward from
(\ref{eqn;StandardYangHopf}) by $f_{BC}^A =
f^{AB}_{C}\tilde{K}_{DB}$ and gives (after including braiding
factors)
\begin{align}
\Delta(\hat{\mathbb{L}}^{a}_{b}) &=
\hat{\mathbb{L}}^{a}_{b}\otimes\mathbbm{1} +
\mathbbm{1}\otimes\hat{\mathbb{L}}^{a}_{b} +
\left.\frac{1}{2}\right[-(\mathbb{L}^{c}_{b}\otimes\mathbb{L}^{a}_{c}-
\mathbb{L}^{a}_{c}\otimes\mathbb{L}^{c}_{b})+\\
&\ \left.-
\mathbb{Q}^{\dag\gamma}_{b}\otimes\mathbb{U}^{-1}\mathbb{Q}^{a}_{\gamma}-\mathbb{Q}^{a}_{\gamma}\otimes\mathbb{U}\mathbb{Q}^{\dag
\gamma}_{b}+\frac{\delta^{a}_{b}}{2}(\mathbb{Q}^{\dag\gamma}_{c}\otimes\mathbb{U}^{-1}\mathbb{Q}^{c}_{\gamma}+
\mathbb{Q}^{c}_{\gamma}\otimes\mathbb{U}\mathbb{Q}^{\dag\gamma}_{c})~\right]\nonumber\\
\Delta(\hat{\mathbb{R}}^{\alpha}_{\beta}) &=
\hat{\mathbb{R}}^{\alpha}_{\beta}\otimes\mathbbm{1} + \mathbbm{1}\otimes\hat{\mathbb{R}}^{\alpha}_{\beta}
+ \left.\frac{1}{2}\right[ (1+\varepsilon)(\mathbb{R}^{\gamma}_{\beta}\otimes\mathbb{R}^{\alpha}_{\gamma} - \mathbb{R}^{\alpha}_{\gamma}\otimes
\mathbb{R}^{\gamma}_{\beta}) + \\
&\ \left.+\mathbb{Q}^{\dag\alpha}_{c}\otimes\mathbb{U}^{-1}\mathbb{Q}^{c}_{\beta}+\mathbb{Q}^{c}_{\beta}\otimes
\mathbb{U}\mathbb{Q}^{\dag\alpha}_{c} - \frac{\delta^{\alpha}_{\beta}}{2}(\mathbb{Q}^{\dag\gamma}_{c}\otimes\mathbb{U}^{-1}\mathbb{Q}^{c}_{\gamma} +  \mathbb{Q}^{c}_{\gamma}\otimes\mathbb{U}\mathbb{Q}^{\dag\gamma}_{c})\right]\nonumber\\
\Delta(\hat{\mathbb{Q}}^{a}_{\alpha}) &= \hat{\mathbb{Q}}^{a}_{\alpha}\otimes\mathbbm{U} + \mathbbm{1}\otimes\hat{\mathbb{Q}}^{a}_{\alpha} +
\left.\frac{1}{2}\right[
(1-\varepsilon)(\mathbb{Q}^{a}_{\gamma}\otimes\mathbb{U}\mathbb{R}^{\gamma}_{\alpha} - \mathbb{R}^{\gamma}_{\alpha}\otimes\mathbb{Q}^{a}_{\gamma} )+\\
&\ -\mathbb{L}^{a}_{1;c}\otimes\mathbb{Q}^{c}_{\alpha} + \mathbb{Q}^{c}_{\alpha}\otimes\mathbb{U}\mathbb{L}^{a}_{c} -\frac{1}{2}\mathbb{H}_{1}\otimes\mathbb{Q}^{a}_{\alpha} + \frac{1}{2}\mathbb{Q}^{a}_{\alpha}\otimes\mathbb{U}\mathbb{H} + \nonumber\\
&\ +\left.\epsilon_{\alpha\gamma}\epsilon^{ac}\mathbb{C}\otimes\mathbb{U}^2\mathbb{Q}^{\dag\gamma}_{c}-\epsilon_{\alpha\gamma}\epsilon^{ac}
\mathbb{Q}^{\dag\gamma}_{c}\otimes\mathbb{U}\mathbb{C}\right],\nonumber\\
\Delta(\hat{\mathbb{Q}}^{\dag\alpha}_{a}) &= \hat{\mathbb{Q}}^{\dag\alpha}_{a}\otimes\mathbb{U}^{-1} + \mathbbm{1}\otimes\hat{\mathbb{Q}}^{\dag\alpha}_{a} +
\left.\frac{1}{2}\right[ \mathbb{L}^{c}_{a}\otimes\mathbb{Q}^{\dag\alpha}_{a}-
\mathbb{Q}^{\dag\alpha}_{c}\otimes\mathbb{U}^{-1}\mathbb{L}^{c}_{a}+\\
&\ +(1-\varepsilon)(\mathbb{R}^{\alpha}_{\gamma}\otimes\mathbb{Q}^{\dag\gamma}_{a} - \mathbb{Q}^{\dag\gamma}_{a}\otimes\mathbb{U}^{-1}\mathbb{R}^{\alpha}_{\gamma}) + \frac{1}{2}\mathbb{H}\otimes\mathbb{Q}^{\dag\alpha}_{a}-\frac{1}{2}\mathbb{Q}^{\dag\alpha}_{a}\otimes\mathbb{U}^{-1}\mathbb{H}+\nonumber\\
&\ -\epsilon_{ac}\epsilon^{\alpha\gamma}\mathbb{C}^{\dag}\otimes\mathbb{U}^{-2}\mathbb{Q}^{c}_{\gamma}+ \epsilon_{ac}\epsilon^{\alpha\gamma}\mathbb{Q}^{c}_{\gamma}\otimes\mathbb{U}\mathbb{C}^{\dag}\mbox{\Huge]}.\nonumber
\end{align}
The coproducts of the central charges become
\begin{align}
\Delta(\hat{\mathbb{H}}) &= \hat{\mathbb{H}}\otimes\mathbbm{1} + \mathbbm{1}\otimes\hat{\mathbb{H}} + \mathbb{C}\otimes\mathbb{U}^{2}\mathbb{C}^{\dag} - \mathbb{C}^{\dag}\otimes\mathbb{U}^{-2}\mathbb{C}+ \\
&\ + \frac{\varepsilon}{2}(\mathbb{Q}^a_{\alpha}\otimes\mathbb{U}\mathbb{Q}^{\dag\alpha}_a +
\mathbb{Q}^{\dag\alpha}_a\otimes\mathbb{U}^{-1}\mathbb{Q}^a_{\alpha}),\nonumber\\
\Delta(\hat{\mathbb{C}}) &= \hat{\mathbb{C}}\otimes\mathbb{U}^2 + \mathbbm{1}\otimes\hat{\mathbb{C}} +
\frac{1}{2}[ \mathbb{H}\otimes\mathbb{C} - \mathbb{C}\otimes\mathbb{U}^2\mathbb{H} + \\
&\ + \frac{\varepsilon}{2}(\epsilon_{ab}\epsilon^{\alpha\beta}\mathbb{Q}^a_{\alpha}\otimes\mathbb{U}\mathbb{Q}^b_{\beta})],\nonumber\\
\Delta(\hat{\mathbb{C}}^{\dag}) &=
\hat{\mathbb{C}}^{\dag}\otimes\mathbb{U}^{-2} +
\mathbbm{1}\otimes\hat{\mathbb{C}}^{\dag} -\frac{1}{2}[
\mathbb{H}\otimes\mathbb{C}^{\dag} - \mathbb{C}^{\dag}\otimes\mathbb{U}^{-2}\mathbb{H} +\\
& \
+\frac{\varepsilon}{2}(\epsilon^{ab}\epsilon_{\alpha\beta}\mathbb{Q}_a^{\dag\alpha}\otimes\mathbb{U}^{-1}\mathbb{Q}_b^{\dag\beta})]\nonumber.
\end{align}
Upon taking the limit $\varepsilon\rightarrow 0$ one reproduces
the $\alg{su}(2|2)$ Yangian coproducts presented above.

\addcontentsline{toc}{section}{Appendix B: Long Representation}

\section*{Appendix B: Long Representation}

We list in this appendix the generators of centrally extended
$\alg{su}(2|2)$ in the long representation. We only report
explicitly the simple roots for a distinguished Dynkin diagram,
the remainder of the algebra being generated via commutation
relations. We present the roots in a unitary representation. To
achieve this, we perform a similarity transformation on the
generators constructed directly from the oscillator basis of
\cite{Zhang:2004qx}, in order to obtain hermitean matrices. First,
the bosonic $\alg{su}(2)\oplus \su (2)$ roots are given by
\begin{align}
\setcounter{MaxMatrixCols}{16} \mathbb{L}^1_2 =
\left(\begin{smallmatrix}
0 & 0 & 0 & 0 & 0 & 0 & 0 & 0 & 0 & 0 & 0 & 0 & 0 & 0 & 0 & 0 \\
 0 & 0 & 0 & 0 & 0 & 0 & 0 & 0 & 0 & 0 & 0 & 0 & 0 & 0 & 0 & 0 \\
 0 & 1 & 0 & 0 & 0 & 0 & 0 & 0 & 0 & 0 & 0 & 0 & 0 & 0 & 0 & 0 \\
 0 & 0 & 0 & 0 & 0 & 0 & 0 & 0 & 0 & 0 & 0 & 0 & 0 & 0 & 0 & 0 \\
 0 & 0 & 0 & 1 & 0 & 0 & 0 & 0 & 0 & 0 & 0 & 0 & 0 & 0 & 0 & 0 \\
 0 & 0 & 0 & 0 & 0 & 0 & 0 & 0 & 0 & 0 & 0 & 0 & 0 & 0 & 0 & 0 \\
 0 & 0 & 0 & 0 & 0 & \sqrt{2} & 0 & 0 & 0 & 0 & 0 & 0 & 0 & 0 & 0 & 0 \\
 0 & 0 & 0 & 0 & 0 & 0 & \sqrt{2} & 0 & 0 & 0 & 0 & 0 & 0 & 0 & 0 & 0 \\
 0 & 0 & 0 & 0 & 0 & 0 & 0 & 0 & 0 & 0 & 0 & 0 & 0 & 0 & 0 & 0 \\
 0 & 0 & 0 & 0 & 0 & 0 & 0 & 0 & 0 & 0 & 0 & 0 & 0 & 0 & 0 & 0 \\
 0 & 0 & 0 & 0 & 0 & 0 & 0 & 0 & 0 & 0 & 0 & 0 & 0 & 0 & 0 & 0 \\
 0 & 0 & 0 & 0 & 0 & 0 & 0 & 0 & 0 & 0 & 0 & 0 & 0 & 0 & 0 & 0 \\
 0 & 0 & 0 & 0 & 0 & 0 & 0 & 0 & 0 & 0 & 0 & 1 & 0 & 0 & 0 & 0 \\
 0 & 0 & 0 & 0 & 0 & 0 & 0 & 0 & 0 & 0 & 0 & 0 & 0 & 0 & 0 & 0 \\
 0 & 0 & 0 & 0 & 0 & 0 & 0 & 0 & 0 & 0 & 0 & 0 & 0 & 1 & 0 & 0 \\
 0 & 0 & 0 & 0 & 0 & 0 & 0 & 0 & 0 & 0 & 0 & 0 & 0 & 0 & 0 & 0
\end{smallmatrix}\right),\quad
\mathbb{L}^2_1 = \left(\begin{smallmatrix}
 0 & 0 & 0 & 0 & 0 & 0 & 0 & 0 & 0 & 0 & 0 & 0 & 0 & 0 & 0 & 0 \\
 0 & 0 & 1 & 0 & 0 & 0 & 0 & 0 & 0 & 0 & 0 & 0 & 0 & 0 & 0 & 0 \\
 0 & 0 & 0 & 0 & 0 & 0 & 0 & 0 & 0 & 0 & 0 & 0 & 0 & 0 & 0 & 0 \\
 0 & 0 & 0 & 0 & 1 & 0 & 0 & 0 & 0 & 0 & 0 & 0 & 0 & 0 & 0 & 0 \\
 0 & 0 & 0 & 0 & 0 & 0 & 0 & 0 & 0 & 0 & 0 & 0 & 0 & 0 & 0 & 0 \\
 0 & 0 & 0 & 0 & 0 & 0 & \sqrt{2} & 0 & 0 & 0 & 0 & 0 & 0 & 0 & 0 & 0 \\
 0 & 0 & 0 & 0 & 0 & 0 & 0 & \sqrt{2} & 0 & 0 & 0 & 0 & 0 & 0 & 0 & 0 \\
 0 & 0 & 0 & 0 & 0 & 0 & 0 & 0 & 0 & 0 & 0 & 0 & 0 & 0 & 0 & 0 \\
 0 & 0 & 0 & 0 & 0 & 0 & 0 & 0 & 0 & 0 & 0 & 0 & 0 & 0 & 0 & 0 \\
 0 & 0 & 0 & 0 & 0 & 0 & 0 & 0 & 0 & 0 & 0 & 0 & 0 & 0 & 0 & 0 \\
 0 & 0 & 0 & 0 & 0 & 0 & 0 & 0 & 0 & 0 & 0 & 0 & 0 & 0 & 0 & 0 \\
 0 & 0 & 0 & 0 & 0 & 0 & 0 & 0 & 0 & 0 & 0 & 0 & 1 & 0 & 0 & 0 \\
 0 & 0 & 0 & 0 & 0 & 0 & 0 & 0 & 0 & 0 & 0 & 0 & 0 & 0 & 0 & 0 \\
 0 & 0 & 0 & 0 & 0 & 0 & 0 & 0 & 0 & 0 & 0 & 0 & 0 & 0 & 1 & 0 \\
 0 & 0 & 0 & 0 & 0 & 0 & 0 & 0 & 0 & 0 & 0 & 0 & 0 & 0 & 0 & 0 \\
 0 & 0 & 0 & 0 & 0 & 0 & 0 & 0 & 0 & 0 & 0 & 0 & 0 & 0 & 0 & 0
\end{smallmatrix}\right)
\end{align}
and
\begin{align}
\setcounter{MaxMatrixCols}{16} \mathbb{R}^3_4 =
\left(\begin{smallmatrix}
 0 & 0 & 0 & 0 & 0 & 0 & 0 & 0 & 0 & 0 & 0 & 0 & 0 & 0 & 0 & 0 \\
 0 & 0 & 0 & 0 & 0 & 0 & 0 & 0 & 0 & 0 & 0 & 0 & 0 & 0 & 0 & 0 \\
 0 & 0 & 0 & 0 & 0 & 0 & 0 & 0 & 0 & 0 & 0 & 0 & 0 & 0 & 0 & 0 \\
 0 & 1 & 0 & 0 & 0 & 0 & 0 & 0 & 0 & 0 & 0 & 0 & 0 & 0 & 0 & 0 \\
 0 & 0 & 1 & 0 & 0 & 0 & 0 & 0 & 0 & 0 & 0 & 0 & 0 & 0 & 0 & 0 \\
 0 & 0 & 0 & 0 & 0 & 0 & 0 & 0 & 0 & 0 & 0 & 0 & 0 & 0 & 0 & 0 \\
 0 & 0 & 0 & 0 & 0 & 0 & 0 & 0 & 0 & 0 & 0 & 0 & 0 & 0 & 0 & 0 \\
 0 & 0 & 0 & 0 & 0 & 0 & 0 & 0 & 0 & 0 & 0 & 0 & 0 & 0 & 0 & 0 \\
 0 & 0 & 0 & 0 & 0 & 0 & 0 & 0 & 0 & 0 & 0 & 0 & 0 & 0 & 0 & 0 \\
 0 & 0 & 0 & 0 & 0 & 0 & 0 & 0 & \sqrt{2} & 0 & 0 & 0 & 0 & 0 & 0 & 0 \\
 0 & 0 & 0 & 0 & 0 & 0 & 0 & 0 & 0 & \sqrt{2} & 0 & 0 & 0 & 0 & 0 & 0 \\
 0 & 0 & 0 & 0 & 0 & 0 & 0 & 0 & 0 & 0 & 0 & 0 & 0 & 0 & 0 & 0 \\
 0 & 0 & 0 & 0 & 0 & 0 & 0 & 0 & 0 & 0 & 0 & 0 & 0 & 0 & 0 & 0 \\
 0 & 0 & 0 & 0 & 0 & 0 & 0 & 0 & 0 & 0 & 0 & 1 & 0 & 0 & 0 & 0 \\
 0 & 0 & 0 & 0 & 0 & 0 & 0 & 0 & 0 & 0 & 0 & 0 & 1 & 0 & 0 & 0 \\
 0 & 0 & 0 & 0 & 0 & 0 & 0 & 0 & 0 & 0 & 0 & 0 & 0 & 0 & 0 & 0
\end{smallmatrix}\right),\quad
\mathbb{R}^4_3 = \left(\begin{smallmatrix}
 0 & 0 & 0 & 0 & 0 & 0 & 0 & 0 & 0 & 0 & 0 & 0 & 0 & 0 & 0 & 0 \\
 0 & 0 & 0 & 1 & 0 & 0 & 0 & 0 & 0 & 0 & 0 & 0 & 0 & 0 & 0 & 0 \\
 0 & 0 & 0 & 0 & 1 & 0 & 0 & 0 & 0 & 0 & 0 & 0 & 0 & 0 & 0 & 0 \\
 0 & 0 & 0 & 0 & 0 & 0 & 0 & 0 & 0 & 0 & 0 & 0 & 0 & 0 & 0 & 0 \\
 0 & 0 & 0 & 0 & 0 & 0 & 0 & 0 & 0 & 0 & 0 & 0 & 0 & 0 & 0 & 0 \\
 0 & 0 & 0 & 0 & 0 & 0 & 0 & 0 & 0 & 0 & 0 & 0 & 0 & 0 & 0 & 0 \\
 0 & 0 & 0 & 0 & 0 & 0 & 0 & 0 & 0 & 0 & 0 & 0 & 0 & 0 & 0 & 0 \\
 0 & 0 & 0 & 0 & 0 & 0 & 0 & 0 & 0 & 0 & 0 & 0 & 0 & 0 & 0 & 0 \\
 0 & 0 & 0 & 0 & 0 & 0 & 0 & 0 & 0 & \sqrt{2} & 0 & 0 & 0 & 0 & 0 & 0 \\
 0 & 0 & 0 & 0 & 0 & 0 & 0 & 0 & 0 & 0 & \sqrt{2} & 0 & 0 & 0 & 0 & 0 \\
 0 & 0 & 0 & 0 & 0 & 0 & 0 & 0 & 0 & 0 & 0 & 0 & 0 & 0 & 0 & 0 \\
 0 & 0 & 0 & 0 & 0 & 0 & 0 & 0 & 0 & 0 & 0 & 0 & 0 & 1 & 0 & 0 \\
 0 & 0 & 0 & 0 & 0 & 0 & 0 & 0 & 0 & 0 & 0 & 0 & 0 & 0 & 1 & 0 \\
 0 & 0 & 0 & 0 & 0 & 0 & 0 & 0 & 0 & 0 & 0 & 0 & 0 & 0 & 0 & 0 \\
 0 & 0 & 0 & 0 & 0 & 0 & 0 & 0 & 0 & 0 & 0 & 0 & 0 & 0 & 0 & 0 \\
 0 & 0 & 0 & 0 & 0 & 0 & 0 & 0 & 0 & 0 & 0 & 0 & 0 & 0 & 0 & 0
\end{smallmatrix}\right)~.
\end{align}
Next, if we define $q_\pm = \sqrt{q\pm1}$, then the fermionic simple roots are given by
\begin{align}
\setcounter{MaxMatrixCols}{16} \mathbb{Q}^1_3 =
\left(\begin{smallmatrix}
   0 &   0 &   0 &   -b \sqrt{q} &   0 &   0 &   0 &   0 &   0 &   0 &   0 &   0 &   0 &   0 &   0 &   0 \\
   0 &   0 &   0 &   0 &   0 &   b q_- &   0 &   0 &   0 &   0 &   0 &   0 &   0 &   0 &   0 &   0 \\
   a \sqrt{q} &   0 &   0 &   0 &   0 &   0 & \scriptstyle \frac{b q_-}{\sqrt{2}} &   0 &   0 & \scriptstyle \frac{b q_+}{\sqrt{2}} &   0 &   0 &   0 &   0 &   0 &   0 \\
   0 &   0 &   0 &   0 &   0 &   0 &   0 &   0 &   0 &   0 &   0 &   0 &   0 &   0 &   0 &   0 \\
   0 &   0 &   0 &   0 &   0 &   0 &   0 &   0 &   0 &   0 &   b q_+ &   0 &   0 &   0 &   0 &   0 \\
   0 &   0 &   0 &   0 &   0 &   0 &   0 &   0 &   0 &   0 &   0 &   0 &   0 &   0 &   0 &   0 \\
   0 &   0 &   0 & \scriptstyle \frac{a q_-}{\sqrt{2}} &   0 &   0 &   0 &   0 &   0 &   0 &   0 &   0 &   0 & \scriptstyle -\frac{b q_+}{\sqrt{2}} &   0 &   0 \\
   0 &   0 &   0 &   0 &   a q_- &   0 &   0 &   0 &   0 &   0 &   0 &   0 &   0 &   0 &   -b q_+ &   0 \\
   0 &   a q_+ &   0 &   0 &   0 &   0 &   0 &   0 &   0 &   0 &   0 &   b q_- &   0 &   0 &   0 &   0 \\
   0 &   0 &   0 & \scriptstyle \frac{a q_+}{\sqrt{2}} &   0 &   0 &   0 &   0 &   0 &   0 &   0 &   0 &   0 & \scriptstyle \frac{b q_-}{\sqrt{2}} &   0 &   0 \\
   0 &   0 &   0 &   0 &   0 &   0 &   0 &   0 &   0 &   0 &   0 &   0 &   0 &   0 &   0 &   0 \\
   0 &   0 &   0 &   0 &   0 &   -a q_+ &   0 &   0 &   0 &   0 &   0 &   0 &   0 &   0 &   0 &   0 \\
   0 &   0 &   0 &   0 &   0 &   0 & \scriptstyle -\frac{a q_+}{\sqrt{2}} &   0 &   0 & \scriptstyle \frac{a q_-}{\sqrt{2}} &   0 &   0 &   0 &   0 &   0 &   -b \sqrt{q} \\
   0 &   0 &   0 &   0 &   0 &   0 &   0 &   0 &   0 &   0 &   0 &   0 &   0 &   0 &   0 &   0 \\
  0 &   0 &   0 &   0 &   0 &   0 &   0 &   0 &   0 &   0 &   a q_- &   0 &   0 &   0 &   0 &   0 \\
   0 &   0 &   0 &   0 &   0 &   0 &   0 &   0 &   0 &   0 &   0 &   0 &   0 & \scriptstyle a \sqrt{q} &   0 &   0
\end{smallmatrix}\right),
\end{align}
and
\begin{align}
\setcounter{MaxMatrixCols}{16} \mathbb{Q}^{\dag4}_2 =
\left(\begin{smallmatrix}
  0 &   0 &   0 &   -d \sqrt{q} &   0 &   0 &   0 &   0 &   0 &   0 &   0 &   0 &   0 &   0 &   0 &   0 \\
  0 &   0 &   0 &   0 &   0 &   d q_- &   0 &   0 &   0 &   0 &   0 &   0 &   0 &   0 &   0 &   0 \\
  c \sqrt{q} &   0 &   0 &   0 &   0 &   0 &   \frac{d q_-}{\sqrt{2}} &   0 &   0 &   \frac{d q_+}{\sqrt{2}} &   0 &   0 &   0 &   0 &   0 &   0 \\
  0 &   0 &   0 &   0 &   0 &   0 &   0 &   0 &   0 &   0 &   0 &   0 &   0 &   0 &   0 &   0 \\
  0 &   0 &   0 &   0 &   0 &   0 &   0 &   0 &   0 &   0 &   d q_+ &   0 &   0 &   0 &   0 &   0 \\
  0 &   0 &   0 &   0 &   0 &   0 &   0 &   0 &   0 &   0 &   0 &   0 &   0 &   0 &   0 &   0 \\
  0 &   0 &   0 &   \frac{c q_-}{\sqrt{2}} &   0 &   0 &   0 &   0 &   0 &   0 &   0 &   0 &   0 &   -\frac{d q_+}{\sqrt{2}} &   0 &   0 \\
  0 &   0 &   0 &   0 &   c q_- &   0 &   0 &   0 &   0 &   0 &   0 &   0 &   0 &   0 &   -d q_+ &   0 \\
  0 &   c q_+ &   0 &   0 &   0 &   0 &   0 &   0 &   0 &   0 &   0 &   d q_- &   0 &   0 &   0 &   0 \\
  0 &   0 &   0 &   \frac{c q_+}{\sqrt{2}} &   0 &   0 &   0 &   0 &   0 &   0 &   0 &   0 &   0 &   \frac{d q_-}{\sqrt{2}} &   0 &   0 \\
  0 &   0 &   0 &   0 &   0 &   0 &   0 &   0 &   0 &   0 &   0 &   0 &   0 &   0 &   0 &   0 \\
  0 &   0 &   0 &   0 &   0 &   -c q_+ &   0 &   0 &   0 &   0 &   0 &   0 &   0 &   0 &   0 &   0 \\
  0 &   0 &   0 &   0 &   0 &   0 &   -\frac{c q_+}{\sqrt{2}} &   0 &   0 &   \frac{c q_-}{\sqrt{2}} &   0 &   0 &   0 &   0 &   0 &   -d \sqrt{q} \\
  0 &   0 &   0 &   0 &   0 &   0 &   0 &   0 &   0 &   0 &   0 &   0 &   0 &   0 &   0 &   0 \\
  0 &   0 &   0 &   0 &   0 &   0 &   0 &   0 &   0 &   0 &   c q_- &   0 &   0 &   0 &   0 &   0 \\
  0 &   0 &   0 &   0 &   0 &   0 &   0 &   0 &   0 &   0 &   0 &   0 &   0 &   c \sqrt{q} &   0 &   0
\end{smallmatrix}\right).
\end{align}
For completeness let us also explicitly give the similarity transformation that relates the unitary
representation to the one from \cite{Zhang:2004qx}
\begin{align}
 \mathcal{V} =
\mathrm{diag}(\sqrt{q^3-q},q_+q_-,q_+q_-,q_+q_-,q_+q_-,2q_+,\sqrt{2}q_+,2q_+,2q_-,\sqrt{2}q_-,2q_-,1,1,1,1,\frac{1}{\sqrt{q}})
\end{align}
We notice that this transformation is singular for $q^2=1$, where
the representation becomes reducible but indecomposable.

%% file: BoundSmatrix2.tex
\chapter{Bound State S-Matrices}\label{chap;BoundSmat}

The light cone gauge-fixed Hamiltonian of the $\ads$ superstring
admits two copies of the centrally extended $\alg{su}(2|2)$
($\alg{h}$) as a symmetry algebra \cite{Arutyunov:2006ak}.
World-sheet excitations transform in the fundamental
representation of this algebra, hence their scattering data is
encoded in the S-matrix of $\alg{h}$ in this representation. The
same algebra emerges in $\mathcal{N}=4$ SYM, where it appears as
the algebra governing a spin chain whose energies encode the
anomalous dimensions of large operators \cite{Beisert:2005tm}.

Now that we have discussed the symmetry algebra in detail and also
studied its Yangian in the previous chapter, we can put it to use
in the computation of S-matrices. First we will discuss the
fundamental S-matrix. We will give its explicit form and indicate
the pole structure that gives rise to bound states. We will also
discuss some of its properties including Yangian symmetry. By
assuming Yangian symmetry we will be able to give a complete
derivation of the S-matrix that describes scattering between
arbitrary bound states, reproducing all data known so far.

\section{The Fundamental S-matrix}\label{sec:FundSmat}

The S-matrix describing the scattering of fundamental excitations
is given by the S-matrix of $\alg{h}$ seen as a Hopf algebra.
Since we explicitly know the fundamental representation it is
straightforward to compute the $16\times16$ dimensional matrix
that intertwines the coproduct and the opposite coproduct.
Consider the fundamental representation $V^F(p)$ depending on the
momentum $p$ and coupling constant $g$. The S-matrix $\S^F :
V^F(p_1)\otimes V^F(p_2) \rightarrow V^F(p_1)\otimes V^F(p_2)$ is
given by
\begin{align}\label{eqn;Sfund}
\S^{F}=
\left(\begin{smallmatrix}
 a_1 & 0 & 0 & 0 & \vline &  0 & 0 & 0 & 0 & \vline  & 0 & 0 & 0 & 0 & \vline & 0 & 0 & 0 & 0 \\
 0 & a_1+a_2 & 0 & 0 & \vline & -a_2 & 0 & 0  & 0 & \vline & 0 & 0 & 0  & -a_7 & \vline& 0 & 0 & a_7 & 0 \\
 0 & 0 & a_5 & 0 & \vline & 0 & 0 & 0 & 0 & \vline & a_9 & 0 & 0 & 0& \vline  & 0 & 0 & 0 & 0 \\
 0 & 0 & 0 & a_5 & \vline & 0 & 0 & 0 & 0 & \vline & 0 & 0 & 0 & 0 & \vline & a_9 & 0 & 0 & 0 \\
 - & - & - & - & - & - & - & - & - & - & - & - & - & - & - & - & - & - & - \\
 0 & -a_2 & 0 & 0 & \vline & a_1+a_2 & 0 & 0 & 0 & \vline & 0 & 0 & 0 & a_7 & \vline & 0 & 0 & -a_7 & 0 \\
 0 & 0 & 0 & 0 & \vline & 0 & a_1 & 0 & 0 & \vline & 0 & 0 & 0 & 0 & \vline & 0 & 0 & 0 & 0 \\
 0 & 0 & 0 & 0 & \vline & 0 & 0 & a_5 & 0 & \vline & 0 & a_9 & 0 & 0 & \vline & 0 & 0 & 0 & 0 \\
 0 & 0 & 0 & 0 & \vline & 0 & 0 & 0 & a_5 & \vline & 0 & 0 & 0 & 0 & \vline & 0 & a_9 & 0 & 0 \\
 - & - & - & - & - & - & - & - & - & - & - & - & - & - & - & - & - & - & - \\
 0 & 0 & a_10 & 0 & \vline & 0 & 0 & 0 & 0 & \vline & a_6 & 0 & 0 & 0 & \vline & 0 & 0 & 0 & 0 \\
 0 & 0 & 0 & 0 & \vline & 0 & 0 & a_10 & 0 & \vline & 0 & a_6 & 0 & 0 & \vline & 0 & 0 & 0 & 0 \\
 0 & 0 & 0 & 0 & \vline & 0 & 0 & 0 & 0 & \vline & 0 & 0 & a_3 & 0 & \vline & 0 & 0 & 0 & 0 \\
 0 & -a_8 & 0 & 0 & \vline & a_8 & 0 & 0 & 0 & \vline & 0 & 0 & 0 & a_3+a_4 & \vline & 0 & 0 & a_4 & 0 \\
 - & - & - & - & - & - & - & - & - & - & - & - & - & - & - & - & - & - & - \\
 0 & 0 & 0 & a_{10} & \vline & 0 & 0 & 0 & 0 & \vline & 0 & 0 & 0 & 0 & \vline & a_6 & 0 & 0 & 0 \\
 0 & 0 & 0 & 0 & \vline & 0 & 0 & 0 & a_{10} & \vline & 0 & 0 & 0 & 0 & \vline & 0 & a_6 & 0 & 0 \\
 0 & a_8 & 0 &  0 &\vline & -a_8 & 0 & 0 & 0 & \vline & 0 & 0 & 0 & -a_4 & \vline & 0 & 0 & a_3+a_4 & 0 \\
 0 & 0 & 0 & 0 & \vline & 0 & 0 & 0 & 0 & \vline & 0 & 0 & 0 & 0 & \vline & 0 & 0 & 0 & a_3
\end{smallmatrix}\right)
\end{align}
with coefficients
\begin{align}
& a_1 = 1 \nonumber\\
& a_2 = 2\frac{(x^+_2-x^+_1)(x^-_1 x^+_2-1)x^-_2}{(x^-_2-x^+_1)(x^-_1 x^-_2-1)x^+_2}-1\nonumber\\
& a_3=\frac{x^+_2-x^-_1}{x^-_2-x^+_1}\sqrt{\frac{x^+_1}{x^-_1}}\sqrt{\frac{x^-_2}{x^+_2}}\nonumber\\
& a_{4}=\frac{x^-_1-x^+_2}{x^-_2-x^+_1}\sqrt{\frac{x^+_1}{x^-_1}}\sqrt{\frac{x^-_2}{x^+_2}}-2\frac{(x^-_2 x^+_1-1)(x^+_1-x^+_2)x^-_1}{(x^-_1 x^-_2-1) (x^-_2-x^+_1) x^+_1}\sqrt{\frac{x^+_1}{x^-_1}}\sqrt{\frac{x^-_2}{x^+_2}}\nonumber\\
& a_5 = \frac{x^-_1-x^-_2}{x^+_1-x^-_2}\sqrt{\frac{x^+_1}{x^-_1}}\nonumber\\
& a_{6}=\frac{x^+_2-x^+_1}{x^-_2-x^+_1}\sqrt{\frac{x^-_2}{x^+_2}}\nonumber\\
& a_7= -\frac{i (x^-_1-x^+_1) (x^-_2-x^+_2) (x^+_1-x^+_2)}{(x^-_1x^-_2-1)(x^-_2-x^+_1)\eta(p_1)\eta(p_2)}\sqrt{\frac{x^-_2}{x^+_2}}\nonumber\\
& a_8 =\frac{i (x^+_1-x^+_2)\eta(p_1)\eta(p_2)x^-_1 x^-_2}{(x^-_1 x^-_2-1)(x^-_2-x^+_1) x^+_1 x^+_2}\sqrt{\frac{x^+_1}{x^-_1}}\nonumber\\
& a_9 = \frac{x^-_1-x^+_1}{x^-_2-x^+_1}\sqrt{\frac{x^+_1}{x^-_1}}\sqrt{\frac{x^-_2}{x^+_2}}\frac{\eta(p_2)}{\eta(p_1)}\nonumber\\
&a_{10}=\frac{x^-_2-x^+_2}{x^-_2-x^+_1}\frac{\eta(p_1)}{\eta(p_2)},\nonumber
\end{align}
where the parameters $x^{\pm}_i$ are related to the momentum $p_i$
via (\ref{eqn;ParamtersPandG}). We have normalized the S-matrix in
such a way that $a_1 =1$. The S-matrix satisfies the identities
common to S-matrices to two-dimensional integrable systems
\begin{description}
    \item[Unitarity:] $\qquad \, \, \, \, \, \, \, \, \, \, \, \, \S^F_{12}(z_1,z_2)\S^F_{21}(z_2,z_1)=\mathbbm{1}$.
    \item[Hermiticity:] $\, \, \, \, \, \, \qquad \S^F_{12}(z_1,z_2)\S^F_{12}(z_1^*,z_2^*)^{\dag} = \mathbbm{1}$.
    \item[CPT Invariance:] $\, \, \, \, \S^F_{12}=(\S^F_{12})^t$.
    \item[Yang-Baxter:] $\, \, \, \qquad \S^F_{12}\S^F_{13}\S^F_{23}=\S^F_{23}\S^F_{13}\S^F_{12}$.
\end{description}
The full S-matrix includes an overall scalar factor
$S_0(p_1,p_2)$, which is not fixed by the invariance condition
(\ref{eqn;SymmProp}). However crossing symmetry puts restrictions
on it \cite{Janik:2006dc}. The problem of finding the appropriate
overall scalar factor which reproduces the correct anomalous
dimensions has been extensively studied in the literature
\cite{Arutyunov:2004vx,Beisert:2006ib,Hernandez:2006tk,Beisert:2007hz,Eden:2006rx}.
An exact conjecture has been put forward in \cite{Beisert:2006ez}
\begin{align}
S_{F}(p_{1},p_{2})=
\left(\frac{x_{1}^{-}}{x_{1}^{+}}\right)^{\frac{1}{2}}\left(\frac{x_{2}^{+}}{x_{2}^{-}}\right)^{\frac{1}{2}}\sigma(x_{1},x_{2})\sqrt{G(2)}.
\end{align}
where
\begin{align}\label{fattoreG}
&G(n) = \frac{u_1 - u_2 + \frac{n}{2}}{u_1 - u_2 - \frac{n}{2}},&&
\sigma(p_1,p_2) = e^{\frac{i}{2}\theta(p_1,p_2)}.
\end{align}
The function $\theta(p_1,p_2)$ is called the dressing phase and it
is defined in terms of conserved charges \cite{Arutyunov:2004vx}
\begin{align}
q_{n}(x_{i})=\frac{i}{n-1}\left(\frac{1}{(x_{i}^{+})^{n-1}}-\frac{1}{(x_{i}^{-})^{n-1}}\right).
\end{align}
as follows:
\begin{eqnarray}
\theta(p_1,p_2) =\sum_{r=2}^{\infty}\sum_{n=0}^{\infty}
c_{r,r+1+2n} (q_r(p_1) q_{r+1+2n}(p_2t)-q_r(p_2) q_{r+1+2n}(p_1)),
\end{eqnarray}
for some coefficients $c$. This solution satisfies crossing
symmetry and agrees with all data known so far from string and
gauge perturbation theory.

As is not uncommon in integrable field theories, the S-matrix
actually respects a bigger symmetry group, namely the Yangian of
$\alg{h}$. It is readily checked that this matrix intertwines the
different coproducts of the Yangian (\ref{eqn;YangianCoprod}) in
the evaluation representation \cite{Beisert:2007ds}.

Finally, the S-matrix has a pole at $x_1^+=x_2^-$. This indicates
the presence of bound states \cite{Arutyunov:2007tc}. The
S-matrices corresponding to scattering of bound states of up to
two fundamental particles have been computed in
\cite{Arutyunov:2008zt}. It appears \cite{deLeeuw:2008dp} that in
these cases the requirement of the Yang-Baxter equation is
equivalent to the presence of Yangian symmetry (see also
\cite{Arutyunov:2006yd}). In what follows we will make use of this
fact and assume Yangian symmetry to be present for all bound state
S-matrices.

\section{Kinematical Structure of the S-Matrix}

It is useful to take a closer look at the kinematical structure of
the S-matrix. In particular, we will use $\alg{h}$ invariance to
show that the S-matrix is of block diagonal form. The bound state
S-matrices should respect the $\alg{h}$ symmetry, which is imposed
by requiring invariance under the coproducts of the generators
(\ref{eqn;SymmProp})
\begin{eqnarray}\label{eqn;SymmPropSU22}
\S~\Delta(\mathbb{J}^{A})&=&\Delta^{op}(\mathbb{J}^{A})~\S.
\end{eqnarray}
This formula will be our starting point. We consider for the
generators the bound state representations in the superspace
formalism discussed in the previous chapter.

\subsection{Invariant subspaces}

Consider two short symmetric solutions with bound state numbers
$\ell_1,\ell_2$ respectively (cf. section
\ref{sec;boundstaterep}). The tensor product of the corresponding
bound state representations in superspace \cite{Arutyunov:2008zt}
is given by:
\begin{eqnarray}
\Phi(w,\theta)\Phi(v,\vartheta),
\end{eqnarray}
where $w,\theta$ denote the superspace variables of the first
particle and $v,\vartheta$ describe the representation of the
second particle.

The S-matrix acts on this tensor space and should, according to
(\ref{eqn;SymmPropSU22}), commute in particular with
$\Delta\mathbb{L}^1_1 = \Delta^{op}\mathbb{L}^1_1$ and
$\Delta\mathbb{R}^3_3 = \Delta^{op}\mathbb{R}^3_3$. From this, it
is easily deduced that the numbers
\begin{eqnarray}
K^{\rm{II}} &\equiv&
\#\theta_3+\#\theta_4+\#\vartheta_3+\#\vartheta_4+2\#w_2+2\#v_2,\nonumber\\
K^{\rm{III}} &\equiv& \#\theta_3+\#\vartheta_3+\#w_2+\#v_2
\end{eqnarray}
are conserved. Here $\# w_2$ means the number of $w_2$'s, i.e.
$\#w_2$ of the state $w_2^k$ is $k$, etc. More precisely, for any
state with bound state number $\ell$ we have
\begin{align}
&(\ell - \fL^1_1)\theta_3^k \theta_4^l w_1^m w_2^n = (2 n + k + l ) \theta_3^k \theta_4^l w_1^m w_2^n \\
&(\ell - \fL^1_1 + \fR^3_3)\theta_3^k \theta_4^l w_1^m w_2^n = (2
n + 2k) \theta_3^k \theta_4^l w_1^m w_2^n
\end{align}
Applying these formulas to the coproducts projected into the
tensor product of two bound states we obtain the above expressions
for $K^{\rm{II}},K^{\rm{III}}$.

The variables $w_2,v_2$ can be interpreted as being a combined
state of two fermions of different type \cite{Beisert:2005tm}.
Hence, the number $K^{\rm{II}}$ corresponds to the total
number of fermions, and the number $K^{\rm{III}}$ counts the
number of fermions of type 3. The fact that
these numbers are conserved allows us to define subspaces that the
S-matrix has to leave invariant. For each of these subspaces we
will derive the corresponding S-matrix.\smallskip

Let us write out the tensor product more explicitly. Since we are
considering bound states with bound state number $\ell_1,\ell_2$
we restrict to
\begin{eqnarray}
&&( w_1^{\ell_1-k} w_2^k +  \theta_{3}w_1^{\ell_1-k-1} w_2^k +
\theta_{4}w_1^{\ell_1-k-1} w_2^k +
 \theta_{3}\theta_{4}w_1^{\ell_1-k-1} w_2^{k-1})\times\nonumber\\
&&\times~( v_1^{\ell_2-l} v_2^l + \vartheta_{3}v_1^{\ell_2-l-1}
v_2^l +  \vartheta_{4}v_1^{\ell_2-l-1} v_2^l +
\vartheta_{3}\vartheta_{4}v_1^{\ell_2-l-1} v_2^{l-1}).
\end{eqnarray}
The ranges over which the labels $k,l$ are allowed to vary can be
straightforwardly read off for each term. By multiplying
everything out, we reproduce the basis vectors that span the
tensor product representation of these two bound states. One can
compute the quantum numbers $K^{\rm{II}},K^{\rm{III}}$ for any of
these basis vectors. The results are listed in Table
\ref{Tab;Tensor}.
\begin{table}[htbp]
    \centering
        \begin{tabular}{|l|l||l|l||l||l|}
            \hline
            Space 1 & Space 2 &  $K^{\rm{II}}$ & $K^{\rm{III}}$ & $N$ & Case\\
            \hline \hline
            $\theta_{3}w_1^{\ell_1-k-1} w_2^k$ & $\vartheta_{3}v_1^{\ell_2-l-1} v_2^l$ & $2(k+l)+2$ & $k+l+2$ & $k+l$ & Ia\\
            & & & & &\\
            $\theta_{4}w_1^{\ell_1-k-1} w_2^k$ & $\vartheta_{4}v_1^{\ell_2-l-1} v_2^l$ & $2(k+l)+2$ & $k+l$ & $k+l$ & Ib\\
            & & & & &\\
            $\theta_{3}w_1^{\ell_1-k-1} w_2^k$ & $v_2^{\ell_2-l} v_2^l$ & $2(k+l)+1$ & $k+l+1$ & $k+l$ & IIa\\
            $w_1^{\ell_1-k} w_2^k$ & $\vartheta_{3}v_1^{\ell_2-l-1} v_2^l$ & $2(k+l)+1$ & $k+l+1$ & $k+l$ & IIa\\
            $\theta_{3}w_1^{\ell_1-k-1} w_2^k$ & $\vartheta_{3}\vartheta_{4} v_1^{\ell_2-l-1} v_2^{l-1}$ & $2(k+l)+1$ & $k+l+1$ & $k+l$ & IIa\\
            $\theta_{3}\theta_{4}w_1^{\ell_1-k-1} w_2^{k-1}$ & $\vartheta_{3}v_1^{\ell_2-l-1} v_2^l$ & $2(k+l)+1$ & $k+l+1$ & $k+l$ & IIa\\
            & & & & &\\
            $\theta_{4}w_1^{\ell_1-k-1} w_2^k$ & $v_2^{\ell_2-l} v_2^l$ & $2(k+l)+1$ & $k+l$ & $k+l$ & IIb\\
            $w_1^{\ell_1-k} w_2^k$ & $\vartheta_{4}v_1^{\ell_2-l-1} v_2^l$ & $2(k+l)+1$ & $k+l$ & $k+l$ & IIb\\
            $\theta_{4}w_1^{\ell_1-k-1} w_2^k$ & $\vartheta_{3}\vartheta_{4} v_1^{\ell_2-l-1} v_2^{l-1}$ & $2(k+l)+1$ & $k+l$ & $k+l$ & IIb\\
            $\theta_{3}\theta_{4}w_1^{\ell_1-k-1} w_2^{k-1}$ & $\vartheta_{4}v_1^{\ell_2-l-1} v_2^l$ & $2(k+l)+1$ & $k+l$ & $k+l$ & IIb\\
            & & & & &\\
            $w_1^{\ell_1-k} w_2^k$ & $v_2^{\ell_2-l} v_2^l$ & $2(k+l)$ & $k+l$ & $k+l$ & III\\
            $w_1^{\ell_1-k} w_2^k$ & $\vartheta_{3}\vartheta_{4} v_1^{\ell_2-l-1} v_2^{l-1}$ & $2(k+l)$ & $k+l$ & $k+l$ & III\\
            $\theta_{3}\theta_{4}w_1^{\ell_1-k-1} w_2^{k-1}$ & $v_2^{\ell_2-l} v_2^l$ & $2(k+l)$ & $k+l$ & $k+l$ & III\\
            $\theta_{3}\theta_{4}w_1^{\ell_1-k-1} w_2^{k-1}$ & $\vartheta_{3}\vartheta_{4} v_1^{\ell_2-l-1} v_2^{l-1}$ & $2(k+l)$ & $k+l$ & $k+l$ & III\\
            $\theta_{3}w_1^{\ell_1-k-1} w_2^k$ & $\vartheta_{4}v_1^{\ell_2-l-1} v_2^l$ & $2(k+l+1)$ & $k+l+1$ & $k+l+1$ & III\\
            $\theta_{4}w_1^{\ell_1-k-1} w_2^k$ & $\vartheta_{3}v_1^{\ell_2-l-1} v_2^l$ & $2(k+l+1)$ & $k+l+1$ & $k+l+1$ & III\\
            \hline
        \end{tabular}
    \caption{The $16\ell_1\ell_2$ vectors from the tensor product and their $\alg{su}(2)\times\alg{su}(2)$ quantum numbers.}
    \label{Tab;Tensor}
\end{table}
When we take a closer look at the result, we see that ordering the
states by the quantum numbers $K^{\rm{II}},K^{\rm{III}}$, there
are exactly five different types of states:
\begin{description}
  \item[Case Ia:] $K^{\mathrm{II}}=2 N+2, K^{\mathrm{III}}=N+2$,
  \item[Case Ib:] $K^{\mathrm{II}}=2 N+2, K^{\mathrm{III}}=N$,
  \item[Case IIa:] $K^{\mathrm{II}}=2 N+1, K^{\mathrm{III}}=N+1$,
  \item[Case IIb:] $K^{\mathrm{II}}=2 N+1, K^{\mathrm{III}}=N$,
  \item[Case III:] $K^{\mathrm{II}}=2 N, K^{\mathrm{III}}=N$,
\end{description}
for some integer $N$. For fixed $N$, each of these states has
different quantum numbers $K^{\rm{II}},K^{\rm{III}}$ and hence the
states belonging to each of these cases form an invariant
subspace under the action of the S-matrix. \smallskip

Clearly, vectors from Case Ia and Case Ib only differ by the
exchange of the fermionic index $3\leftrightarrow 4$, which is
easily realized in terms of the (fermionic) $\mathfrak{su}(2)$
symmetry generators of type $\mathbb{R}$. Hence, the subspaces
spanned by the two types of states are isomorphic, and scatter via
the same S-matrix. An analogous relationship connects Case IIa and
IIb. Thus, there are only three non-equivalent cases:
\begin{description}
  \item[Case I:] $K^{\mathrm{II}}=2 N+2, K^{\mathrm{III}}=N+2$,
  \item[Case II:] $K^{\mathrm{II}}=2 N+1, K^{\mathrm{III}}=N+1$,
  \item[Case III:] $K^{\mathrm{II}}=2 N, K^{\mathrm{III}}=N$.
\end{description}
For fixed $N$ (i.e. for fixed $K^{\rm{II}},K^{\rm{III}}$) we
denote the vector spaces spanned by vectors from each of the
inequivalent cases by $V_N^{\rm{I}},V_N^{\rm{II}},V_N^{\rm{III}}$
respectively. In what follows we will compute the S-matrix for
each of these invariant subspaces. For this we will first need to
study these invariant spaces in more detail.

\subsection{Basis and relations}

Let us give a complete description of the bases of the invariant
subspaces. Later on we will introduce different choices of basis
for the different cases, but in this section we will discuss the
bases as obtained simply by multiplying out the tensor product as
seen from Table \ref{Tab;Tensor}. We will call this type of basis
the standard one.

\subsection*{Case I, $K^{\mathrm{II}}=2 N +2, K^{\mathrm{III}}=N+2$.}

For fixed $N$, the vector space of states $V^{\rm{I}}_N$ is
$N+1$-dimensional. The standard basis for this vector space is
\begin{eqnarray}\label{eqn;BasisCase1}
\stateA{k,l}\equiv\underbrace{\theta_{3}w_1^{\ell_1-k-1}w_2^{k}}_{\rm{Space
1}}~\underbrace{\vartheta_{3}v_1^{\ell_2-l-1}v_2^{l}}_{\rm{Space
2}},
\end{eqnarray}
for all $k+l=N$. These indeed give $N+1$ different vectors.

\subsection*{Case II, $K^{\mathrm{II}}=2 N+1, K^{\mathrm{III}}=N+1$.}

For fixed $N$, the dimension of this vector space is $4N+2$. The
standard basis is
\begin{eqnarray}\label{eqn;BasisCase2}
\stateB{k,l}_1&\equiv& \underbrace{\theta_{3}w_1^{\ell_1-k-1}w_2^{k}}~\underbrace{v_1^{\ell_2-l}v_2^{l}},\nonumber\\
\stateB{k,l}_2&\equiv&\underbrace{w_1^{\ell_1-k}w_2^{k}}~\underbrace{\vartheta_{3}v_1^{\ell_2-l-1}v_2^{l}},\\
\stateB{k,l}_3&\equiv&\underbrace{\theta_{3}w_1^{\ell_1-k-1}w_2^{k}}~\underbrace{\vartheta_{3}\vartheta_{4}v_1^{\ell_2-l-1}v_2^{l-1}},\nonumber\\
\stateB{k,l}_4&\equiv&\underbrace{\theta_{3}\theta_{4}w_1^{\ell_1-k-1}w_2^{k-1}}~\underbrace{\vartheta_{3}v_1^{\ell_2-l-1}v_2^{l}},\nonumber
\end{eqnarray}
where $k+l=N$. As a lighter notation, we will from now on, with no
risk of confusion, omit indicating ``Space 1" and ``Space 2" under
the curly brackets. The ranges of $k,l$ are clear from the
explicit expressions and it is easily seen that we get $4N+2$
states.

\subsection*{Case III: $K^{\rm{II}}=2N, K^{\rm{III}}=N$}

For fixed $N=k+l$, the dimension of this vector space is $6N$. The
standard basis is
\begin{eqnarray}\label{eqn;BasisCase3}
\stateC{k,l}_1&\equiv&\underbrace{w_1^{\ell_1-k}w_2^{k}}~\underbrace{v_1^{\ell_2-l}v_2^{l}},\nonumber\\
\stateC{k,l}_2&\equiv&\underbrace{w_1^{\ell_1-k}w_2^{k}}~\underbrace{\vartheta_{3}\vartheta_{4}v_1^{\ell_2-l-1}v_2^{l-1}},\nonumber\\
\stateC{k,l}_3&\equiv&\underbrace{\theta_{3}\theta_{4}w_1^{\ell_1-k-1}w_2^{k-1}}~\underbrace{v_1^{\ell_2-l}v_2^{l}},\nonumber\\
\stateC{k,l}_4&\equiv&\underbrace{\theta_{3}\theta_{4}w_1^{\ell_1-k-1}w_2^{k-1}}~\underbrace{\vartheta_{3}\vartheta_{4}v_1^{\ell_2-l-1}v_2^{l-1}},\\
\stateC{k,l}_5&\equiv&\underbrace{\theta_{3}w_1^{\ell_1-k-1}w_2^{k}}~\underbrace{\vartheta_{4}v_1^{\ell_2-l}v_2^{l-1}},\nonumber\\
\stateC{k,l}_6&\equiv&\underbrace{\theta_{4}w_1^{\ell_1-k}w_2^{k-1}}~\underbrace{\vartheta_{3}v_1^{\ell_2-l-1}v_2^{l}}\nonumber.
\end{eqnarray}
Note that our numbering slightly differs from the one used in
Table \ref{Tab;Tensor}, in the sense that $\stateC{k,l}_{5,6}$ are
rescaled for convenience in such a way that they also have
$N=k+l$, instead of $k+l+1$ as in Table \ref{Tab;Tensor}.

It is useful to supply all these spaces with a canonical inner
product:
\begin{eqnarray}
 ~^{\rm{A}}_j\langle k,l|m,n\rangle^{\rm{B}}_i =
 \delta_{ij}\delta_{km}\delta_{ln}\delta_{AB}.
\end{eqnarray}
Actually, for the sake of our arguments, orthogonality of these
vectors will always be sufficient.

For later convenience we also introduce the vector spaces
$V^{\rm{A}}_{k,l} = {\rm span}\{|k,l\rangle^{\rm{A}}_{i}\}$, for
$\rm{A}= \rm{I,II,III}$. These subspaces are generated by the
basis elements for fixed $k,l$ and together build up the full
invariant subspace
\begin{align}
V^A_N = \bigoplus_{k+l=N}V^A_{k,l}.
\end{align}
The dimensions of these spaces for generic $k,l$ are $\dim
V^{\rm{I}}_{k,l} = 1, \dim V^{\rm{II}}_{k,l} = 4, \dim
V^{\rm{III}}_{k,l} = 6$. For specific values of $k,l$ they can be
lower dimensional, e.g. $\dim V^{\rm{II}}_{0,0} = 2$.

\begin{figure}
  \centering
  \includegraphics[scale=.5]{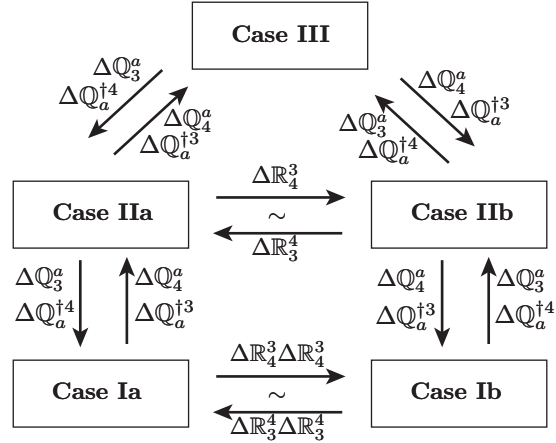}
  \caption{Schematic representation of the relations between the invariant subspaces.
  The opposite coproducts also respect the above diagram, as well as all their Yangian counterparts.}\label{Fig;Cases}
\end{figure}

So far we have only used the bosonic part of the algebra to
determine the invariant subspaces. The fermionic generators will
give maps between the different cases. One can use the (opposite)
coproducts of the symmetry generators to move between the
different subspaces. In particular, the cases are distinguished by
their quantum numbers $K^{\rm{II}},K^{\rm{III}}$ and acting with
supersymmetry generators will change these numbers. How this works
is schematically depicted in figure \ref{Fig;Cases}.

These relations between the different cases will play an important
role in the derivation of the full S-matrix. We can employ the
different arrows in Figure \ref{Fig;Cases} (and their Yangian
counterparts) to relate the different S-matrices to the Case I
S-matrix. In the next section, we will introduce two different
sets of bases which allow for a natural interpretation of the
S-matrix. These bases will make use of the full Yangian symmetry
rather than just $\alg{h}$. Since we will be able to uniquely
determine the form of the S-matrix reduced to Case I states, by
applying the aforementioned maps we can use this to compute the
S-matrix also in the other cases.

Summarizing, we find that the S-matrix is of block-diagonal form
\begin{eqnarray}\label{eqn;GeneralBoundStateMatrix}
\S=\begin{pmatrix}
  \fbox{\small{$\mathscr{X}$}} & ~ & ~ & ~ & ~ \\
  ~ & \fbox{\LARGE{$\mathscr{Y}$}} & ~ & \mbox{\Huge{$0$}} & ~ \\
  ~ & ~ & \fbox{\Huge{$\mathscr{Z}$}} & ~ & ~ \\
  ~ & \mbox{\Huge{$0$}} & ~ & \fbox{\LARGE{$\mathscr{Y}$}} & ~ \\
  ~ & ~ & ~ & ~ & \fbox{\small{$\mathscr{X}$}}
\end{pmatrix}.
\end{eqnarray}
The outer blocks scatter states from $V^{\rm{I}}$
\begin{eqnarray}
&&\mathscr{X}:V^{\rm{I}}_N\longrightarrow V^{\rm{I}}_N\\
&&\stateA{k,l}\mapsto \sum_{m=0}^{k+l}
\mathscr{X}^{k,l}_m\stateA{m,k+l-m},
\end{eqnarray}
where $k+l=N$ and $\mathscr{X}^{k,l}_m$ will be given by (\ref{eqn;SCase1}).
The blocks $\mathscr{Y}$ describe the scattering of states from
$V^{\rm{II}}$
\begin{eqnarray}
&&\mathscr{Y}:V^{\rm{II}}_N\longrightarrow V^{\rm{II}}_N\\
&&\stateB{k,l}_j\mapsto \sum_{m=0}^{k+l}\sum_{j=1}^{4}
\mathscr{Y}^{k,l;j}_{m;i}\stateB{m,k+l-m}_j.
\end{eqnarray}
These S-matrix elements are given in (\ref{eqn;SCase2}). Finally,
the middle block deals with the third case
\begin{eqnarray}
&&\mathscr{Z}:V^{\rm{III}}_N\longrightarrow V^{\rm{III}}_N\\
&&\stateC{k,l}_j\mapsto \sum_{m=0}^{k+l}\sum_{j=1}^{6}
\mathscr{Z}^{k,l;j}_{m;i}\stateC{m,k+l-m}_j,
\end{eqnarray}
with $\mathscr{Z}^{k,l;j}_{m;i}$ from (\ref{eqn;SCase3}).

\section{Yangian Symmetry and Coproducts}\label{sect;Bases}

Up to now, we have only used $\alg{su}(2|2)$ symmetry to study the
bound state S-matrix. This, however, is not enough to fix the
tensor structure of the S-matrix. In particular, it was found that
one needs to impose the Yang-Baxter equation by hand to attain
this \cite{Arutyunov:2008zt}. An alternative to this method was
shown to come from Yangian symmetry \cite{deLeeuw:2008dp}. We will
follow the latter approach and employ Yangian symmetry to fully
fix the bound state S-matrix.

\subsection{(Opposite) coproduct basis}

Let us turn back to the invariant subspaces. We define different
bases for each case in addition to the standard basis, which is
the one commonly used in the literature. These bases are more
convenient for the computation of the bound state S-matrix, and
they will be called the coproduct basis and the opposite coproduct
basis. The basis transformation between the coproduct (opposite
coproduct) basis and the standard one will be denoted by $\Lambda$
($\Lambda^{op}$, respectively).

The (opposite) coproduct basis will be constructed by using
Yangian generators to create states out of a chosen vacuum. This
is similar to \cite{deLeeuw:2008ye} where it was used to study the
Bethe Ansatz. We define our vacuum to be
\begin{eqnarray}\label{eqn;BoundStateVacuum}
|0\rangle \equiv w_1^{\ell_1}\ v_1^{\ell_2}.
\end{eqnarray}
Note that this
state is from $V_{0}^{\rm{III}}$, which is a one dimensional
space. The S-matrix maps this space onto itself, and we normalize
our S-matrix in such a way that $\mathbb{S}|0\rangle=|0\rangle$.
The (opposite) coproduct basis will consist of states created by
the (opposite) coproducts of various symmetry generators acting on
this vacuum.

Clearly, the S-matrix has a natural interpretation in these bases,
and can be formulated in terms of $\Lambda$ and $\Lambda^{op}$, as
will be explained below in section \ref{sect;SmatCoProd}. We will
now list the explicit formulae for the different bases.

\subsection*{Case I, $K^{\mathrm{II}}=2 N +2, K^{\mathrm{III}}=N+2$.}

The coproduct basis is given by
\begin{eqnarray}\label{eqb;CoProdBasisCase1}
\Delta(\mathbb{Q}^{1}_{3})\Delta(\mathbb{Q}_{2}^{\dag4})\prod_{i=q+1}^{N}\Delta(\mathbb{L}^{1}_{2})
\prod_{j=1}^{q}\Delta(\hat{\mathbb{L}}^{1}_{2})|0\rangle, \qquad
q=0,1,\ldots N,
\end{eqnarray}
and the opposite coproduct basis is given by
\begin{eqnarray}
\Delta^{op}(\mathbb{Q}^{1}_{3})\Delta^{op}(\mathbb{Q}_{2}^{\dag4})\prod_{i=k+1}^{N}\Delta^{op}(\mathbb{L}^{1}_{2})
\prod_{j=1}^{k}\Delta^{op}(\hat{\mathbb{L}}^{1}_{2})|0\rangle,
\qquad k=0,1,\ldots N.
\end{eqnarray}
Each of these two bases is indeed composed of $N+1$ different
vectors. By explicitly working out the coproducts one can see that
these vectors form a basis for Case I. One could also consider an
alternative choice, like for instance
\begin{eqnarray}
\Delta(\mathbb{Q}^{1}_{3})\Delta(\hat{\mathbb{Q}}^{1}_{3})\prod_{i=k+1}^{N}\Delta(\mathbb{L}^{1}_{2})
\prod_{j=1}^{k}\Delta(\hat{\mathbb{L}}^{1}_{2})|0\rangle,
\end{eqnarray}
but these vectors are readily seen to be proportional to
(\ref{eqb;CoProdBasisCase1}).

It is also straightforwardly seen why (\ref{eqb;CoProdBasisCase1})
actually describes Case I from the point of view of the quantum
numbers $K^{\rm{II}}, K^{\rm{III}}$. The operators
$\Delta\mathbb{L}^1_2,\Delta\hat{\mathbb{L}}^1_2$ create a boson
of type $2$ out of the vacuum and the supersymmetry generators
$\Delta\mathbb{Q}^{1}_{3},\Delta\hat{\mathbb{Q}}^{1}_{3}$ create a
fermion of type $3$. Hence we find that $K^{\rm{II}} = 2\#
\mathbb{L}^1_2 + 2 \# \Delta\hat{\mathbb{L}}^1_2 + \#
\Delta\mathbb{Q}^{1}_{3} + \# \Delta\hat{\mathbb{Q}}^{1}_{3}$ and
$K^{\rm{III}} = \# \mathbb{L}^1_2 + \# \Delta\hat{\mathbb{L}}^1_2
+ \# \Delta\mathbb{Q}^{1}_{3} + \#\Delta\hat{\mathbb{Q}}^{1}_{3}$.
This indeed coincides with $K^{\mathrm{II}}=2 N +2,
K^{\mathrm{III}}=N+2$.

\subsection*{Case II, $K^{\mathrm{II}}=2 N+1, K^{\mathrm{III}}=N+1$.}

The coproduct basis is given by
\begin{align}
&\Delta(\mathbb{Q}^{1}_{3})\prod_{i=q+1}^{N}\Delta(\mathbb{L}^{1}_{2})
\prod_{j=1}^{q}\Delta(\hat{\mathbb{L}}^{1}_{2})|0\rangle,&&\Delta(\hat{\mathbb{Q}}^{1}_{3})\prod_{i=q+1}^{N}\Delta(\mathbb{L}^{1}_{2})
\prod_{j=1}^{q}\Delta(\hat{\mathbb{L}}^{1}_{2})|0\rangle,\\
&\Delta(\mathbb{Q}^{1}_{3})\Delta(\hat{\mathbb{Q}}^{1}_{3})\Delta(\mathbb{Q}^{1}_{4})\prod_{i=q+1}^{N-1}\Delta(\mathbb{L}^{1}_{2})
\prod_{j=1}^{q}\Delta(\hat{\mathbb{L}}^{1}_{2})|0\rangle,&&\Delta(\mathbb{Q}^{1}_{3})\Delta(\hat{\mathbb{Q}}^{1}_{3})\Delta(\hat{\mathbb{Q}}^{1}_{4})\prod_{i=q+1}^{N-1}\Delta(\mathbb{L}^{1}_{2})
\prod_{j=1}^{q}\Delta(\hat{\mathbb{L}}^{1}_{2})|0\rangle,\nonumber
\end{align}
and similar expressions hold for the opposite coproduct basis. One
can again compute $K^{\rm{II}}, K^{\rm{III}}$ for these states and
see explicitly that they describe Case II. Also in this case one
could use for example the operator
$\Delta(\mathbb{Q}_{2}^{\dag4})$ to create the coproduct basis.

\subsection*{Case III, $K^{\mathrm{II}}=2 N, K^{\mathrm{III}}=N$.}

The coproduct basis is
\begin{eqnarray}
&&\prod_{i=q+1}^{N}\Delta(\mathbb{L}^{1}_{2})
\prod_{j=1}^{q}\Delta(\hat{\mathbb{L}}^{1}_{2})|0\rangle,\nonumber\\
&&\Delta(\mathbb{Q}^{1}_{3})\Delta(\mathbb{Q}^{1}_{4})\prod_{i=q+1}^{N-1}\Delta(\mathbb{L}^{1}_{2})
\prod_{j=1}^{q}\Delta(\hat{\mathbb{L}}^{1}_{2})|0\rangle,\nonumber\\
&&\Delta(\mathbb{Q}^{1}_{3})\Delta(\hat{\mathbb{Q}}^{1}_{4})\prod_{i=q+1}^{N-1}\Delta(\mathbb{L}^{1}_{2})
\prod_{j=1}^{q}\Delta(\hat{\mathbb{L}}^{1}_{2})|0\rangle,\nonumber\\
&&\Delta(\hat{\mathbb{Q}}^{1}_{3})\Delta(\mathbb{Q}^{1}_{4})\prod_{i=q+1}^{N-1}\Delta(\mathbb{L}^{1}_{2})
\prod_{j=1}^{q}\Delta(\hat{\mathbb{L}}^{1}_{2})|0\rangle,\nonumber\\
&&\Delta(\hat{\mathbb{Q}}^{1}_{3})\Delta(\hat{\mathbb{Q}}^{1}_{4})\prod_{i=q+1}^{N-1}\Delta(\mathbb{L}^{1}_{2})
\prod_{j=1}^{q}\Delta(\hat{\mathbb{L}}^{1}_{2})|0\rangle,\nonumber\\
&&\Delta(\mathbb{Q}^{1}_{3})\Delta(\mathbb{Q}^{1}_{4})\Delta(\hat{\mathbb{Q}}^{1}_{3})
\Delta(\hat{\mathbb{Q}}^{1}_{4})\prod_{i=q+1}^{N-2}\Delta(\mathbb{L}^{1}_{2})
\prod_{j=1}^{q}\Delta(\hat{\mathbb{L}}^{1}_{2})|0\rangle.\nonumber
\end{eqnarray}
These are readily seen to be $6N$ states and their quantum numbers
are of the form $K^{\mathrm{II}}=2 N, K^{\mathrm{III}}=N$.

The Yangian generators also provide maps between the different
cases. In particular, one finds that figure \ref{Fig;Cases} also
holds for Yangian generators. In this basis the arrows between the
different cases are obvious. One important thing to notice is the
following. Even though, for example, $\Delta\mathbb{Q}^2_{3}$ maps
Case II onto Case I, this does not automatically give a
straightforward map between the vector spaces $V_{k,l}^{\rm{A}}$.
For instance, one has
\begin{eqnarray}
&&\Delta\mathbb{Q}^1_{3}:V_{k,l}^{\rm{II}}\longrightarrow V_{k,l-1}^{\rm{I}}\oplus V_{k-1,l}^{\rm{I}},\\
&&\Delta\hat{\mathbb{Q}}^2_{3}:V_{k,l}^{\rm{II}}\longrightarrow
V_{k+1,l-1}^{\rm{I}}\oplus V_{k,l}^{\rm{I}}\oplus
V_{k-1,l+1}^{\rm{I}}.
\end{eqnarray}
This provides an additional complication we will have to deal with
in the computation of the S-matrix on Case II states.

\subsection{S-matrix in coproduct basis}\label{sect;SmatCoProd}

The fact that the coproduct basis is well suited for computing the
S-matrix can be seen from (\ref{eqn;SymmPropSU22}). One sees that
the S-matrix directly maps the coproduct basis onto the opposite
coproduct basis. In particular, since we normalize the S-matrix in
such a way that $\mathbb{S}|0\rangle=|0\rangle$, we see that the
S-matrix, when written as a map between these two bases, is just
the identity matrix.\smallskip

In other words, one can obtain the general formula for the
S-matrix in the standard basis (which is ultimately the basis we
are interested in) just by applying the appropriate basis
transformations. Let us denote the S-matrix written in the
standard basis as $\mathbb{S}$. One then finds:
\begin{eqnarray}\label{eqn;Smat}
\mathbb{S} = \Lambda^{op} \Lambda^{-1}.
\end{eqnarray}
Since $\Lambda_{12}^{op}(p_1,p_2) = \Lambda_{21}(p_2,p_1)$, this
is reminiscent of a Drinfel'd twist \cite{twi}. Unfortunately
however, the matrix $\Lambda$ is not of upper triangular form.

Note that the explicit matrices $\Lambda$ and $\Lambda^{op}$ just
consist of the coproduct vectors written in the standard basis, and
that these matrices trivially have the same block structure as the
S-matrix with respect to the quantum numbers
$K^{\rm{II}},K^{\rm{III}}$. The above discussion can be summarized
in the following commutative diagram:
\begin{eqnarray}
\begin{CD}
\{\rm{coproduct\ basis}\} @>\mathbbm{1}>>
\{\rm{opposite\ coproduct\ basis}\}\\
@V\Lambda VV @V\Lambda^{op}VV\\
\{\rm{standard\ basis}\} @>\mathbb{S}>> \{\rm{standard\ basis}\}.
\end{CD}
\end{eqnarray}
The computationally hard part is finding the explicit inverse of
$\Lambda$. For any concrete case at hand this can be done by
simple linear algebra, but the expressions become rather involved.
However, we will be able to carry out this procedure in full
generality for the S-matrix of Case I, and use this result to find
the S-matrix for all the other states.

\section{Fundamental S-matrix revisited}

To illustrate the above discussion, we will give a full derivation
of the fundamental S-matrix $\S^{F}$ in this formalism. As we saw
earlier, the fundamental S-matrix can be completely fixed without
usage of Yangian symmetry and we will indeed see this reflected
throughout the derivation.

Let us start by explicitly giving the coproduct basis and the
standard basis. The vacuum is given by
\begin{eqnarray}\label{eqn;FundVacuum}
|0\rangle\in V^{\rm{III}}_0, \qquad |0\rangle = w_1 v_1.
\end{eqnarray}
We choose the normalization $\S^F|0\rangle= |0\rangle$.

It is readily seen that there is only one Case I state, namely
\begin{eqnarray}
\theta_3\vartheta_3.
\end{eqnarray}
The corresponding state from Case Ib is found by replacing
$3\leftrightarrow 4$ and does not give anything new. The
(opposite) coproduct basis for this case is given by
\begin{eqnarray}
\Delta \mathbb{Q}^1_3 \Delta \mathbb{Q}_2^{\dag4}|0\rangle, \qquad
\Delta^{op} \mathbb{Q}^1_3 \Delta^{op}
\mathbb{Q}_2^{\dag4}|0\rangle.
\end{eqnarray}
One can explicitly work out the above to find
\begin{eqnarray}
\Delta \mathbb{Q}^1_3 \Delta \mathbb{Q}_2^{\dag4}|0\rangle =(a_1
c_2-a_2 c_1)\theta_3\vartheta_3.
\end{eqnarray}
Hence, the piece of $\Lambda$ that describes the basis
transformation of Case III states is given by
\begin{align}
\Lambda_{\mathscr{X}} &= a_1 c_2-a_2 c_1.\\
\Lambda^{op}_{\mathscr{X}} &= a_3 c_4-a_4 c_3.
\end{align}
To avoid cluttered notation, we introduce the following quantities
\begin{align}
\mathscr{Q}_{ij}&=a_i c_j-a_j c_i,\nonumber\\
\overline{\mathscr{Q}}_{ij}&=b_i d_j-d_j b_i,\\
\mathscr{I}_{ij}&=a_i d_j-b_j c_i\nonumber.
\end{align}
These coefficients satisfy the following identity
\begin{eqnarray}
\mathscr{Q}_{ij}\overline{\mathscr{Q}}_{ij} =
1-\mathscr{I}_{ij}\mathscr{I}_{ji}.
\end{eqnarray}
We also work with coefficients that carry the explicit
braiding factors
(\ref{eqn;LabelWithBrainding1},\ref{eqn;LabelWithBrainding2}) to
keep the notation light. The S-matrix restricted to Case I follows
readily from this,
\begin{eqnarray}
\S^F\cdot\theta_3\theta_3 = \mathscr{X} \theta_3\theta_3,
\end{eqnarray}
with
\begin{eqnarray}
\mathscr{X} = \frac{a_3 c_4-a_4 c_3}{a_1 c_2-a_2 c_1} =
\frac{\mathscr{Q}_{34}}{\mathscr{Q}_{12}}.
\end{eqnarray}
Next, we consider Case II. There are four different invariant
subspaces, but upon changing $3\leftrightarrow 4, 1\leftrightarrow
2$ we find that they are all isomorphic. We will therefore
restrict to $V^{\rm{II}}_0$, which has standard basis
\begin{eqnarray}
\{\theta_3 v_1,w_1\vartheta_3\}.
\end{eqnarray}
The coproduct basis is easily seen to be given by
\begin{eqnarray}
\{\Delta \mathbb{Q}^1_3|0\rangle, \Delta \mathbb{Q}_2^{\dag
4}|0\rangle\}.
\end{eqnarray}
Again,we would like to point out that
\begin{eqnarray}
\{\Delta \mathbb{Q}^1_3|0\rangle, \Delta
\hat{\mathbb{Q}}^1_3|0\rangle\}
\end{eqnarray}
is also a valid basis, and both bases will necessarily result in
the same S-matrix. The coproduct-to-standard basis transformation
can be worked out to find, in the above notation,
\begin{align}
\Lambda_{\mathscr{Y}} =
\begin{pmatrix}
  a_{1} & c_{1} \\
  a_{2} & c_{2}
\end{pmatrix}.
\end{align}
The S-matrix for Case II is given by
\begin{eqnarray}
\S^F\cdot \theta_3v_1 &=& \mathscr{Y}^1_1 \theta_3v_1 + \mathscr{Y}^2_1
w_1\vartheta_3\nonumber\\
\S^F\cdot w_1\vartheta_3 &=& \mathscr{Y}^1_2 \theta_3v_1 +
\mathscr{Y}^2_2 w_1\vartheta_3,
\end{eqnarray}
with
\begin{eqnarray}
\mathscr{Y} =
\begin{pmatrix}
  \mathscr{Y}^1_{1} & \mathscr{Y}^2_{1} \\
  \mathscr{Y}^1_{2} & \mathscr{Y}^2_{2}
\end{pmatrix}=\left(
\begin{array}{ll}
 \frac{a_2 c_3-a_3 c_2}{a_2 c_1-a_1 c_2} & \frac{a_3 c_1-a_1
   c_3}{a_2 c_1-a_1 c_2} \\
 \frac{a_2 c_4-c_2 a_4}{a_2 c_1-a_1 c_2} & \frac{a_4 c_1-a_1
   c_4}{a_2 c_1-a_1 c_2}
\end{array}
\right)=
\left(
\begin{array}{ll}
 \frac{\mathscr{Q}_{23}}{\mathscr{Q}_{21}} & \frac{\mathscr{Q}_{31}}{\mathscr{Q}_{21}} \\
 \frac{\mathscr{Q}_{24}}{\mathscr{Q}_{21}} & \frac{\mathscr{Q}_{41}}{\mathscr{Q}_{21}}
\end{array}
\right).
\end{eqnarray}
Finally, for Case III one finds three different subspaces
$V^{\rm{III}}_0,V^{\rm{III}}_1,V^{\rm{III}}_2$. The subspace
$V^{\rm{III}}_0$ contains only the vacuum (\ref{eqn;FundVacuum}),
and $V^{\rm{III}}_2$ is isomorphic to $V^{\rm{III}}_0$. The only
non-trivial piece is $V^{\rm{III}}_1$ which is spanned by the
standard basis
\begin{eqnarray}
\{w_1v_2,w_2v_1,\theta_3\vartheta_4,\theta_4\vartheta_3\}.
\end{eqnarray}
The coproduct basis is given by
\begin{eqnarray}
\{\Delta \mathbb{L}^1_2|0\rangle,
\Delta\mathbb{Q}^{1}_3\Delta\mathbb{Q}^{1}_4|0\rangle,
\Delta\mathbb{Q}^{1}_3\Delta \mathbb{Q}^{\dag3}_2|0\rangle,
\Delta\mathbb{Q}^{\dag4}_2\Delta \mathbb{Q}^{\dag3}_2|0\rangle \}
\end{eqnarray}
and from this one finds
\begin{eqnarray}
\Lambda_{\mathscr{Z}} = \left(
\begin{array}{llll}
 1 & -a_2 b_2 & a_2 d_2 & -c_2 d_2 \\
 1 & -a_1 b_1 & a_1 d_1 & -c_1 d_1 \\
 0 & -a_1 a_2 & a_1 c_2 & -c_1 c_2 \\
 0 & -a_1 a_2 & a_2 c_1 & -c_1 c_2
\end{array}
\right).
\end{eqnarray}
The resulting S-matrix for Case III is
\begin{eqnarray}
\mathscr{Z} =
\Lambda_{\mathscr{Z}}^{op}(\Lambda_{\mathscr{Z}})^{-1} = \left(
\begin{array}{llll}
 \frac{\mathscr{Q}_{14}\mathscr{I}_{14}}{\mathscr{Q}_{12}\mathscr{I}_{12}} & \frac{\mathscr{Q}_{24}\mathscr{I}_{24}}{\mathscr{Q}_{21}\mathscr{I}_{21}} & \frac{\mathscr{Q}_{24}\overline{\mathscr{Q}}_{14}}{\mathscr{Q}_{12}\mathscr{I}_{21}} & \frac{\mathscr{Q}_{14}\overline{\mathscr{Q}}_{42}}{\mathscr{Q}_{21}\mathscr{I}_{12}} \\
 \frac{\mathscr{Q}_{13}\mathscr{I}_{13}}{\mathscr{Q}_{12}\mathscr{I}_{12}} & \frac{\mathscr{Q}_{23}\mathscr{I}_{23}}{\mathscr{Q}_{21}\mathscr{I}_{21}} & \frac{\mathscr{Q}_{23}\overline{\mathscr{Q}}_{13}}{\mathscr{Q}_{12}\mathscr{I}_{21}} & \frac{\mathscr{Q}_{13}\overline{\mathscr{Q}}_{32}}{\mathscr{Q}_{21}\mathscr{I}_{12}} \\
 \frac{\mathscr{Q}_{13}\mathscr{Q}_{14}}{\mathscr{Q}_{12}\mathscr{I}_{12}} & \frac{\mathscr{Q}_{23}\mathscr{Q}_{24}}{\mathscr{Q}_{21}\mathscr{I}_{21}} & -\frac{\mathscr{Q}_{23}\mathscr{I}_{41}}{\mathscr{Q}_{12}\mathscr{I}_{21}} & \frac{\mathscr{Q}_{13}\mathscr{I}_{42}}{\mathscr{Q}_{21}\mathscr{I}_{12}} \\
 -\frac{\mathscr{Q}_{13}\mathscr{Q}_{14}}{\mathscr{Q}_{12}\mathscr{I}_{12}} & -\frac{\mathscr{Q}_{23}\mathscr{Q}_{24}}{\mathscr{Q}_{21}\mathscr{I}_{21}}& \frac{\mathscr{Q}_{24}\mathscr{I}_{31}}{\mathscr{Q}_{12}\mathscr{I}_{21}} & -\frac{\mathscr{Q}_{14}\mathscr{I}_{32}}{\mathscr{Q}_{21}\mathscr{I}_{12}}
\end{array}
\right).
\end{eqnarray}
This is a nice expression which only depend on the representation
parameters $a,b,c,d$ which automatically incorporate the braiding
factors. If one plugs in the explicit parameterizations in terms
of $x^{\pm}$
(\ref{eqn;LabelWithBrainding1},\ref{eqn;LabelWithBrainding2}) one
recovers the fundamental S-matrix (\ref{eqn;Sfund}). It is also
readily seen that using Yangian generators to construct the
(opposite) coproduct basis leads to the same S-matrix.

Writing $\Lambda$ as a $16\times16$-matrix one finds a
decomposition of $\S^F$ which reminds of a Drinfel'd twist:
\begin{eqnarray}
\S^F = \Lambda^{op}\Lambda^{-1}.
\end{eqnarray}
The inverse S-matrix is easily found by interchanging the
opposite and normal coproduct basis. This just amounts to changing
\begin{align}
(a_1,b_1,c_1,d_1,a_2,b_2,c_2,d_2) \leftrightarrow (a_3,b_3,c_3,d_3,a_4,b_4,c_4,d_4),
\end{align}
in the above formulae. In the previous discussion there was no
need to use the Yangian generators, however they will prove
crucial for general bound states.

\section{Complete Solution of Case I}

We will now move on to generic bound state representations. As
mentioned before we will first derive the S-matrix for states from
Case I. To this end, we employ Yangian symmetry. We will work in
the evaluation representation. This S-matrix proves to be the
building block out of which the S-matrices for both Case II and
Case III can be constructed\footnote{Our procedure will somehow be
reminiscent of employing highest weight states of Yangians.}.

Our starting point is the coproduct basis
(\ref{eqb;CoProdBasisCase1}).  It is convenient to reorder the
products in the following way:
\begin{eqnarray}
\left\{\prod_{i=q+1}^{N}\Delta(\mathbb{L}^{1}_{2})
\prod_{j=1}^{q}\Delta(\hat{\mathbb{L}}^{1}_{2})\right\}\Delta(\mathbb{Q}^{1}_{3})\Delta(\mathbb{Q}^{\dag4}_{2})|0\rangle.
\end{eqnarray}
The action of the susy generators on the vacuum is of the form
\begin{eqnarray}
\Delta(\mathbb{Q}^{1}_{3})\Delta(\mathbb{Q}^{\dag4}_{2})|0\rangle
= (a_2c_1-a_1c_2)\ell_1\ell_2|0,0\rangle^{\rm{I}},
\end{eqnarray}
with a similar expression for the opposite version. In complete
analogy with the fundamental S-matrix, this defines the coordinate
transformation $\Lambda$ of Case I states to the standard basis,
from which one can straightforwardly read off the action of the
S-matrix on $|0,0\rangle^{\rm{I}}$:
\begin{eqnarray}\label{eqn;Sfor00}
\mathbb{S}|0,0\rangle^{\rm{I}}&=&
\frac{\mathbb{S}\Delta(\mathbb{Q}^{1}_{3})\Delta(\mathbb{Q}^{\dag4}_{2})|0\rangle}{(a_2c_1-a_1c_2)\ell_1\ell_2}\nonumber\\
&=&\frac{\Delta^{op}(\mathbb{Q}^{1}_{3})\Delta^{op}(\mathbb{Q}^{\dag4}_{2})\mathbb{S}|0\rangle}{(a_2c_1-a_1c_2)\ell_1\ell_2}\nonumber\\
&=&\frac{a_4c_3-a_3c_4}{a_2c_1-a_1c_2}|0,0\rangle^{\rm{I}}.
\end{eqnarray}
In other words, the S-matrix multiplies $|0,0\rangle^{\rm{I}}$ by
a scalar. We will denote this scalar by $\mathcal{D}$. In terms of
$x^{\pm}$, it is given by
\begin{eqnarray}
\label{D} \mathcal{D}\equiv\frac{a_4c_3-a_3c_4}{a_2c_1-a_1c_2} =
\frac{x^+_2-x^-_1}{x^-_2-x^+_1}\sqrt{\frac{x^+_1}{x^-_1}}\sqrt{\frac{x^-_2}{x^+_2}}.
\end{eqnarray}
This factor coincides with the one found for the fundamental
S-matrix on the corresponding state.\smallskip

One can now use the generators
$\mathbb{L}^1_2,\hat{\mathbb{L}}^1_2$ to construct a generic Case
I state $|k,l\rangle^{\rm{I}}$ from $|0,0\rangle^{\rm{I}}$, for
arbitrary $k,l$. This can be seen by considering the following
identities:
\begin{align}\label{eqn;Lplusmin}
&(\mathbb{L}^1_2\otimes\mathbbm{1})(\delta
u+\Delta\mathbb{L}^1_1)|k,l\rangle^{\rm{I}}=
\left\{\Delta\hat{\mathbb{L}}^1_2-u_2 \Delta\mathbb{L}^1_2+
\Delta\mathbb{L}^1_2\circ(\mathbb{L}^1_1\otimes
\mathbbm{1})\right\}|k,l\rangle^{\rm{I}},\\
&(\mathbbm{1}\otimes\mathbb{L}^1_2)(\delta
u+\Delta\mathbb{L}^1_1)|k,l\rangle^{\rm{I}}=
-\left\{\Delta\hat{\mathbb{L}}^1_2-u_1 \Delta\mathbb{L}^1_2-
\Delta\mathbb{L}^1_2\circ(\mathbbm{1}\otimes\mathbb{L}^1_1)\right\}|k,l\rangle^{\rm{I}},\nonumber
\end{align}
where
\begin{eqnarray}
\delta u = u_1-u_2.
\end{eqnarray}
Since $\Delta(\mathbb{L}^1_1)|k,l\rangle^{\rm{I}} =
\frac{\ell_1+\ell_2-2(k+l+1)}{2}|k,l\rangle^{\rm{I}}$, it is
obvious that the left hand side of (\ref{eqn;Lplusmin}) is
proportional to $|k+1,l\rangle$ (first line) and $|k,l+1\rangle$
(respectively, second line). By applying the right hand side
operators in (\ref{eqn;Lplusmin}) inductively to $|0,0\rangle$,
one finds
\begin{align}
&\left\{\prod_{m=1}^{k}(\ell_1-m)\prod_{n=1}^{l}(\ell_2-n)\prod_{q=1}^{k+l}\left(\delta
u+\frac{\ell_1+\ell_2}{2}-q\right)\right\}|k,l\rangle =\nonumber\\
&\qquad\qquad\qquad\left[(\mathbb{L}^1_2\otimes\mathbbm{1})(\delta
u+\Delta\mathbb{L}^1_1)\right]^k\left[(\mathbbm{1}\otimes\mathbb{L}^1_2)(\delta
u+\Delta\mathbb{L}^1_1)\right]^l|0, 0\rangle.
\end{align}
Then, by (\ref{eqn;Lplusmin}),
\begin{align}
|k,l\rangle^{\rm{I}}=
\frac{\prod_{i=1}^k\left[\Delta\hat{\mathbb{L}}^1_2-\frac{\ell_1+2u_2-2i+1}{2}
\Delta\mathbb{L}^1_2\right]\prod_{j=1}^l\left[\frac{1+2j+2u_1-\ell_2}{2}\Delta\mathbb{L}^1_2-\Delta\hat{\mathbb{L}}^1_2
\right]}
{\prod_{m=1}^{k}(\ell_1-m)\prod_{n=1}^{l}(\ell_2-n)\prod_{q=1}^{k+l}\left(\delta
u+\frac{\ell_1+\ell_2}{2}-q\right)}|0,0\rangle^{\rm{I}}.\nonumber
\end{align}
This exactly tells us how to write a state in the standard basis
in terms of the coproduct basis. In other words, this explicitly
indicates how to construct $\Lambda^{-1}$. We would also like to
point out that the presence of the full Yangian symmetry is
crucial here. It is not possible to construct the operators that
link the vectors $\stateA{k,l}$ to $\stateA{0,0}$ without the
Yangian coproducts.

It is now straightforward to obtain the action of the S-matrix on
Case I states from the above. The symmetry properties of the
S-matrix, together with (\ref{eqn;Sfor00}), now imply
\begin{align}\label{eqn;EasyCaseSL}
\mathbb{S}|k,l\rangle^{\rm{I}} =&\mathcal{D}\times\\
&
\frac{\prod_{i=1}^k\left[\Delta^{op}\hat{\mathbb{L}}^1_2-\frac{2u_2-\ell_1+2i-1}{2}
\Delta^{op}\mathbb{L}^1_2\right]\prod_{j=1}^l\left[\frac{2u_1+\ell_2-1-2j}{2}
\Delta^{op}\mathbb{L}^1_2-\Delta^{op}\hat{\mathbb{L}}^1_2\right]}
{\prod_{m=1}^{k}(\ell_1-m)\prod_{n=1}^{l}(\ell_2-n)\prod_{q=1}^{k+l}\left(\delta
u+\frac{\ell_1+\ell_2}{2}-q\right)}|0,0\rangle^{\rm{I}}.\nonumber
\end{align}
By explicitly computing the right hand side, one finds that
$\mathbb{S}|k,l\rangle^{\rm{I}}$ is of the form
\begin{eqnarray}\label{transit}
\mathbb{S}|k,l\rangle^{\rm{I}} =
\sum_{n=0}^{k+l}\mathscr{X}^{k,l}_n|n,k+l-n\rangle^{\rm{I}},
\end{eqnarray}
with
\begin{align}\label{eqn;SCase1}
&\mathscr{X}^{k,l}_{n} =
\mathcal{D}\frac{\prod_{i=1}^{n}(\ell_1-i)\prod_{i=1}^{k+l-n}(\ell_2-i)}{\prod_{p=1}^{k}(\ell_1-p)\prod_{p=1}^{l}(\ell_2-p)\prod_{p=1}^{k+l}(\delta u +\frac{\ell_1+\ell_2}{2}-p)}\times \\
& \ \ \times \sum_{m=0}^{k}\left\{ {k\choose k-m }{l\choose n-m
}\prod_{p=1}^{m}\mathfrak{c}^+_p
\prod_{p=1-m}^{l-n}\mathfrak{c}^-_p
\prod_{p=1}^{k-m}\mathfrak{d}_{\frac{k-p+2}{2}}
\prod_{p=1}^{n-m}\tilde{\mathfrak{d}}_{\frac{k+l-m-p+2}{2}}\right\}.\nonumber
\end{align}
The coefficients are given by
\begin{eqnarray}
\mathfrak{c}^{\pm}_m&=&\delta u \pm \frac{\ell_1-\ell_2}{2} -m+1,\nonumber\\
\mathfrak{d}_i&=&\ell_1+1-2i,\nonumber\\
\tilde{\mathfrak{d}}_i&=&\ell_2+1-2i.\nonumber
\end{eqnarray}
It is worthwhile noticing that in the special case $l=0$ (and
similarly for $k=0$) this expression reduces considerably. For
later use, we can write it in the following way:
\begin{eqnarray}
\label{primalzero} \mathscr{X}^{k,0}_{k-n} = \mathcal{D}D{k\choose n}
\frac{\prod_{p=1}^{n}(\ell_2-p)\prod_{p=1}^{k-n}(\delta u
+\frac{\ell_1-\ell_2}{2}-p+1)}{\prod_{p=1}^{k}(\delta
u+\frac{\ell_1+\ell_2}{2}-p)}.
\end{eqnarray}
In all of the above expressions it is understood that products are
set to 1 whenever they run over negative integers, i.e.
$\prod_a^b=1$ if $b< a$, and the binomial ${x\choose y}$ is taken
to be zero if $y>x$ and if $y<0$.\smallskip

We can see how the formula we have found bears a rational
dependence on the difference of the spectral parameters, as
typical of Yangian universal R-matrices in evaluation
representations cf. e.g. \cite{Molev,MacKay:2004tc}. {\rm The
following function, meromorphic in all the parameters, coincides
with (\ref{eqn;SCase1}) in the appropriate domain of integer
values:}
\begin{align}\label{hypergeom}
&\mathscr{X}^{k,l}_n =
\frac{(-1)^{k+n}  \pi \mathcal{D} \sin [(k-\ell_1) \pi ] \, \Gamma (l+1)}{\sin [\ell_1 \pi]
\sin [(k +l -\ell_2-n) \pi ] \, \Gamma (l-\ell_2+1) \Gamma
   (n+1)} \times \nonumber\\
&\ \  \frac{\Gamma
   (n+1-\ell_1) \Gamma
   \left(l+\frac{\ell_1-\ell_2}{2}-n-\delta u
   \right) \Gamma \left(1-\frac{\ell_1+\ell_2}{2}-\delta u \right)}{\Gamma
   \left(k+l-\frac{\ell_1+\ell_2}{2}-\delta u +1\right) \Gamma \left(\frac{\ell_1-\ell_2}{2}- \delta u \right)} \times \\
&\ _4\tilde{F}_3
   \left[-k,-n,\frac{2\delta u
   +2-\ell_1+\ell_2}{2} ,\frac{\ell_2-\ell_1-2\delta u}{2}; 1-\ell_1,\ell_2-k-l,l-n+1;1 \right],\nonumber
\end{align}
where one has defined the regularized hypergeometric function
$_4\tilde{F}_3 (a_1,a_2,a_3,a_4;b_1,b_2,b_3;\tau) = {_4F_3}
(a_1,a_2,a_3,a_4;b_1,b_2,b_3;\tau)/[\Gamma (b_1) \Gamma (b_2)
\Gamma (b_3)]$.

Moreover, we can easily see that we are in a special situation,
since the parameters entering the hypergeometric function $_4F_3
(a_1,a_2,a_3,a_4;b_1,b_2,b_3;1)$ satisfy $\sum_{i=1}^4 a_i -
\sum_{j=1}^3 b_j=-1$. When this happens, the hypergeometric
function reduces to a $6j$-symbol, according to the following
formula (see for example \cite{shbook}):
\begin{align}
\label{6jsy}
&_4F_3\left(a_1,a_2,a_3,a_4;b_1,b_2,b_3;1\right)=\\
&\quad\frac{(-1)^{b_1+1}\Gamma
\left(b_2\right) \Gamma \left(b_3\right)\sqrt{\Gamma
\left(1-a_1\right)\Gamma \left(1-a_2\right)\Gamma
   \left(1-a_3\right)} }{\Gamma \left(1-b_1\right) \sqrt{\Gamma
\left(b_2-a_1\right)\Gamma \left(b_2-a_2\right)}}\times \nonumber\\
&\times \frac{\sqrt{\Gamma \left(1-a_4\right)\Gamma
\left(a_1-b_1+1\right)\Gamma \left(a_2-b_1+1\right)\Gamma
\left(a_3-b_1+1\right)\Gamma \left(a_4-b_1+1\right)}}{\sqrt{\Gamma
\left(b_2-a_3\right)\Gamma
   \left(b_2-a_4\right)\Gamma \left(b_3-a_1\right)\Gamma \left(b_3-a_2\right)\Gamma \left(b_3-a_3\right)\Gamma
   \left(b_3-a_4\right)}}\times \nonumber \\
&\times\left\{\begin{array}{ccc}
 \frac{1}{2} \left(-a_1-a_4+b_3-1\right) & \frac{1}{2} \left(-a_1-a_3+b_2-1\right) & \frac{1}{2} \left(a_1+a_2-b_1-1\right) \\
 \frac{1}{2} \left(-a_2-a_3+b_3-1\right) & \frac{1}{2} \left(-a_2-a_4+b_2-1\right) & \frac{1}{2} \left(a_3+a_4-b_1-1\right)
\end{array}
\right\}.
\end{align}
By identifying the parameters we see that the relevant $6j$-symbol
\begin{eqnarray}
\left\{ \begin{array}{ccc}
 j_1 & j_2 & j_3 \\
 j_4 & j_5 & j_6
\end{array}\right\}
\end{eqnarray}
has coefficients
\begin{align}
\label{co6j}
&j_1 = \frac{1}{2}\left(k+l-n+ \frac{\ell_1-\ell_2}{2} +\delta u\right),& &j_2 = \frac{1}{2}\left( \frac{\ell_1+\ell_2}{2} -2 - l  -
\delta u\right) ,\nonumber\\
&j_3 = \frac{1}{2}\left( \ell_1 - 2 - k - n\right),&& j_4 = \frac{1}{2}\left( \frac{\ell_1-\ell_2}{2}- 1 + l-\delta u\right),\nonumber\\
&j_5 = \frac{1}{2}\left( \frac{\ell_1+\ell_2}{2}-1 - k - l + n + \delta u\right), && j_6=  \frac{1}{2}\left(\ell_2-1\right).
\end{align}
For generic values of $\delta u$, the $6j$-symbol is understood
{\rm in the same sense as in the comment above formula
(\ref{hypergeom}).}
However, one can prove that, for values of $\delta u$
corresponding to the physical poles, the entries of the $6j$-symbol are indeed half-integer, as
one may expect from the fusion rules of $\alg{su}(2)$
representations. One expects this because the action of the
bosonic $\alg{su}(2)$ generators $\fL^a_b$ on Case I states forms a
$\alg{su}(2)$ algebra.

In the special case $l=0$ (a similar argument would hold for
$k=0$), we can go back to expression (\ref{eqn;SCase1}), and see
that it can be casted in the following form:
\begin{align}\label{dopolzero2}
&\mathscr{X}^{k,0}_{k-n} =  \frac{\mathcal{D}\Gamma (k+1) \Gamma
(1+n-\ell_2) \Gamma \left(1-\frac{\ell_1+\ell_2}{2}-\delta u
\right) \Gamma
   \left(k+\frac{\ell_2}{2}-\frac{\ell_1}{2}-n-\delta u \right)}{\Gamma (1-\ell_2) \Gamma (k-n+1) \Gamma (n+1) \, \Gamma
   \left(k-\frac{\ell_1+\ell_2}{2}-\delta u +1\right) \Gamma
   \left(\frac{\ell_2-\ell_1}{2}-\delta u \right)}.
\end{align}

\section{The S-matrix for Case II}\label{sect;SmatCaseII}

As explained in the previous sections, $\Delta\mathbb{Q}^1_{3},
\Delta\mathbb{Q}^{\dag4}_2$ and their Yangian counterparts map
Case II states onto Case I states. We introduce the Case II
S-matrix in the following way
\begin{eqnarray}
\mathbb{S}\stateB{k,l}_i =
\sum_{j=1}^4\sum_{m=0}^{k+l}\mathscr{Y}^{k,l;j}_{m;i}\stateB{m,N-m}_j,
\end{eqnarray}
where again $N=k+l$. This means that the coefficients
$\mathscr{Y}^{k,l;j}_{m;i}$ actually correspond to the S-matrix
restricted to the following spaces
\begin{eqnarray}
\mathscr{Y}^{k,l;j}_{n;i}: V^{\rm{II}}_{k,l}\longrightarrow
V^{\rm{II}}_{n,N-n}.
\end{eqnarray}
Generically, both spaces are 4 dimensional, and
$\mathscr{Y}^{k,l;j}_{m;i}$ correspond to the coefficients of a
$4\times4$ matrix. One might wonder what happens for special
values of $k,l,n,N$ since $V^{\rm{II}}_0$ is lower dimensional. It
turns out that the $4\times4$ matrix actually contains these
non-generic cases. This will be explained later on in Section
\ref{sect;Reduction} and we will continue with deriving the
generic $4\times4$ matrix.
\smallskip

By considering the action of $\Delta\mathbb{Q}^1_{3}$, we can
relate the Case II S-matrix to (\ref{eqn;SCase1}). It is easily
checked that
\begin{eqnarray}
\Delta\mathbb{Q}^1_{3}\stateB{k,l}_j = Q_j(k,l)\stateA{k,l},
\end{eqnarray}
with
\begin{eqnarray}
\begin{array}{lll}
  Q_1(k,l) = a_2(l-\ell_2), &\qquad & Q_2(k,l) = a_1(\ell_1-k), \\
  Q_3(k,l) = b_2, &\qquad & Q_4(k,l) = -b_1.
\end{array}
\end{eqnarray}
Similar expressions are of course obtained for
$\Delta^{op}\mathbb{Q}^1_{3},\Delta^{op}\mathbb{Q}^{\dag4}_2,\Delta\mathbb{Q}^{\dag4}_2$.
We can now apply our general strategy in the following fashion:
\begin{align}\label{eqn;LHSderivationCase2}
\costateA{n,N-n}|\Delta^{op}\mathbb{Q}^1_{3}\S\stateB{k,l}_i
&=\sum_{j=1}^4\sum_{m=0}^{k+l}\mathscr{Y}^{k,l;j}_{m;i}\
\costateA{n,N-n}
|\Delta^{op}\mathbb{Q}^1_{3}\stateB{m,N-m}_j\nonumber\\
&=\sum_{j,m}\mathscr{Y}^{k,l;j}_{m;i}Q^{op}_j(m,N-m)
\costateA{n,N-n}\stateA{m,N-m}\nonumber\\
&=\sum_{j=1}^4\mathscr{Y}^{k,l;j}_{n;i}Q^{op}_j(n,N-n).
\end{align}
On the other hand, we can use the symmetry properties of the
S-matrix to obtain
\begin{eqnarray}\label{eqn;RHSderivationCase2}
\costateA{n,N-n}|~\Delta^{op}\mathbb{Q}^1_{3}\mathbb{S}~\stateB{k,l}_i
&=&\costateA{n,N-n}|~\mathbb{S}\Delta\mathbb{Q}^1_{3}~\stateB{k,l}_i\nonumber\\
&=&Q_{i}(k,l)\costateA{n,N-n}|~\mathbb{S}~\stateA{k,l}\nonumber\\
&=&Q_{i}(k,l)\sum_{m=0}^{N}\mathscr{X}^{k,l}_{m}\ \costateA{n,N-n}\stateA{m,N-m}\nonumber\\
&=&Q_{i}(k,l)\mathscr{X}^{k,l}_{n}.
\end{eqnarray}
Clearly, this gives us four linear equations relating the S-matrix
from Case II to the S-matrix of Case I. A similar computation can
be worked out using $\Delta^{op}\mathbb{Q}_2^{\dag4}$, giving four
additional equations. We can cast the above formulae in a
convenient matrix form:
\begin{eqnarray}\label{eqn;SmatCase2easypart}
\left(\begin{smallmatrix}
  a_{4}(N-n-\ell_2) & a_{3}(\ell_1-n) & b_{4} & -b_{3} \\
  c_{4}(N-n-\ell_2) & c_{3}(\ell_1-n) & d_{4} & -d_{3} \\
\end{smallmatrix}\right)
\mathscr{Y}^{k,l}_n = \mathscr{X}^{k,l}_n
\left(\begin{smallmatrix}
  a_{2}(l-\ell_2) & a_{1}(\ell_1-k) & b_{2} & -b_{1} \\
  c_{2}(l-\ell_2) & c_{1}(\ell_1-k) & d_{2} & -d_{1} \\
\end{smallmatrix}\right),
\end{eqnarray}
with
\begin{eqnarray}
\mathscr{Y}^{k,l}_n &\equiv& \begin{pmatrix}
  \mathscr{Y}^{k,l;1}_{n;1} & \mathscr{Y}^{k,l;1}_{n;2} & \mathscr{Y}^{k,l;1}_{n;3} & \mathscr{Y}^{k,l;1}_{n;4} \\
  \mathscr{Y}^{k,l;2}_{n;1} & \mathscr{Y}^{k,l;2}_{n;2} & \mathscr{Y}^{k,l;2}_{n;3} & \mathscr{Y}^{k,l;2}_{n;4} \\
  \mathscr{Y}^{k,l;3}_{n;1} & \mathscr{Y}^{k,l;3}_{n;2} & \mathscr{Y}^{k,l;3}_{n;3} & \mathscr{Y}^{k,l;3}_{n;4} \\
  \mathscr{Y}^{k,l;4}_{n;1} & \mathscr{Y}^{k,l;4}_{n;2} & \mathscr{Y}^{k,l;4}_{n;3} & \mathscr{Y}^{k,l;4}_{n;4}
\end{pmatrix}.
\end{eqnarray}
Written in this way, the relation to (\ref{eqn;SymmPropSU22})
becomes apparent. However, because the matrix
$\mathscr{Y}^{k,l}_n$ has 16 unknown coefficients it is clear that
in order to fully determine $\mathscr{Y}^{k,l}_n$ (and therefore
the full Case II S-matrix ) one needs more equations in
addition to (\ref{eqn;SmatCase2easypart}).

\begin{figure}
  \centering
  \includegraphics[scale=1]{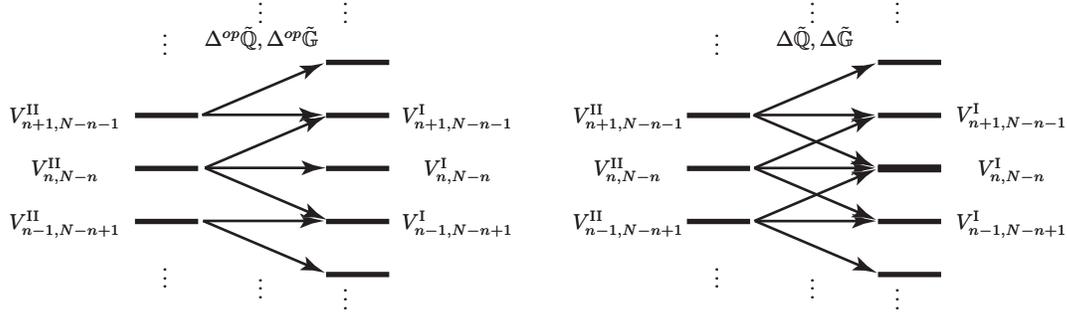}
  \caption{\label{Fig;Case2}Action of $\Delta^{op}\tilde{\mathbb{Q}},\Delta^{op}\tilde{\mathbb{G}}$ and
   $\Delta\tilde{\mathbb{Q}},\Delta\tilde{\mathbb{G}}$. They map Case II states
   (on the left) to Case I states (on the right).}
\end{figure}

These equations can be obtained via the Yangian generators.
Consider the following operators:
\begin{align}
&\Delta\tilde{\mathbb{Q}}=\Delta\hat{\mathbb{Q}}^1_{3}+\frac{\Delta\hat{\mathbb{L}}^1_2\Delta\mathbb{Q}^2_{3}}{\frac{\ell_1+\ell_2}{2}-(N+1+\delta
u
)}-\frac{\frac{\ell_1-\ell_2}{2}+N-2n+u_1+u_2}{\ell_1+\ell_2-(N+1+\delta
u )}\Delta\mathbb{L}^1_2\Delta\mathbb{Q}^2_{3},\nonumber\\
~\\
&\Delta\tilde{\mathbb{G}}=\Delta\hat{\mathbb{Q}}^{\dag4}_2+\frac{\Delta\hat{\mathbb{L}}^1_2\Delta\mathbb{Q}^{\dag4}_1}{\frac{\ell_1+\ell_2}{2}-(N+1+\delta
u)}+\frac{\frac{\ell_1-\ell_2}{2} +
N-2n+u_1+u_2}{\ell_1+\ell_2-2(N+1+\delta u
)}\Delta\mathbb{L}^1_2\Delta\mathbb{Q}^{\dag4}_1.\nonumber
\end{align}
These operators are chosen in such a way that \emph{only} states
of the form $\stateB{n,N-n}_i$ are mapped to $\stateA{n,N-n}_i$.
When we follow the same derivation as before, we see that this
fact is important in (\ref{eqn;LHSderivationCase2}) in order to be
able to factorize the matrix $\mathscr{Y}^{k,l}_n$ in front of the
final expression, and be therefore able to solve for it. In fact,
$\Delta\tilde{\mathbb{Q}}$ generically maps
\begin{eqnarray}
\Delta\tilde{\mathbb{Q}}:V^{\rm{II}}_{k,l} \longrightarrow
V^{\rm{I}}_{k+1,l-1}\oplus V^{\rm{I}}_{k,l}\oplus
V^{\rm{I}}_{k-1,l+1},
\end{eqnarray}
or, more precisely, we can write
\begin{align}
&\Delta\tilde{\mathbb{Q}}\stateB{k,l}_i =\\
&\ \ \tilde{Q}_i(k,l)\stateA{k,l}+\tilde{Q}^+_i(k+1,l-1)\stateA{k+1,l-1}+\tilde{Q}^-_i(k-1,l+1)\stateA{k-1,l+1}.\nonumber
\end{align}
This means that, if one follows (\ref{eqn;LHSderivationCase2}),
one obtains
\begin{align}
&\costateA{n,N-n}|~\Delta^{op}\tilde{\mathbb{Q}}\mathbb{S}~\stateB{k,l}_i
=\\
&\quad\sum_{j=1}^4\mathscr{Y}^{k,l;j}_{n;i}\tilde{Q}^{op}_j(n,N-n)+\mathscr{Y}^{k,l;j}_{n+1;i}\tilde{Q}^{op,+}_j(n,N-n)+\mathscr{Y}^{k,l;j}_{n-1;i}\tilde{Q}^{op,-}_j(n,N-n).\nonumber
\end{align}
However, the specific choice we made for
$\Delta\tilde{\mathbb{Q}}$ means that
$\tilde{Q}^{op,+}_j(n,N-n)=\tilde{Q}^{op,-}_j(n,N-n)=0$. In other
words, we can again put in evidence the matrix factor
$\mathscr{Y}^{k,l}_n$ on the left hand side of the final equation.
Since this is specifically tuned to work for the opposite
coproducts, the right hand side of the equation will not have this
property, and $\tilde{Q}^{\pm}$ will contribute there. This is
exemplified in figure \ref{Fig;Case2}.

For compactness, let us define $M\equiv N-2n$. By
combining all the equations one is lead to the following matrix
equation:
\begin{align}
&\left(
\begin{smallmatrix}
  a_4 & a_3 & 0 & 0 \\
  c_4 & c_3 & 0 & 0 \\
  0 & 0 & a_4 & a_3 \\
  0 & 0 & c_4 & c_3
\end{smallmatrix}\right)
A\mathscr{Y}^{k,l}_n =
\left(\begin{smallmatrix}
  a_2 & a_1 & 0 & 0 \\
  c_2 & c_1 & 0 & 0 \\
  0 & 0 & a_2 & a_1 \\
  0 & 0 & c_2 & c_1
\end{smallmatrix}\right)
\left[B^+\mathscr{X}^{k+1,l-1}_n + B^-\mathscr{X}^{k-1,l+1}_n +B
\mathscr{X}^{k,l}_n \right]
\end{align}
where the matrix on the left hand side is given by
\begin{align}
&A=
\begin{pmatrix}
  \scriptstyle{N-n-\ell_2} & \scriptstyle{0} & \scriptstyle{\frac{\mathscr{I}_{34}}{\mathscr{Q}_{34}}} & \scriptstyle{\frac{1}{\mathscr{Q}_{43}}} \\
  \scriptstyle{0} & \scriptstyle{\ell_1-n} & \scriptstyle{\frac{1}{\mathscr{Q}_{43}}} & \scriptstyle{\frac{\mathscr{I}_{43}}{\mathscr{Q}_{34}}} \\
  \scriptstyle{(N-n-\ell_2)(M-\delta u)} & \scriptstyle{(n-\ell_1)\ell_2\mathscr{I}_{34}} & \scriptstyle{\frac{(\delta u-M+\ell_2)\mathscr{I}_{34}}{\mathscr{Q}_{43}}} & \scriptstyle{\frac{\delta u+M+\ell_1-\ell_2\mathscr{Q}_{34}\overline{\mathscr{Q}}_{34}}{\mathscr{Q}_{43}}} \\
  \scriptstyle{(N-n-\ell_2)(\ell_1\mathscr{I}_{43})} & \scriptstyle{(\ell_1-n)(\delta u+M)} & \scriptstyle{\frac{M-\delta u-\ell_2+\ell_1\mathscr{Q}_{34}\overline{\mathscr{Q}}_{34}}{\mathscr{Q}_{43}}} & \scriptstyle{\frac{(\delta u+M+\ell_1)\mathscr{I}_{43}}{\mathscr{Q}_{34}}}
\end{pmatrix}
\end{align}
and the matrices on the right hand side by
\begin{align}
&B^+=
{\textstyle{\frac{2(\ell_1-k-1)\mathfrak{c}^-_{l-n}}{\tilde{\mathfrak{c}}^-_{-N}}}}
\left(
\begin{smallmatrix}
 0 & 0 & 0 & 0 \\
 0 & 0 & 0 & 0 \\
 l & 0 & \frac{\mathscr{I}_{12}}{\mathscr{Q}_{12}} & 0 \\
 0 & 0 & \frac{1}{\mathscr{Q}_{21}} & 0
\end{smallmatrix}
\right),& B^-=
{\textstyle{\frac{2(\ell_2-l-1)\mathfrak{c}^+_{n-l}}{\tilde{\mathfrak{c}}^-_{-N}}}}
\left(
\begin{smallmatrix}
 0 & 0 & 0 & 0 \\
 0 & 0 & 0 & 0 \\
 0 & 0 & 0 & \frac{1}{\mathscr{Q}_{12}} \\
 0 & k & 0 & \frac{\mathscr{I}_{21}}{\mathscr{Q}_{21}}
\end{smallmatrix}
\right),
\end{align}
\begin{align}
&B=
\begin{pmatrix}
  \scriptstyle{l-\ell_2} & \scriptstyle{0} & \scriptstyle{\frac{\mathscr{I}_{12}}{\mathscr{Q}_{12}}} & \scriptstyle{\frac{1}{\mathscr{Q}_{21}}} \\
  \scriptstyle{0} & \scriptstyle{\ell_1-k} & \scriptstyle{\frac{1}{\mathscr{Q}_{21}}} & \scriptstyle{\frac{\mathscr{I}_{21}}{\mathscr{Q}_{12}}} \\
  \scriptstyle{(l-\ell_2)(N-\delta u)} & \scriptstyle{(\ell_1-k)\ell_2\mathscr{I}_{12}} & \scriptstyle{\frac{(N-\delta u-\ell_2)\mathscr{I}_{12}}{\mathscr{Q}_{12}}} & \scriptstyle{\frac{N-\delta u-\ell_1-\ell_2\mathscr{Q}_{12}\overline{\mathscr{Q}}_{12}}{\mathscr{Q}_{12}}} \\
  \scriptstyle{(\ell_2-l)(\ell_1\mathscr{I}_{21})} & \scriptstyle{(\ell_1-k)(\delta u-N)} & \scriptstyle{\frac{\delta u-N+\ell_1\mathscr{Q}_{12}\overline{\mathscr{Q}}_{12}+\ell_2}{\mathscr{Q}_{12}}} & \scriptstyle{\frac{(\delta u-N+\ell_1)\mathscr{I}_{21}}{\mathscr{Q}_{12}}}
\end{pmatrix}&&\nonumber\\
&\qquad -2
\begin{pmatrix}
  \scriptstyle{0} & \scriptstyle{0} & \scriptstyle{0} & \scriptstyle{0} \\
  \scriptstyle{0} & \scriptstyle{0} & \scriptstyle{0} & \scriptstyle{0} \\
  \scriptstyle{\frac{l(1+n+k-\ell_1)(l-\ell_2)}{\tilde{\mathfrak{c}}^-_{-N}}} & \scriptstyle{0} & \scriptstyle{\frac{(l-\ell_2)(1+n+k-\ell_1)\mathscr{I}_{12}}{\tilde{\mathfrak{c}}^-_{-N}\mathscr{Q}_{12}}} & \scriptstyle{\frac{(\ell_1-k)(1+N-n+l-\ell_2)}{\tilde{\mathfrak{c}}^-_{-N}\mathscr{Q}_{21}}} \\
  \scriptstyle{0} & \scriptstyle{\frac{k(1+N-n+l-\ell_2)(k-\ell_1)}{\tilde{\mathfrak{c}}^-_{-N}}} & \scriptstyle{\frac{(l-\ell_2)(1+n+k-\ell_1)}{\tilde{\mathfrak{c}}^-_{-N}\mathscr{Q}_{21}}} & \scriptstyle{\frac{(\ell_1-k)(1+N-n+l-\ell_2)\mathscr{I}_{21}}{\tilde{\mathfrak{c}}^-_{-N}\mathscr{Q}_{12}}}
\end{pmatrix}&&\nonumber,
\end{align}
where we defined
\begin{align}
\tilde{\mathfrak{c}}^{\pm}_m=\delta u \pm \frac{\ell_1+\ell_2}{2}
-m+1.
\end{align}
Notice the similarities between the matrices $A,B$ and $B^+,B^-$.
From this, it is now straightforward to extract
$\mathscr{Y}^{k,l}_n$ by simple linear algebra
\begin{align}\label{eqn;SCase2}
\mathscr{Y}^{k,l}_n = A^{-1}\left(
\begin{smallmatrix}
  \frac{\mathscr{Q}_{32}}{\mathscr{Q}_{34}} & \frac{\mathscr{Q}_{31}}{\mathscr{Q}_{34}} & 0 & 0 \\
  \frac{\mathscr{Q}_{42}}{\mathscr{Q}_{43}} & \frac{\mathscr{Q}_{41}}{\mathscr{Q}_{43}} & 0 & 0 \\
  0 & 0 & \frac{\mathscr{Q}_{32}}{\mathscr{Q}_{34}} & \frac{\mathscr{Q}_{31}}{\mathscr{Q}_{34}} \\
  0 & 0 & \frac{\mathscr{Q}_{42}}{\mathscr{Q}_{43}} & \frac{\mathscr{Q}_{41}}{\mathscr{Q}_{43}}
\end{smallmatrix}\right)
\left\{\mathscr{X}^{k+1,l-1}_n B^+ + \mathscr{X}^{k-1,l+1}_n B^- +
\mathscr{X}^{k,l}_n B \right\}.
\end{align}
Note that the final result for $\mathscr{Y}^{k,l}_n$ purely
depends on the spectral parameters through their difference
$\delta u$, and the representation parameters only appear in the
combinations $\mathscr{Q}_{ij},\mathcal{I}_{ij}$ (modulo perhaps
the overall scalar factor, which, as usual, has to be determined
separately). The rest of the formula is taken care of purely by
combinatorial factors involving the integer bound state
components.

\section{Complete Solution of Case III}

We will perform here a similar construction as done in the
previous section, in order to solve Case III in terms of Case II.
Let us first set few additional notations. We introduce the
S-matrix at this level in the following way:
\begin{eqnarray}
\mathbb{S}\stateC{k,l}_i \equiv
\sum_{m=0}^{k+l}\sum_{j=1}^{6}\mathscr{Z}^{k,l;j}_{m;i}\stateC{m,k+l-m}_j.
\end{eqnarray}
It is clear that one can repeat a very similar derivation as
performed in (\ref{eqn;LHSderivationCase2}) and
(\ref{eqn;RHSderivationCase2}), where, instead of $\mathscr{X}$,
one has to think of having $\mathscr{Y}$ (and indices running over
the appropriate domains). This time, one again considers the
action of $\Delta\mathbb{Q}^1_{3},\Delta\mathbb{Q}_2^{\dag4}$. The
result is now the following matrix equations:
\begin{align}
&\left(\begin{smallmatrix}
  (n-\ell_1)a_3 & 0 & b_3 & 0 & b_4 & 0 \\
  (N-n-\ell_2)a_4 & b_4 & 0 & 0 & 0 & -b_3 \\
  0 & (n-\ell_1)a_3 & 0 & b_3 & (N-n-\ell_2)a_4 & 0 \\
  0 & 0 & (N-n-\ell_2)a_4 & b_4 & 0 & (n-\ell_1)a_3
\end{smallmatrix}\right)\mathscr{Z}^{k,l}_n
=\\
& \qquad\mathscr{Y}^{k,l}_n \left(\begin{smallmatrix}
  (k-\ell_1)a_1 & 0 & b_1 & 0 & b_2 & 0 \\
  (l-\ell_2)a_2 & b_2 & 0 & 0 & 0 & -b_1 \\
  0 & (k-\ell_1)a_1 & 0 & b_1 & (l-\ell_2)a_2 & 0 \\
  0 & 0 & (l-\ell_2)a_2 & b_2 & 0 & (k-\ell_1)a_1
\end{smallmatrix}\right)\nonumber,
\end{align}
and
\begin{align}
&\left(\begin{smallmatrix}
  (n-\ell_1)c_3 & 0 & d_3 & 0 & d_4 & 0 \\
  (N-n-\ell_2)c_4 & d_4 & 0 & 0 & 0 & -d_3 \\
  0 & (n-\ell_1)c_3 & 0 & d_3 & (N-n-\ell_2)d_4 & 0 \\
  0 & 0 & (N-n-\ell_2)c_4 & d_4 & 0 & (n-\ell_1)c_3
\end{smallmatrix}\right)\mathscr{Z}^{k,l}_n
=\\
& \qquad \mathscr{Y}^{k,l}_n \left(\begin{smallmatrix}
  (k-\ell_1)c_1 & 0 & d_1 & 0 & d_2 & 0 \\
  (l-\ell_2)c_2 & d_2 & 0 & 0 & 0 & -d_1 \\
  0 & (k-\ell_1)c_1 & 0 & d_1 & (l-\ell_2)c_2 & 0 \\
  0 & 0 & (l-\ell_2)c_2 & d_2 & 0 & (k-\ell_1)c_1
\end{smallmatrix}\right)\nonumber,
\end{align}
where
\begin{eqnarray}
\mathscr{Z}^{k,l}_n &\equiv& \begin{pmatrix}
  \scriptstyle{\mathscr{Z}^{k,l;1}_{n;1}} & \scriptstyle{\mathscr{Z}^{k,l;1}_{n;2}} & \scriptstyle{\mathscr{Z}^{k,l;1}_{n;3}} & \scriptstyle{\mathscr{Z}^{k,l;1}_{n;4}} & \scriptstyle{\mathscr{Z}^{k,l;1}_{n;5}} & \scriptstyle{\mathscr{Z}^{k,l;1}_{n;6}} \\
  \scriptstyle{\mathscr{Z}^{k,l;2}_{n;1}} & \scriptstyle{\mathscr{Z}^{k,l;2}_{n;2}} & \scriptstyle{\mathscr{Z}^{k,l;2}_{n;3}} & \scriptstyle{\mathscr{Z}^{k,l;2}_{n;4}} & \scriptstyle{\mathscr{Z}^{k,l;2}_{n;5}} & \scriptstyle{\mathscr{Z}^{k,l;2}_{n;6}} \\
  \scriptstyle{\mathscr{Z}^{k,l;3}_{n;1}} & \scriptstyle{\mathscr{Z}^{k,l;3}_{n;2}} & \scriptstyle{\mathscr{Z}^{k,l;3}_{n;3}} & \scriptstyle{\mathscr{Z}^{k,l;3}_{n;4}} & \scriptstyle{\mathscr{Z}^{k,l;3}_{n;5}} & \scriptstyle{\mathscr{Z}^{k,l;3}_{n;6}} \\
  \scriptstyle{\mathscr{Z}^{k,l;4}_{n;1}} & \scriptstyle{\mathscr{Z}^{k,l;4}_{n;2}} & \scriptstyle{\mathscr{Z}^{k,l;4}_{n;3}} & \scriptstyle{\mathscr{Z}^{k,l;4}_{n;4}} & \scriptstyle{\mathscr{Z}^{k,l;4}_{n;5}} & \scriptstyle{\mathscr{Z}^{k,l;4}_{n;6}} \\
  \scriptstyle{\mathscr{Z}^{k,l;5}_{n;1}} & \scriptstyle{\mathscr{Z}^{k,l;5}_{n;2}} & \scriptstyle{\mathscr{Z}^{k,l;5}_{n;3}} & \scriptstyle{\mathscr{Z}^{k,l;5}_{n;4}} & \scriptstyle{\mathscr{Z}^{k,l;5}_{n;5}} & \scriptstyle{\mathscr{Z}^{k,l;5}_{n;6}} \\
  \scriptstyle{\mathscr{Z}^{k,l;6}_{n;1}} & \scriptstyle{\mathscr{Z}^{k,l;6}_{n;2}} & \scriptstyle{\mathscr{Z}^{k,l;6}_{n;3}} & \scriptstyle{\mathscr{Z}^{k,l;6}_{n;4}} & \scriptstyle{\mathscr{Z}^{k,l;6}_{n;5}} & \scriptstyle{\mathscr{Z}^{k,l;6}_{n;6}}
\end{pmatrix}.
\end{eqnarray}
Once again, the relation with (\ref{eqn;SymmPropSU22}) is
apparent.

It is readily checked that in this case these equations are not
all independent. Hence, one similarly needs additional equations,
very much like in the previous section in order to compute
$\mathscr{Y}$. In that case, these additional equations were
provided by Yangian generators. In this case we are more fortunate
and do not need the Yangian, since one can consider the action of
$\Delta\mathbb{Q}^2_4$ and $\Delta\mathbb{Q}^{\dag3}_1$. It is
easy to check that, by repeating the above procedure using these
additional symmetries, one arrives at the following matrix
equations:
\begin{align}
&\left(\begin{smallmatrix}
  n a_3 & 0 & b_3 & 0 & 0 & -b_4 \\
  (N-n)a_4 & b_4 & 0 & 0 & b_3 & 0 \\
  0 & n a_3 & 0 & b_3 & 0 & (N-n) a_4 \\
  0 & 0 & (N-n)a_4 & b_4 & -n a_3 & 0
\end{smallmatrix}\right)\mathscr{Z}^{k,l}_n
= \tilde{\mathscr{Y}}^{k,l}_n \left(\begin{smallmatrix}
  k a_1 & 0 & b_1 & 0 & 0 & -b_2 \\
  l a_2 & b_2 & 0 & 0 & b_1 & 0 \\
  0 & k a_1 & 0 & b_1 & 0 & l a_2 \\
  0 & 0 & l a_2 & b_2 & -k a_1 & 0
\end{smallmatrix}\right)
\end{align}
and
\begin{align}
&\left(\begin{smallmatrix}
  n c_3 & 0 & d_3 & 0 & 0 & -d_4 \\
  (N-n)c_4 & d_4 & 0 & 0 & d_3 & 0 \\
  0 & n c_3 & 0 & d_3 & 0 & (N-n) c_4 \\
  0 & 0 & (N-n)c_4 & d_4 & -n c_3 & 0
\end{smallmatrix}\right)\mathscr{Z}^{k,l}_n
= \tilde{\mathscr{Y}}^{k,l}_n \left(\begin{smallmatrix}
  k c_1 & 0 & d_1 & 0 & 0 & -d_2 \\
  l c_2 & d_2 & 0 & 0 & d_1 & 0 \\
  0 & k c_1 & 0 & d_1 & 0 & l c_2 \\
  0 & 0 & l c_2 & d_2 & -k c_1 & 0
\end{smallmatrix}\right),
\end{align}
where we have defined
\begin{eqnarray}
\tilde{\mathscr{Y}}^{k,l}_n &\equiv& \begin{pmatrix}
  \mathscr{Y}^{k-1,l;1}_{n-1;1} & \mathscr{Y}^{k,l-1;1}_{n-1;2} & \mathscr{Y}^{k-1,l;1}_{n-1;3} & \mathscr{Y}^{k,l-1;1}_{n-1;4}  \\
  \mathscr{Y}^{k-1,l;2}_{n;1} & \mathscr{Y}^{k,l-1;2}_{n;2} & \mathscr{Y}^{k-1,l;2}_{n;3} & \mathscr{Y}^{k,l-1;2}_{n;4}  \\
  \mathscr{Y}^{k-1,l;3}_{n-1;1} & \mathscr{Y}^{k,l-1;3}_{n-1;2} & \mathscr{Y}^{k-1,l;3}_{n-1;3} & \mathscr{Y}^{k,l-1;3}_{n-1;4}  \\
  \mathscr{Y}^{k-1,l;4}_{n;1} & \mathscr{Y}^{k,l-1;4}_{n;2} & \mathscr{Y}^{k-1,l;4}_{n;3} & \mathscr{Y}^{k,l-1;4}_{n;4}
\end{pmatrix}.
\end{eqnarray}
Combining all of the above equations is sufficient in order to
solve for $\mathscr{Z}$. To be more precise, one can write the
equation for $\mathscr{Z}^{k,l}_{n}$ in the following way:
\begin{align}
&\left( \begin{smallmatrix}
 \scriptstyle{(n-\ell_1)\mathscr{Q}_{43}} & \scriptstyle{0} & \scriptstyle{\mathscr{I}_{43}} & \scriptstyle{0} & \scriptstyle{1} & \scriptstyle{0} \\
 \scriptstyle{0} & \scriptstyle{1} & \scriptstyle{0} & \scriptstyle{0} & \scriptstyle{0} & \scriptstyle{-\mathscr{I}_{43}} \\
 \scriptstyle{0} & \scriptstyle{(n-\ell_1)\mathscr{Q}_{43}} & \scriptstyle{0} & \scriptstyle{\mathscr{I}_{43}} & \scriptstyle{0} & \scriptstyle{0} \\
 \scriptstyle{-n \mathscr{Q}_{43}} & \scriptstyle{0} & \scriptstyle{-\mathscr{I}_{43}} & \scriptstyle{0} & \scriptstyle{0} & \scriptstyle{1} \\
 \scriptstyle{0} & \scriptstyle{-1} & \scriptstyle{0} & \scriptstyle{0} & \scriptstyle{-\mathscr{I}_{43}} & \scriptstyle{0} \\
 \scriptstyle{0} & \scriptstyle{-n \mathscr{Q}_{43}} & \scriptstyle{0} & \scriptstyle{-\mathscr{I}_{43}} & \scriptstyle{0} & \scriptstyle{0}
\end{smallmatrix}
\right) \mathscr{Z}^{k,l}_n =\\
&\qquad \qquad\qquad\qquad\qquad \check{\mathscr{Y}}^{k,l}_n
\left(
\begin{smallmatrix}
  \scriptstyle{(\ell_1-k ) \mathscr{Q}_{14}} & \scriptstyle{0} & \scriptstyle{\mathscr{I}_{41}} & \scriptstyle{0} & \scriptstyle{\mathscr{I}_{42}} & \scriptstyle{0} \\
  \scriptstyle{(l-\ell_2 ) \mathscr{Q}_{42}} & \scriptstyle{\mathscr{I}_{42}} & \scriptstyle{0} & \scriptstyle{0} & \scriptstyle{0} & \scriptstyle{-\mathscr{I}_{41}} \\
 \scriptstyle{0} &  \scriptstyle{(\ell_1-k ) \mathscr{Q}_{14}} & \scriptstyle{0} & \scriptstyle{\mathscr{I}_{41}} &  \scriptstyle{(\ell_2-l ) \mathscr{Q}_{42}} & \scriptstyle{0} \\
 \scriptstyle{0} & \scriptstyle{0} &  \scriptstyle{(l-\ell_2 ) \mathscr{Q}_{42}} & \scriptstyle{\mathscr{I}_{42}} & \scriptstyle{0} &  \scriptstyle{(\ell_1-k ) \mathscr{Q}_{14}} \\
 \scriptstyle{k \mathscr{Q}_{14}} & \scriptstyle{0} & \scriptstyle{-\mathscr{I}_{41}} & \scriptstyle{0} & \scriptstyle{0} & \scriptstyle{\mathscr{I}_{42}} \\
 \scriptstyle{-l \mathscr{Q}_{42}} & \scriptstyle{-\mathscr{I}_{42}} & \scriptstyle{0} & \scriptstyle{0} & \scriptstyle{-\mathscr{I}_{41}} & \scriptstyle{0} \\
 \scriptstyle{0} & \scriptstyle{k \mathscr{Q}_{14}} & \scriptstyle{0} & \scriptstyle{-\mathscr{I}_{41}} & \scriptstyle{0} & \scriptstyle{-l \mathscr{Q}_{42}} \\
 \scriptstyle{0} & \scriptstyle{0} & \scriptstyle{-l \mathscr{Q}_{42}} & \scriptstyle{-\mathscr{I}_{42}} & \scriptstyle{-k \mathscr{Q}_{14}} & \scriptstyle{0}
\end{smallmatrix}
\right)\nonumber,
\end{align}
with
\begin{eqnarray}
\check{\mathscr{Y}}^{k,l}_n\equiv
\left(\begin{array}{cccccccccccc}
  \scriptstyle{\mathscr{Y}^{k,l;1}_{n;1}} & \scriptstyle{\mathscr{Y}^{k,l;1}_{n;2}} & \scriptstyle{\mathscr{Y}^{k,l;1}_{n;3}} & \scriptstyle{\mathscr{Y}^{k,l;1}_{n;4}} & \scriptstyle{0} & \scriptstyle{0} & \scriptstyle{0} & \scriptstyle{0}  \\
  \scriptstyle{\mathscr{Y}^{k,l;2}_{n;1}} & \scriptstyle{\mathscr{Y}^{k,l;2}_{n;2}} & \scriptstyle{\mathscr{Y}^{k,l;2}_{n;3}} & \scriptstyle{\mathscr{Y}^{k,l;2}_{n;4}} & \scriptstyle{0} & \scriptstyle{0} & \scriptstyle{0} & \scriptstyle{0}  \\
  \scriptstyle{\mathscr{Y}^{k,l;3}_{n;1}} & \scriptstyle{\mathscr{Y}^{k,l;3}_{n;2}} & \scriptstyle{\mathscr{Y}^{k,l;3}_{n;3}} & \scriptstyle{\mathscr{Y}^{k,l;3}_{n;4}} & \scriptstyle{0} & \scriptstyle{0} & \scriptstyle{0} & \scriptstyle{0}  \\
  \scriptstyle{0} & \scriptstyle{0} & \scriptstyle{0} & \scriptstyle{0} & \scriptstyle{\mathscr{Y}^{k-1,l;1}_{n-1;1}} & \scriptstyle{\mathscr{Y}^{k,l-1;1}_{n-1;2}} & \scriptstyle{\mathscr{Y}^{k-1,l;1}_{n-1;3}} & \scriptstyle{\mathscr{Y}^{k,l-1;1}_{n-1;4}} \\
  \scriptstyle{0} & \scriptstyle{0} & \scriptstyle{0} & \scriptstyle{0} & \scriptstyle{\mathscr{Y}^{k-1,l;2}_{n;1}} & \scriptstyle{\mathscr{Y}^{k,l-1;2}_{n;2}} & \scriptstyle{\mathscr{Y}^{k-1,l;2}_{n;3}} & \scriptstyle{\mathscr{Y}^{k,l-1;2}_{n;4}} \\
  \scriptstyle{0} & \scriptstyle{0} & \scriptstyle{0} & \scriptstyle{0} & \scriptstyle{\mathscr{Y}^{k-1,l;3}_{n-1;1}} & \scriptstyle{\mathscr{Y}^{k,l-1;3}_{n-1;2}} & \scriptstyle{\mathscr{Y}^{k-1,l;3}_{n-1;3}} & \scriptstyle{\mathscr{Y}^{k,l-1;3}_{n-1;4}}
\end{array}\right).
\end{eqnarray}
The explicit matrix inversion gives
\begin{eqnarray}\label{eqn;SCase3}
&&\mathscr{Z}=\left(
\begin{smallmatrix}
 \frac{1}{\ell_1 \mathscr{Q} _{34}} & \frac{1}{\ell_1 \mathscr{Q} _{34} \mathscr{I} _{43}} & \scriptstyle{0} & \frac{1}{\ell_1 \mathscr{Q} _{34}} & \frac{1}{\ell_1 \mathscr{Q} _{34} \mathscr{I} _{43}} & \scriptstyle{0} \\
 \scriptstyle{0} & \scriptstyle{0} & \frac{1}{\ell_1 \mathscr{Q} _{34}} & \scriptstyle{0} & \scriptstyle{0} & \frac{1}{\ell_1 \mathscr{Q} _{34}} \\
 \frac{n}{\ell_1 \mathscr{I} _{43}} & \frac{n-\ell_1}{\ell_1 \mathscr{I} _{43}^2} & \frac{1}{\ell_1 \mathscr{Q} _{34} \mathscr{I} _{43}^2} & \frac{n-\ell_1}{\ell_1 \mathscr{I} _{43}} & \frac{n}{\ell_1 \mathscr{I}
   _{43}^2} & \frac{1}{\ell_1 \mathscr{Q} _{34} \mathscr{I} _{43}^2} \\
 \scriptstyle{0} & \scriptstyle{0} & \frac{n}{\ell_1 \mathscr{I} _{43}} & \scriptstyle{0} & \scriptstyle{0} & \frac{n-\ell_1}{\ell_1 \mathscr{I} _{43}} \\
 \scriptstyle{0} & \scriptstyle{0} & \frac{1}{\ell_1 \mathscr{Q} _{43} \mathscr{I} _{43}} & \scriptstyle{0} & \scriptstyle{-}\frac{1}{\mathscr{I} _{43}} & \frac{1}{\ell_1 \mathscr{Q} _{43} \mathscr{I} _{43}} \\
 \scriptstyle{0} & \scriptstyle{-}\frac{1}{\mathscr{I} _{43}} & \frac{1}{\ell_1 \mathscr{Q} _{34} \mathscr{I} _{43}} & \scriptstyle{0} & \scriptstyle{0} & \frac{1}{\ell_1 \mathscr{Q} _{34} \mathscr{I} _{43}}
\end{smallmatrix}
\right)\check{\mathscr{Y}}^{k,l}_n\times\\
&&\qquad\qquad\qquad\qquad\times \left(
\begin{smallmatrix}
  \scriptstyle{(\ell_1-k ) \mathscr{Q}_{14}} & \scriptstyle{0} & \scriptstyle{\mathscr{I}_{41}} & \scriptstyle{0} & \scriptstyle{\mathscr{I}_{42}} & \scriptstyle{0} \\
  \scriptstyle{(l-\ell_2 ) \mathscr{Q}_{42}} & \scriptstyle{\mathscr{I}_{42}} & \scriptstyle{0} & \scriptstyle{0} & \scriptstyle{0} & \scriptstyle{-\mathscr{I}_{41}} \\
 \scriptstyle{0} &  \scriptstyle{(\ell_1-k ) \mathscr{Q}_{14}} & \scriptstyle{0} & \scriptstyle{\mathscr{I}_{41}} &  \scriptstyle{(\ell_2-l ) \mathscr{Q}_{42}} & \scriptstyle{0} \\
 \scriptstyle{0} & \scriptstyle{0} &  \scriptstyle{(l-\ell_2 ) \mathscr{Q}_{42}} & \scriptstyle{\mathscr{I}_{42}} & \scriptstyle{0} &  \scriptstyle{(\ell_1-k ) \mathscr{Q}_{14}} \\
 \scriptstyle{k \mathscr{Q}_{14}} & \scriptstyle{0} & \scriptstyle{-\mathscr{I}_{41}} & \scriptstyle{0} & \scriptstyle{0} & \scriptstyle{\mathscr{I}_{42}} \\
 \scriptstyle{-l \mathscr{Q}_{42}} & \scriptstyle{-\mathscr{I}_{42}} & \scriptstyle{0} & \scriptstyle{0} & \scriptstyle{-\mathscr{I}_{41}} & \scriptstyle{0} \\
 \scriptstyle{0} & \scriptstyle{k \mathscr{Q}_{14}} & \scriptstyle{0} & \scriptstyle{-\mathscr{I}_{41}} & \scriptstyle{0} & \scriptstyle{-l \mathscr{Q}_{42}} \\
 \scriptstyle{0} & \scriptstyle{0} & \scriptstyle{-l \mathscr{Q}_{42}} & \scriptstyle{-\mathscr{I}_{42}} & \scriptstyle{-k \mathscr{Q}_{14}} & \scriptstyle{0}
\end{smallmatrix}
\right).\nonumber
\end{eqnarray}
It is now straightforward to do the matrix multiplication. This
solves the final case. Once again, the dependence of the entries
solely on the difference of the spectral parameters, and on the
characteristic combinations of representation labels already
observed in Case II, is a noticeable feature of the result.

\section{Reduction and Comparison}\label{sect;Reduction}

Let us now compare our formulae with the known S-matrices. Here,
one runs into potential difficulties. The formulae from the
previous sections were derived for generic bound states, and one
might wonder whether there could be obstructions for small bound
states. A first problem arises when $n$ is comparable to
$\ell_1,\ell_2$. A second problem is encountered for $n=0,n=k+l$,
since the basis of two-particle states in these two cases is
lower-dimensional. One may wonder whether our formulae
\begin{eqnarray}
\S\stateA{k,l} &=& \sum_{n=0}^{k+l}\mathscr{X}^{k,l}_{n}\stateA{n,N-n}\\
\S\stateB{k,l}_i &=& \sum_{n=0}^{k+l}\sum_{j=1}^4\mathscr{Y}^{k,l;j}_{n;i}\stateB{n,N-n}_j\label{eqn;reduction}\\
\S\stateC{k,l}_i &=&
\sum_{n=0}^{k+l}\sum_{j=1}^6\mathscr{Z}^{k,l;j}_{n;i}\stateC{n,N-n}_j,
\end{eqnarray}
with $N=k+l$ and $\mathscr{Y},\mathscr{Z}$ given by
(\ref{eqn;SCase2}) and (\ref{eqn;SCase3}), remain valid also for
these particular values.

It turns out that this is indeed the case. Let us deal with the
first problem. One can see from (\ref{eqn;SCase1}) that, when
$n>\ell_1$, precisely the unwanted S-matrix elements vanish,
basically thanks to the vanishing of the correspondent
coefficients $\mathscr{X}^{k,l}_{n}$.

Concerning the second potential problem, we notice that the issue
arises only for Case II and III states. In Case II, the
corresponding sum on the right hand side of (\ref{eqn;reduction})
contains terms like
\begin{eqnarray}
\mathscr{Y}^{k,l;4}_{0;i}\stateB{0,N}_4.
\end{eqnarray}
But, as seen from (\ref{eqn;BasisCase2}), $\stateB{0,N}_4$ is not
well-defined (actually it is not part of our bound state
representation). Hence, the S-matrix transition amplitudes toward
these states, $\mathscr{Y}^{k,l;4}_{0;i}$, should vanish
identically. We verified that this indeed turns out to be the
case, which means that these states completely decouple.

More specifically, from (\ref{eqn;SCase1}) it can be shown that
\begin{eqnarray}
\mathscr{X}^{k+1,l-1}_0 &=&\frac{\ell_2-l}{\delta u-\frac{\ell_1-\ell_2}{2}-l+1} \mathscr{X}^{k,l}_0\\
\mathscr{X}^{k-1,l+1}_0 &=&\frac{\delta
u-\frac{\ell_1-\ell_2}{2}-l}{\ell_2-l-1} \mathscr{X}^{k,l}_0.
\end{eqnarray}
This means that in (\ref{eqn;SCase2}) one can pull out a factor
$\mathscr{X}^{k,l}_0$. The remaining matrix part is
straightforwardly seen to have zeroes for the states corresponding
to the amplitudes $\mathscr{Y}^{k,l;4}_{0;i}$, for all $i=1,2,3,4$
as it should indeed be the case. In other words, one can
unambiguously write
\begin{eqnarray}
\S \stateB{k,l}_{i} = \sum_{n=0}^{k+l}\sum_{j=1}^{4}
\mathscr{Y}^{k,l;j}_{n;i}\stateB{n,N-n}_j,
\end{eqnarray}
where $\mathscr{Y}^{k,l;j}_{n;i}$ is given by the complete
$4\times4$ matrix from (\ref{eqn;SCase2}). The same is true for
Case III states.


One can now compare our coefficients against the known S-matrices.
Complete agreement is found with $\S^{AA},\S^{AB},\S^{BB}$ from
\cite{Beisert:2005tm,Arutyunov:2006yd,Arutyunov:2008zt}. We also
checked several coefficients of the S-matrix $\S^{1\ell}$ from
\cite{Bajnok:2008bm}, and also in that case we find agreement with
our results.

To make this comparison more explicit, let us list some explicit
entries from the S-matrix. These entries can be directly compared
against the coefficients from the known S-matrices like the
fundamental S-matrix (\ref{eqn;Sfund}). The coefficients we list
here will also be needed in later chapters.

The lowest entries of the Case II S-matrix are given
by\footnote{We suppress the dependence on momenta in order to have
a lighter notation. All functions appearing in this section have
to be understood as $\mathscr{X} \equiv \mathscr{X}(p_1, p_2)$,
$\mathscr{Y} \equiv \mathscr{Y}(p_1, p_2)$, $\mathscr{Z} \equiv
\mathscr{Z}(p_1, p_2)$ (indices are omitted here for simplicity).}
\begin{eqnarray}
\mathscr{Y}^{k,0;1}_{k;1} &=&
\frac{x^+_1-x^+_2}{x^-_1-x^+_2}\sqrt{\frac{x^-_1}{x^+_1}}\left[1-\frac{k}{\delta
u+\frac{\ell_1-\ell_2}{2} }\right]\mathscr{X}^{k,0}_{k},\\
\mathscr{Y}^{k,0;2}_{k;2}
&=&\frac{x^-_1-x^-_2}{x^-_1-x^+_2}\sqrt{\frac{x^+_2}{x^-_2}}\mathscr{X}^{k,0}_{k},\\
\mathscr{Y}^{k,0;1}_{k;2}
&=&\frac{x^-_2-x^+_2}{x^-_1-x^+_2}\sqrt{\frac{x^-_1 x^+_2}{x^+_1
x^-_2}}\frac{\sqrt{\ell_1}\eta(p_1)}{\sqrt{\ell_2}\eta(p_2)}\frac{k-\ell_1}{\ell_1}\mathscr{X}^{k,0}_{k},\\
\mathscr{Y}^{k,0;2}_{k;1}
&=&\frac{x^+_1-x^-_1}{x^-_1-x^+_2}\frac{\sqrt{\ell_2}\eta(p_2)}{\sqrt{\ell_1}\eta(p_1)}\mathscr{X}^{k,0}_{k},\\
\mathscr{Y}^{k,0;4}_{k;1} &=&\frac{\sqrt{\ell_1\ell_2}\eta(p_1)\eta(p_2)}{x^+_1 x^+_2-1}\frac{k}{i\ell_1}\mathscr{X}^{k,0}_{k},\\
\mathscr{Y}^{k,0;4}_{k;4}
&=&\sqrt{\frac{x^+_2}{x^-_2}}\frac{x^+_1x^-_2-1}{x^+_1x^+_2-1}\mathscr{X}^{k,0}_{k},\\
\mathscr{Y}^{k,0;1}_{k;4} &=&
\frac{i}{\sqrt{\ell_1\ell_2}\eta(p_1)\eta(p_2)}\sqrt{\frac{x^+_1x^+_2}{x^-_1x^-_2}}\frac{(x^-_1-x^+_1)(x^-_2-x^+_2)}{x^+_1x^+_2-1}\mathscr{X}^{k,0}_{k},
\end{eqnarray}
From the Case III S-matrix we will encounter
\begin{eqnarray}
\mathscr{Z}^{k,0;1}_{k;1} &=&\left[1-\frac{2ik}{g}\frac{x^+_1
(x^-_2-x^-_1 x^+_1
x^+_2)}{(x^-_2-x^+_1)(1-x^-_1x^+_1)(1-x^+_1x^+_2)}\right]\frac{\mathscr{X}^{k,0}_{k}}{\mathcal{D}},\\
\mathscr{Z}^{k,0;1}_{k;3} &=& \frac{2
(x^-_1-x^+_1)(x^+_1)^2(x^-_2-x^+_2)} {g
(x^+_1-x^-_2)(1-x^+_1x^-_1) (1-x^+_1
x^+_2)\eta(p_1)^2}\frac{\mathscr{X}^{k,0}_{k}}{\mathcal{D}},\\
\mathscr{Z}^{k,0;3}_{k;1} &=& \frac{i k (\ell_1-k)}{\ell_1}\frac{
(x^-_2-x^+_2) \eta(p_1)^2}{(x^+_1-x^-_2)(1-x^+_1
x^+_2)}\frac{\mathscr{X}^{k,0}_{k}}{\mathcal{D}},\\
\mathscr{Z}^{k,0;3}_{k;3} &=&\left[
\frac{(x^+_1-x^+_2)(1-x^-_2x^+_1)}{(x^+_1-x^-_2)(1-x^+_1 x^+_2)} +
\frac{2ik}{g}\frac{x^+_1
(x^+_2-x^-_1x^+_1x^-_2)}{(x^+_1-x^-_2)(1-x^-_1
x^+_1)(1-x^+_1x^+_2)}\right]\frac{\mathscr{X}^{k,0}_{k}}{\mathcal{D}}\quad
\quad
\end{eqnarray}
and
\begin{eqnarray}
\mathscr{Z}^{k,0;6}_{k;1} &=&
\frac{ik\sqrt{\ell_2}\eta(p_1)\eta(p_2)}{\sqrt{\ell_1}}\frac{
x^-_1-x^-_2 }{(x^-_1-x^+_2)
(1-x^+_1x^+_2)}\sqrt{\frac{x^+_2}{x^-_2}}\mathscr{X}^{k,0}_k,\\
\mathscr{Z}^{k,0;6}_{k;3}
&=&\sqrt{\frac{\ell_2}{\ell_1}}\frac{\eta(p_2)}{\eta(p_1)} \frac{
(x^-_1-x^+_1)(x^-_2
x^+_1-1)}{(x^-_1-x^+_2)(x^+_1x^+_2-1)}\sqrt{\frac{x^+_2}{x^-_2}}
\mathscr{X}^{k,0}_k,\\
\mathscr{Z}^{k,0;6}_{k;6}&=&\frac{(x^-_1-x^-_2)(x^-_2x^+_1-1)x^+_2}
{(x^-_1-x^+_2)(x^+_1 x^+_2-1)x^-_2}\mathscr{X}^{k,0}_k,\\
\mathscr{Z}^{k,0;3}_{k;6}&=&\frac{(\ell_1-k)\eta
(p_1)}{\sqrt{\ell_1\ell_2}\eta
(p_2)}\frac{(x^-_2x^+_1-1)(x^-_2-x^+_2) x^+_2}{
(x^-_1-x^+_2)(x^+_1x^+_2-1)x^-_2}\sqrt{\frac{x^-_1}{x^+_1}}\mathscr{X}^{k,0}_k,\\
\mathscr{Z}^{k,0;1}_{k;6}&=&\frac{i}{\sqrt{\ell_1\ell_2}\eta(p_1)
\eta(p_2)}\frac{(x^-_1-x^-_2) (x^-_1-x^+_1)(x^-_2-x^+_2) x^+_2}{
(x^-_1-x^+_2)(x^+_1 x^+_2-1)x^-_2}
\sqrt{\frac{x^+_1}{x^-_1}}\mathscr{X}^{k,0}_k.
\end{eqnarray}

\section{Summary}\label{sec;ConclSmat}

In this chapter we explicitly constructed S-matrix that
intertwines two symmetric short representations of $\alg{h}$ by
using Yangian symmetry.

Because of ${\alg{su}(2)}\times {\alg{su}(2)}$ invariance, when
this S-matrix acts on such a (tensor-product) bound state
representation space, it leaves five different subspaces
invariant. Each of these subspaces is characterized by a specific
assignment of ${\alg{su}(2)}\times {\alg{su}(2)}$ Dynkin labels,
which are quantum numbers that are trivially conserved during the
scattering. We found that two pairs of these subspaces are simply
related to each other by exchanging the type of fermions.
Therefore, we found only three non-equivalent cases. The S-matrix
has the following block-diagonal form:
\begin{eqnarray}
\S=\begin{pmatrix}
  \fbox{\small{$\mathscr{X}$}} & ~ & ~ & ~ & ~ \\
  ~ & \fbox{\LARGE{$\mathscr{Y}$}} & ~ & \mbox{\Huge{$0$}} & ~ \\
  ~ & ~ & \fbox{\Huge{$\mathscr{Z}$}} & ~ & ~ \\
  ~ & \mbox{\Huge{$0$}} & ~ & \fbox{\LARGE{$\mathscr{Y}$}} & ~ \\
  ~ & ~ & ~ & ~ & \fbox{\small{$\mathscr{X}$}}
\end{pmatrix}.
\end{eqnarray}
The outer blocks scatter states from Case I (\ref{eqn;BasisCase1})
\begin{eqnarray}
\mathscr{X}:\stateA{k,l}\mapsto \sum_{m=0}^{k+l}
\mathscr{X}^{k,l}_m\stateA{m,k+l-m},
\end{eqnarray}
where $\mathscr{X}^{k,l}_m$ is given by equation
(\ref{eqn;SCase1}). The blocks $\mathscr{Y}$ describe the
scattering of states from Case II (\ref{eqn;BasisCase2})
\begin{eqnarray}
\mathscr{Y}:\stateB{k,l}_j\mapsto \sum_{m=0}^{k+l}\sum_{j=1}^{4}
\mathscr{Y}^{k,l;j}_{m;i}\stateB{m,k+l-m}_j.
\end{eqnarray}
These S-matrix elements are given in equation (\ref{eqn;SCase2}).
Finally, the middle block deals with the third case
(\ref{eqn;BasisCase3})
\begin{eqnarray}
\mathscr{Z}:\stateC{k,l}_j\mapsto \sum_{m=0}^{k+l}\sum_{j=1}^{6}
\mathscr{Z}^{k,l;j}_{m;i}\stateC{m,k+l-m}_j,
\end{eqnarray}
where $\mathscr{Z}^{k,l;j}_{m;i}$ can be found in
(\ref{eqn;SCase3}).

This S-matrix is canonically normalized, namely, it leaves the
vacuum state (\ref{eqn;BoundStateVacuum}) exactly invariant. The
full $\ads$ string bound state S-matrix is obtained by taking two
copies of the above derived S-matrix and multiplying \emph{each
one of them} with the phase factor \cite{Arutyunov:2008zt}
\begin{align}\label{eqn;FullPhase}
S_{0}(p_{1},p_{2})=
\left(\frac{x_{1}^{-}}{x_{1}^{+}}\right)^{\frac{\ell_2}{2}}\left(\frac{x_{2}^{+}}{x_{2}^{-}}\right)^{\frac{\ell_1}{2}}\sigma(x_{1},x_{2})\sqrt{G(\ell_2-\ell_1)G(\ell_2+\ell_1)}\prod_{q=1}^{\ell_1-1}G(\ell_2-\ell_1+2q),
\end{align}
where $G$ has been given in (\ref{fattoreG}). Concretely the
complete S-matrix is given by
\begin{align}
\S_{{\rm Full}}(p_1,p_2) = S_{0}(p_{1},p_{2})^2
~\S(p_1,p_2)\otimes \S(p_1,p_2),
\end{align}
which for explicit bound state numbers $\ell_1,\ell_2$ is a
$(16\ell_1\ell_2)^2$ by $(16\ell_1\ell_2)^2$ dimensional matrix.

%% file: RMatrix2.tex
\chapter{The Classical r-Matrix}\label{chap;Rmat}

For many integrable models the semiclassic limit of the S-matrix
is described by a special object, called classical $r$-matrix.
This matrix is a universal structure in the sense that it can be
defined purely in algebraic terms and hence it is representation
independent. Its possible forms have been classified for simple
Lie algebras \cite{BelDrin} and have also been studied for certain
superalgebras \cite{Zhang:1990du,Leites:1984pt,Karaali200483}.
However, those results do not apply to $\alg{h}$.

Consider a generic Yangian evaluation representation and rescale
$u\rightarrow u/\hbar$. It is readily seen that the coproduct
of the Yangian generators (\ref{eqn;StandardYangHopf}) becomes of
the form
\begin{align}
\Delta\hat{\mathbb{J}}^A = \hat{\mathbb{J}}^A\otimes\mathbbm{1} +
\mathbbm{1}\otimes\hat{\mathbb{J}}^A + \hbar
f^A_{BC}\mathbb{J}^B\otimes\mathbb{J}^C.
\end{align}
Obviously, this induces an $\hbar$ dependence in the S-matrix, and
one generically finds that for small $\hbar$
\begin{align}
\S = \mathbbm{1} + \hbar r + \mathcal{O}(\hbar^2).
\end{align}
The matrix $r$ is called the classical $r$-matrix. For the Yangian
of $\alg{h}$ the situation is non-generic
since $u$ is actually fixed in terms of the underlying
representation. The conventions used here are such that a rescaling by
$\hbar$ is not needed, instead one can define a classical limit by
identifying $\hbar = g^{-1}$.

Generically, the classical $r$-matrix for Yangians in the
evaluation representation can be given purely in terms of algebra
generators
\begin{align}\label{eqn;standardClassicalR}
r = \frac{K_{AB}\mathbb{J}^A\otimes\mathbb{J}^B}{u_1-u_2},
\end{align}
where $K_{AB}$ is the Killing form. It satisfies the classical
Yang-Baxter equation
\begin{align}
[r_{12},r_{13}] + [r_{12},r_{23}] + [r_{13},r_{23}]=0.
\end{align}
However, since the Killing form $K_{AB}$ of $\alg{h}$ is zero, this formula is not applicable.
Nevertheless, the classical limit of the S-matrix discussed in the previous chapter was studied in
\cite{Torrielli:2007mc} and shortly after a proposal of the
classical $r$-matrix for the $\ads$ superstring was put forward in
\cite{Moriyama:2007jt}. A different proposal closer to
(\ref{eqn;standardClassicalR}) was made in \cite{Beisert:2007ty}. In
this chapter we will discuss both proposals and show that the
second indeed agrees with the (semi-)classical limit of the bound
state S-matrices.

\section{The near plane-wave limit}

As was discussed in  Section \ref{sect;Parameterization}, in order
to describe bound states with momentum $p$ and bound state number $\ell$ it is convenient to
introduce parameters $x^{\pm}$ such that
\begin{eqnarray}
\frac{x^+}{x^-} = e^{ip}, \qquad x^+ + \frac{1}{x^+} -x^-
-\frac{1}{x^-} = \frac{2i\ell}{g},
\end{eqnarray}
where $g = \sqrt{\lambda}/2\pi$ is related to the 't Hooft
coupling $\lambda$. A convenient parameterization of $x^\pm$ that
can be used to study the large coupling limit $g\rightarrow\infty$
was introduced in \cite{Arutyunov:2006iu}.
\begin{eqnarray}\label{eqn;ParameterXpm}
x^{\pm}_{i}& =&
x_{i}\left(\sqrt{1-\frac{(\ell_i/g)^{2}}{(x_{i}-\frac{1}{x_{i}})^{2}}}\pm
\frac{i \ell_i/g}{x_{i}-\frac{1}{x_{i}}}\right).
\end{eqnarray}
In this parametrization, the Hamiltonian is given by
\begin{eqnarray}
H = \ell \frac{x^2+1}{x^2-1}
\end{eqnarray}
and the momentum is related to the parameter $x$ by
\begin{eqnarray}
\sin \frac{p}{2} = \frac{\ell}{g} \frac{x}{x^2-1}.
\end{eqnarray}
Notice that the energy does not depend explicitly on the coupling
$g$.

The large coupling limit corresponds to the semiclassical limit of
spinning strings \cite{Frolov:2002av} or to the near plane-wave
limit \cite{Berenstein:2002jq}. In other words, we can indeed
identify the inverse coupling $1/g$ with the $\hbar$ parameter
discussed above. In the remainder of the chapter we will study the
classical r-matrix by taking the $g\rightarrow\infty$ limit of the
S-matrix (\ref{eqn;GeneralBoundStateMatrix}). To this end we will
need the expansion of the parameters $a,b,c,d$
(\ref{eqn;ABCDparameters}) describing the bound state
representation. We denote the near-classical limit of these
parameters as $\alg{a},\alg{b},\alg{c},\alg{d}$:
\begin{eqnarray}\label{eqn;classicalCoeffi}
\begin{array}{lll}
    \alg{a}= \frac{x}{\sqrt{x^2-1}}, &\ &
    \alg{b}= -\frac{1}{\sqrt{x^2-1}}, \\
    \alg{c}= -\frac{1}{\sqrt{x^2-1}}, &\ &
    \alg{d}= \frac{x}{\sqrt{x^2-1}}.
\end{array}
\end{eqnarray}
The square root factors are coming from expanding $\eta$
(\ref{eqn;DefEta}). The other parameter that appears in the
S-matrices is the Yangian evaluation parameter $u$
(\ref{eqn;defU}) and this reduces to
\begin{eqnarray}\label{eqn;classicalU}
u =
-\frac{ig}{2}\left(x+\frac{1}{x}\right)+\frac{i\ell^2}{4g}\frac{x+\frac{1}{x}}{\left(x-\frac{1}{x}\right)^2}
+ \mathcal{O}\left(g^{-3}\right).
\end{eqnarray}
Finally the braiding factor $\mathbb{U}$ is related to the
momentum $p$ and its semiclassical expansion is easily found to be
\begin{align}
\sqrt{\frac{x^+}{x^-}} = 1 + \frac{i\ell}{g} \frac{x}{x^2-1}.
\end{align}

\section{The Moriyama-Torrielli proposal}

The first proposal for the classical r-matrix has been made in
\cite{Moriyama:2007jt} and is given by
\begin{align}\label{eqn;MT-Rmat}
r = \sum_{n=0}^{\infty}
&\left[(\mathbb{L}^a_b)_n\otimes(\tilde{\mathbb{L}}^b_a)_{-n-1}-
\left(\mathbb{L}^a_b\right)_{-n-1}\otimes\left(\tilde{\mathbb{L}}^b_a\right)_{n}-\right.\nonumber\\
&\left(\mathbb{R}^{\alpha}_{\beta}\right)_{n}\otimes\left(\tilde{\mathbb{R}}^{\beta}_{\alpha}\right)_{-n-1}+
\left(\mathbb{R}^{\alpha}_{\beta}\right)_{-n-1}\otimes\left(\tilde{\mathbb{R}}^{\beta}_{\alpha}\right)_{n}+\\
& \left.+ \left(\mathbb{Q}^a_{\alpha}\right)_n\otimes\left(\mathbb{Q}^{\dag\alpha}_{a}\right)_{-n-1}-
\left(\mathbb{Q}^a_{\alpha}\right)_{n}\otimes\left(\mathbb{Q}^{\dag\alpha}_{a}\right)_{-n-1} +
\fHH_n\otimes\mathbb{B}_{-n-1} +
\mathbb{B}_{n}\otimes\fHH_{-n-1}\right].\nonumber
\end{align}
where the bosonic generators are given by
\begin{align}
& (\mathbb{L}^{a}_{b})_n  = \frac{x^{n+1}-x^{-n-1}}{x-\frac{1}{x}}
\mathbb{L}^{a}_{b}, && (\tilde{\mathbb{L}}^{a}_{b})_n  =
\frac{x^{n-1}-x^{1-n}}{x-\frac{1}{x}}
\mathbb{L}^{a}_{b},\nonumber\\
&(\mathbb{R}^{\alpha}_{\beta})_n  =
\frac{x^{1+n}-x^{-1-n}}{x-\frac{1}{x}}
\mathbb{R}^{\alpha}_{\beta}, &&
(\tilde{\mathbb{R}}^{\alpha}_{\beta})_n  =
\frac{x^{n-1}-x^{1-n}}{x-\frac{1}{x}}
\mathbb{R}^{\alpha}_{\beta}\\
&\mathbb{H}_n = \tilde{\mathbb{H}}_n =
\frac{x^{n+1}+x^{-n-1}}{x+\frac{1}{x}}\mathbb{H}, &&
\mathbb{B}_n = \tilde{\mathbb{B}}_n =
\frac{x^{n}-x^{-n}}{2}\frac{x+\frac{1}{x}}{x-\frac{1}{x}}\mathbb{B}\nonumber.
\end{align}
The operator $\mathbb{B}$ is the extra generator that makes
$\alg{su}(2|2)$ into $\alg{u}(2|2)$ and is, in the operator
language given by (\ref{eqn;defBoperator})
\begin{align}
\mathbb{B} = \frac{1}{2(a d+ b
c)}\left(w_a\frac{\partial}{\partial w_a}
-\theta_\alpha\frac{\partial}{\partial \theta_\alpha}\right).
\end{align}
The supersymmetry generators are given by
\begin{align}
& (\mathbb{Q}^{a}_{\alpha})_n  = (\tilde{\mathbb{Q}}^{a}_{\alpha})_n  =
\mathbb{Q}^{a}_{\alpha}(x^n\Pi_b + x^{-n}\Pi_f),\nonumber\\
& (\mathbb{Q}_{a}{^\dag\alpha})_n  = (\tilde{\mathbb{Q}}_{a}^{\dag\alpha})_n  =
\mathbb{Q}_{a}^{\dag\alpha}(x^{-n}\Pi_b + x^{n}\Pi_f).
\end{align}
In the above expression $\Pi_b$ and $\Pi_f$ are the projectors on
the bosonic and fermionic subspace, respectively.

The dependence of the generators on the parameter $x$ is quite
different from the standard evaluation representation, especially
because of the presence of the bosonic and fermionic projectors.

One can check that this classical $r$-matrix matches with the
semiclassical limit of the fundamental S-matrix. Nevertheless,
when one considers bound state S-matrices disagreement is found.
This shows that this proposal for classical $r$-matrix is not
universal.

\section{The Beisert-Spill proposal}

The other conjecture for the classical $r$-matrix was put forward
in \cite{Beisert:2007ty}. This is again done by introducing the
extra generator $\mathbb{B}$ from (\ref{eqn;defBoperator}). This
proposal for the classical $r$-matrix was made in terms of algebra
generators in the evaluation representation and its form is close
to the standard classical $r$-matrix for Yangians
(\ref{eqn;standardClassicalR}).

The $r$-matrix is given in terms of algebra generators and
evaluation parameters $u_{1},u_{2}$. Consider the following
two-site operator
\begin{eqnarray}
\mathcal{T}_{12}=2\left(\mathbb{R}^{
\alpha}_{\beta}\otimes\mathbb{R}^{\beta}_{\alpha}- \mathbb{L}^{
a}_{b}\otimes\mathbb{L}^{b}_{a}+ \mathbb{Q}^{\dag
\alpha}_{a}\otimes\mathbb{Q}^{a}_{\alpha}-
\mathbb{Q}^{a}_{\alpha}\otimes\mathbb{Q}^{\dag \alpha}_{a}\right).
\end{eqnarray}
In terms of the operator $\mathbb{B}$, the proposed classical
$r$-matrix is \cite{Beisert:2007ty}
\begin{align}\label{eqn;RmatBeisert}
&r_{12} = \frac{g}{2}\left[\frac{\mathcal{T}_{12}-\mathbb{B}\otimes
\mathbb{H}-\mathbb{H}\otimes
\mathbb{B}}{u_{1}-u_{2}}-\frac{\mathbb{B}\otimes \mathbb{H}
}{u_{2}} +\frac{\mathbb{H}\otimes \mathbb{B}}{u_{1}}+
\left(\frac{1}{u_{1}}-\frac{1}{u_{2}}\right)\mathbb{H}\otimes \mathbb{H}\right].
\end{align}
All the operators and $u$ in the above expression are understood
in the strict classical limit, i.e. the lowest order terms in
(\ref{eqn;classicalCoeffi}),(\ref{eqn;classicalU}). The last term
is proportional to the identity operator and is related to the
phase factor of the S-matrix. It was shown in
\cite{Beisert:2007ty} that $r$ satisfies the classical Yang-Baxter
equation.\smallskip

Via (\ref{eqn;AlgDiff}) it is straightforward to put $r$ into
differential operator form since it is completely defined in terms
of the algebra generators and central elements. Upon taking the
near plane-wave limit discussed above we can then compare the
action of this operator to the semiclassical limit of the bound state S-matrix.

For completeness, we give here the operator form of $\mathcal{T}_{12}$. The operator
$\mathcal{T}_{12}$ is composed of two operators acting in
different spaces, whose superspace variables are again denoted by
$w_a,\theta_{\alpha}$ and $v_a,\vartheta_{\alpha}$ respectively.
Writing it out is straightforward:
\begin{align}
\mathcal{T}_{12} =&
(-2w_{b}v_{a}+w_{a}v_{b})\frac{\partial^{2}}{\partial
w_{a}\partial v_{b}} +
(2\theta_{\beta}\vartheta_{\alpha}-\theta_{\alpha}\vartheta_{\beta})\frac{\partial^{2}}{\partial \theta_{\alpha}\partial \vartheta_{\beta}} +\nonumber\\
&
2(\mathfrak{a}_{1}\mathfrak{d}_{2}-\mathfrak{b}_{2}\mathfrak{c}_{1})v_{a}\theta_{\alpha}
\frac{\partial^{2}}{\partial w_{a}\partial \vartheta_{\alpha}} +
 2(\mathfrak{a}_{2}\mathfrak{d}_{1}-\mathfrak{b}_{1}\mathfrak{c}_{2})w_{a}\vartheta_{\alpha} \frac{\partial^{2}}{\partial v_{a}\partial \theta_{\alpha}} +\\
&
2(\mathfrak{a}_{2}\mathfrak{c}_{1}-\mathfrak{b}_{1}\mathfrak{d}_{2})\theta_{\alpha}\vartheta_{\beta}
\epsilon_{ab}\epsilon^{\alpha\beta}\frac{\partial^{2}}{\partial
w_{a}\partial v_{b}} +
2(\mathfrak{a}_{1}\mathfrak{c}_{2}-\mathfrak{b}_{2}\mathfrak{d}_{1})w_{a}v_{b}
\epsilon^{ab}\epsilon_{\alpha\beta}\frac{\partial^{2}}{\partial
\theta_{\alpha}\partial \vartheta_{\beta}}.\nonumber
\end{align}
The coefficients
$\mathfrak{a},\mathfrak{b},\mathfrak{c},\mathfrak{d}$ are the
classical limits of $a,b,c,d$ defined above
(\ref{eqn;classicalCoeffi}).

\section{The semi-classical limit of the S-matrix}

We will now concentrate on the plane-wave limit of the bound state
S-matrices. In this limit it should agree with the universal
classical $r$-matrix. There are two parts to the S-matrix, the
matrix part and the overall scalar factor. We will start with a
discussion of the latter.

\subsection*{The dressing phase}

By making use of fusion techniques, the scalar factor of the bound
state S-matrix scattering bound states of length $\ell_1,\ell_2$
respectively was found in \cite{Arutyunov:2008zt}. It was shown
that this factor respects crossing symmetry. Recall that, if one defines
\begin{eqnarray}
G(n):=\frac{u_{1}-u_{2}+\frac{n}{2}}{u_{1}-u_{2}-\frac{n}{2}},
\end{eqnarray}
the explicit form of the scalar factor is given by (\ref{eqn;FullPhase})
\begin{align}
S_{0}^{\ell_1\ell_2}(p_{1},p_{2})=
\left(\frac{x_{1}^{-}}{x_{1}^{+}}\right)^{\frac{\ell_2}{2}}\left(\frac{x_{2}^{+}}{x_{2}^{-}}\right)^{\frac{\ell_1}{2}}\sigma(x_{1},x_{2})
\sqrt{G(\ell_2-\ell_1)G(\ell_1+\ell_2)}\prod_{k=1}^{\ell_1-1}G(\ell_2-\ell_1+2k)\nonumber,
\end{align}
where $\sigma(x_{2},x_{1})$ is the dressing factor
\cite{Beisert:2006ez}. For comparison with the classical
$r$-matrix this has to be evaluated to order
$\mathcal{O}(g^{-1})$, which will then be combined with the matrix
part later on. First of all, the functions $G(n)$ and the factors
proportional to the momenta are easily expanded around
$g\rightarrow \infty$ by using (\ref{eqn;ParameterXpm}). We find
\begin{eqnarray}
G(n) &=& 1+\frac{2 i n}{g}\frac{ x_1 x_2}{(x_1-x_2) (x_1
x_2-1)}+\mathcal{O}(g^{-2}).
\end{eqnarray}
To examine the dressing factor $\sigma(x_{1},x_{2})$, we first
introduce the conserved charges
\begin{eqnarray}
q_{n}(x_{i}) &\equiv&
\frac{i}{n-1}\left(\frac{1}{(x_{i}^{+})^{n-1}}-\frac{1}{(x_{i}^{-})^{n-1}}\right)\nonumber\\
&=&\frac{2\ell_{i}}{g}\frac{x_{i}^{2-n}}{x_{i}^2-1} +
\mathcal{O}(g^{-2}).
\end{eqnarray}
The dressing phase $\theta$ is related to the conserved charges as
follows
\begin{eqnarray}
\sigma(x_{1},x_{2}) = e^{\frac{i}{2}\theta(x_{1},x_{2})},
\end{eqnarray}
where
\begin{eqnarray}
\theta_{12} =g\sum_{r=2}^{\infty}\sum_{n=0}^{\infty} c_{r,r+1+2n}
\left(q_r\left(x_1\right)
q_{r+1+2n}\left(x_2\right)-q_r\left(x_2\right)
q_{r+1+2n}\left(x_1\right)\right),
\end{eqnarray}
with \cite{Arutyunov:2006iu}
\begin{eqnarray}
c_{r,s} =
\delta_{r+1,s}-g^{-1}\frac{4}{\pi}\frac{(r-1)(s-1)}{(r+s-2)(s-r)}+\mathcal{O}(g^{-2}).
\end{eqnarray}
Since in the near plane-wave limit $q_{n}\sim g^{-1}$, we see that
if we work to order $\mathcal{O}(g^{-1})$, it suffices to take
$c_{r,s} = \delta_{r+1,s}$. Hence, the dressing phase reduces to
\begin{eqnarray}
\theta_{12} &=&g\sum_{r=2}^{\infty}\sum_{n=0}^{\infty}
\delta_{n,0} \left(q_r\left(x_1\right)
q_{r+1+2n}\left(x_2\right)-q_r\left(x_2\right)
q_{r+1+2n}\left(x_1\right)\right)+\mathcal{O}(g^{-2})\nonumber\\
&=&g\sum_{r=2}^{\infty} \left(q_r\left(x_1\right)
q_{r+1}\left(x_2\right)-q_r\left(x_2\right)
q_{r+1}\left(x_1\right)\right)+\mathcal{O}(g^{-2})\nonumber\\
&=&\frac{4\ell_1\ell_2}{g}\frac{x_{1}^{2}x_{2}^{2}(x_{1}-x_{2})}{(x_{1}^2-1)(x_{2}^2-1)}\sum_{r=2}^{\infty}
\left(\frac{1}{x_{1}x_{2}}\right)^{r+1}+\mathcal{O}(g^{-2})\nonumber\\
&=&\frac{4\ell_1\ell_2}{g}\frac{(x_{1}-x_{2})}{(x_{1}^2-1)(x_{1}x_{2}-1)(x_{2}^2-1)}+\mathcal{O}(g^{-2}).
\end{eqnarray}
This gives
\begin{eqnarray}
\sigma(x_{1},x_{2}) = 1 + \frac{2 i
\ell_1\ell_2}{g}\frac{(x_{1}-x_{2})}{(x_{1}^2-1)(x_{1}x_{2}-1)(x_{2}^2-1)}+\mathcal{O}(g^{-2}).
\end{eqnarray}
From this expression it is easy to see that, at least to first
order, the dressing phases of bound states indeed respect fusion.

The remainder of $S_{0}^{\ell_1\ell_2}$ is easily found to give
\begin{align}
&\left(\frac{x_{1}^{-}}{x_{1}^{+}}\right)^{\frac{\ell_2}{2}}\left(\frac{x_{2}^{+}}{x_{2}^{-}}\right)^{\frac{\ell_1}{2}}\sqrt{G(\ell_2-\ell_1)G(\ell_1+\ell_2)}\prod_{k=1}^{\ell_1-1}G(\ell_2-\ell_1+2k)=\\
&\qquad\qquad 1-\frac{i \ell_1 \ell_2}{g} \left(
\frac{x_1}{x_1^2-1}-\frac{2 x_1 x_2}{(x_1-x_2) (x_1
x_2-1)}-\frac{x_2}{x_2^2-1}\right) + \mathcal{O}(g^{-2}).\nonumber
\end{align}
Combining this with the dressing phase, we obtain in the near
plane-wave limit
\begin{eqnarray}
S_{0}^{\ell_1\ell_2}(p_{1},p_{2})={ 1+ \frac{i \ell_1\ell_2}{g}
\frac{(x_1 x_2-1) (x_1^2+x_2^2)
 }{(x_1^2-1) (x_1-x_2) (x_2^2-1)}} + \mathcal{O}(g^{-2}).
\end{eqnarray}

\subsection*{Matrix Part}

Let us now turn to the matrix part. We will first study the
S-matrix $\mathscr{X}$ (\ref{eqn;SCase1}) from Case I in detail.

When looking at formula (\ref{eqn;SCase1}), one can see that,
besides expanding the factor $\mathcal{D}$, one needs to expand
the remaining expression, depending only on the difference $\delta
u$ of the spectral parameters, for large values of $\delta u$. The
terms relevant to the classical limit of (\ref{eqn;SCase1}) are
given by the following expansion:
\begin{align}\label{eqn;SCase3class}
&\mathscr{X}^{k,l}_n \sim
(1+\mathcal{D}_{cl})\frac{\bigg( 1 - \frac{1}{\delta u}\sum_{p=1}^{k+l} (\frac{\ell_1+\ell_2}{2}-p)\bigg) \, \prod_{i=1}^{n}(\ell_1-i)\prod_{i=1}^{k+l-n}(\ell_2-i)}{\prod_{p=1}^{k}(\ell_1-p)\prod_{p=1}^{l}(\ell_2-p)}\times \nonumber \\
&\ \ \times \sum_{m=0}^{k}\Bigg\{ \bigg( 1+ \frac{1}{\delta u} \, \sum_{p=1}^m (\frac{\ell_1-\ell_2}{2}+1-p)+\frac{1}{\delta u} \sum_{p=1-m}^{l-n} (\frac{\ell_2-\ell_1}{2}+1-p) \bigg)\times \nonumber\\
&\ \  \, \times {\delta u}^{2m-k-n} \, {k\choose k-m }{l\choose
n-m } \prod_{p=1}^{k-m}\mathfrak{d}_{\frac{k-p+2}{2}}
\prod_{p=1}^{n-m}\tilde{\mathfrak{d}}_{\frac{k+l-m-p+2}{2}}
\Bigg\} ,
\end{align}
where $\mathcal{D}_{cl}$ denotes the first order in $1/g$ of
$\mathcal{D}$. Here, we have used the fact that the binomials
enforce $l\geq n-m$, in order to obtain the power of ${\delta
u}^{2m-k-n}$. Let us start by considering non-diagonal amplitudes,
namely, $n$ different from $k$ (cfr. (\ref{transit})). In order to
do that, let us first reduce the above formula for the case $n
\geq k$. In this case, the leading piece in the above expression
is given by the term in the sum with $m=k$ (the binomials are in
this case non-zero, since, from (\ref{transit}), one has $l\geq
n-k$). The amplitude tends to
\begin{align}\label{eqn;SCase3classngeqk}
\mathscr{X}^{k,l}_n \sim \frac{1}{{\delta u}^{n-k}} \,
\frac{\prod_{i=1}^{n}(\ell_1-i)\prod_{i=1}^{k+l-n}(\ell_2-i)}{\prod_{p=1}^{k}(\ell_1-p)\prod_{p=1}^{l}(\ell_2-p)}
\, {l\choose n-k }
\prod_{p=1}^{n-k}\tilde{\mathfrak{d}}_{\frac{l-p+2}{2}} .
\end{align}
As one can see, in the non-diagonal case only one of these
amplitudes actually contributes to the classical limit
(corresponding to the order $1/g$ of the scattering matrix).
Namely, only the transition from a state characterized by quantum
number $k$ to one with corresponding quantum number $n=k+1$ has
the right order, the other ones being suppressed by higher powers
of $\delta u$. In this situation, the classical amplitudes reads
\begin{eqnarray}\label{eqn;SCase3classneqkp1}
\mathscr{X}^{k,l}_{k+1} &\sim& \frac{1}{{\delta u}} \, l (\ell_1 -
k - 1) .
\end{eqnarray}
Next, let us consider $k \geq n$. In this case the binomials force
the leading piece in the sum to be the one with $m=n$. This reads
(quite symmetrically w.r.t the previous case)
\begin{eqnarray}\label{eqn;SCase3classkgeqn}
\mathscr{X}^{k,l}_n &\sim& \frac{1}{{\delta u}^{k-n}} \,
\frac{\prod_{i=1}^{n}(\ell_1-i)\prod_{i=1}^{k+l-n}(\ell_2-i)}{\prod_{p=1}^{k}(\ell_1-p)\prod_{p=1}^{l}(\ell_2-p)}
\, {k\choose k-n } \prod_{p=1}^{k-n}
\mathfrak{d}_{\frac{k-p+2}{2}} .
\end{eqnarray}
Analogously, only one of the non-diagonal terms has the right
falloff to be able to contribute to the classical r-matrix, namely
the amplitude for quantum numbers $k$ to $n=k-1$. The contribution
is given by
\begin{eqnarray}\label{eqn;SCase3classneqkp2}
\mathscr{X}^{k,l}_{k-1} &\sim& \frac{1}{{\delta u}} \, k (\ell_2 -
l - 1) .
\end{eqnarray}
The diagonal part, for $n=k$, is slightly more complicated. The
leading term can be obtained by restricting to $k=n$ either of the
two formulas (\ref{eqn;SCase3classngeqk}) or
(\ref{eqn;SCase3classkgeqn}), and is easily seen to be equal to
$1$. The S-matrix tends in fact to the identity in the strict
classical limit. The next to leading term of order $1/\delta u$
contributes to the classical r-matrix, and can be
straightforwardly obtained from (\ref{eqn;SCase3class}) as
\begin{align}\label{eqn;SCase3classngeqn}
&\mathscr{X}^{k,l}_k  \sim 1 +  D_{cl} + \frac{1}{{\delta u}}
\Bigg[ \sum_{p=1}^{k+l} (\frac{\ell_1+\ell_2}{2}-p) +
\sum_{p=1}^{k} (\frac{\ell_1-\ell_2}{2}+1-p)+ \sum_{p=1-k}^{l-k}
(\frac{\ell_2-\ell_1}{2}+1-p)\Bigg].\nonumber
\end{align}
Having now determined the semi-classical limit of the Case I
S-matrix, one can easily produce the semi-classical limit of the
Case II S-matrix (\ref{eqn;SCase2}). By expanding the matrices
$A,B,B^\pm$ to order $g^{-1}$ one again finds that the only terms
$\mathscr{Y}^{k,l;i}_{n;j}$ that survive are those with $k=n$ and $k=n\pm 1$. A
similar discussion also holds for the Case III S-matrix.

\section{Comparison in the near plane-wave limit}

We will now compare the classical $r$-matrix
(\ref{eqn;RmatBeisert}) against the semi-classical limit of the
bound state S-matrix. Let us first look at the dressing phase. To
this end, we recall that the bound state S-matrices
$\S^{\ell_1\ell_2}$ are canonically normalized by setting
\begin{eqnarray}
\mathbb{S}_{can}^{\ell_1\ell_2} w_{1}^{\ell_1}v_{1}^{\ell_2} =
w_{1}^{\ell_1}v_{1}^{\ell_2}.
\end{eqnarray}
For the fully dressed S-matrix we therefore  obtain
\begin{eqnarray}
\mathbb{S}^{\ell_1\ell_2} w_{1}^{\ell_1}v_{1}^{\ell_2} =
S_{0}^{\ell_1\ell_2}w_{1}^{\ell_1}v_{1}^{\ell_2},
\end{eqnarray}
where $S_{0}^{}$ is the scalar factor given by
(\ref{eqn;FullPhase}). On the other hand, assuming that the
classical $r$-matrix is universal, we can easily compute its
action on the state $w_{1}^{\ell_1}v_{1}^{\ell_2}$. One finds
\begin{align}
&(1+g^{-1}r) w_{1}^{\ell_1}v_{1}^{\ell_2} =
S_{0}^{\ell_1\ell_2}(x_{1},x_{2}) w_{1}^{\ell_1}v_{1}^{\ell_2}.
\end{align}
This means that the phase factor (\ref{eqn;FullPhase}) derived in
\cite{Arutyunov:2008zt} is indeed compatible with $r$.

The matrix structure is now easily compared by acting with $r$ as
an operator on states $\stateA{k,l}$,
$\stateB{k,l}_i,\stateC{k,l}_i$ and comparing the coefficients of
the resulting states against the S-matrix elements. Doing this
leads to perfect agreement.

As a curiosity, we note that the classical r-matrix actually
can be used to describe the S-matrices up to second order (apart from the overall
factor)
\begin{eqnarray}
\S \sim 1 +\frac{r}{g} + \frac{r^2}{2g^2} + \mathcal{O}(g^{-3}).
\end{eqnarray}
It is easily checked that this exponential pattern breaks down at third order.


\section{Summary}

In this chapter we compared the classical limit of the bound state
S-matrix against two different algebraic expressions for the
classical $r$-matrix. We find that the universal $r$-matrix
(\ref{eqn;MT-Rmat}) from \cite{Moriyama:2007jt} agrees with the
semi-classical limit of the fundamental S-matrix but this
agreement breaks down for higher bound state numbers. However, the
universal $r$-matrix (\ref{eqn;RmatBeisert}) put forward in
\cite{Beisert:2007ty} correctly describes the semi-classical limit
of all bound state S-matrices. Moreover, it even captures the
matrix structure at one order higher.

%% file: UniversalBlocks4.tex
\chapter{Universal Blocks}\label{chap;UnivBlocks}

In chapter \ref{chap;IntegrabilityAndHopf} the notion of a
quasi-triangular Hopf algebra was introduced. These Hopf algebras
admit an R-matrix, which is a purely algebraic object intertwining
the coproduct and the opposite coproduct (for the relevant
definitions we remind to section \ref{sec;HopfAlgebra}). When
evaluated in explicit representations the R-matrix gives rise to
the S-matrix in that representation.


Of course, in a concrete representation one can compute the
S-matrix without knowing the universal R-matrix. This is what was
done in chapter \ref{chap;BoundSmat}, where bound state S-matrices
were computed. One might then wonder if this S-matrix (or parts
thereof) has an algebraic origin. The existence of a universal
R-matrix would be interesting from a mathematical point of view
and useful for computing the S-matrix in representations for which
a direct derivation may be cumbersome.

In this chapter we will explore this universality. The bound state
representations we have discussed in section
\ref{sec;boundstaterep} contain representations of subalgebras of
$\alg{h}$. We can identify two such subspaces, namely a
$\alg{gl}(1|1)$ and an $\su(2)$. The $\su(2)$ is particulary
interesting since it describes Case I states, which provided the
starting point for our construction in chapter
\ref{chap;BoundSmat}. The (double) Yangian of both subalgebras
admits a universal R-matrix and one can ask whether the
restriction of the S-matrix to these states corresponds to this
universal R-matrix. We will see that this is indeed the case.


\section{The $\alg{su}(2)$ subspace}

The first subspace is given by states that span Case I. We remind
that the $\alg{psu}(2|2)$ algebra has two (``bosonic" and
``fermionic", according to the indices they transform)
$\alg{su}(2)$ subalgebras,  with generators $\mathbb{L}^a_b$ and
$\mathbb{R}^{\alpha}_{\beta}$, respectively. The first ones
satisfy the following commutation relations:
\begin{align}
&[\mathbb{L}_{a}^{\ b},\mathbb{J}_{c}] = \delta_{c}^{b}\mathbb{J}_{a}-\frac{1}{2}\delta_{a}^{b}\mathbb{J}_{c},\nonumber\\
&[\mathbb{L}_{a}^{\ b},\mathbb{J}^{c}] = -\delta_{a}^{c}\mathbb{J}^{b}+\frac{1}{2}\delta_{a}^{b}\mathbb{J}^{c}.
\end{align}
 The states
\begin{align}
\theta_3 w_1^{\ell-k-1}w_2^k,
\end{align}
form a natural representation on which the ``bosonic"
$\alg{su}(2)$ subalgebra of $\mathbb{L}^a_b$'s acts. They form an
$\ell-1$ dimensional representation. Obviously any vector
$\stateA{k,l}$ from Case I (\ref{eqn;BasisCase1}) originates from
the tensor product of two such states. It is easy to see that the
coproducts of the Yangian generators on theses states coincide
with the truncation to the $\alg{su}(2)$ generators of the general
expressions (\ref{eqn;YangianCoprod}). Furthermore, the Case I
S-matrix satisfies the Yang-Baxter equation by itself, and it is
of difference form. This means that such S-matrix should naturally
come from the universal R-matrix of the $\alg{su}(2)$ Yangian
double \cite{Khoroshkin:1994uk}.

\subsection*{Drinfeld II for $\alg{su}(2)$}

In \cite{Khoroshkin:1994uk}, the universal R-matrix for Yangian
doubles has been constructed using Drinfeld's second realization
of the Yangian. The discussion there can straightforwardly be
applied to the bound state representations in the superspace
formalism.

The map between the first and the second realization becomes
\begin{align}\label{def:isomsu2}
&\kappa_{0}=2\fL^2_2,&& \xi^+_{0}=\fL^1_2,&& \xi^-_{0}=\fL^2_1,\nonumber\\
&\kappa_{1}=2\hat{\fL}^2_2-v,&& \xi^+_{1}=\hat{\fL}^1_2-w,&&
\xi^-_{1}=\hat{\fL}^2_1 -z,
\end{align}
where
\begin{eqnarray}
v = \frac{1}{2} (\{ \fL^2_1,\fL^1_2\} - (\fL^2_2)^2 ), \qquad w =
- \frac{1}{4} \{ \fL^1_2,\fL^2_2\}, \qquad z = - \frac{1}{4} \{
\fL^2_1,\fL^2_2\}.
\end{eqnarray}
The operators $\mathbb{L}^a_b$ are realized as in
(\ref{eqn;AlgDiff}). The higher level generators are given by
\begin{eqnarray}\label{eqn;UnivSU2Generators}
\xi^-_n &=& \xi^-_0 (u +\frac{\kappa_0-1}{2} )^n,\nonumber\\
\xi^+_n &=& \xi^+_0 (u +\frac{\kappa_0+1}{2} )^n,\\
\kappa_n &=& \xi^+_0 \xi^-_n - \xi^-_0 \xi^+_n.\nonumber
\end{eqnarray}
The parameter $u$ corresponds to (\ref{eqn;defUparameter}). The
generators (\ref{eqn;UnivSU2Generators}) coincide with those
obtained in \cite{Khoroshkin:1994uk} for generic highest-weight
representations of $Y(\alg{su}(2))$. It is easy to check that
these generators satisfy the correct relations
\begin{align}
\label{eqn;DrinIIsu2}
&[\kappa_{m},\kappa_{n}]=0,\quad \, \, \, \, \, \, \, [\xi^+_{m},\xi^-_{n}]=\, \kappa_{n+m},\nonumber\\
&[\kappa_{0},\xi^+_{m}]= 2\,\xi^+_{m},\quad [\kappa_{0},\xi^-_{m}]=-  2\,\xi^-_{m},\nonumber\\
&[\kappa_{m+1},\xi^+_{n}]-[\kappa_{m},\xi^+_{n+1}] =   \{\kappa_{m},\xi^+_{n}\},\nonumber\\
&[\kappa_{m+1},\xi^-_{n}]-[h_{m},\xi^-_{n+1}] = -  \{\kappa_{m},\xi^-_{n}\},\nonumber\\
&[\xi^+_{m+1},\xi^+_{n}]-[\xi^+_{m},\xi^+_{n+1}] =   \{\xi^+_{m},\xi^+_{n}\},\nonumber\\
&[\xi^-_{m+1},\xi^-_{n}]-[\xi^-_{m},\xi^-_{n+1}] = -
\{\xi^-_{m},\xi^-_{n}\}
\end{align}
that define Drinfeld's second realization of $Y(\alg{su}(2))$.

\subsection*{The universal R-matrix}

We will now proceed to compute the universal R-matrix for the
$\alg{su}(2)$ block of our bound state S-matrix, following
\cite{Khoroshkin:1994uk}. The derivation is split up into three
parts, corresponding to the factorization
\begin{eqnarray}
R = R_+ R_0 R_-,
\end{eqnarray}
$R_+$ and $R_-$ being ``root" factors, while $R_0$ is a purely
diagonal ``Cartan" factor. The different terms are
\begin{align}
& R_+=\prod_{n\geq 0}^{\rightarrow}\exp(- \xi^+_n\otimes \xi^-_{-n-1}),  \\
& R_-=\prod_{n\geq 0}^{\leftarrow}\exp(- \xi^-_n\otimes \xi^+_{-n-1}),  \\
& R_0=\prod_{n\geq 0} \exp \left\{ {\rm Res}_{u=v}\left[ \frac{\rm
d}{{\rm d}u}({ \log }H^+(u))\otimes { \log
}H^-(v+2n+1)\right]\right\}.
\end{align}
One has defined
\begin{eqnarray}
\label{eqn;Res}
&&{\rm Res}_{u=v}\left(A(u)\otimes B(v)\right)=\sum_k a_k\otimes b_{-k-1}
\end{eqnarray}
for $A(u)=\sum_k a_k u^{-k-1}$ and $B(u)=\sum_k b_k u^{-k-1}$, and the
so-called Drinfeld's currents are given by
\begin{eqnarray}
\label{curr} &&E^{\pm}(u)=\pm \sum_{n \ge 0 \atop n<0} \xi^+_n
u^{-n-1} ~,~~~~~~~~
F^{\pm}(u)=\pm \sum_{n \ge 0 \atop n<0} \xi^-_n u^{-n-1} \nonumber \\
&&H^{\pm}(u)=1\pm \sum_{n \ge 0 \atop n<0} \kappa_n u^{-n-1}
~.~~~~
\end{eqnarray}
The arrows on the products indicate the ordering one has to follow
in the multiplication, and are a consequence of the normal
ordering prescription for the root factors in the universal
R-matrix \cite{Khoroshkin:1994uk}. For the generic bound state
representations which we have described above, the ordering will
be essential to get the correct result. To keep notation concise we introduce
\begin{align}
&\langle A,B \rangle\langle C,D \rangle = \theta_3w_1^A w_2^B \vartheta_3v_1^C v_2^D,&&\langle A,B \rangle = \theta_3w_1^A w_2^B.
\end{align}
In this notation we have for the state $\stateA{k,l}$, $A=\ell_1-k-1,B=k,C=\ell_2-l-1,D=l$.
Let us first compute how $R_-$ acts on an arbitrary Case I state.
We find
\begin{eqnarray}\label{eqn;RE2}
\prod^{\leftarrow}_{n\geq0} \exp[-\xi^-_n\otimes \xi^+_{-1-n}]|k,l\rangle &=& \sum_{m} A_m |k-m,l+m\rangle.
\end{eqnarray}
The term $A_m$ is built up out of $m$ copies of $-\xi^-\otimes\xi^+$
acting on the state $\langle A,B\rangle\langle C,D\rangle$, which
is made of an $A$ number of $w_1$'s, a $B$ number of $w_2$'s in
the first space, and analogously $C$ and $D$ for $v_1$, $v_2$ in
the second space. In view of (\ref{eqn;RE2}), we find that such
terms can come from different exponentials, i.e. with different
$n$'s, or from the same exponential. One first needs to know how the
product of $m$ $\xi^+$'s acts on the state $\langle A,B\rangle$. We conveniently define
\begin{align}\label{cd}
&c_i = u_1-\frac{A-B+1}{2} -i, & & d_i = u_2-\frac{C-D-1}{2} +i,\nonumber \\
&\tilde{c}_i = u_2-\frac{C-D+1}{2} -i,& &\tilde{d}_i =
u_1-\frac{A-B-1}{2} +i,
\end{align}
and
\begin{align}
\delta u = u_1 - u_2.
\end{align}
In general one has
\begin{eqnarray}
\xi^-_{n_m}\ldots \xi^-_{n_2} \xi^-_{n_1}\langle A,B\rangle &=&\xi^-_{n_m}\ldots \xi^-\left(u+\frac{h-1}{2}\right)^{n_2}\xi^-\left(u+\frac{h-1}{2}\right)^{n_1}\langle A,B\rangle \nonumber\\
&=& \xi^-_{n_m}\ldots \xi^-\left(u+\frac{h-1}{2}\right)^{n_2} \xi^-\left(c_0\right)^{n_1}\langle A,B\rangle \nonumber\\
&=& B\left(c_0\right)^{n_1}\xi^-_{n_m}\ldots \xi^-\left(u+\frac{h-1}{2}\right)^{n_2}\langle A+1,B-1\rangle \nonumber\\
&=& B(B-1)\left(c_0\right)^{n_1} \left(c_1\right)^{n_2}\xi^-_{n_m}\ldots \xi^-_{n_3}\langle A+2,B-2\rangle \nonumber\\
&=& \frac{B!}{(B-m)!} c_0^{n_1}\ldots c_{m-1}^{n_m}\langle A+m,B-m\rangle.
\end{eqnarray}
Similar expressions hold for $\xi^+_n$ acting on $\langle C,D\rangle$,
but with $d_i$ instead of $c_i$, and producing the state $\langle
C-m,D+m\rangle$. When we consider terms like this coming from the
ordered exponential (\ref{eqn;RE2}), we always have that $n_i\geq
n_{i-1}$. In case $n_i = n_{i+1}$, we also pick up a combinatorial
factor coming from the series of the exponential. Putting all of
this together, we find
\begin{eqnarray}
&&A_m = (-)^m \frac{B!}{(B-m)!}\frac{C!}{(C-m)!}\left\{ \sum_{n_1\leq\ldots\leq n_m} \frac{1}{N(\{n_1,\ldots,n_m\})}\frac{c_0^{n_1}}{d_0^{n_1+1}}\ldots \frac{c_{m-1}^{n_m}}{d_{m-1}^{n_m+1}} \right\}, \nonumber\\
&&N(\{n_1,\ldots,n_m\})=\frac{1}{{\rm ord} S(\{n_1,\ldots,n_m\})}.
\end{eqnarray}
$N$ is a combinatorial factor which is defined as the inverse of
the order of the permutation group of the set
$\{n_1,\ldots,n_m\}$. For example, $N(\{1,1,2\})=\frac{1}{2}$
and\\ \noindent
$N(\{1,1,1,2,3,3,4,5\})=\frac{1}{3!}\frac{1}{2!}=\frac{1}{12}$. By
using the fact that $c_i= c_{i+1}+1, d_i= d_{i+1}-1$, one can
evaluate this sum explicitly and find
\begin{eqnarray}
A_m (A,B,C,D)= m! {B\choose m}{C\choose m} \prod_{i=0}^{m-1}\frac{1}{c_0-d_0-i-m+1},
\end{eqnarray}
where we have indicated the dependence on the parameters $A,B,C,D$
of the state we are acting on. As one can easily see using
(\ref{cd}), the resulting expression is manifestly of difference
form, i.e. it only depends on $\delta u$.

A similar consideration works for $R_+$. One has
\begin{eqnarray}\label{eqn;RF2}
\prod^{\rightarrow}_{n\geq0} \exp[-\xi^+_n\otimes \xi^-_{-1-n}]|k,l\rangle
&=& \sum_{m} B_m |k+m,l-m\rangle.
\end{eqnarray}
where
\begin{eqnarray}
B_m (A,B,C,D) = m! {A\choose m}{D\choose m}
\prod_{i=0}^{m-1}\frac{1}{\tilde{d}_0-\tilde{c}_0-i+m-1}.
\end{eqnarray}
Finally, we turn to the Cartan part. First, we work out
\begin{eqnarray}\label{eqn;kappaexplicit}
\kappa_n\langle A,B\rangle = \left\{(A+1)B\left[u-\frac{A-B+1}{2}\right]^n - (B+1)A\left[u-\frac{A-B-1}{2}\right]^n \right\}\langle A,B\rangle \nonumber.
\end{eqnarray}
We then recall the definition of $H_\pm$ from (\ref{curr}). From
the explicit realization (\ref{eqn;kappaexplicit}) it follows that
\begin{align}
H_+(t)\langle A,B\rangle&=H_-(t)\langle A,B\rangle\nonumber\\
 &=\left\{1-\frac{(A+1)B}{u-t-\frac{1}{2}(A - B +1)}+\frac{A(B+1)}{u-t-\frac{1}{2}(AB-1)}\right\}\langle A,B\rangle.
\end{align}
Defining $K_{\pm} = \log H_{\pm}$, the Cartan part of the
universal R-matrix can be written as
\begin{eqnarray}
R_0 = \prod_{n\geq0} \exp\left[{\rm Res}_{t=x}\left(\frac{d}{dt}K_+(t)\otimes K_-(x+2n+1)\right) \right],
\end{eqnarray}
where the residue is defined in (\ref{eqn;Res}). We have to find
the suitable series representations corresponding to
$\frac{d}{dt}K_+(t)$ and $K_-(x+2n+1)$. With an appropriate choice
of domains for the variables $t$ and $x$, one can write in
particular
\begin{eqnarray}
\frac{d}{dt}K_+(t) &=& \sum_{m\geq 1}\left\{ \alpha_{1}^{m} + \alpha_{2}^{m} - \alpha_{3}^{m} - \alpha_{4}^{m}\right\}t^{-m-1},\\
K_-(x+2n+1) &=& K_-(0) + \sum_{m\geq 1}\left\{ \beta_{1}^{-m} + \beta_{2}^{-m} - \beta_{3}^{-m} - \beta_{4}^{-m}\right\}\frac{x^{m}}{m},
\end{eqnarray}
where
\begin{eqnarray}
\begin{array}{lcl}
    \alpha_1 = u_1+ \frac{1}{2}(A+B+1), & ~ & \alpha_2 = u_1- \frac{1}{2}(A+B+1),\\
    \alpha_3 = u_1- \frac{1}{2}(A-B+1), & ~ & \alpha_4 = u_1- \frac{1}{2}(A-B-1),
\end{array}
\end{eqnarray}
and
\begin{eqnarray}
\begin{array}{lcl}
    \beta_1 = u_2-2n + \frac{1}{2}(D-C-1), & ~ & \beta_2 = u_2-2n+ \frac{1}{2}(D-C-3),\\
    \beta_3 = u_2-2n+ \frac{1}{2}(D+C-1), & ~ & \beta_4 = u_2-2n- \frac{1}{2}(D+C+3),
\end{array}\end{eqnarray}
This leads to
\begin{align}\label{eqn;RH2}
R_0 \langle A,B\rangle\langle C,D\rangle =&
\frac{2^{1-2 \delta u} \, \pi  \, \, \Gamma  \big(\frac{2 \delta u
+A+B+C-D+2}{2} \big) \, \, \Gamma  \big(\frac{2 \delta u+B-
A+C+D+2}{2} \big)}{\Gamma  (\frac{2 \delta u+ A+B-C-D+2}{4} ) \Gamma
(\frac{\delta u- A+B+C-D+2}{2} )\Gamma(\frac{2 \delta u +A+B+C+D+4}{4} )}\times \nonumber\\
&\times\frac{ \Gamma  \big(\frac{2 \delta u-A+B-C-D}{2} \big) \, \, \, \, \, \Gamma  \big( \frac{2 \delta u -A-B+C-D}{2} \big)}{\Gamma  (\frac{2\delta u -A-B+C+D +2}{4}) \Gamma  (\frac{\delta u- A+B+C-D}{2} )\Gamma  (\frac{2\delta u
-A-B-C-D}{4}) } \langle A,B\rangle\langle C,D\rangle \nonumber\\
&\nonumber\\
& \equiv {\cal{H}}(A,B,C,D) \, \langle A,B\rangle\langle C,D\rangle.
\end{align}
We are now ready to put things together and evaluate the action of
the universal R-matrix of $\alg{su}(2)$ on Case I states. We
obtain
\begin{align}
R |k,l\rangle =& \sum_{m=0}^{min(B,C)} \, \sum_{n=0}^{min(A,D)+m} \, B_n (A+m,B-m,C-m,D+m) \\
&\times \, {\cal{H}} (A+m,B-m,C-m,D+m) \, A_m (A,B,C,D) \, |k-m+n,l+m-n\rangle,\nonumber
\end{align}
where
\begin{eqnarray}
&&A=\ell_1 - k - 1, \qquad B=k,\nonumber\\
&&C=\ell_2 - l - 1, \qquad D=l,
\end{eqnarray}
and the various factors are given by formulas
(\ref{eqn;RE2}),(\ref{eqn;RF2}) and (\ref{eqn;RH2}). It is now
easy to convert the above expression into
\begin{eqnarray}
R |k,l\rangle = \sum_{n=0}^{k+l} R_n \, |n,k+l-n\rangle.
\end{eqnarray}
In order to find the amplitudes $R_n$, we proceed as follows.
Taking into account the presence of binomial factors in the
expressions for $A_m$ and $B_n$, which naturally truncate the sum
when $m,n$ lie outside the correct intervals, we can extend the
summation indices to run from $-\infty$ to $\infty$. In this way,
manipulations of the above sums are easier, and one ends up with
\begin{eqnarray}
\label{Rn}
R_n &=& \sum_{m=-n+k}^{\infty} A_m (\ell_1 - k - 1,k,\ell_2-l-1,l)\, \nonumber\\
&&\times \, {\cal{H}} (\ell_1 - k - 1+m,k-m,\ell_2-l-1-m,l+m)\nonumber\\
&&\times \, B_{n-k+m} (\ell_1 - k - 1+m,k-m,\ell_2-l-1-m,l+m).
\end{eqnarray}
We have checked that this coincides with the r.h.s. of
(\ref{hypergeom}) for a large selection of choices of the integer
parameters, keeping $\delta u$ arbitrary, when taking into account
the proper normalization. We have in fact, with the notations of
\cite{Arutyunov:2009mi},
\begin{eqnarray}
\label{macc} {\cal{D}} \, \frac{\Gamma \left( \frac{1}{4}(2 +
\ell_1 - \ell_2 + 2 \delta u)\right) \, \Gamma\left(\frac{1}{4}(2
+ \ell_2 - \ell_1 + 2 \delta u)\right)}{\Gamma\left( \frac{1}{4}(4
- \ell_1 - \ell_2 + 2 \delta u)\right) \,
\Gamma\left(\frac{1}{4}(\ell_1 + \ell_2 + 2 \delta u)\right)} \,
R_n = \mathscr{X}^{k,l}_n.
\end{eqnarray}
The ratio of gamma functions appearing in the above formula is the
(inverse of the) so-called ``character" of the universal R-matrix
in evaluation representations \cite{Khoroshkin:1994uk}, namely its
action on states of highest-weight $\lambda_i = l_i - 1$.

\section{Universal R-matrix for ${\alg{gl}}(1|1)$}

The other subspace we will consider is obtained by restricting the
bound states to having bosonic and fermionic indices of only one
respective type. For definiteness, we will take the bosonic index
to be $1$ and the fermionic index to be $3$. There are four copies
of this subspace, corresponding to the four different choices of
these indices we can make. The embedding of this subspace in the
full bound state representation is spanned by the vectors
\begin{eqnarray}
\left\{\stateC{0,0}_1,~\stateB{0,0}_1,~\stateB{0,0}_2,~\stateA{0,0}\right\}.
\end{eqnarray}
As one can see, this subspace takes particular states from all
three Cases listed earlier, yet being closed under the action of
the S-matrix. This means that the S-matrix for this subsector
corresponds to a block-diagonal $4\times4$ matrix. Its form can be
readily found from the explicit expressions in section
\ref{sect;Reduction}. Putting this together, one obtains
\begin{eqnarray}\label{eqn;4x4Smat}
\mathbb{S} = \begin{pmatrix}
 1 ~&~ 0 ~&~ 0 ~&~ 0\\
 0 ~&~ e^{-i\frac{p_2}{2}}\frac{x^{+}_{1}-x^{+}_{2}}{x^{+}_{1}-x^{-}_{2}} ~&~ \frac{\sqrt{\ell_1}\eta(p_{1})}{\sqrt{\ell_2}\eta(p_{2})}\frac{x^{+}_{2}-x^{-}_{2}}{x^{+}_{1}-x^{-}_{2}} ~&~ 0\\
 0 ~&~ \frac{e^{i\frac{p_1}{2}}}{e^{i\frac{p_2}{2}}}\frac{\sqrt{\ell_2}\eta(p_{2})}{\sqrt{\ell_1}\eta(p_{1})}\frac{x^{+}_{1}-x^{-}_{1}}{x^{+}_{1}-x^{-}_{2}} ~&~ e^{i\frac{p_1}{2}}\frac{x^{-}_{1}-x^{-}_{2}}{x^{+}_{1}-x^{-}_{2}} ~&~ 0\\
 0 ~&~ 0 ~&~ 0 ~&~ \frac{x_1^-
-x_2^+}{x_1^+ -x_2^-}\frac{e^{i\frac{p_1}{2}}}{e^{i\frac{p_2}{2}}}
\end{pmatrix}.
\end{eqnarray}
We remark that, taken in the fundamental representation, and
suitably un-twisted in order to eliminate the braiding factors
coming from the nontrivial coproduct
\cite{Gomez:2006va,Plefka:2006ze,Arutyunov:2006yd}, this matrix
coincides with the S-matrix of \cite{Beisert:2005wm}. It is
readily checked that this matrix satisfies the Yang-Baxter
equation by itself, therefore it is natural to ask whether it is
the representation of a known (Yangian) universal R-matrix.

\subsection*{Drinfeld II}

The algebra transforming the states inside these sectors is an
${\alg{sl}}(1|1)$. As it is known, this type of superalgebras
(with a degenerate Cartan matrix) do not admit a universal
R-matrix, therefore we will introduce an extra Cartan generator
\cite{KT} and study the Yangian of the algebra ${\alg{gl}}(1|1)$
instead\footnote{For the purposes of the universal R-matrix, it
will not make any difference to consider real forms of the
algebras when needed.}. Let us start with the canonical
derivation, and adapt the representation later in order to exactly
match with our S-matrix.  We will follow
\cite{Khoroshkin:1994uk,Cai:q-alg9709038}. The super Yangian
double $DY\left(gl(1|1)\right)$ is the Hopf algebra generated by
the elements $\xi^+_n$, $\xi^-_n$, $\kappa_{1;n}$, $\kappa_{2;n}$,
with $n$ an integer number, satisfying (Drinfeld's second
realization)
\begin{align}\label{LieGL}
&[\kappa_{i;m} ~,~ \kappa_{j;n}]=0, \nonumber \\
&[\kappa_{2;m} ~,~ \xi^+_n]=[\kappa_{2;m} ~,~ \xi^-_n] =0, \nonumber \\
&[\kappa_{1;0} ~,~ \xi^+_n]=-2\xi^+_n ~,~ [\kappa_{1;0}~,~\xi^-_n] = 2\xi^-_n, \nonumber \\
&[\kappa_{1;m+1} ~,~\xi^+_n]-[\kappa_{1;m} ~,~ \xi^+_{n+1}]+\{\kappa_{1;m}~,~\xi^+_n\}=0, \nonumber\\
&[\kappa_{1;m+1} ~,~\xi^-_n]-[\kappa_{1;m} ~,~ \xi^-_{n+1}]-\{\kappa_{1;m}~,~\xi^-_n\}=0, \nonumber \\
&\{\xi^+_m ~,~ \xi^+_n\}=\{\xi^-_m ~,~\xi^-_n\}=0, \nonumber \\
&\{\xi^+_m ~,~ \xi^-_n\}=-\kappa_{2;m+n}.
\end{align}
Drinfeld's currents are given by
\begin{eqnarray}
&&E^{\pm}(t)=\pm \sum_{n \ge 0 \atop n<0} \xi^+_n t^{-n-1}
~,~~~~~~~~
F^{\pm}(t)=\pm \sum_{n \ge 0 \atop n<0} \xi^-_n t^{-n-1}, \\
&&H^{\pm}(t)=1\pm \sum_{n \ge 0 \atop n<0} \kappa_{1;n} t^{-n-1}
~,~~~~ K^{\pm}(t)=1\pm \sum_{n \ge 0 \atop n<0}
\kappa_{2;n}t^{-n-1}\label{eqn;DrinfeldCurrentsGL11}.
\end{eqnarray}
One can show that the following bound state representation, acting
on monomials made of a generic bosonic state $v$ and a generic
fermionic state $\theta$, satisfies all the defining relations of
the second realization (\ref{LieGL}):
\begin{align}\label{eqn;EvRepDrinfeldIIforGL11}
&\xi^+_n = \lambda^n \fQ^1_3, &&  \xi^-_n = \lambda^n \fQ^{\dag3}_1, \nonumber\\
&\kappa_{1;n} = 2(\lambda+\ell-1)^n(\fL^1_1 - \fR^3_3), &&
\kappa_{2;n} = -\lambda^n
\left(\frac{1}{2}\fHH+\fL^1_1+\fR^3_3\right).
\end{align}
As usual, we denote by $\ell$ the number of components of the
bound state. At this stage, $\lambda$ is a generic spectral
parameter and we leave the parameterization of $\fQ,\fQ^{\dag}$
unspecified. One can see that the coproducts of the Yangian
generators does not truncate nicely as in the $\alg{su}(2)$ case.
Because of this reason, we do not expect $\lambda$ to agree with
$u$. We will later specify the value $\lambda$ has to take in
order to match with the bound state S-matrix in these subsectors.


\subsection*{Universal R-matrix}

The universal $R$-matrix reads
\begin{eqnarray}
\label{univ11} &&{\cal R}={\cal R}_+{\cal R}_1{\cal R}_2{\cal
R}_-, \label{rmatrix}
\end{eqnarray}
where
\begin{eqnarray}
&&{\cal R}_+=\prod_{n\ge 0}^{\rightarrow}\exp(- \xi^+_n\otimes \xi^-_{-n-1}),  \\
&&{\cal R}_-=\prod_{n\ge 0}^{\leftarrow}\exp(\xi^-_n\otimes \xi^+_{-n-1}),  \\
&&{\cal R}_1=\prod_{n\ge 0} \exp \left\{ {\rm Res}_{t=z}\left[(-1)
\frac{ d}{{ d}t}({ \log }H^+(t))\otimes
{\rm ln}K^-(z+2n+1)\right]\right\}, \\
&&{\cal R}_{2}=\prod_{n\ge 0} \exp \left\{ {\rm Res}_{t=z}\left[(-1)
\frac{ d}{{ d}t}({\log }K^+(t))\otimes
{\rm ln}H^-(z+2n+1)\right]\right\},
\end{eqnarray}
and again
\begin{eqnarray}
&&{\rm Res}_{t=z}\left(A(t)\otimes B(z)\right)=\sum_k a_k\otimes b_{-k-1}
\end{eqnarray}
for $A(t)=\sum_k a_k t^{-k-1}$, $B(z)=\sum_k b_k z^{-k-1}$. We
first compute
\begin{eqnarray}
\mathcal{R}_-=\prod_{n\geq0}^{\leftarrow} \exp [ \xi^-_n\otimes
\xi^+_{-n-1}]
\end{eqnarray}
in our representation (\ref{eqn;EvRepDrinfeldIIforGL11}). Because
of the fermionic nature of the operators $\xi^-_n\otimes
\xi^+_{-n-1}$, the above expression simplifies to
\begin{eqnarray}
\mathcal{R}_-&=&1+\sum_{n\geq0}\xi^-_n\otimes \xi^+_{-n-1}\nonumber\\
&=&1+\sum_{n\geq0} \frac{\lambda_1^n}{\lambda_2^{n+1}}\xi^-_0\otimes\xi^+_0\nonumber\\
&=&1-\frac{\xi^-\otimes \xi^+}{\delta \lambda}
\end{eqnarray}
Considering that this term will act non-trivially only on states
with a fermion in the first space, we easily obtain
\begin{eqnarray}
\mathcal{R}_-=
\begin{pmatrix}
  1 & 0 & 0 & 0 \\
  0 & 1 & 0 & 0 \\
  0 & \frac{a_2d_1\ell_2}{\delta \lambda} & 1 & 0 \\
  0 & 0 & 0 & 1
\end{pmatrix}.
\end{eqnarray}
We have defined
\begin{eqnarray}
\delta \lambda = \lambda_1 - \lambda_2.
\end{eqnarray}
Similarly, one finds
\begin{eqnarray}
\mathcal{R}_+&=&1-\sum_{n\geq0}\xi^+_n\otimes\xi^-_{-n-1}\nonumber\\
&=&1+\frac{\xi^+_0\otimes \xi^-_0}{\delta \lambda},
\end{eqnarray}
which, written as a matrix, takes the form
\begin{eqnarray}
\mathcal{R}_+=
\begin{pmatrix}
  1 & 0 & 0 & 0 \\
  0 & 1 & \frac{a_1d_2\ell_1}{\delta \lambda} & 0 \\
  0 & 0 & 1 & 0 \\
  0 & 0 & 0 & 1
\end{pmatrix}.
\end{eqnarray}
Let us now turn to the Cartan part. For this, we first need to
compute the currents (\ref{eqn;DrinfeldCurrentsGL11}). They are found to be
\begin{eqnarray}
H^\pm &=& 1-\frac{\kappa_{1;0}}{1+\lambda-\ell-t},\\
K^\pm &=& 1-\frac{\kappa_{2;0}}{\lambda-t},
\end{eqnarray}
where we used the fact that both $\kappa_{1;0}$ and $\kappa_{2;0}$
are diagonal operators. In appropriate domains of convergence of
the series one then has in particular
\begin{eqnarray}
-\frac{d}{dt}\log H^+ = \sum_{m=1}^{\infty}
\left\{{(\lambda+\ell-1)^m}
-{(\lambda+\ell-1-\kappa_{1;0})^m}\right\} t^{-m-1}
\end{eqnarray}
and
\begin{eqnarray}
\log K^-(z+2n+1) &=& \log K^-(2n+1) + \\
&& +\sum_{m=1}^{\infty} \left\{\frac{1}{(\lambda-1-2n)^m}
-\frac{1}{(\lambda-1-2n-\kappa_{2;0})^m}\right\}
\frac{z^{m}}{m}\nonumber.
\end{eqnarray}
Straightforwardly computing the residue and performing the sum
yields, in matrix form,
\begin{eqnarray}
\mathcal{R}_1 =
\frac{\Gamma \left(\frac{\delta \lambda + \ell_1}{2} \right) \Gamma \left(\frac{\delta \lambda - a_2d_2\ell_2}{2}\right)}{\Gamma \left(\frac{\delta \lambda}{2}\right) \Gamma \left(\frac{\delta \lambda + \ell_1-a_2d_2\ell_2}{2}\right)}
\begin{pmatrix}
  1 & 0 & 0 & 0 \\
  0 & \frac{\delta \lambda-a_2 d_2 \ell_2}{\delta \lambda} & 0 & 0 \\
  0 & 0 & 1 & 0 \\
  0 & 0 & 0 &  \frac{\delta \lambda-a_2 d_2 \ell_2}{\delta \lambda}
\end{pmatrix}.
\end{eqnarray}
One can perform an analogous derivation for $R_2$ and find
\begin{eqnarray}
\mathcal{R}_2 =
\frac{\Gamma \left(\frac{\delta \lambda + a_1d_1\ell_1+2}{2}\right) \Gamma \left(\frac{\delta \lambda-\ell_2+2}{2}\right)}{\Gamma \left(\frac{\delta \lambda+2}{2}\right) \Gamma \left(\frac{\delta \lambda + a_1d_1\ell_1-\ell_2+2}{2}\right)}
\begin{pmatrix}
  1 & 0 & 0 & 0 \\
  0 & 1 & 0 & 0 \\
  0 & 0 & \frac{\delta \lambda}{\delta \lambda + a_1d_1 \ell_1} & 0 \\
  0 & 0 & 0 &  \frac{\delta \lambda}{\delta \lambda + a_1 d_1 \ell_1}
\end{pmatrix}.
\end{eqnarray}
Multiplying everything out finally gives us the universal R-matrix
in our bound state representation:
\begin{eqnarray}
\label{uni}
\mathcal{R} = A
\left(
\begin{array}{cccc}
 1 & 0 & 0 & 0 \\
 0 & 1-\frac{a_2 d_2 \ell_2}{\delta \lambda+a_1 d_1 \ell_1} & \frac{a_1 d_2 \ell_1}{\delta \lambda+a_1
   d_1 \ell_1} & 0 \\
 0 & \frac{a_2 d_1 \ell_2}{\delta \lambda+a_1 d_1 \ell_1} & \frac{\delta \lambda}{\delta \lambda+a_1 d_1 \ell_1} &
   0 \\
 0 & 0 & 0 & \frac{\delta \lambda-a_2 d_2 \ell_2}{\delta \lambda+a_1 d_1 \ell_1}
\end{array}
\right),
\end{eqnarray}
where
\begin{eqnarray}
\label{normA} A=\frac{\Gamma \left(\frac{\delta
\lambda+\ell_1}{2}\right) \Gamma \left(\frac{\delta \lambda+a_1 d_1
   \ell_1+2}{2}\right) \Gamma \left(\frac{\delta \lambda-\ell_2+2}{2}\right) \Gamma \left(\frac{\delta \lambda-a_2
   d_2 \ell_2}{2} \right)}{\Gamma \left(\frac{\delta \lambda}{2}\right) \Gamma \left(\frac{\delta \lambda+2}{2}\right) \Gamma \left(\frac{\delta \lambda+a_1 d_1 \ell_1-\ell_2+2}{2} \right) \Gamma \left(\frac{\delta \lambda+\ell_1-a_2 d_2\ell_2}{2}\right)}.
\end{eqnarray}
For $a_i=d_i=\ell_i=1$ this reduces to the formula in \cite{Cai:q-alg9709038},
\begin{eqnarray}
\mathcal{R} \propto \mathbbm{1} + \frac{P}{\delta \lambda},
\end{eqnarray}
where $P$ is the graded permutation matrix.

But we can also take $a,d$ to be the representation labels of the
supercharges in the centrally extended ${\alg{psu}}(2|2)$
superalgebra, i.e.
\begin{eqnarray}
a = \sqrt{\frac{g}{2\ell}}\eta, \qquad
d=\sqrt{\frac{g}{2\ell}}\frac{x^{+}-x^-}{i\eta}.
\end{eqnarray}
This corresponds to considering the generators $\xi^{\pm}$ as the
restriction to this subsector of the two supercharges
$\mathbb{Q}_{1}^{3}$ and $\mathbb{Q}_{3}^{\dag 1}$. It is now readily
seen that by choosing $\lambda$ to be $\frac{g}{2i} x^-$, we can
exactly reproduce\footnote{This is similar to the observation in
\cite{Beisert:2005wm} for the the case of the fundamental
representation.} the $4\times4$ block (\ref{eqn;4x4Smat}) from
(\ref{uni}), after we properly normalize it and introduce the
appropriate braiding factors. To normalize, we simply divide the
formula coming from the universal R-matrix by $A$ (\ref{normA}).
To introduce the braiding factors, we need to twist it by
\cite{Arutyunov:2008zt}
$$
U_2^{-1} (p_1) \, \mathcal{R} \, U_1 (p_2),
$$
with $U(p) = diag(1,e^{- i p/2})$.

One can also restrict the supercharges $\mathbb{Q}_{2}^{4}$ and
$\mathbb{Q}_{4}^{\dag2}$ to this sector and repeat the procedure.
Remarkably, in order to match with (\ref{eqn;4x4Smat}), one has to
choose $\lambda = \frac{i g}{2 x^-}$. The correct braiding factors
can be incorporated by means of the inverse of the above mentioned
twist \cite{Arutyunov:2008zt}.

A similar argument can finally be seen to hold for all the other
subsectors corresponding to different choices of the fixed bosonic
and fermionic indices.

While it is likely that in the full universal R-matrix (where one
is supposed to have at once all generators of ${\alg{psu}}(2|2)$)
some kind of ``average" of the two situations will
occur\footnote{In the fundamental representation, this is
exemplified by some of the formulas in \cite{Torrielli:2008wi}.},
we have shown here that the S-matrix in these subspaces can be
``effectively" described by the universal R-matrix of
$DY(\alg{gl}(1|1))$ taken in (two inequivalent choices of)
evaluation representations.

\section{Summary}

In this chapter we explored some universal structures underlying
the bound state S-matrix that was derived in chapter
\ref{chap;BoundSmat}. We found two representations that are
contained in the bound state representation of $\alg{h}$, namely a
$\alg{su}(2)$ and a $\alg{gl}(1|1)$. For these algebras the
universal R-matrix is known
\cite{Khoroshkin:1994uk,Cai:q-alg9709038} and one finds that the
restriction of the bound state S-matrix to these subspaces indeed
agrees with these universal R-matrices.

%% file: BetheAnsatz.tex
\chapter{The Coordinate Bethe Ansatz}\label{chap;BetheAnsatz}

The spectrum in integrable models can often be computed exactly by
a technique called the Bethe ansatz. It was first used in the
context of the one-dimensional XXX Heisenberg model in 1931
\cite{Bethe:1931hc}. In this approach one makes an explicit
plane-wave type ansatz for the eigenvectors of the Hamiltonian. On
these eigenstates one can then impose periodic boundary
conditions, which result in a quantization condition of the
particle momenta. From this ansatz one computes the exact
eigenvalues of the Hamiltonian for the system. This version of the
Bethe ansatz is commonly called the coordinate Bethe ansatz.

The coordinate Bethe ansatz procedure for theories with multiple
species of particles was first solved for a system with a
repulsive $\delta$-interaction \cite{Yang:1967bm}. As we will see,
the fact that there are different types of particles in the model
results in a certain matrix structure. To deal with this, one
makes repeated use of a Bethe ansatz. For this reason this
approach is referred to as the nested Bethe ansatz.

In this chapter we will apply the nested Bethe ansatz to bound
states of the $\ads$ superstring. However, the Hamiltonian is not
explicitly known. Nevertheless, one can apply the Bethe ansatz
procedure by using the explicit S-matrix and the dispersion
relation. We will first exemplify the nested Bethe ansatz
procedure in the nonlinear Schr\"odinger model, highlighting some
of its features, before moving to the string model.

\section{Formalism}

We will set up the machinery of the Bethe ansatz in the context of
field theories. We start by discussing the nonlinear Schr\"odinger
model which is a theory without internal degrees of freedom. After
this we will add internal degrees of freedom and discuss the
nesting procedure.

\subsection{Nonlinear Schr\"odinger Model}

Let us first discuss the Bethe ansatz in the nonlinear
Schr\"odinger model. An excellent review on this subject is given
in \cite{Thacker}. The nonlinear Schr\"odinger model is defined by
the following Hamiltonian operator
\begin{align}\label{eqn;HamiltonianNLSE}
H = \int dx \left[\partial_x \phi^* \partial_x \phi + c
\phi^*\phi^*\phi\phi\right],
\end{align}
where $\phi$ is a bosonic field with canonical equal time
commutation relations
\begin{align}\label{eqn:CommRulesNLSE}
[\phi(x),\phi^*(y)] = \delta(x-y).
\end{align}
This model is integrable and the Hamiltonian commutes with the
number operator $N = \int dx \phi^*\phi$ (the number of particles
is preserved). This means that we can consider the different
$N$-body states of the Hilbert space separately. For simplicity we
will restrict to two-particle wave functions.

The Bethe ansatz is a plane-wave type ansatz for the eigenstates
of the Hamiltonian. Consider the system on a line. Suppose we have
two particles with momenta $k_1,k_2$ at positions $x_1,x_2$ that
are well-separated. If $x_1<x_2$, the particles do not interact
and the natural guess for the wave function of this configuration
would be
\begin{align}
|\Psi(k_1,k_2)\rangle_{1} = \int dx_1 dx_2
\theta(x_1<x_2)e^{i(k_1x_1+k_2x_2)}\phi^*(x_1)\phi^*(x_2)|0\rangle,
\end{align}
where $\theta$ is the heaviside step function
\begin{align}
\theta(x<y)\equiv \theta(y-x)= \left\{ \begin{array}{l} 1  \mathrm{\ if\ }x<y,\\
\frac{1}{2} \mathrm{\ if\ }x=y,\\ 0 \mathrm{\ else.}
\end{array}\right.
\end{align}
Similarly for $x_1>x_2$ one would write
\begin{align}
|\Psi(k_1,k_2)\rangle_{2} = \int dx_1 dx_2
\theta(x_2<x_1)e^{i(k_1x_1+k_2x_2)}\phi^*(x_1)\phi^*(x_2)|0\rangle.
\end{align}
Of course when traversing from $x_1<x_2$ to $x_1<x_2$ the
particles come within interaction range and should interact
according to the S-matrix. This leads us to the following ansatz
for the total eigenstate
\begin{align}\label{eqn;NLSMeigenstate}
|\Psi(k_1,k_2)\rangle =& |\Psi(k_1,k_2)\rangle_{1} + \mathcal{A}|\Psi(k_1,k_2)\rangle_{2}\nonumber\\
 = &\int dx_1 dx_2
[\theta(x_1<x_2)+\theta(x_2<x_1)\mathcal{A}]e^{i(k_1x_1+k_2x_2)}\phi^*(x_1)\phi^*(x_2)|0\rangle,
\end{align}
where the coefficient $\mathcal{A}$ should capture the scattering
information.

Since $|\Psi(k_1,k_2)\rangle$ should be an eigenstate of the
Hamiltonian, we will find restrictions on $\mathcal{A}$. By the
canonical commutation rules (\ref{eqn:CommRulesNLSE}) one finds
\begin{align}
[H,\phi^*] = -\partial^2 \phi^* + 2c \phi^*\phi^*\phi.
\end{align}
From this it is easily deduced that
\begin{align}\label{eqn;NLSMintermediateEigenstate}
H |\Psi(k_1,k_2)\rangle = &\int dx_1 dx_2
[\theta(x_2-x_1)+\theta(x_1-x_2)\mathcal{A}]e^{i(k_1x_1+k_2x_2)}\times\\
&\times \left(2 c \phi^*(x_1) \phi^*(x_2)\delta(x_1-x_2) -
\partial^2\phi^*(x_1) \phi^*(x_2) - \phi^*(x_1)
\partial^2\phi^*(x_2)
\right)|0\rangle.\nonumber
\end{align}
Let us first work out the term proportional to $c$. By using the
relation $\delta(x)\theta(x)= \frac{1}{2}\delta(x)$ this term
reduces to
\begin{align}
(1+\mathcal{A})c\int dx~ e^{i(k_1+k_2)x}
\phi^*(x)\phi^*(x)|0\rangle.
\end{align}
The derivative term in the above expression can be evaluated by
repeated partial integrating. Let us focus on the term
$\partial^2\phi^*(x_1) \phi^*(x_2)$
\begin{align}
&\int dx_1 dx_2
[\theta(x_2-x_1)+\theta(x_1-x_2)\mathcal{A}]e^{i(k_1x_1+k_2x_2)}
\partial^2\phi^*(x_1) \phi^*(x_2)|0\rangle\nonumber\\
&\qquad\qquad=-\int dx_1 dx_2
ik_1[\theta(x_2-x_1)+\theta(x_1-x_2)\mathcal{A}]e^{i(k_1x_1+k_2x_2)}
\partial\phi^*(x_1) \phi^*(x_2)|0\rangle + \nonumber\\
&\qquad\qquad\ \ \  - \int dx_1 dx_2
(\mathcal{A}-1)\delta(x_1-x_2)e^{i(k_1x_1+k_2x_2)}
\partial\phi^*(x_1) \phi^*(x_2)|0\rangle\nonumber \\
&\qquad\qquad=\int dx_1 dx_2
-k_1^2[\theta(x_2-x_1)+\theta(x_1-x_2)\mathcal{A}]e^{i(k_1x_1+k_2x_2)}
\phi^*(x_1) \phi^*(x_2)|0\rangle + \nonumber\\
&\qquad\qquad \ \ \ +\int dx_1 dx_2
ik_1(\mathcal{A}-1)\delta(x_2-x_1)e^{i(k_1x_1+k_2x_2)}
\phi^*(x_1) \phi^*(x_2)|0\rangle + \\
&\qquad\qquad \ \ \  - \int dx (\mathcal{A}-1)e^{i(k_1+k_2)x}
\frac{1}{2}\partial(\phi^*(x))^2 |0\rangle\nonumber\\
&\qquad\qquad=-k_1^2\int dx_1 dx_2
[\theta(x_2-x_1)+\theta(x_1-x_2)\mathcal{A}]e^{i(k_1x_1+k_2x_2)}
\phi^*(x_1) \phi^*(x_2)|0\rangle + \nonumber\\
&\qquad\qquad\ \ \  + \int dx i(k_1 +
\frac{k_1+k_2}{2})(\mathcal{A}-1)e^{i(k_1+k_2)x}
\phi^*(x)\phi^*(x)|0\rangle\nonumber
\end{align}
where we used $\partial_x \theta(x) = \delta (x)$ and applied
partial integration in all the steps. The other derivative term in
(\ref{eqn;NLSMintermediateEigenstate}) can be computed analogously
and we obtain
\begin{align}
H |\Psi(k_1,k_2)\rangle =& (k_1^2+k_2^2) |\Psi(k_1,k_2)\rangle +\\
&+\left\{c (1+\mathcal{A})+i(k_1-k_2)(1-\mathcal{A} )\right\}\int
dx~ e^{i(k_1+k_2)x} \phi^*(x)\phi^*(x)|0\rangle.\nonumber
\end{align}
From this one finds that $|\Psi(k_1,k_2)\rangle$ is an eigenstate
of the Hamiltonian provided that
\begin{align}
i(k_1-k_2)(1-\mathcal{A}) + c (1+\mathcal{A}) = 0,
\end{align}
which is solved by
\begin{align}\label{eqn;SmatNLSE}
\mathcal{A}(k_2,k_1) = \frac{k_2-k_1-ic}{k_2-k_1+ic}.
\end{align}
Concluding, with this choice for $\mathcal{A}$ we have that
$|\Psi(k_1,k_2)\rangle$ is an eigenstate of the Hamiltonian with
eigenvalue
\begin{align}\label{eqn;NLSMeigenvalues}
H |\Psi(k_1,k_2)\rangle = (k_1^2+k_2^2)|\Psi(k_1,k_2)\rangle.
\end{align}
The value for $\mathcal{A}$, (\ref{eqn;SmatNLSE}) is actually the
two-particle S-matrix. This can be seen in a very intuitive way.
Consider our ansatz with $k_1>k_2$, then
$|\Psi(k_1,k_2)\rangle_{1}$ has the interpretation of two
particles that are going to collide. This means that for $k_1>k_2$
one would consider $|\Psi(k_1,k_2)\rangle_1$ to be an in-state.
Equivalently, for $k_2>k_1$ this would be an out-state, from which
the relation of $\mathcal{A}$ with the S-matrix becomes apparent.
It is now straightforward to extend this ansatz to more than two
particles. Notice also that in case of vanishing interaction
($c=0$) one finds $\mathcal{A} = 1$ and the eigenstate reduces to
a genuine sum of plane-waves.

The above discussion is valid on an infinite line. There is no
restriction on the values of the momenta $k_1,k_2$ and the
spectrum is continuous. The next step is to consider periodic
boundary conditions. Consider the system on a line of large length
$L\rightarrow\infty$, then the wave function
\begin{align}
\psi(x_1,x_2):=\langle0|\phi(x_1)\phi(x_2)|\Psi(k_1,k_2)\rangle
\end{align}
needs to be periodic
\begin{align}
&\psi(0,x_2) = \psi(L,x_2), && \psi(x_1,0) = \psi(x_1,L).
\end{align}
One immediately sees that this reduces to the following equations
\begin{align}\label{eqn;BAE-NLSM}
e^{i k_j L} = \prod_{i\neq j}\mathcal{A}(k_j,k_i),
\end{align}
which are the Bethe equations. They capture the $\frac{1}{L}$
corrections to the momenta. This can be seen by considering for
example the case $c=0$. Solving $k$ will then give $k=2n\pi/L$ for
some integer $n$. One now obtains the eigenvalues of the
Hamiltonian by solving (\ref{eqn;BAE-NLSM}) and plugging the
solutions in equation (\ref{eqn;NLSMeigenvalues}).

We work in the regime $L\rightarrow\infty$ because we have freely
made use of partial integration discarding all boundary terms. To
this end we tacitly assumed that we worked with rapidly decreasing
fields.

\begin{figure}[h]
  \centering
\parbox[b]{5in}{  \includegraphics[scale = .7]{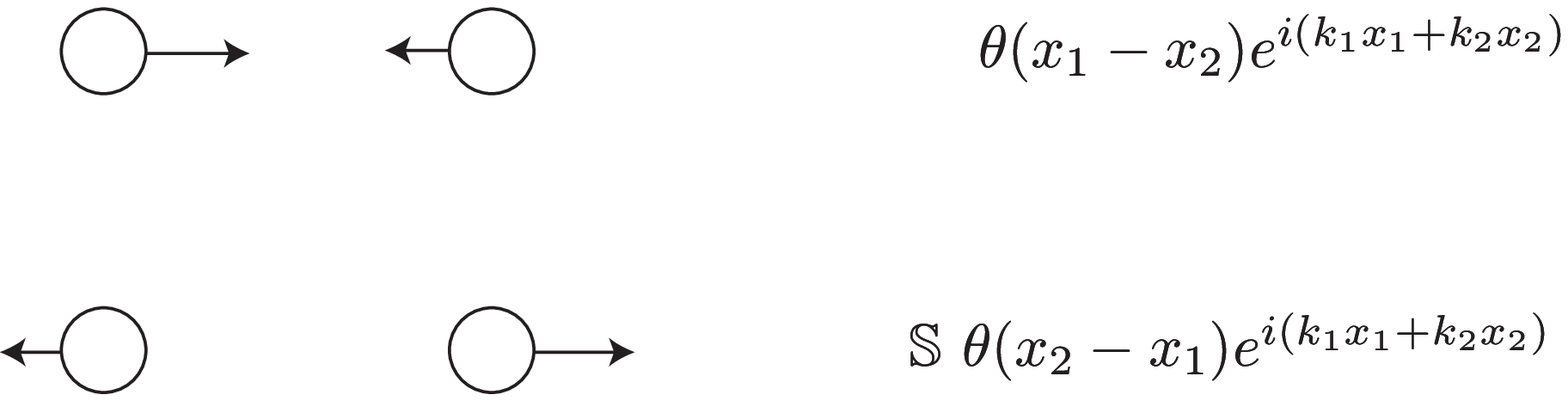}
  \caption{Schematic overview of the Bethe ansatz for two-particles. There are two regions for which one makes a plane-wave ansatz.
  The regions are denoted by the theta-functions and related via the S-matrix.}\label{Fig;BetheA}}
\end{figure}

In short, in the coordinate Bethe ansatz one builds up an
eigenstate of the Hamiltonian of the system by considering regions
where the particles are separated. In each of these regions one
makes an ansatz for the wave function consisting of a sum of plane
waves. The coefficients relating the different regions are
described by the S-matrix of the system. This is exemplified in
figure \ref{Fig;BetheA}. The reasoning behind this ansatz is that
if the particles are well-separated they do not feel interactions
and behave as plane-waves \cite{Zamolodchikov:1978xm}. When
particles scatter they will cross regions and pick up the factor
from the S-matrix. Since the system is integrable, only
two-particle interactions will play a role. Periodicity is then
imposed on these coefficients, resulting in a set of Bethe
equations for the particle momenta $k_i$.

\subsection{Adding color}

We now consider a system with internal degrees of freedom, i.e.
particles have color. As an example we will consider the system
described by the Hamiltonian \cite{Thacker}
\begin{align}
H = \int dx~ \sum_{a=1,2}\partial_x \phi^*_a(x)\partial_x
\phi_a(x) +
\sum_{a,b=1,2}\phi^*_a(x)\phi^*_b(x)\phi_a(x)\phi_b(x).
\end{align}
The natural way to generalize (\ref{eqn;NLSMeigenstate}) would be
to consider all possible types of orderings of momenta and
positions of the particles. For two particles this becomes
\begin{align}
|\Psi(k_1,k_2)\rangle = \int dx_1 dx_2 ~
\psi(x_1,x_2)\phi^*_{a_1}(x_1)\phi^*_{a_2}(x_1)|0\rangle,
\end{align}
with
\begin{align}
\psi(x_1,x_2) =\sum_{\mathcal{P},\mathcal{Q}}
\mathcal{A}^{\mathcal{P}|\mathcal{Q}}\theta(x_{\mathcal{Q}_1}<x_{\mathcal{Q}_2})
e^{i(k_{\mathcal{P}_1}x_{\mathcal{Q}_1}+k_{\mathcal{P}_2}x_{\mathcal{Q}_2})},
\end{align}
where $\mathcal{A}^{\mathcal{P}|\mathcal{Q}}$ are constants, and
$\mathcal{P},\mathcal{Q}$ are permutations of $\{1,2\}$.
Explicitly, we get
\begin{align}
\psi(x_1,x_2) =& \theta(x_1<x_2)\left\{\mathcal{A}^{12|12}
e^{i(k_1 x_1 + k_2 x_2)} + \mathcal{A}^{12|21} e^{i(k_2 x_1 + k_2
x_1)}\right\} +\nonumber\\
&+ \theta(x_2<x_1)\left\{\mathcal{A}^{21|12} e^{i(k_1 x_2 + k_2
x_1)} + \mathcal{A}^{21|21} e^{i(k_1 x_1 + k_2 x_2)}\right\}
\end{align}
This wave function is continuous if we require that
\begin{align}
\mathcal{A}^{12|12} +  \mathcal{A}^{12|21} = \mathcal{A}^{21|12} +
\mathcal{A}^{21|21}.
\end{align}
By performing a similar computation as in the previous section,
one finds that $|\Psi(k_1,k_2)\rangle$ is a continuous eigenstate
of the Hamiltonian provided the coefficients satisfy
\begin{align}\label{eqn;SmatNLSMNested}
&{\mathcal{A}^{12|21} \choose \mathcal{A}^{21|21}} =-
\frac{\lambda_{12}}{ 1 + \lambda_{12}} {\mathcal{A}^{12|12}
\choose \mathcal{A}^{21|12}} + \frac{1}{ 1 + \lambda_{12}}
{\mathcal{A}^{21|12} \choose \mathcal{A}^{12|12}}, &&\lambda_{ij}
= \frac{ic}{k_i-k_j}.
\end{align}
The eigenvalue of $|\Psi(k_1,k_2)\rangle$ is
\begin{align}
H|\Psi(k_1,k_2)\rangle = (k_1^2 + k_2^2)|\Psi(k_1,k_2)\rangle.
\end{align}
We would once again like to interpret the different terms in the
wave function as in- and out-states. To this end it is convenient
to make the change of variables $x_1\leftrightarrow x_2$ in the
terms proportional to $\mathcal{A}^{12|21},\mathcal{A}^{21|12}$ to
obtain
\begin{align}\label{eqn;NLSMinandoutNested}
|\Psi(k_1,k_2)\rangle =& \int dx_1 dx_2\\
& e^{i(k_1 x_1 + k_2 x_2)}\theta(x_1<x_2)\left\{
\mathcal{A}^{12|12} \phi^*_{a_1}(x_1)\phi^*_{a_2}(x_2) +
\mathcal{A}^{21|12} \phi^*_{a_2}(x_1)\phi^*_{a_1}(x_2)
\right\}|0\rangle +\nonumber\\
&+ e^{i(k_1 x_1 + k_2 x_2)}\theta(x_2<x_1)\left\{
\mathcal{A}^{21|21} \phi^*_{a_1}(x_1)\phi^*_{a_2}(x_2) +
\mathcal{A}^{12|21} \phi^*_{a_2}(x_1)\phi^*_{a_1}(x_2)
\right\}|0\rangle\nonumber.
\end{align}
In this expression on recognizes the first term as an in-state,
which is a superposition of a wave where the first particle has
color $a_1$ and the second $a_2$ and of a wave with the color
indices interchanged. The different coefficients from
(\ref{eqn;SmatNLSMNested}) then admit an interpretation as being
coefficients of the S-matrix. The permutation $\mathcal{Q}$ in
coefficients $\mathcal{A}^{\mathcal{P}|\mathcal{Q}}$ actually
labels the color distribution. To make this concrete let us
explicitly include these indices, i.e. we write
\begin{align}\label{eqn;CoefficientWithIndicesNLSM}
&\mathcal{A}^{12}_{a_1 a_2} \equiv \mathcal{A}^{12|12}, && \mathcal{A}^{12}_{a_2 a_1} \equiv \mathcal{A}^{21|12}, \nonumber\\
&\mathcal{A}^{21}_{a_1 a_2} \equiv \mathcal{A}^{21|21}, &&
\mathcal{A}^{21}_{a_2 a_1} \equiv \mathcal{A}^{12|21}.
\end{align}
There are two distinct cases to be considered, namely $a_1=a_2$
and $a_1\neq a_2$. For the case $a_1 = a_2$ it is readily seen
that both $\mathcal{A}^{12|12}$ and $\mathcal{A}^{21|12}$ describe
the same state, which is made particulary clear by our notation
above (\ref{eqn;CoefficientWithIndicesNLSM}). Hence in this case
our state is described by just two components
$\mathcal{A}^{12|12}=\mathcal{A}^{12}_{a_1 a_1}$ and
$\mathcal{A}^{21|21}=\mathcal{A}^{21}_{a_1 a_1}$. One finds that
(\ref{eqn;SmatNLSMNested}) implies
$\mathcal{A}^{21|21}=\frac{1-\lambda_{12}}{1+\lambda_{12}}\mathcal{A}^{12|12}$,
which agrees with the result (\ref{eqn;SmatNLSE}) from the
previous section. Define the following S-matrix on
$\mathbb{C}^2\otimes\mathbb{C}^2$
\begin{align}
&\S = \frac{\mathbbm{1}  -  \lambda_{12} P }{1+\lambda_{12}}, &&
\S = \S^{ij}_{kl}E^{i}_k\otimes E^j_l,
\end{align}
where $P$ is the permutation operator and $E^i_j$ are the standard
matrix unities, that have all zeroes except for a $1$ at position
$(j,i)$. The S-matrix $\S$ can easily be shown to satsify
unitarity and the Yang-Baxter equation. From
(\ref{eqn;SmatNLSMNested}) we find that $|\Psi(k_1,k_2)\rangle$ is
an eigenfunction of the Hamiltonian provided that
\begin{align}\label{eqn;SmatNLSMNested2}
\mathcal{A}^{21}_{a_1 a_2} = \S\cdot\mathcal{A}^{12}_{a_1 a_2}= \S^{b_1 b_2}_{a_1 a_2} \mathcal{A}^{12}_{b_1 b_2},
\end{align}
where repeated indices are summed over. Because of this,
specifying the vector ${ \mathcal{A}^{12}_{a_1a_2} \choose
\mathcal{A}^{12}_{a_2a_1} }$ fixes the wave function completely.

The generalization to $N$ particles is now obvious; one takes
\begin{align}\label{eqn;AnsatzNLSMcolor}
\psi = \sum_{\mathcal{P},\mathcal{Q}}
\mathcal{A}^{\mathcal{P}|\mathcal{Q}}\theta(x_{\mathcal{Q}_1}<\ldots<x_{\mathcal{Q}_N})
e^{i\sum_i k_{\mathcal{P}_i}x_{\mathcal{Q}_i}}.
\end{align}
The coefficients of the vectors $\xi_{\mathcal{P}}\equiv
(\mathcal{A}^{\mathcal{P}}_{a_1\ldots a_N})$ are then related by
(\ref{eqn;SmatNLSMNested2})
\begin{align}
\mathcal{A}^{\mathcal{P}}_{a_1\ldots a_N} = \S_{i i+1}\cdot
\mathcal{A}^{(ii+1)\mathcal{P}}_{a_1\ldots a_N},
\end{align}
where $(ii+1)$ is the 2-cycle that permutes $i$ and $i+1$. Hence
the state is fixed by specifying the vector $\xi_0 \equiv
(\mathcal{A}^{\mathbbm{1}}_{a_1\ldots a_N})$, where $\mathbbm{1}$
stands for the trivial permutation. The periodic boundary
conditions now become the following matrix equation
\begin{align}\label{eqn;NestedNSLMequation}
e^{ik_j L}\xi_0 = \vec{\S}_j \xi_0 \equiv
\S_{j+1,j}\S_{j+2,j}\ldots \S_{N,j}\S_{1,j}\S_{2,j}
\ldots\S_{j-1,j} \xi_0.
\end{align}
From the Yang-Baxter equation it is readily checked that
$[\vec{\S}_i,\vec{\S}_j]=0$. This means that we can simultaneously
diagonalize these matrix operators. To cope with the matrix
structure in (\ref{eqn;NestedNSLMequation}) one introduces a
so-called nested structure. We will do this by finding the
eigenvectors $\xi(N,M)$ of $\vec{\S}_j$ and then require that
their eigenvalues $\Lambda(N,M)$ satisfy $\Lambda(N,M) = e^{ik_j
L}$.

In our system we consider particles of two different species for
simplicity, created by $\phi^*_1,\phi^*_2$ respectively (the
indices label the color). It is readily seen that the S-matrix
preserves the number of particles of type 2 (and hence also of
type 1). This means that there are two numbers that determine an
eigenspace of $\vec{\S}_j$, namely the total number of particles
$N$ and the number of particles of type 2, denoted by $M$. This
reminds of the XXX spin chain were $M$ would be the number of spin
up particles in a background of spin down particles. In analogy to
this, we will identify $\xi_0$ with states of the XXX spin chain
and treat $M$ as excitations thereon.

Consider a spin chain with $N$ sites $\bigotimes_{i=1}^N
\mathbb{C}^2$. To any distribution of particles, i.e. any
component of $\xi_0$, we can associate a state of this chain and
vice versa
\begin{align}\label{eqn;IdentificationSpinChain}
&\mathcal{A}^{\mathbbm{1}}_{a_1\ldots a_N} =
 \mathcal{A}^{\mathbbm{1}}_{a_1\ldots a_N} e_{a_1}\otimes\ldots\otimes e_{a_N}, && e_1 = {1\choose0},&&e_2 =
{0\choose1},
\end{align}
here repeated indices are not summed over. By
(\ref{eqn;SmatNLSMNested2}) we see that the action of the S-matrix
on $\xi_0$ is exactly the same as the action of the S-matrix on
the spin chain. Thus, the diagonalization of $\vec{\S}_j$ is
reduced to a problem on the spin chain. Therefore we will solve
(\ref{eqn;NestedNSLMequation}) on the chain introduced above. Let
us stress that the introduced $N$-site spin chain is more of a
practical bookkeeping device to deal with the index structure than
a real physical system.

We will first first deal with the cases $M=0,1,2$ before
presenting the general ansatz. For $M=0$, all the particles are of
the same color and we are back in the situation discussed in
previous section. There is only one independent component
$\mathcal{A}^{\mathbbm{1}}_{1\ldots 1}$ and the rest are related
to that via the scattering amplitude $\S^{11}_{11}$, e.g.
$\mathcal{A}^{(21)}_{1\ldots 1} =
S_0(k_1,k_2)\mathcal{A}^{\mathbbm{1}}_{1\ldots 1}$, where we
defined
\begin{align}
S_0(k_i,k_j) = \S^{11}_{11}(k_i,k_j)
=\frac{1-\lambda_{ij}}{1+\lambda_{ij}}.
\end{align}
Then (\ref{eqn;NestedNSLMequation}) becomes
\begin{align}
e^{ik_jL} &= \prod_{i\neq j}^N S_0(k_i,k_j),
\end{align}
which indeed agrees with (\ref{eqn;BAE-NLSM}).

For $M=1$ we make the following ansatz
\begin{align}\label{eqn;NLSM-OneExcitation}
&\xi(N,1) = \sum_i \Phi_i(y) \underbrace{e_1 \otimes \ldots \otimes e_{1}}_{i-1} \otimes
 e_2 \otimes \underbrace{e_{1} \ldots \otimes e_1}_{N-i},&&
\Phi_i(y) = f(y,k_i)\prod_{n=1}^{i-1}S(y,k_n),
\end{align}
where $f,S$ are functions that are to be determined. The vector
$\xi(N,1)$ is then related to the actual ansatz
(\ref{eqn;AnsatzNLSMcolor}) via equation
(\ref{eqn;IdentificationSpinChain}). The function $\Phi_i(y)$ in
the above ansatz describes an excitation with momentum $y$ being
moved to the place $i$ on the chain. With a modest amount of
foresight we have put in a parameter $y$ explicitly as it will
appear in the solutions of $f,S$ later. The interpretation is that
when it is moved there, it scatters with the different particles
along the way, producing $i-1$ scattering terms $S(y,k)$ and then
it is inserted at position $i$ that is described by the term
$f(y,k_i)$.

Returning back to the two-particle case, let us see what
restrictions one can put on the functions $f,S$. By relabelling
$x_1 \leftrightarrow x_2$ in (\ref{eqn;NLSMinandoutNested}) one
can map the in-state to the out-state. From this one sees it is
natural to take
\begin{align}
&\mathcal{A}^{21|21}(k_1,k_2) \sim \mathcal{A}^{21|12}(k_2,k_1),&&
\mathcal{A}^{12|21}(k_1,k_2) \sim \mathcal{A}^{12|12}(k_2,k_1).
\end{align}
On the other hand, when $a_1 = a_2$ we know that the coefficients
are related via $S_0(k_1,k_2)$. From this we find that one should
require
\begin{align}
&\mathcal{A}^{21|21}(k_1,k_2) = S_0
\mathcal{A}^{21|12}(k_2,k_1),&& \mathcal{A}^{12|21}(k_1,k_2) =S_0
 \mathcal{A}^{12|12}(k_2,k_1).
\end{align}
In general for $N$ sites, this is formulated as
\begin{align}
\mathcal{A}^{(ii+1)}_{a_1\ldots a_N}(k_1,\ldots,k_N) =
S_0(k_i,k_{i+1})\mathcal{A}^{\mathbbm{1}}_{a_1\ldots
a_{i+1}a_i\ldots a_N}(k_1,\ldots,k_{i+1},k_i,\ldots,k_N).
\end{align}
Via (\ref{eqn;SmatNLSMNested}) one can then relate this to the
S-matrix.
\begin{align}\label{eqn;compatiNLSM}
\S_{ii+1}\mathcal{A}^{\mathbbm{1}}_{a_1\ldots a_N}(k_1,\ldots,k_N)
= S_0(k_i,k_{i+1})\mathcal{A}^{\mathbbm{1}}_{a_1\ldots
a_{i+1}a_i\ldots a_N}(k_1,\ldots,k_{i+1},k_i,\ldots,k_N).
\end{align}
We see that vectors satisfying the above condition diagonalize
$\vec{\S}_j$ `up to a permutation'. Restricting
(\ref{eqn;NLSM-OneExcitation}) to two particles ($N=2$) then
(\ref{eqn;compatiNLSM}) amounts to the following equations
\begin{align}
- \frac{\lambda_{12}}{ 1 + \lambda_{12}} f(k_1,y) + \frac{1}{ 1 +
\lambda_{12}} f(k_2,y)S(k_1,y) &= S_0(k_1,k_2)f(k_2,y) \\
\frac{1}{ 1 + \lambda_{12}} f(k_1,y) - \frac{\lambda_{12}}{ 1 +
\lambda_{12}} f(k_2,y)S(k_1,y) &= S_0(k_1,k_2)f(k_1,y)S(k_2,y).
\end{align}
These equations are uniquely solved by
\begin{align}
&f(k,y) = \frac{y}{k-y+\frac{ic}{2}}, && S(k,y) =
\frac{k-y-\frac{ic}{2}}{k-y+\frac{ic}{2}},
\end{align}
where $y$ arises as an integration constant. Returning back to our
original matrix equation (\ref{eqn;NestedNSLMequation}) one can
check that $\xi(N,1)$ is an eigenstate provided
\begin{align}
1&=\prod_{i=1}^N
\frac{k_i-y_{\alpha}-\frac{ic}{2}}{k_i-y_{\alpha}+\frac{ic}{2}}.
\end{align}
The eigenvalue equation becomes
\begin{align}
e^{ik_jL} &=
\frac{k_i-y-\frac{ic}{2}}{k_i-y+\frac{ic}{2}}\prod_{i\neq j}
S_0(k_i,k_j).
\end{align}
Finally let us treat the case $M=2$. In this case we introduce an
auxiliary S-matrix $\S^{\mathrm{II}}$ that deals with the
scattering of two auxiliary excitations on the spin chain with
momenta $y_1,y_2$. Explicitly for two excitations one makes the
following ansatz
\begin{align}
\xi(N,2) =& \sum_{j<i} \Phi_i(y_1)\Phi_j(y_2) \underbrace{e_1 \otimes \ldots \otimes e_{1}}_{j-1} \otimes
 e_2 \otimes \underbrace{e_{1} \ldots \otimes e_1}_{i-j-1}\otimes e_2\otimes \underbrace{e_{1} \ldots \otimes e_1}_{N-i}+\nonumber\\
& \S^{\mathrm{II}}\sum_{j<i} \Phi_i(y_2)\Phi_j(y_1) \underbrace{e_1 \otimes \ldots \otimes e_{1}}_{j-1} \otimes
 e_2 \otimes \underbrace{e_{1} \ldots \otimes e_1}_{i-j-1}\otimes e_2\otimes \underbrace{e_{1} \ldots \otimes e_1}_{N-i}.
\end{align}
We again require (\ref{eqn;compatiNLSM}). By considering the
restricting of the expression for $\xi(N,2)$ to two sites one
finds a unique solution $\S^{\mathrm{II}}$ of
(\ref{eqn;compatiNLSM})
\begin{align}
\S^{\mathrm{II}}(y_1,y_2) = \frac{y_1-y_2-ic}{y_1-y_2+ic}.
\end{align}
In general we make the following ansatz for $\xi(N,M)$
\begin{align}
\xi(N,M) =& \sum_{i_1<\ldots<i_M}
\Phi_{i_1}(y_1)\ldots\Phi_{i_M}(y_M) \underbrace{e_1 \otimes
\ldots \otimes e_{1}}_{i_1-1} \otimes
 e_2 \otimes e_1\ldots \otimes \ldots + \nonumber\\
& + \S_{12}^{\mathrm{II}}  \sum_{i_1<\ldots<i_M}
\Phi_{i_1}(y_2)\Phi_{i_2}(y_1)\ldots\Phi_{i_M}(y_M)
\underbrace{e_1 \otimes \ldots \otimes e_{1}}_{i_1-1} \otimes
 e_2 \otimes e_1\otimes \ldots +\nonumber\\
&+\ldots.
\end{align}
where $\ldots$ stands for all terms with the auxiliary rapidities
$y_i$ permuted and multiplied with the appropriate factor. By
using the explicit expression for $\xi(N,M)$, one can see that it
is an eigenstate of $\vec{\S}_i$ from
(\ref{eqn;NestedNSLMequation}) if
\begin{align}\label{eqn;AuxBAE-NLSM}
1&=\prod_{i=1}^N
\frac{k_i-y_{\alpha}-\frac{ic}{2}}{k_i-y_{\alpha}+\frac{ic}{2}}
\prod_{\beta=1}^M
\frac{y_{\alpha}-y_{\beta}-ic}{y_{\alpha}-y_{\beta}+ic}.
\end{align}
The complete solution of the eigenvalue equation
(\ref{eqn;NestedNSLMequation}) is now encoded in a set of coupled
algebraic equations
\begin{align}
e^{ik_jL} &= \prod_{i\neq j}^N S_0(k_i,k_j) \prod_{\alpha=1}^M
\frac{k_j-y_{\alpha}-\frac{ic}{2}}{k_j-y_{\alpha}+\frac{ic}{2}}\label{eqn;FullBAE-NLSMA}\\
1&=\prod_{i=1}^N
\frac{k_i-y_{\alpha}-\frac{ic}{2}}{k_i-y_{\alpha}+\frac{ic}{2}}
\prod_{\beta=1}^M
\frac{y_{\alpha}-y_{\beta}-ic}{y_{\alpha}-y_{\beta}+ic}\label{eqn;FullBAE-NLSMB}.
\end{align}
Concluding we find that $|\Psi(k_1,k_2)\rangle$ is an
eigenfunction of the Hamiltonian $H$ on a circle of circumference
$L$ with eigenvalue $\sum_{i=1}^N k_i^2$ provided the momenta
$k_i$ satisfy (\ref{eqn;FullBAE-NLSMA}),(\ref{eqn;FullBAE-NLSMB}).

\section{The $\su(2|2)$ (nested) coordinate Bethe Ansatz}

The Hamiltonian for the $\ads$ superstring is not explicitly
known. Therefore, one can not explicitly propose a state
(\ref{eqn;NLSMinandoutNested}) and compute the action of the
Hamiltonian on this. Even so, in the nonlinear Schr\"odinger model,
integrability implies that the coefficients of the eigenstate are
related via the S-matrix and the eigenvalue is described by the
dispersion relation $H(k) = k^2$.

Luckily, for the $\ads$ superstring both the S-matrix and the
dispersion relation are explicitly known from symmetries. Thus we
will use these objects to construct our Bethe ansatz rather than
the explicit Hamiltonian. The dispersion relation is given by
\cite{Beisert:2004hm}
\begin{align}
H(p) = \sqrt{1+4 g^2 \sin^2 \frac{p}{2}}
\end{align}
and the S-matrix was explicitly derived in chapter
\ref{chap;BoundSmat}. The full S-matrix contains two copies of the
$\alg{h}$ invariant S-matrix, cf. section \ref{sec;ConclSmat}. We
will first restrict to just one copy of the invariant S-matrix.

Mimicking the discussion from the previous section, we divide the
space into asymptotic regions $\mathcal{P}|\mathcal{Q}$. In these
regions we make the following ansatz
\begin{align}
\psi = \sum_{\mathcal{P},\mathcal{Q}}
\mathcal{A}^{\mathcal{P}|\mathcal{Q}}\theta(x_{\mathcal{Q}_1}<\ldots<x_{\mathcal{Q}_{K^{\mathrm{I}}}})
e^{i\sum_i p_{\mathcal{P}_i}x_{\mathcal{Q}_i}}.
\end{align}
The wave functions in the different asymptotic regions are related
via the S-matrix. Since we are dealing with closed strings, we
impose periodicity, from which the BAE can be read off:
\begin{align}\label{eqn;NestedEquationSU22}
e^{ik_j L}\xi_0 = \vec{\S}_j \xi_0 \equiv
\S_{j+1,j}\S_{j+2,j}\ldots \S_{K^{\mathrm{I}},j}\S_{1,j}\S_{2,j}
\ldots\S_{j-1,j} \xi_0,
\end{align}
where $\xi_0 = (\mathcal{A}^{\mathbbm{1}}_{a_1\ldots
a_{K^{\mathrm{I}}}})$ is again the vector describing the in-state
and this time the indices $a_i$ run in the $4\ell_i$ dimensional
bound state representation.

The auxiliary spin chain whose states can be identified with
$\xi_0$ is obtained under the identification
\begin{align}
\mathcal{A}^{\mathbbm{1}}_{a_1\ldots a_{K^{\mathrm{I}}}} \sim
e_{a_1}\otimes\ldots\otimes e_{a_{K^{\mathrm{I}}}} \in
 V_{\ell_1}(p_1)\otimes\ldots\otimes
V_{\ell_{K^{\mathrm{I}}}}(p_{K^{\mathrm{I}}}),
\end{align}
where $V_{\ell_i}(p_i)$ is the bound state representation with
bound state number $\ell_i$ and momentum $p_i$.

For simplicity, we will first discuss the Bethe ansatz for
fundamental particles. This discussion is straightforwardly
generalized to bound states. In the second part of this chapter we
will discuss a slightly different approach to apply the Bethe
ansatz procedure; we will employ the Yangian symmetry of the
S-matrix to obtain the BAE rather than the S-matrix itself.

\subsection{Solving for the coefficients}

In chapter \ref{chap;BoundSmat} it was shown that three quantum
numbers $K^{\mathrm{I}},K^{\mathrm{II}}$ and $K^{\mathrm{III}}$
are preserved in scattering processes. $K^{\mathrm{I}}$
corresponds to the number of particles, $K^{\mathrm{II}}$ is the
total number of fermions and $K^{\mathrm{III}}$ labels the number
of fermions of type $4$. This means that the eigenspaces of
$\vec{\S}_j$ are labelled by these numbers. We will now construct
the eigenvectors
$\xi(K^{\mathrm{I}},K^{\mathrm{II}},K^{\mathrm{III}})$ of
$\vec{\S}_j$ in a way analogous to the nonlinear Schr\"odinger
model.

Let us first define a `vacuum' on the spin chain:
\begin{eqnarray}\label{eqn;vacSU22NBA}
\xi(K,0,0) = |0\rangle = w_{1}^{(1)}\ldots w_{1}^{(K)},
\end{eqnarray}
where $w_1^{(i)}$ is the bosonic variable associated to the space
$V_{\ell_i}(p_i)$ in the superspace formalism. We find that
\begin{eqnarray}
\vec{\S}_j|0\rangle = \prod_{i\neq j} S_0(k_i,k_j) |0\rangle.
\end{eqnarray}
In the light of (\ref{eqn;NestedEquationSU22}), this means that
the state (\ref{eqn;vacSU22NBA}) is indeed an eigenstate of
$\vec{\S}_j$ and the corresponding BAE give
\begin{align}
e^{ik_jL} &= \prod_{i\neq j} S_0(k_i,k_j).
\end{align}
The next thing to consider is the case where we have a fermionic
excitation in this vacuum (the case in which the other boson is
inserted is treated later on). We can treat $\xi(K,1,0)$ and
$\xi(K,1,1)$ simultaneously by making an ansatz of the following
form (cf. equation (\ref{eqn;NLSM-OneExcitation})):
\begin{eqnarray}\label{eqn;SU22-OneExcitation}
\xi(K,1)=|\alpha\rangle:= \sum_{i} \Psi_{i}(y) w^{(1)}_{1}\ldots
\theta^{(i)}_{\alpha}\ldots w_{1}^{(K)},\qquad \Psi_{k}(y) =
f(y,p_{k})\prod_{l<k} S^{\mathrm{II,I}}(y,p_{l}),~ ~
\end{eqnarray}
where $\alpha= 3,4$. We will denote $S(y,p)\equiv
S^{\mathrm{II,I}}(y,p)$. We must check whether this construction
is well-defined in the sense that it respects
(\ref{eqn;compatiNLSM}).

Because of the factorization property of the S-matrix, it again
suffices to restrict to a two-particle state. By reducing equation
(\ref{eqn;compatiNLSM}) to this case, one derives from the
explicit form of the S-matrix (cf. equation (\ref{eqn;Sfund})) the
following equations
\begin{align}\label{eqn;level1}
\frac{e^{i\frac{p_1}{2}}}{e^{i\frac{p_2}{2}}}\frac{\eta(p_2)}{\eta(p_{1})}\frac{x^{+}_{1}-x^{-}_{1}}{x^{+}_{1}-x^{-}_{2}}~f(y,p_1)
+
 e^{i\frac{p_1}{2}}\frac{x^{-}_{1}-x^{-}_{2}}{x^{+}_{1}-x^{-}_{2}}~ f(y,p_2)S(y,p_1) =& f(y,p_2)\nonumber\\
e^{-i\frac{p_2}{2}}\frac{x^{+}_{1}-x^{+}_{2}}{x^{+}_{1}-x^{-}_{2}}~f(y,p_1)
+
\frac{\eta(p_{1})}{\eta(p_{2})}\frac{x^{+}_{2}-x^{-}_{2}}{x^{+}_{1}-x^{-}_{2}}~
f(y,p_2)S(y,p_1) =& f(y,p_1) S(y,p_2).
\end{align}
These equations can be solved explicitly and the solution is given
by:
\begin{align}\label{eqn;SolutionfS}
&f(y,p_{k}) = \eta(p_{k}) \sqrt{\frac{x^-_k}{x^+_k}}\frac{y}{y-
x_{k}^{-}}\sqrt{\frac{g \ell_{k}}{2}}, && S(y,p_{k}) =
\sqrt{\frac{x^-_k}{x^+_k}}\frac{y-x^{+}_{k}}{y- x_{k}^{-}}.
\end{align}
With a modest amount of foresight, we fix the overall
normalization of $f$ to be dependent on the bound state number
$\ell_{k}$.

The problem becomes more involved when inserting two excitations.
In this case, one again introduces an auxiliary S-matrix,
$\S^{\mathrm{II}}$, that deals with interchanging excitations. The
ansatz for the coefficient can be written as
\begin{eqnarray}
|\alpha\beta\rangle = |\alpha\beta\rangle_{y_1 y_2} +
\S^{\mathrm{II}}\cdot|\alpha\beta\rangle_{y_1 y_2},
\end{eqnarray}
where we define $\S^{\mathrm{II}}$ as
\begin{eqnarray}
\S^{\mathrm{II}}\cdot|\alpha\beta\rangle_{y_1 y_2} &=&
M(y_1,y_2)|\alpha\beta\rangle_{y_2 y_1} +
N(y_1,y_2)|\beta\alpha\rangle_{y_2 y_1}.
\end{eqnarray}
When $\alpha=\beta=3$, we find
\begin{align}
\xi(K,2,0) =& |33\rangle\nonumber\\
=& \sum_{k<l}\Psi_k(y_1)\Psi_l(y_2)w^{(1)}_{1}\ldots
\theta^{(k)}_{3}\ldots \theta^{(l)}_{3}\ldots
w_{1}^{(K)}+\\
&+(M(y_1,y_2)+N(y_1,y_2))\sum_{k<l}\Psi_k(y_2)\Psi_l(y_1)w^{(1)}_{1}\ldots
\theta^{(k)}_{3}\ldots \theta^{(l)}_{3}\ldots
w_{1}^{(K)},\nonumber
\end{align}
which is compatible with (\ref{eqn;compatiNLSM}) if
$M(y_1,y_2)+N(y_1,y_2) = -1$. In general we write
\begin{eqnarray}
|\alpha\beta\rangle &=&
\sum_{k<l}\Psi_k(y_1)\Psi_l(y_2)w^{(1)}_{1}\ldots
\theta^{(k)}_{\alpha}\ldots \theta^{(l)}_{\beta}\ldots
w_{1}^{(K)}+\nonumber\\
&&+\S^{\mathrm{II}}\cdot\sum_{k<l}\Psi_k(y_1)\Psi_l(y_2)w^{(1)}_{1}\ldots
\theta^{(k)}_{\alpha}\ldots \theta^{(l)}_{\beta}\ldots w_{1}^{(K)}+\\
&&+\epsilon^{\alpha\beta}\sum_{k}\Psi_k(y_1)\Psi_k(y_2)h(y_1,y_2,p_k)w^{(1)}_{1}\ldots
w^{(k)}_{2}\ldots w_{1}^{(K)}.\nonumber
\end{eqnarray}
The term containing the $w_2$ variable needs to be included here
since it has the same quantum numbers
$K^{\mathrm{II}},K^{\mathrm{III}}$ as the state $\theta_3\theta_4$
as was explained in chapter \ref{chap;BoundSmat}. We allow for a
new function $h$, which describes the case when two fermions
occupy the same site. This term also receives a contribution from
$\S^{\mathrm{II}}$, which for simplicity, we have absorbed in $h$.

Because of integrability we can restrict to just two sites. The
above state $|\alpha\beta\rangle$ splits into the sum of a wave
function with either zero, one or two fermions. The only new piece
is the part containing two fermions which is given by
\begin{eqnarray}
|\alpha\beta\rangle &=& \left\{f(y_1,p_1) f(y_2,p_2)S(y_2,p_1) + M f(y_2,p_1) f(y_1,p_2)S(y_1,p_1)\right\} \theta^{(1)}_{\alpha}\theta^{(2)}_{\beta}\nonumber\\
&& + N f(y_2,p_1) f(y_1,p_2)S(y_1,p_1)\theta^{(1)}_{\beta}\theta^{(2)}_{\alpha}\nonumber \\
&& + \epsilon^{\alpha\beta}h(y_1,y_2,p_1)f(y_2,p_1) f(y_1,p_1)w^{(1)}_{2}w^{(2)}_{1}\\
&& + \epsilon^{\alpha\beta}h(y_1,y_2,p_2)f(y_2,p_2)
f(y_1,p_2)S(y_2,p_1) S(y_1,p_1)w^{(1)}_{1}w^{(2)}_{2}\nonumber
\end{eqnarray}
Plugging this into (\ref{eqn;compatiNLSM}) again allows one to
find the explicit (unique) solutions of the unknown functions:
\begin{align}\label{eqn;TwoExcSolution}
&M(y_1,y_2) = \frac{2
i/g}{y_1+\frac{1}{y_1}-y_2-\frac{1}{y_2}-\frac{2 i}{g}},
\qquad\qquad N(y_1,y_2) =
-\frac{y_1+\frac{1}{y_1}-y_2-\frac{1}{y_2}}{y_1+\frac{1}{y_1}-y_2-\frac{1}{y_2}-\frac{2 i}{g}}\\
&h(y_1,y_2,p_k) = \frac{i}{\ell_{k}\eta(p_{k})^2}~ \frac{y_1 y_2 -
x_{k}^{+}x_{k}^{-}}{y_1
y_2}~\frac{x_{k}^{+}-x_{k}^{-}}{x^{-}_{k}}~
\frac{y_1-y_2}{y_1+\frac{1}{y_1}-y_2-\frac{1}{y_2}-\frac{2
i}{g}}\nonumber
\end{align}
Upon defining $v = y + \frac{1}{y}$ one finds
\begin{align}
&M = \frac{2i/g}{v_1-v_2-2i/g} ,&& N =
-\frac{v_1-v_2}{v_1-v_2-2i/g}.
\end{align}
From this we see that (\ref{eqn;SmatNLSMNested2}) and
$\S^{\mathrm{II}}$ are basically the same operator. It is readily
seen that in order for $|\alpha\beta\rangle$ to be an eigenvector
of $\vec{\S}_j$ one has to require that
\begin{align}
\S^{\mathrm{II}}|\alpha\beta\rangle = |\beta\alpha\rangle,
\end{align}
which just indicates we should apply the procedure a second time
to deal with the index structure associated with the fermionic
indices $\alpha\beta$.

One can repeat the discussion for the nonlinear Schr\"odinger
model to deal with this since (\ref{eqn;SmatNLSMNested2}) agrees
with $\S^{\mathrm{II}}$ \cite{Leeuw:2007uf,Beisert:2005tm}. In
this process we are lead to introduce functions
$S^{\mathrm{II,II}},f^{(2)}, S^{\mathrm{III},\mathrm{II}},
S^{\mathrm{III,III}}$ similar to the $f,S$ introduced for the
nonlinear Schr\"odinger model in the previous section. The result
of this consideration is
\begin{align}
&S^{\mathrm{II,II}}=-M-N=1, &
&f^{(2)}(w,y_{k})=\frac{w-\frac{i}{g}}{w-v_{k}-\frac{i}{g}}\nonumber\\
&S^{\mathrm{III},\mathrm{II}}(w,y_{k})=\frac{w-v_{k}+\frac{i}{g}}{w-v_{k}-\frac{i}{g}}
& &S^{\mathrm{III,III}}(w_{1},w_{2}) =
\frac{w_{1}-w_{2}-\frac{2i}{g}}{w_{1}-w_{2}+\frac{2i}{g}}.
\end{align}
By putting all of this together one obtains the Bethe equations
describing the large volume spectrum of the $\ads$ superstring
\cite{Beisert:2005fw,Beisert:2005tm,Martins:2007hb,Leeuw:2007uf}:
\begin{eqnarray}
e^{ip_{k}L}&=& \prod_{l=1,l\neq k
}^{K^{\mathrm{I}}}\left[S_{0}(p_{k},p_{l})\frac{x_k^+-x_l^-}{x_k^--x_l^+}\sqrt{\frac{x_l^+x_k^-}{x_l^-x_k^+}}\right]^2\prod_{\alpha=1}^{2}\prod_{l=1}^{K_{(\alpha)}^{\mathrm{II}}}\frac{{x_{k}^{-}-y^{(\alpha)}_{l}}}{x_{k}^{+}-y^{(\alpha)}_{l}}\sqrt{\frac{x^+_k}{x^-_k}}\nonumber \\
1&=&\prod_{l=1}^{K^{\mathrm{I}}}\frac{y^{(\alpha)}_{k}-x^{+}_{l}}{y^{(\alpha)}_{k}-x^{-}_{l}}\sqrt{\frac{x^-_k}{x^+_k}}
\prod_{l=1}^{K_{(\alpha)}^{\mathrm{III}}}\frac{y_{k}^{(\alpha)}+\frac{1}{y_{k}^{(\alpha)}}-w_{l}^{(\alpha)}+\frac{i}{g}}{y_{k}^{(\alpha)}+\frac{1}{y_{k}^{(\alpha)}}-w_{l}^{(\alpha)}-\frac{i}{g}}\\
1&=&\prod_{l=1}^{K_{(\alpha)}^{\mathrm{II}}}\frac{w_{k}^{(\alpha)}-y_{k}^{(\alpha)}-\frac{1}{y_{k}^{(\alpha)}}+\frac{i}{g}}{w_{k}^{(\alpha)}-y_{k}^{(\alpha)}-\frac{1}{y_{k}^{(\alpha)}}-\frac{i}{g}}
\prod_{l\neq
k}^{K_{(\alpha)}^{\mathrm{III}}}\frac{w_{k}^{(\alpha)}-w_{l}^{(\alpha)}-\frac{2i}{g}}{w_{k}^{(\alpha)}-w_{l}^{\alpha}+\frac{2i}{g}}\nonumber,
\end{eqnarray}
where $\alpha=1,2$ reinstates the two independent copies of
$\mathfrak{su}(2|2)$ and $S_{0}(p_{k},p_{l})$ is the overall phase
of the S-matrix.

One can straightforwardly apply the same procedure to derive the
spectrum for bound states by using the explicit bound state
S-matrix. However, in the next section we choose a slightly
different approach according to which the Bethe equations can be
derived without reference to the explicit S-matrix.

\section{Bethe Ansatz and Yangian Symmetry}

In this section we will generalize the above construction to
arbitrary bound states. We will do this by considering coproducts
of (Yangian) symmetry generators. This formulation allows us to
solve (\ref{eqn;compatiNLSM}) without referring to the explicit
form of the bound state S-matrix, but rather use its underlying
symmetry.

\subsection{Single excitations}

We will again start by considering a single excitation on the
vacuum
\begin{eqnarray}
|0\rangle =( w^{(1)}_{1})^{\ell_1}\ldots (w^{(K)}_{1})^{\ell_K}.
\end{eqnarray}
The natural generalization of a single excitation wave function
(\ref{eqn;SU22-OneExcitation}) is:
\begin{align}\label{eqn;SingleExBound}
&|\alpha\rangle:= \sum_{i} \Psi_{i}(y) (
w^{(1)}_{1})^{\ell_1}\ldots
\theta^{(i)}_{\alpha}(w^{(i)}_{1})^{\ell_i-1}\ldots
(w^{(K)}_{1})^{\ell_K},\nonumber\\
&\Psi_{k}(y) = f(y,p_{k})\prod_{l<k} S^{\mathrm{I,II}}(y,p_{l}),~
\end{align}
As noted above, it suffices to restrict to two bound state
representations, for which the wave function is of the form
\begin{eqnarray}
|\alpha\rangle =
f(p_1)\theta^{(1)}_{\alpha}(w^{(1)}_{1})^{\ell_1-1}(w^{(2)}_{1})^{\ell_2}
+
f(p_2)S(p_1)(w^{(1)}_{1})^{\ell_1}\theta^{(2)}_{\alpha}(w^{(2)}_{1})^{\ell_2-1}.
\end{eqnarray}
The remarkable fact is that one can write this as:
\begin{eqnarray}
\tilde{\Delta}\mathbb{Q}^{1}_{\alpha}|0\rangle:=\left(K_{0}(p_1,p_2)\Delta\mathbb{Q}^{1}_{\alpha}+K_{1}(p_1,p_2)\Delta\hat{\mathbb{Q}}^{1}_{\alpha}\right)|0\rangle,
\end{eqnarray}
with
\begin{eqnarray}
&K_{0}&=
-\sqrt{\frac{2}{g}}\frac{x_2^-\{x_1^-x_2^-(x_1^+x_2^+-1)-x_1^+x_2^+(1+x_1^-[x_1^-+x_1^++x_2^+])\}}{(x_2^-
-x_1^+) (2 x_1^+
   x_2^+x_1^- x_2^--x_1^- x_2^--x_1^+
   x_2^+)}\times\nonumber\\
&&   \times\left[\frac{f(p_2)S(p_1)}{\sqrt{\ell_2} \eta(p_2)}-\frac{e^{-i\frac{p_2}{2}}f(p_1)}{\sqrt{\ell_1} \eta(p_1)}\right]+\frac{e^{-i\frac{p_2}{2}}f(p_1)}{\sqrt{\ell_1}\eta(p_1)}\\
&K_{1}&=\frac{4i\sqrt{2}}{g^{3/2}}\frac{x_1^- x_2^- x_1^+ x_2^+} {
   (x_2^- -x_1^+) (2 x_1^+
   x_2^+x_1^- x_2^--x_1^- x_2^--x_1^+
   x_2^+)}\left[\frac{f(p_2)S(p_1)}{\sqrt{\ell_2} \eta(p_2)}-\frac{e^{-i\frac{p_2}{2}}f(p_1)}{\sqrt{\ell_1}
\eta(p_1)}\right]\nonumber
\end{eqnarray}
For the moment let us keep $f,S$ arbitrary. The invariance of the
S-matrix under Yangian symmetry means that
\begin{eqnarray}
\S\Delta \mathbb{Q}^{1}_{\alpha} = \Delta^{op}
\mathbb{Q}^{1}_{\alpha}\S,\qquad \S\Delta
\hat{\mathbb{Q}}^{1}_{\alpha} =\Delta^{op}
\hat{\mathbb{Q}}^{1}_{\alpha}\S.
\end{eqnarray}
In other words, we find:
\begin{eqnarray}
\S |\alpha\rangle &=& \S
\left(K_{0}(p_1,p_2)\Delta\mathbb{Q}^{1}_{\alpha}+K_{1}(p_1,p_2)\Delta\hat{\mathbb{Q}}^{1}_{\alpha}\right)|0\rangle\nonumber\\
&=&\left(K_{0}(p_1,p_2)\Delta^{op}\mathbb{Q}^{1}_{\alpha}+K_{1}(p_1,p_2)\Delta^{op}\hat{\mathbb{Q}}^{1}_{\alpha}\right)\S|0\rangle\\
&=&S_0\left(K_{0}(p_1,p_2)\Delta^{op}\mathbb{Q}^{1}_{\alpha}+K_{1}(p_1,p_2)\Delta^{op}\hat{\mathbb{Q}}^{1}_{\alpha}\right)|0\rangle.
\end{eqnarray}
since $\S|0\rangle = S_0|0\rangle$. However, on the right hand
side of (\ref{eqn;compatiNLSM}) we find the coefficient with
indices and momenta interchanged, which we denote by
$|\alpha\rangle_{\pi}$. It is readily seen that
\begin{eqnarray}
|\alpha\rangle_{\pi}&=&\left(K_{0}(p_2,p_1)\Delta^{op}\mathbb{Q}^{1}_{\alpha}+K_{1}(p_2,p_1)\Delta^{op}\hat{\mathbb{Q}}^{1}_{\alpha}\right)|0\rangle.
\end{eqnarray}
This means that (\ref{eqn;compatiNLSM}) corresponds to requiring
that $K_0$ and $K_1$ are symmetric under interchanging $p_1
\leftrightarrow p_2$. In other words to find our coefficients, we
have to solve
\begin{eqnarray}\label{eqn;symmCoeff}
K_{0}(p_1,p_2) = K_{0}(p_2,p_1),\qquad K_{1}(p_1,p_2) =
K_{1}(p_2,p_1),
\end{eqnarray}
for the functions $f$ and $S$.\smallskip

From the explicit expressions for $K_0,K_1$ it is straightforward
to prove that (\ref{eqn;symmCoeff}) is equivalent to the
equations:
\begin{eqnarray}
K f(p_1) + G f(p_2)S(p_1) &=& f(p_2)\nonumber\\
L f(p_1) + H f(p_2)S(p_1) &=& f(p_1)S(p_2),
\end{eqnarray}
with
\begin{eqnarray}
\begin{array}{lll}
  K=\frac{e^{i\frac{p_1}{2}}}{e^{i\frac{p_2}{2}}}\frac{\sqrt{\ell_2}\eta(p_{2})}{\sqrt{\ell_1}\eta(p_{1})}\frac{x^{+}_{1}-x^{-}_{1}}{x^{+}_{1}-x^{-}_{2}}&\quad & G=
 e^{i\frac{p_1}{2}}\frac{x^{-}_{1}-x^{-}_{2}}{x^{+}_{1}-x^{-}_{2}} \\
  L=e^{-i\frac{p_2}{2}}\frac{x^{+}_{1}-x^{+}_{2}}{x^{+}_{1}-x^{-}_{2}} &\quad& H=
\frac{\sqrt{\ell_1}\eta(p_{1})}{\sqrt{\ell_2}\eta(p_{2})}\frac{x^{+}_{2}-x^{-}_{2}}{x^{+}_{1}-x^{-}_{2}}
\end{array}.
\end{eqnarray}
These equations are solved by the $f,S$ found before, i.e. we
again find (\ref{eqn;SolutionfS}) as unique solution. Moreover,
notice that from this construction we can read off the elements of
the S-matrix. Namely, we rediscover in this way our coefficients
describing the scattering of Case II states $\mathscr{Y}^{k,0}_k$,
cf. section \ref{sect;Reduction}.

In conclusion, Yangian symmetry uniquely fixes the form of our
wave function. We can now write the wave function, restricted to
two sites, completely in terms of coproducts and, as a
consequence, (\ref{eqn;compatiNLSM}) is automatically satisfied.
Finally, the explicit expressions for $K_0,K_1$ are
\begin{align}
K_{0}(p_{1},p_{2},y)&=\sqrt{\frac{x_1^-}{x_1^+}}
\sqrt{\frac{x_2^-}{x_2^+}}\frac{y}{(y-x_1^-)(y-x_2^-)}\left[y-\frac{
x_1^-x_2^- x_1^+x_2^+(x_1^-+x_2^-+ x_1^++x_2^+)}{2 x_1^- x_2^- x_1^+ x_2^+ -x_1^- x_2^--x_1^+ x_2^+}\right]\nonumber\\
K_{1}(p_{1},p_{2},y)&=\frac{4i}{g}\sqrt{\frac{x_1^-}{x_1^+}}
\sqrt{\frac{x_2^-}{x_2^+}}\frac{y}{(y-x_1^-)(y-x_2^-)}\left[\frac{x_1^-x_2^-
x_1^+x_2^+}{2 x_1^- x_2^- x_1^+ x_2^+ -x_1^-
x_2^--x_1^+x_2^+}\right].
\end{align}
This consideration is valid for \textit{any} bound state numbers
and hence wave function (\ref{eqn;SingleExBound}) is valid for
\textit{any} bound state representations. In particular, all bound
state representations share the same function $S^{\mathrm{I,II}}$.

\subsection{Multiple excitations}

When dealing with two excitations, one needs to introduce a level
$\mathrm{II}$ S-matrix that deals with interchanging $y_1$ and
$y_2$.

\subsubsection*{Fundamental representations}

Let us first restrict to fundamental representations and
reformulate this in terms of coproducts. We will then move on to
generic bound states.

The wave function was of the form
\begin{eqnarray}
|\alpha\beta\rangle = |\alpha\beta\rangle_{y_1 y_2} +
\S^{\mathrm{II}}|\alpha\beta\rangle_{y_1 y_2},
\end{eqnarray}
where
\begin{eqnarray}
\S^{\mathrm{II}}|\alpha\beta\rangle_{y_1 y_2} &=&
M(y_1,y_2)|\alpha\beta\rangle_{y_2 y_1} +
N(y_1,y_2)|\beta\alpha\rangle_{y_2 y_1}.
\end{eqnarray}
This state contained both fermions and bosons $w_2$, so the
natural way to write this would be:
\begin{eqnarray}\label{eqn;TwoExcAnsatz}
|\alpha\beta\rangle_{y_1 y_2} &=&
\left\{(\tilde{\Delta}_{y_{1}}\mathbb{Q}^{1}_{\alpha})(\tilde{\Delta}_{y_{2}}\mathbb{Q}^{1}_{\beta})+
\epsilon_{\alpha\beta}\Delta^{\prime}_{y_1,y_2}\mathbb{L}^{1}_{2}\right\}|0\rangle,
\end{eqnarray}
with
\begin{eqnarray}
\Delta^{\prime}_{y_1,y_2}\mathbb{L}^{1}_{2}&:=&
L_{0}(y_1,y_2,p_1,p_2)\Delta\mathbb{L}^{1}_{2}+L_{1}(y_1,y_2,p_1,p_2)\Delta\hat{\mathbb{L}}^{1}_{2}.
\end{eqnarray}
By taking $\alpha=\beta$, one easily checks that we reproduce the
result from the previous chapter provided $M+N=-1$. Now we have to
solve the coefficients $L_0, L_1$ such that our ansatz agrees with
\begin{eqnarray}
&&\left\{f(y_1,p_1) f(y_2,p_2)S(y_2,p_1) + M f(y_2,p_1) f(y_1,p_2)S(y_1,p_1)\right\} \theta^{(1)}_{\alpha}\theta^{(2)}_{\beta}\nonumber\\
&&+  N f(y_2,p_1) f(y_1,p_2)S(y_1,p_1)\theta^{(1)}_{\beta}\theta^{(2)}_{\alpha}\nonumber \\
&&+  \epsilon^{\alpha\beta}h(y_1,y_2,p_1)f(y_2,p_1) f(y_1,p_1)w^{(1)}_{2}w^{(2)}_{1}\\
&&+  \epsilon^{\alpha\beta}h(y_1,y_2,p_2)f(y_2,p_2)
f(y_1,p_2)S(y_2,p_1) S(y_1,p_1)w^{(1)}_{1}w^{(2)}_{2}\nonumber,
\end{eqnarray}
where we keep the functions $M,N,h$ arbitrary. The above
expression consists of four independent terms which can be shown
to give two equations for $L_0$ and two equations for $L_1$. The
next step is to impose that both $L_0, L_1$ are symmetric under
the interchange $p_1\leftrightarrow p_2$ in order to satisfy
(\ref{eqn;compatiNLSM}). This will give us four equations for the
functions $h,M,N$ which can be shown to be equivalent to the
following set of equations:
\begin{align}\label{eqn;2excSmatrix}
\{f_{12}f_{21}S_{22} + M f_{22}f_{11}S_{12}\} =&
\{f_{11}f_{22}S_{21} + M f_{21}f_{12}S_{11}\}\frac{D+E}{2}+ N f_{21}f_{12}S_{11}\frac{D-E}{2}\nonumber\\
&+ \left( -f_{11}f_{21}h_{121} + f_{12}f_{22}S_{11}S_{21}h_{122}
\right)\frac{C}{2}\nonumber\\
 N f_{22}f_{11}S_{12} =& \{f_{11}f_{22}S_{21} + M f_{21}f_{12}S_{11}\}\frac{D-E}{2}+ N f_{21}f_{12}S_{11}\frac{D+E}{2}\nonumber\\
&- \left( -f_{11}f_{21}h_{121} + f_{12}f_{22}S_{11}S_{21}h_{122}
\right)\frac{C}{2}.\nonumber\\
f_{11}f_{21}S_{12}S_{22}h_{121} =& \{f_{11}f_{22}S_{21} + (M-N)
f_{21}f_{12}S_{11}\}\frac{F}{2}\\
&+f_{11}f_{21}h_{121}\frac{1-B}{2} +
f_{12}f_{22}S_{11}S_{21}h_{122}\frac{1+B}{2}\nonumber\\
f_{12}f_{22}h_{122} =& -\{f_{11}f_{22}S_{21} + (M-N)
f_{21}f_{12}S_{11}\}\frac{F}{2}\nonumber\\
&+f_{11}f_{21}h_{121}\frac{1+B}{2} +
f_{12}f_{22}S_{11}S_{21}h_{122}\frac{1-B}{2},\nonumber
\end{align}
where, for convenience, we introduced the short-hand notation
$f_{kl}:=f(y_{k},p_{l}), S_{kl}:=S^{\mathrm{II,I}}(y_{k},p_{l}),
M:= M(y_{1},y_{2}),N:= N(y_{1},y_{2})$ and $h_{ijk} :=
h(y_i,y_j,p_k)$. The coefficients $B,C,D,E,F$ are given by
\begin{eqnarray}
B&=&\frac{2 x_1^- x_2^- (x_2^+)^2-(x_1^- x_2^-+1)
   (x_2^-+x_1^+) x_2^++2 x_2^- x_1^+}{(1-x_1^-
   x_2^-) (x_1^+-x_2^-) x_2^+}\nonumber\\
C&=&2 i \eta(p_1) \eta(p_2)\frac{x_2^-}{x_2^+}\frac{
e^{-\frac{ip_1}{2}}(x_2^+-x_1^+)}{(1-x_1^-
   x_2^-)(x_1^+-x_2^-) }\nonumber\\
D&=& \frac{ x_1^--x_2^+}{x_2^--x_1^+
  }\frac{e^{\frac{ip_1}{2}}}{e^{\frac{ip_2}{2}}}\\
E&=&\frac{e^{\frac{ip_1}{2}}}{e^{\frac{ip_2}{2}}}\frac{ (x_1^-
(x_2^- (x_1^--2x_1^+)+1)
   x_1^++(x_1^++x_1^- (x_2^- x_1^+-2))
  x_2^+)}{(1-x_1^- x_2^-)(x_1^+-x_2^-)x_1^+ }\nonumber\\
F&=&2 i
\frac{e^{-\frac{ip_1}{2}}}{\eta(p_1)\eta(p_2)}\frac{(x_1^+-x_1^-)
(x_2^+-x_2^-)
   (x_2^+-x_1^+)}{(1-x_1^-
   x_2^-)(x_1^+-x_2^-)}\nonumber.
\end{eqnarray}
It is readily seen that these expressions coincide with elements
from the fundamental S-matrix. Remarkably, these are exactly the
same equations that one encounters in the nested Bethe Ansatz. In
other words, the coefficients $B,C,D,E,F$ indeed correspond to
elements from the fundamental S-matrix and we again find
(\ref{eqn;TwoExcSolution}) as the unique solution for $M,N,h$.

To conclude, let us give the explicit solutions for $L_0,L_1$,
\begin{align}\label{eqn;L0L1}
&L_0 =\frac{ g(y_1-y_2) x^-_1 x^-_2}{2 i (y_1-x^-_1) (y_2-x^-_1)
   (y_1-x^-_2) (y_2-x^-_2)}\times\\
&\times\left[ (y_1 +
   y_2)-\frac{x^-_1 x^-_2 x^+_1 x^+_2
   (x^-_1+x^-_2+x^+_1+x^+_2)}{2x_1^+ x_2^+x_1^- x_2^--x_1^- x_2^--x_1^+
   x_2^+}-\frac{y_1 y_2 x^+_1 x^+_2
   }{2x_1^+ x_2^+x_1^- x_2^--x_1^- x_2^--x_1^+
   x_2^+}\right\{\nonumber\\
&(x^+_1+x^+_2-x^-_1-x^-_2) (x^-_1 x^-_2-x^+_1
   x^+_2)
\left.\left.-\left(\frac{1}{x^-_2}+\frac{1}{x^+_1}+\frac{1}{x^+_2}+\frac{1}{x^-_1}\right)
   (x^-_1 x^-_2+x^+_1 x^+_2)\right\}\right]\nonumber\\
&L_{1} = \frac{y_1 y_2 x^-_1 x^-_2}{(y_1-x^-_1) (y_2-x^-_1)
(y_1-x^-_2)
  (y_2-x^-_2)}\times\\
&\qquad\times\left[
   (y_1-y_2)+\frac{4 i g^{-1}  x^-_1 x^-_2 x^+_1 x^+_2}{2 x_1^+
   x_2^+x_1^- x_2^--x_1^- x_2^--x_1^+
   x_2^+}\right]\nonumber
\end{align}
Note that they are indeed manifestly symmetric under
$p_1\leftrightarrow p_2$.

\subsubsection*{Bound states}

When considering bound states there is a new term proportional to
$\theta_3\theta_4$. To include this term we make the following
generalization of the two excitation ansatz
\begin{eqnarray}
|\alpha\beta\rangle &=&
\sum_{k<l}\Psi_k(y_1)\Psi_l(y_2)(w^{(1)}_{1})^{\ell_1}\ldots
\theta^{(k)}_{\alpha}(w^{(k)}_{1})^{\ell_k-1}\ldots
\theta^{(l)}_{\beta}(w^{(l)}_{1})^{\ell_l-1}\ldots
(w^{(K)}_{1})^{\ell_K}+\nonumber\\
&&+\S^{\mathrm{II}}\cdot\sum_{k<l}\Psi_k(y_1)\Psi_l(y_2)(w^{(1)}_{1})^{\ell_1}\ldots
\theta^{(k)}_{\alpha}(w^{(k)}_{1})^{\ell_k-1}\ldots \theta^{(l)}_{\beta}(w^{(l)}_{1})^{\ell_l-1}\ldots (w_{1}^{(K)})^{\ell_K}+\nonumber\\
&&+\epsilon^{\alpha\beta}\sum_{k}\Psi_k(y_1)\Psi_k(y_2)h(y_1,y_2,p_k)(w^{(1)}_{1})^{\ell_1}\ldots
w^{(k)}_{2}(w_{1}^{(k)})^{\ell_k-1}\ldots (w_{1}^{(K)})^{\ell_K}\\
&&+\sum_{k}\Psi_k(y_1)\Psi_k(y_2)g(y_1,y_2,p_k)(w^{(1)}_{1})^{\ell_1}\ldots
\theta_{\alpha}^{(k)}\theta_{\beta}^{(k)}(w^{(1)}_{2})^{\ell_k-2}\ldots
(w_{1}^{(K)})^{\ell_K},\nonumber
\end{eqnarray}
where the function $g$ is again to be determined. Restricted to
two sites we can try to match this with (\ref{eqn;TwoExcAnsatz}).
This is indeed possible and imposing symmetry of $L_0,L_1$ as
before provides equations that are uniquely solved by
$g(y_1,y_2,p_k)=\frac{\ell_k-1}{2\ell_k}(1+M(y_1,y_2)-N(y_1,y_2))$
and (\ref{eqn;TwoExcSolution}). Plugging these solutions back in
$L_0,L_1$, we find the same functions $L_0,L_1$ as in
(\ref{eqn;L0L1}) but one has to bear in mind that $x^{\pm}$
parameterize bound state solutions; they depend on the bound state
number $\ell$ via equation(\ref{eqn;ParamtersPandG}).

By construction, this wave function satisfies
(\ref{eqn;compatiNLSM}) for any bound state S-matrix. Hence this
solves our two excitation case. In particular one finds that also
the level II S-matrix, $\S^{\mathrm{II}}$ is unchanged for bound
states.

\subsection{Bethe equations}

By making use of coproducts and Yangian symmetry, we have found a
way, independent of the explicit form of the S-matrix, to write
down Bethe wave functions. This allowed us to find
$S^{\mathrm{II,I}}$ and we found that the level two S-matrix,
$\S^{\mathrm{II}}$ remains unchanged. In other words, this yields
that the Bethe equations for any combination of bound states are
given by:
\begin{eqnarray}\label{eqn;FullBAE}
e^{ip_{k}L}&=& \prod_{l=1,l\neq k
}^{K^{\mathrm{I}}}\left[S_{0}(p_{k},p_{l})\frac{x_k^+-x_l^-}{x_k^--x_l^+}\sqrt{\frac{x_l^+x_k^-}{x_l^-x_k^+}}\right]^2\prod_{\alpha=1}^{2}\prod_{l=1}^{K_{(\alpha)}^{\mathrm{II}}}\frac{{x_{k}^{-}-y^{(\alpha)}_{l}}}{x_{k}^{+}-y^{(\alpha)}_{l}}\sqrt{\frac{x^+_k}{x^-_k}}\nonumber \\
1&=&\prod_{l=1}^{K^{\mathrm{I}}}\frac{y^{(\alpha)}_{k}-x^{+}_{l}}{y^{(\alpha)}_{k}-x^{-}_{l}}\sqrt{\frac{x^-_k}{x^+_k}}
\prod_{l=1}^{K_{(\alpha)}^{\mathrm{III}}}\frac{y_{k}^{(\alpha)}+\frac{1}{y_{k}^{(\alpha)}}-w_{l}^{(\alpha)}+\frac{i}{g}}{y_{k}^{(\alpha)}+\frac{1}{y_{k}^{(\alpha)}}-w_{l}^{(\alpha)}-\frac{i}{g}}\\
1&=&\prod_{l=1}^{K_{(\alpha)}^{\mathrm{II}}}\frac{w_{k}^{(\alpha)}-y_{k}^{(\alpha)}-\frac{1}{y_{k}^{(\alpha)}}+\frac{i}{g}}{w_{k}^{(\alpha)}-y_{k}^{(\alpha)}-\frac{1}{y_{k}^{(\alpha)}}-\frac{i}{g}}
\prod_{l\neq
k}^{K_{(\alpha)}^{\mathrm{III}}}\frac{w_{k}^{(\alpha)}-w_{l}^{(\alpha)}-\frac{2i}{g}}{w_{k}^{(\alpha)}-w_{l}^{\alpha}+\frac{2i}{g}}\nonumber,
\end{eqnarray}
with
\begin{eqnarray}
x_{k}^{+}+\frac{1}{x_{k}^+}-x_{k}^{-}-\frac{1}{x_{k}^-} = \frac{2i
\ell_{k}}{g},\qquad \frac{x_{k}^+}{x_{k}^-} = e^{ip_k}.
\end{eqnarray}
However, note that apart from the parameters $x^{\pm}$, the phase
factor $S_{0}(p_{k},p_{l})$ also depends on the bound states
representation one considers (\ref{eqn;FullPhase}).

\section{Summary}

In this chapter we derived the equations that capture the large
volume spectrum of bound states of the $\ads$ superstring. They
were derived by making use of the nested coordinate Bethe ansatz.
In this procedure one makes a plane-wave type ansatz for the
eigenstates of the Hamiltonian.

The coefficients in this ansatz depend crucially on the S-matrix
of the system. The non-trivial matrix structure of the S-matrix
leads to a nested structure which emerges in the form of a set of
auxiliary momenta $y_i,w_i$. One then imposes periodic boundary
conditions on the system which results in a set of coupled
equations (\ref{eqn;FullBAE}) that restrict the particle momenta.
The solutions of these equations describe the large volume
spectrum of $\ads$ bound states.

%% file: Transfer2.tex
\chapter{Algebraic Bethe Ansatz}\label{chap;Transfer}

Apart from the coordinate Bethe ansatz, there exists also another
method of diagonalizing an integrable Hamiltonian. This method
goes under the name of algebraic Bethe ansatz
\cite{Korepin2,Faddeev:1996iy}. They key feature in this method is
the transfer matrix, which is a generator of conserved charges.
The eigenvalues of the transfer matrix are also important since
they encode the asymptotic behavior of the TBA equations. In the
algebraic Bethe ansatz one constructs all eigenvalues of the
transfer matrix by first defining a vacuum eigenstate and then
introducing creation operators which generate excited eigenstates.

In this chapter we will apply this procedure to the AdS/CFT
problem at hand. We will first give the necessary definitions of
monodromy and transfer matrices and introduce the vacuum and find
its eigenvalue. After this we will define the creation operators
and build the excited states and the corresponding eigenvalues.
From these one can again read off the nested Bethe equations that
were derived in the previous chapter. Finally we discuss a fusion
procedure which allows to derive the bound state transfer matrices
from the case where all particles are in the fundamental
representation.

\section{Monodromy and transfer matrices}

Consider $K^{\rm{I}}$ bound state particles with bound state
numbers $\ell_1,\ldots,\ell_{K^{\rm{I}}}$ and momenta
$p_1,\ldots,p_{K^{\rm{I}}}$. To these particles we add an
auxiliary one, with momentum $q$ and bound state number $\ell_0$.
Any state of this system lives in the following tensor
product space
\begin{eqnarray}
\mathcal{V}:=V_{\ell_0}(q)\otimes V_{\ell_1}(p_1)\otimes \ldots
\otimes V_{\ell_{K^{\rm{I}}}}(p_{K^{\rm{I}}}),
\end{eqnarray}
where $V_{\ell_i}$ is the carrier space of the bound state
representation with the number $\ell_i$. We split the states in
the space $\mathcal{V}$ into an auxiliary part and a physical part:
\begin{eqnarray}\label{eqn;TotalStateABA}
|A\rangle_0\otimes|B\rangle_{K^{\rm{I}}} \in \mathcal{V},\nonumber
\end{eqnarray}
where $ |A\rangle_0 \in V_{\ell_0}(q)$ and\footnote{All the tensor
products are defined with increasing order of the index as
$1,2,\ldots max$.} $|B\rangle_{K^{\rm{I}}}
 \in  V_P:=\bigotimes_{i} V_{\ell_i}(p_{i})$. The monodromy matrix
acting in the space $\mathcal{V}$ is defined as follows:
\begin{eqnarray}
\mathcal{T}_{\ell_0}(q|\vec{p}) :=
\prod_{i=1}^{K^{\rm{I}}}\mathbb{S}_{0k}(q,p_i),
\end{eqnarray}
where $\mathbb{S}_{0k}(q,p_k)$ is the bound state S-matrix
describing scattering between the auxiliary particle, with
momentum $q$ and bound state number $\ell_0$, and a `physical'
particle, with momentum $p_k$ and bound state number $\ell_k$. For
convenience we will work with the canonically normalized S-matrix,
meaning that we take $\S~ w_1^{\ell_1}w_2^{\ell_2} =
w_1^{\ell_1}w_2^{\ell_2}$. We will include the appropriate overall
normalization $S_0$ (equation (\ref{eqn;FullPhase})) in the end.

The monodromy matrix can be seen as a $4\ell_0\times 4\ell_0$
dimensional matrix in the auxiliary space $V_{\ell_0}(q)$,  the
corresponding matrix elements  being themselves operators on
$V_P$. Indeed, introducing a basis $|e_{I}\rangle$ for
$V_{\ell_0}(q)$, with the index $I$ labelling a $4
\ell_0$-dimensional space, and a basis $|f_{A}\rangle$ for $V_P$,
the action of the monodromy matrix $\mathcal{T}\equiv
\mathcal{T}_{\ell_0}(q|\vec{p})$ on the total space $\mathcal{V}$
can be written as
\begin{eqnarray}\label{eqn;ActionMonodromy}
\mathcal{T}(|e_{I}\rangle\otimes |f_A\rangle) = \sum_{J,B}
T_{IA}^{JB}|e_{J}\rangle\otimes |f_B\rangle.
\end{eqnarray}
The matrix entries of the monodromy matrix can then be denoted as
\begin{eqnarray}
\mathcal{T}|e_{I}\rangle = \sum_{J} \mathcal{T}^{J}_{I}
|e_{J}\rangle\,  ,
\end{eqnarray}
while the action of the matrix elements $\mathcal{T}^{J}_{I}$ as
operators on $V_P$ can easily be read off:
\begin{eqnarray}
\mathcal{T}^{J}_{I}|f_A\rangle = \sum_{B} T_{IA}^{JB}|f_B\rangle.
\end{eqnarray}
The operators $\mathcal{T}^{J}_{I}$ have non-trivial commutation
relations among themselves. Consider two different auxiliary
spaces $V_{\ell_0}(q),V_{\tilde{\ell}_0}(\tilde{q})$. The
Yang-Baxter equation for $\S$ implies that
\begin{eqnarray}\label{eqn;YBE-operators}
\mathbb{S}(q,\tilde{q})\mathcal{T}_{\ell_0}(q|\vec{p})\mathcal{T}_{\tilde{\ell}_0}(\tilde{q}|\vec{p})
=\mathcal{T}_{\tilde{\ell}_0}(\tilde{q}|\vec{p})\mathcal{T}_{\ell_0}(q|\vec{p})\mathbb{S}(q,\tilde{q}),
\end{eqnarray}
where $\mathbb{S}(q,\tilde{q})$ is the S-matrix describing the
scattering between two bound state particles with bound state
numbers $\ell_0,\tilde{\ell}_0$ and momenta $q,\tilde{q}$
respectively. By explicitly working out these relations, one finds
the commutation relations between the different matrix elements of
the monodromy matrix. The fundamental commutation relations
(\ref{eqn;YBE-operators}) constitute the cornerstone of the
Algebraic Bethe Ansatz \cite{Faddeev:1996iy}.


It is convenient to pick up the following explicit basis
$|e_{I}\rangle$ in the space $V_{\ell_0}(q)$
\begin{equation}
\label{basisV0}
\begin{aligned}
e_{\alpha;k} &:= \theta_{\alpha} w_1^{\ell_0-k-1}w_{2}^{k},\\
e_{k} &:= w_1^{\ell_0-k}w_{2}^{k},\\
e_{34;k} &:= \theta_{3}\theta_{4} w_1^{\ell_0-k-1}w_{2}^{k-1}.
\end{aligned}
\end{equation}
The transfer matrix is then defined as
\begin{eqnarray}
\mathscr{T}_0(q|\vec{p}):={\rm{str}}_0\mathcal{T}_{\ell_0}(q|\vec{p}),
\end{eqnarray}
and it can be viewed as an operator acting on the physical space
$V_P$. In terms of the operator entries of the monodromy matrix,
the transfer matrix is written as
\begin{eqnarray}\label{eqn;Transfer}
\mathscr{T}_0(q|\vec{p})=\sum_{k=0}^{\ell_0} \mathcal{T}^{k}_{k} + \sum_{k=1}^{\ell_0-1} \mathcal{T}^{34;k}_{34;k} -
\sum_{k=0}^{\ell_0-1} \, \sum_{\alpha=3,4} \mathcal{T}^{\alpha;k}_{\alpha;k}.
\end{eqnarray}
Let us now briefly indicate why the transfer matrix is important
in integrable models. From (\ref{eqn;YBE-operators}) one can
easily deduce that
\begin{align}
[\mathscr{T}_0(q|\vec{p}),\mathscr{T}_0(\tilde{q}|\vec{p})] = 0.
\end{align}
Hence by expanding $\mathscr{T}_0(q|\vec{p})$ in $q$ one generates
an infinite set of commuting charges.

In the remainder of this paper we will study the action of
$\mathscr{T}_0(q|\vec{p})$ on the physical space $V_P$ in detail
and derive its eigenvalues.


\section{Diagonalization of the transfer matrix}

We start by defining a vacuum state
\begin{eqnarray}\label{eqn;BosonicVacuum}
|0\rangle_P = w_1^{\ell_1}\otimes\ldots\otimes w_1^{\ell_{K^{\rm{I}}}}.
\end{eqnarray}
We then compute the action of the transfer matrix on this state,
which appears to be one of its eigenstates, and afterwards use specific
elements of the monodromy matrix to generate the whole spectrum of
eigenvalues. Imposing the eigenstate condition should result in the
determination of the full set of eigenvalues and associated Bethe
equations, therefore providing the complete solution of the
asymptotic spectral problem.

\subsection{Eigenvalue of the transfer matrix on the vacuum}

As promised, we first deduce the action of the transfer matrix on
the vacuum. We will do this for each of the separate sums in
(\ref{eqn;Transfer}). Let us start with the fermionic part, {\it
i.e.} we want to compute
\begin{eqnarray}
\sum_{k=0}^{\ell_0-1}
\mathcal{T}^{\alpha;k}_{\alpha;k}|0\rangle_P, \qquad \alpha=3,4.
\end{eqnarray}
Taking into account the explicit form of the S-matrix elements
entering the monodromy matrix, we find that the only contribution
to $\mathcal{T}^{\alpha;k}_{\alpha;k}|0\rangle$ comes from
diagonal scattering elements. To be precise, one finds
\begin{eqnarray}
\mathcal{T}^{\alpha;k}_{\alpha;k}|0\rangle_P=\prod_{i}\mathscr{Y}^{k,0;1}_{k;1}(q,p_i)|0\rangle_P,
\end{eqnarray}
where $\mathscr{Y}^{k,0;1}_{k;1}(q,p_i)$ are Case II S-matrix
elements (see section \ref{sect;Reduction} for explicit
expressions). By explicitly working out this expression, one finds
\begin{eqnarray}\label{eqn;VacTransferFermion}
\mathscr{Y}^{k,0;1}_{k;1}(q,p_i) &=&
\frac{x^+_0-x^+_i}{x^-_0-x^+_i}\sqrt{\frac{x^-_0}{x^+_0}}\left[1-\frac{k}{u_0-u_i+\frac{\ell_0-\ell_i}{2}
}\right]\mathscr{X}^{k,0}_k (q, p_i),
\end{eqnarray}
where $x^{\pm}_0$ are defined in terms of the momentum $q$ as in
(\ref{eqn;ParamtersPandG}), and one uses equation
(\ref{dopolzero2}):
\begin{alignat}{3}
\mathscr{X}^{k,0}_{k} (q,p_i) &=
\mathcal{D} \, \frac{\prod_{j=0}^{k - 1} u_0 - u_i + \frac{\ell_0 - \ell_i - 2j}{2}}{\prod_{j=1}^{k} u_0 - u_i + \frac{\ell_0 + \ell_i - 2j}{2}} \ad{\qquad k =1,\cdots,\ell_0 -1}, \notag \\[1mm]
\ad{\mathscr{X}^{0,0}_0 (q,p_i)} & = \mathcal{D} = \frac{x_0^- - x_i^+}{x_0^+ - x_i^-} \sqrt{\frac{x_0^+}{x_0^-} \, \frac{x_i^-}{x_i^+} }\,.
\label{xlolook}
\end{alignat}
Obviously, the contribution of $\mathcal{T}^{\alpha;k}_{\alpha;k}$
is the same for $\alpha=3,4$. Here $x^{\pm}_m$, with
$m=0,1,...K^I$, are the constrained parameters ($\lambda$ is the
't Hooft coupling)
$$
x_m^++\frac{1}{x^+_m}-x_m^--\frac{1}{x^-_m}=2\ell_m\frac{i}{g}\, ,
~~~~~g=\frac{\sqrt{\lambda}}{2\pi}
$$
related to the particle momenta as $p_m=\frac{1}{i}\log
\frac{x^+_m}{x^-_m}$, cf. equation (\ref{eqn;ParamtersPandG}).
Also, $u_m$ represents the corresponding rapidity variable given
by (see equation (\ref{eqn;defUparameter}))
\begin{eqnarray}
\label{uvar} x_m^{\pm}+\frac{1}{x^{\pm}_m}=\frac{2 i}{g}u_m\pm
\frac{i}{g}\ell_m\, .
\end{eqnarray}
Next, we consider the more involved bosonic part. This can be
written as
\begin{eqnarray}
\mathcal{T}^{0}_{0} + \mathcal{T}^{\ell_0}_{\ell_0} + \sum_{k=1}^{\ell_0-1}\left\{\mathcal{T}^{k}_{k} + \mathcal{T}^{34;k}_{34;k}\right\}.
\end{eqnarray}
We first determine $\mathcal{T}^{0}_{0}|0\rangle_P$ and
$\mathcal{T}^{\ell_0}_{\ell_0}|0\rangle_P$. For these operators,
one again finds that only diagonal scattering elements of the
S-matrices contribute, which leads to
\begin{eqnarray}
\begin{aligned}
\mathcal{T}^{0}_{0}|0\rangle_P &=\prod_{i} \mathscr{Z}^{0,0;1}_{0;1}(q,p_i)  \, |0\rangle_P,\\
\mathcal{T}^{\ell_0}_{\ell_0}|0\rangle_P&= \prod_{i}
\mathscr{Z}^{\ell_0,0;1}_{\ell_0;1}(q,p_i) |0\rangle_P.
\end{aligned}
\end{eqnarray}
These matrix elements can be computed explicitly and give
\begin{eqnarray}
\mathcal{T}^{0}_{0}|0\rangle_P &=&|0\rangle_P,\label{eqn;TrivBosTransfer1}\\
\mathcal{T}^{\ell_0}_{\ell_0}|0\rangle_P&=&
\left\{\prod_{i=1}^{K^{\rm{I}}} \frac{(x^-_0-x^-_i)(1-x^-_0
x^+_i)}{(x^-_0-x^+_i)(1-x^+_0
x^+_i)}\sqrt{\frac{x^+_0x^+_i}{x^-_0x^-_i}}\mathscr{X}^{\ell_0,0}_{\ell_0}(q,p_i)\right\}
|0\rangle_P \ad{\,.}\label{eqn;TrivBosTransfer2}
\end{eqnarray}
\ad{where we define
\begin{equation}
\label{xloloo}
\mathscr{X}^{\ell_0,0}_{\ell_0} (q,p_i)=
\mathcal{D} \, \frac{\prod_{j=0}^{\ell_0 - 1} u_0 - u_i + \frac{\ell_0 - \ell_i - 2j}{2}}{\prod_{j=1}^{\ell_0} u_0 - u_i + \frac{\ell_0 + \ell_i - 2j}{2}} \,.
\end{equation}}
The next thing to consider is the sum
\begin{eqnarray}
\sum_{k=1}^{\ell_0-1}\left\{\mathcal{T}^{k}_{k} + \mathcal{T}^{34;k}_{34;k}\right\}.
\end{eqnarray}
While in the previous computations one could simply restrict to
the diagonal elements, one obtains instead a matrix structure for
this last piece. This is due to the fact that there are scattering
processes that relate $w_2 \leftrightarrow
\theta_{\alpha}\theta_{\beta}$. To be more precise, for the action
of $\mathcal{T}^k_k$ and $\mathcal{T}^{34,k}_{34,k}$ one finds
\begin{eqnarray}
\mathcal{T}^{k}_{k}|0\rangle_P &=&\sum_{a_1\ldots
a_{K^{\rm{I}}}=1,3}\mathscr{Z}^{k,0;a_1}_{k;1}(q,p_1) \, \mathscr{Z}^{k,0;a_2}_{k;a_1}(q,p_2)\ldots\mathscr{Z}^{k,0;1}_{k;a_{K^{\rm{I}}}}(q,p_{K^{\rm{I}}})|0\rangle_P,\\
\mathcal{T}^{34,k}_{34,k}|0\rangle_P&=&
\sum_{a_1\ldots
a_{K^{\rm{I}}}=1,3}\mathscr{Z}^{k,0;a_1}_{k;3}(q,p_1)\, \mathscr{Z}^{k,0;a_2}_{k;a_1}(q,p_2)\ldots\mathscr{Z}^{k,0;3}_{k;a_{K^{\rm{I}}}}(q,p_{K^{\rm{I}}})|0\rangle_P.
\end{eqnarray}
In order to evaluate the above expressions explicitly, it proves
useful to use a slightly more general reformulation\footnote{We
remark that this computation has been performed at weak coupling
in \cite{Bajnok:2008bm,Bajnok:2008qj}.}. One can reintroduce the
elements $\mathcal{T}^{34,k}_{k}$ and $\mathcal{T}^{k}_{ 34,k}$
from the monodromy matrix. Their action on the vacuum is
\begin{eqnarray}
\mathcal{T}^{34,k}_{k}|0\rangle_P &=&\sum_{a_1\ldots
a_{K^{\rm{I}}}=1,3}\mathscr{Z}^{k,0;a_1}_{k;1}(q,p_1)\, \mathscr{Z}^{k,0;a_2}_{k;a_1}(q,p_2)\ldots\mathscr{Z}^{k,0;3}_{k;a_{K^{\rm{I}}}}(q,p_{K^{\rm{I}}})|0\rangle_P,\\
\mathcal{T}^{k}_{34,k}|0\rangle_P&=& \sum_{a_1\ldots
a_{K^{\rm{I}}}=1,3}\mathscr{Z}^{k,0;a_1}_{k;3}(q,p_1)\, \mathscr{Z}^{k,0;a_2}_{k;a_1}(q,p_2)\ldots\mathscr{Z}^{k,0;1}_{k;a_{K^{\rm{I}}}}(q,p_{K^{\rm{I}}})|0\rangle_P.
\end{eqnarray}
They describe the mixing between the states
$|e_{34,k}\rangle$ and $|e_{k}\rangle$. If we consider
the two-dimensional vector space spanned by
$|e_{34,k}\rangle$ and $|e_{k}\rangle$ for fixed $k \in \{1,...,\ell_0 -1\}$, we
see that the above elements define a $2\times2$ dimensional matrix
\begin{eqnarray}
\mathcal{T}_{2\times2}=
\begin{pmatrix}
    \mathcal{T}^{k}_{k} & \mathcal{T}^{34,k}_{k}\\
    \mathcal{T}^{k}_{34,k} & \mathcal{T}^{34,k}_{34,k}
\end{pmatrix},
\end{eqnarray}
and the bosonic part, $\mathcal{T}^k_k+\mathcal{T}^{34,k}_{34,k}$,
of the transfer matrix is just the trace of this matrix. Moreover,
it is easily seen from the definition of the transfer matrix that
this matrix factorizes
\begin{eqnarray}
\mathcal{T}_{2\times2} = \prod_{i=1}^{K}
\begin{pmatrix}
    \mathscr{Z}^{k,0;1}_{k;1}(q,p_i) & \mathscr{Z}^{k,0;3}_{k;1}(q,p_i)\\
    \mathscr{Z}^{k,0;1}_{k;3}(q,p_i) & \mathscr{Z}^{k,0;3}_{k;3}(q,p_i)
\end{pmatrix}.
\end{eqnarray}
The trace of this matrix is given by the sum of its eigenvalues,
hence it remains to find the eigenvalues of this matrix.
Actually, it is easily checked that the eigenvectors of
\begin{eqnarray}
\begin{pmatrix}
    \mathscr{Z}^{k,0;1}_{k;1}(q,p_i) & \mathscr{Z}^{k,0;3}_{k;1}(q,p_i)\\
    \mathscr{Z}^{k,0;1}_{k;3}(q,p_i) & \mathscr{Z}^{k,0;3}_{k;3}(q,p_i)
\end{pmatrix}
\end{eqnarray}
are independent of $p_i$. In other words, these are automatically
eigenvectors of $\mathcal{T}_{2\times2}$, and the corresponding eigenvalues are
the product of the eigenvalues of the above matrices. The
individual eigenvalues are given by
\begin{align}\label{eqn;lambda-pm}
\lambda_\pm(q,p_i,k)=&\frac{\mathscr{X}^{k,0}_k}{2\mathcal{D}}\left[1-\frac{(x^-_ix^+_0-1)
   (x^+_0-x^+_i)}{(x^-_i-x^+_0)
   (x^+_0x^+_i-1)}+\frac{2ik}{g}\frac{x^+_0
   (x^-_i+x^+_i)}{(x^-_i-x^+_0)
   (x^+_0x^+_i-1)}\right.\\
&\left.\pm\frac{i x^+_0
   (x^-_i-x^+_i)}{(x^-_i-x^+_0)
 (x^+_0x^+_i-1)}\sqrt{\left(\frac{2k}{g}\right)^2+2i
\left[x^+_0+\frac{1}{x^+_0}\right]
\frac{2k}{g}-\left[x^+_0-\frac{1}{x^+_0}\right]^2}\right]\nonumber.
\end{align}
The action of the transfer matrix on the vacuum is now given by
the summing of all the above terms
(\ref{eqn;VacTransferFermion},\ref{eqn;TrivBosTransfer1},\ref{eqn;TrivBosTransfer2},\ref{eqn;lambda-pm}).
From this it is easily seen that the vacuum is indeed an
eigenvector of the transfer matrix with the following eigenvalue
\begin{eqnarray}\label{transfer-rankone}
\Lambda(q|\vec{p})&=&1+\prod_{i=1}^{K^{\rm{I}}}
\left[\frac{(x^-_0-x^-_i)(1-x^-_0 x^+_i)}{(x^-_0-x^+_i)(1-x^+_0
x^+_i)}\sqrt{\frac{x^+_0x^+_i}{x^-_0x^-_i}}\mathscr{X}^{\ell_0,0}_{\ell_0}\right]\nonumber\\
&&-2\sum_{k=0}^{\ell_0-1}\prod_{i=1}^{K^{\rm{I}}}\left(\frac{x^+_0-x^+_i}{x^-_0-x^+_i}
\sqrt{\frac{x^-_0}{x^+_0}}\left[1-\frac{k}{u_0-u_i+\frac{\ell_0-\ell_i}{2}
}\right]\mathscr{X}^{k,0}_k\right) \nonumber\\
&& +
\sum_{k=1}^{\ell_0-1}\prod_{i=1}^{K^{\rm{I}}}\lambda_+(q,p_i,k)
+\sum_{k=1}^{\ell_0-1} \prod_{i=1}^{K^{\rm{I}}}\lambda_-(q,p_i,k).
\end{eqnarray}
For the fundamental case ($\ell_0=\ell_i=1 \, \, \forall i$), this reduces to
\begin{eqnarray}\label{T fundamental-all}\nonumber
\mathcal{T}_0(q|\vec{p})|0\rangle_P
&=& \left\{\prod_{i}\mathscr{Z}^{0,0;1}_{0;1}(q,p_i) +
\prod_{i}\mathscr{Z}^{1,0;1}_{1;1}(q,p_i) -2
\prod_{i}\mathscr{Y}^{0,0;1}_{0;1}(q,p_i)\right\}|0\rangle_P\\
&=&\left\{1 + \prod_{i=1}^{K^{\rm{I}}}
\frac{1-\frac{1}{x^-_0x^+_i}}{1-\frac{1}{x^-_0x^-_i}}\frac{x^+_0-x^+_i}{x^+_0-x^-_i}
-2
\prod_{i=1}^{K^{\rm{I}}}\frac{x^+_0-x^+_i}{x^+_0-x^-_i}\sqrt{\frac{x^-_i}{x^+_i}}\right\}|0\rangle_P.
\end{eqnarray}

We would like to point out that the square roots
in the eigenvalues $\lambda_{\pm}$ never appear in the vacuum
eigenvalue. This is because the square root part only depends on
the auxiliary momentum $q$, and it can be seen that, after summing
the contribution from $\lambda_+$ and $\lambda_-$, only even powers
of this square root piece survive.

\subsection{Creation operators and excited states}
The next step in the algebraic Bethe ansatz is to introduce
creation operators. These operators will be certain entries from
our monodromy matrix. By acting with these operators on the vacuum
one creates new (excited) states, which, under conditions to be
determined, will be eigenstates of the transfer matrix. We will
need to specify which monodromy matrix entries correspond to
creation operators for our purposes.

Recall that, from the symmetry invariance of the S-matrix, one can
deduce that the quantum numbers $K^{\rm{II}}$ (total number of
fermions) and $K^{\rm{III}}$ (total number of fermions of a
definite species, say, $3$) are conserved upon acting with the
monodromy matrix on a state (\ref{eqn;TotalStateABA}). Any element
$\mathcal{T}^{J}_{I}$ is called a creation operator if
$K^{\rm{II}}(|e_I\rangle_0)>K^{\rm{II}}(|e_J\rangle_0)$, it is
called an annihilation operator if
$K^{\rm{II}}(|e_I\rangle_0)<K^{\rm{II}}(|e_J\rangle_0)$ and
diagonal if
$K^{\rm{II}}(|e_I\rangle_0)=K^{\rm{II}}(|e_J\rangle_0)$. The
reason for this assignment is the following. Consider a creation
operator $\mathcal{T}^{J}_{I}$ and any physical state
$|A\rangle_P$. The action of a creation operator is defined via
(\ref{eqn;ActionMonodromy}). Since the total number $K^{\rm{II}}$
is preserved, and the $K^{\rm{II}}$ charge in the auxiliary space
has decreased, it has necessarily increased in the physical space.
The number $K^{\rm{II}}$ corresponds to the number of fermions in
the system, hence, by acting with $\mathcal{T}^{J}_{I}$ on
$|A\rangle_P$, one creates extra fermions in the physical space.
Notice that this also implies that acting with an annihilation
operator on the vacuum gives zero, whence the name.

We will create excited states by considering auxiliary fundamental
representations with momenta $\lambda_i$. We will use these to
define creation operators $B_{\alpha}(\lambda_i),F(\lambda_i)$.
Since these representations  are fundamental, their monodromy
matrices are only $4\times4$-dimensional. Our discussion will be
very similar to the treatment of the algebraic Bethe ansatz for
the Hubbard model which was first performed in
\cite{martins-1997,Ramos:1996us}. In order to make contact with
the treatment of \cite{Martins:2007hb} and with the standard
notation used for the Hubbard model, we parameterize this
monodromy matrix as
\begin{eqnarray}
\begin{pmatrix}
B & B_{3} & B_{4} & F\\
C_{3} & A^{3}_{3} & A^{3}_{4} & B^{*}_{3}\\
C_{4} & A^{4}_{3} & A^{4}_{4} & B^{*}_{4}\\
C & C^{*}_{3} & C^{*}_{4} & D\\
\end{pmatrix}.
\end{eqnarray}
One finds two seemingly different sets of creation operators
$B_{3}(\lambda_i),B_{4}(\lambda_i),F(\lambda_i)$ and
$B^*_{3}(\lambda_i),B^*_{4}(\lambda_i)$, $F(\lambda_i)$. As
discussed in \cite{martins-1997}, it is enough to restrict to one
set. In what follows, we will use the operators
$B_{3}(\lambda_i),B_{4}(\lambda_i),F(\lambda_i)$ to create
fermionic excitations out of the vacuum.

A generic excited state will now be formed by acting with a number
of those operators on the vacuum, e.g. one can consider states
like
\begin{eqnarray}
B_{3}(\lambda_1)B_{4}(\lambda_2)|0\rangle.
\end{eqnarray}
To find out whether this is an eigenstate of the transfer matrix,
one has to commute the diagonal elements of the transfer matrices
through the creation operators and let them act on the vacuum.
Imposing the eigenstate condition will in general give constraints
on the momenta $\lambda_i$. The explicit commutation relations
will be the subject of the next section.

\subsection{Commutation relations}

In order to compute the action of the transfer matrix on an
excited state, we need to compute the commutation relations
between the diagonal elements $\mathcal{T}^A_A$ and the
aforementioned creation operators. While we have to use creation
operators in a fundamental auxiliary representation, the diagonal
elements are to be taken in the bound state representation with
generic $\ell_0$. The commutation relations follow from
(\ref{eqn;YBE-operators}). We will report the complete derivation
of one specific commutation relation, and only give the final
result for the remaining ones. In the derivation, one has to pay
particular attention to the fermionic nature of the operators.

Consider the
operator $B_{3}(\lambda)$ and the element
$\mathcal{T}^{3,k}_{3,k}$ from the transfer matrix. From
(\ref{eqn;YBE-operators}), one finds
\begin{eqnarray}
\mathbb{P}_{3,k|0}\mathbb{S}(q,\lambda)\mathcal{T}(q)\mathcal{T}(\lambda)e_{3,k}\tilde{e}_{3,0}
=
\mathbb{P}_{3,k|0}\mathcal{T}(\lambda)\mathcal{T}(q)\mathbb{S}(q,\lambda)e_{3,k}\tilde{e}_{3,0},
\end{eqnarray}
where we have dropped the index $\ell_0$ and chosen
$\tilde{\ell}_0=1$, and where the tilde on $\tilde{e}_{3,0}$
denotes a basis element in the second auxiliary space. The
operator $\mathbb{P}_{A|B}$ is the projection operator onto the
subspace generated by the basis element $e_A\tilde{e}_B$. The
right hand side of the above equation gives
\begin{eqnarray}
\mathbb{P}_{3,k|0}\mathcal{T}(\lambda)\mathcal{T}(q)\mathbb{S}(q,\lambda)e_{3,k}\tilde{e}_{3,0}&=&
\mathbb{P}_{3,k|0}\mathscr{X}^{k,0}_k\mathcal{T}(\lambda)\mathcal{T}(q)e_{3,k}\tilde{e}_{3,0}\nonumber\\
&=&\mathbb{P}_{3,k|0}\sum_{A,B}\mathscr{X}^{k,0}_k(-1)^{F_A}(\mathcal{T}_{3}^B\tilde{e}_B)(\lambda)(\mathcal{T}_{3,k}^{A}(q)e_{A})\nonumber\\
&=&\mathscr{X}^{k,0}_k(-1)^{F_{(3,k)}}(\mathcal{T}_{3}^0\tilde{e}_0)(\lambda)(\mathcal{T}_{3,k}^{3,k}(q)e_{3,k})\nonumber\\
&=&-\mathscr{X}^{k,0}_k
B_{3}^0(\lambda)\mathcal{T}_{3,k}^{3,k}(q)e_{3,k}\tilde{e}_0.
\end{eqnarray}
The left hand side reduces to
\begin{align}
\mathbb{P}_{3,k|0}\mathbb{S}(q,\lambda)\mathcal{T}(q)\mathcal{T}(\lambda)e_{3,k}\tilde{e}_{3,0}&=
-\mathbb{P}_{3,k|0}\mathbb{S}(q,\lambda)\mathcal{T}(q)(\mathcal{T}(\lambda)_{3}^B\tilde{e}_{B})e_{3,k}\\
&=-\mathbb{P}_{3,k|0}\mathbb{S}(q,\lambda)\mathcal{T}(q)
\left\{\mathcal{T}_{3}^0(\lambda)\tilde{e}_{0}+\mathcal{T}_{3}^{3}(\lambda)\tilde{e}_{3}+\mathcal{T}_{3}^1(\lambda)\tilde{e}_{1}\right\}e_{3,k}\nonumber\\
&=\mathbb{P}_{3,k|0}\mathbb{S}(q,\lambda)(\mathcal{T}_{3,k}^A(q)e_A)
\left\{\mathcal{T}_{3}^0(\lambda)\tilde{e}_{0}+\mathcal{T}_{3}^{3}(\lambda)\tilde{e}_{3}+\mathcal{T}_{3}^1(\lambda)\tilde{e}_{1}\right\}.\nonumber
\end{align}
Because of the projection, we only need to take into account terms
that are mapped onto $e_{3,k}\tilde{e}_0$ by the action of the
S-matrix. These are given by
\begin{eqnarray}
&&\mathbb{P}_{3,k|0}\mathbb{S}(q,\lambda)\left\{\mathcal{T}_{3,k}^{3,k}(q)e_{3,k}\mathcal{T}_{3}^0(\lambda)\tilde{e}_{0}
+
\mathcal{T}_{3,k}^{3,k-1}(q)e_{3,k-1}\mathcal{T}_{3}^1(\lambda)\tilde{e}_{1}
+\mathcal{T}_{3,k}^{k}(q)e_{k}\mathcal{T}_{3}^{3}(\lambda)\tilde{e}_{3,0}\right\}=\nonumber\\
&&\qquad\mathbb{P}_{3,k|0}\mathbb{S}(q,\lambda)\left\{-\mathcal{T}_{3,k}^{3,k}(q)\mathcal{T}_{3}^0(\lambda)e_{3,k}\tilde{e}_{0}
-
\mathcal{T}_{3,k}^{3,k-1}(q)\mathcal{T}_{3}^1(\lambda)e_{3,k-1}\tilde{e}_{1}
+ \right.\nonumber\\
&&\qquad\qquad\left.+\mathcal{T}_{3,k}^{k}(q)\mathcal{T}_{3}^{3}(\lambda)e_{k}\tilde{e}_{3,0}+\mathcal{T}_{3,k}^{34,k-1}(q)\mathcal{T}_{3}^{3}(\lambda)e_{34,k-1}\tilde{e}_{3,0}\right\}.~~~~
\end{eqnarray}
Working this out explicitly yields
\begin{eqnarray}
&&\mathbb{P}_{3,k|0}\mathbb{S}(q,\lambda)\left\{\mathcal{T}_{3,k}^{3,k}(q)e_{3,k}\mathcal{T}_{3}^0(\lambda)\tilde{e}_{0}
+
\mathcal{T}_{3,k}^{3,k-1}(q)e_{3,k-1}\mathcal{T}_{3}^1(\lambda)\tilde{e}_{1}
+\mathcal{T}_{3,k}^{k}(q)e_{k}\mathcal{T}_{3}^{3}(\lambda)\tilde{e}_{3,0}\right\}=\nonumber\\
&&\qquad\left\{-\mathcal{T}_{3,k}^{3,k}(q)\mathcal{T}_{3}^0(\lambda)\mathscr{Y}^{k,0;1}_{k;1}-
\mathscr{Y}^{k-1,1;1}_{k;1}\mathcal{T}_{3,k}^{3,k-1}(q)\mathcal{T}_{3}^1(\lambda)
+\right.\nonumber\\
&&\qquad\qquad\left.+\mathscr{Y}^{k,1;1}_{k;2}\mathcal{T}_{3,k}^{k}(q)\mathcal{T}_{3}^{3}(\lambda)
+\mathscr{Y}^{k,1;1}_{k;4}\mathcal{T}_{3,k}^{34,k-1}(q)\mathcal{T}_{3}^{3}(\lambda)
\right\}e_{3,k}\tilde{e}_{0}.
\end{eqnarray}
From this we now read off the final commutation
relation\footnote{Throughout the rest of this section 3.3, if not
otherwise indicated, the coefficient functions appearing have to
be understood as $\mathscr{X} \equiv \mathscr{X}(q,\lambda)$,
$\mathscr{Y} \equiv \mathscr{Y}(q,\lambda)$, $\mathscr{Z} \equiv
\mathscr{Z}(q,\lambda)$ (indices are omitted here for
simplicity).}
\begin{eqnarray}\label{eqn;CommRel}
\mathscr{X}^{k,0}_k B_{3}(\lambda)\mathcal{T}_{3,k}^{3,k}(q)
&=&\mathscr{Y}^{k,0;1}_{k;1}\mathcal{T}_{3,k}^{3,k}(q)B_{3}(\lambda)+\mathscr{Y}^{k-1,1;1}_{k;1}\mathcal{T}_{3,k}^{3,k-1}(q)C^*_{3}(\lambda)+\\
&&-\mathscr{Y}^{k,1;1}_{k;2}\mathcal{T}_{3,k}^{k}(q)A_{3}^{3}(\lambda)-\mathscr{Y}^{k,1;1}_{k;4}\mathcal{T}_{3,k}^{34,k-1}(q)A_{3}^{3}(\lambda)\nonumber.
\end{eqnarray}
Notice that in the above relation the operators are ordered in
such a way that all annihilation and diagonal elements are on the
right. This is done because the action of those elements on the
vacuum is known. We would also like to compare these commutation
relations with \cite{martins-1997,Ramos:1996us} for the Hubbard
model. We see that the first and third term are also present in
the Hubbard model. However, due to the fact that we are dealing
with bound state representation, we also obtain {\it two
additional terms}.

Generically, the commutation relations produce ``wanted" terms,
which are those which directly contribute to the eigenvalue, and
other ``unwanted" terms. The latter terms are those which need to
vanish, in order for the state of our ansatz to be an eigenstate.
In (\ref{eqn;CommRel}), one can easily see by acting on the vacuum
that the wanted term is the first term on the right hand side,
while the other terms are unwanted. The cancellation of the
unwanted terms will give rise to certain constraints, which are
precisely the auxiliary Bethe equations.

The other commutation relations one needs to compute are those
with $\mathcal{T}^{k}_{k},\mathcal{T}^{34,k}_{34,k}$ and
$\mathcal{T}^{4,k}_{4,k}$. Their derivation is considerably more
involved, especially the procedure of reordering them according to
the above ``annihilation and diagonal to the right" prescription.
We will present the commutation relations we will actually need in
the coming sections. We will give the wanted terms, and focus on
one specific type of unwanted terms. Schematically, we will focus
on the following structure:
\begin{align}\label{eqn;CommRel2}
\left[\mathcal{T}^k_k(q)
+\mathcal{T}^{\alpha\beta,k}_{\alpha\beta,k}(q)
\right]B_{\alpha}(\lambda) =&
\frac{\mathscr{X}^{k,0}_k}{\mathscr{Y}^{k,0;1}_{k;1}}B_{\alpha}(\lambda)\left[\mathcal{T}^k_k(q)
+\mathcal{T}^{\alpha\beta,k}_{\alpha\beta,k}(q) \right] +\\
&+
\frac{\mathscr{Y}^{k,0;1}_{k;2}}{\mathscr{Y}^{k,0;1}_{k;1}}\mathcal{T}^{k}_{\alpha,k}(q)B(\lambda)
+ \ldots\nonumber\\
\mathcal{T}^{\gamma,k}_{\alpha_1,k}(q)B_{\alpha_2}(\lambda) =&
\frac{\mathscr{X}^{k,0}_k}{\mathscr{Y}^{k,0;1}_{k;1}}B_{\beta_2}(\lambda)\mathcal{T}^{\gamma,k}_{\beta_1,k}(q)
r_{\alpha_1\alpha_2}^{\beta_1\beta_2}(u_0+{\textstyle{\frac{\ell_0-1}{2}}}-k,u_{\lambda}) + \\
&+
\frac{\mathscr{Y}^{k,0;1}_{k;2}}{\mathscr{Y}^{k,0;1}_{k;1}}\mathcal{T}^{k}_{\beta_1,k}(q)A^{\gamma}_{\beta_2}(\lambda)
r_{\alpha_1\alpha_2}^{\beta_1\beta_2}(u_{\lambda},u_{\lambda})+\nonumber\ldots
\end{align}
Here, $u_\lambda$ is given by (cf. (\ref{uvar}))

\begin{eqnarray}
u_\lambda \, = \, \frac{g}{2 i} \bigg( x^+ (\lambda) +
\frac{1}{x^+ (\lambda)} - \frac{i}{g} \bigg)\nonumber
\end{eqnarray}
and $r^{\gamma\delta}_{\alpha\beta}(u_\lambda,u_\mu)$ are the
components of the 6-vertex model S-matrix
(\ref{eqn;6vertexComponents}) with $U=-1$. We would like to point
out that, when comparing this structure against formulas (34-36)
of \cite{martins-1997}, one immediately recognizes a similarity
between the commutation relations. As was shown in
\cite{Martins:2007hb}, for the case in which all representations
are taken to be fundamental, the commutation relations do agree.
The additional contributions coming from the fact that we are
dealing with bound states in e.g. (\ref{eqn;CommRel}), will only
generate a new class of unwanted terms. Hence, these new terms
will not contribute to the eigenvalues.

Let us mention one commutation relation which is particularly
straightforward to derive, namely, the one between two fermionic
creation operators, as found from (\ref{eqn;YBE-operators}) with
$\ell_0 = \tilde{\ell}_0 =1$. This relation reads
\begin{eqnarray}\label{eqn;CommRelCreation}
B_{\alpha}(\lambda)B_{\beta}(\mu) &=& -\mathscr{X}^{0,0}_0 (\lambda,\mu) \, B_{\delta}(\mu)B_{\gamma}(\lambda)r^{\gamma\delta}_{\alpha\beta}(u_\lambda,u_\mu)\nonumber\\
&&\qquad \qquad + \, \frac{\mathscr{Z}^{1,0;1}_{1;6}(\lambda,\mu)}{\mathscr{Z}^{1,0;1}_{1;1}(\lambda,\mu)}\left[F(\lambda)B(\mu)-F(\mu)B(\lambda)\right]\epsilon_{\alpha\beta}.\quad
~
\end{eqnarray}
This reproduces the result of \cite{Martins:2007hb}, and, in this
way, one can see the emergence of nesting. As a matter of fact, in
\cite{Ramos:1996us,martins-1997,Martins:2007hb} the appearance of
the 6-vertex model S-matrix was used to completely fix the form of
the excited eigenstates, and this can also be done in our case.

\subsection{First excited state}

The first excited state is of the form
\begin{eqnarray}
|1\rangle = \mathcal{F}^{\alpha}B_{\alpha}(\lambda)|0\rangle_P,
\end{eqnarray}
where we sum over the repeated fermionic index. This state has
$K^{\rm{II}}=1$. As previously discussed, all the
commutation relations are ordered in such a way that all
annihilation and diagonal operators are on the right. From the
commutation relations (\ref{eqn;CommRel2}) one finds
\begin{align}
&\left[\mathcal{T}^k_k(q)
+\mathcal{T}^{\alpha\beta,k}_{\alpha\beta,k}(q) \right]
\mathcal{F}^{\alpha}B_{\alpha}(\lambda)|0\rangle_P =
\frac{\mathscr{X}^{k,0}_k}{\mathscr{Y}^{k,0;1}_{k;1}}
\mathcal{F}^{\alpha}B_{\alpha}(\lambda)\left[\mathcal{T}^k_k(q)
+\mathcal{T}^{\alpha\beta,k}_{\alpha\beta,k}(q)
\right]|0\rangle_P\nonumber\\
&\qquad \qquad \qquad \qquad \qquad \qquad \qquad \qquad +\frac{\mathscr{Y}^{k,0;1}_{k;2}}{\mathscr{Y}^{k,0;1}_{k;1}} \mathcal{F}^{\alpha}\mathcal{T}^{k}_{\alpha,k}(q)B(\lambda)|0\rangle_P+\ldots,\\
&\left[\mathcal{T}^{\alpha_1,k}_{\alpha_1,k}(q)\right]
\mathcal{F}^{\alpha_2}B_{\alpha_2}(\lambda)|0\rangle_P =
\frac{\mathscr{X}^{k,0}_k}{\mathscr{Y}^{k,0;1}_{k;1}}
\mathcal{F}^{\alpha_2}r_{\alpha_1\alpha_2}^{\beta_1\beta_2}(u_0+{\textstyle{\frac{\ell_0-1}{2}}}-k,u_{\lambda})B_{\beta_2}(\lambda)\left[\mathcal{T}^{\beta_1,k}_{\alpha_1,k}(q)\right]
|0\rangle_P\nonumber \\
& \qquad \qquad \qquad \qquad \qquad \qquad +
\frac{\mathscr{Y}^{k,0;1}_{k;2}}{\mathscr{Y}^{k,0;1}_{k;1}}\mathcal{F}^{\alpha_2}r_{\alpha_1\alpha_2}^{\beta_1\beta_2}(u_{\lambda},u_{\lambda})\mathcal{T}^{k}_{\beta_2,k}(q)A^{\alpha_1}_{\beta_1}(\lambda)
|0\rangle_P +\ldots,
\end{align}
where we remind that we concentrate on only one type of unwanted
terms, for the sake of clarity. Notice the appearance of
six-vertex model S-matrix $r$ (see appendix A) in
the commutation relations. The coefficient functions appearing in
the above two formulas have to be understood as $\mathscr{X}
\equiv \mathscr{X}(q,\lambda)$, $\mathscr{Y} =
\mathscr{Y}(q,\lambda)$ (indices are omitted here for simplicity).

Since $\mathcal{T}^{\alpha,k}_{\beta,k}|0\rangle_P \sim
\delta^{\alpha}_{\beta}|0\rangle_P $, we find that $|1\rangle$ can
only be an eigenstate of the transfer matrix if
\begin{align}
\mathcal{F}^{\alpha}r_{\gamma\alpha}^{\gamma\beta}(u_0+{\textstyle{\frac{\ell_0-1}{2}}}-k,u_{\lambda})
\sim \mathcal{F}^{\beta}.
\end{align}
This means that $\mathcal{F}^{\alpha}$ is an eigenvector of the
transfer matrix of the 6-vertex model. Luckily, one finds that the
eigenstates of the 6-vertex model are independent of the auxiliary
momentum. The $k$ dependence in the above r-matrix appears in the
eigenvalue $\Lambda^{(6v)}$, where $\Lambda^{(6v)}$ is the
eigenvalue of the auxiliary 6-vertex model. From
(\ref{eqn;eigenvalue6V}) we find ($K=K^{\rm{II}}=1$)
\begin{align}
\Lambda^{(6v)}(u_0 |u_\lambda ) &=& \prod_{i=1}^{K^{\rm{III}}}
\frac{1}{b(w_i,u_0+{\textstyle{\frac{\ell_0-1}{2}}}-k)}+b(u_0+{\textstyle{\frac{\ell_0-1}{2}}}-k,u_\lambda )\prod_{i=1}^{K^{\rm{III}}}
\frac{1}{b(u_0+{\textstyle{\frac{\ell_0-1}{2}}}-k,w_i)} ,\quad \nonumber
\end{align}
together with the auxiliary equation (\ref{eqn;AuxEqns6V})
\begin{eqnarray}
b(w_j,u_\lambda )=\prod_{i=1,i\neq j}^{K^{\rm{III}}}
\frac{b(w_j,w_i)}{b(w_i,w_j)}.
\end{eqnarray}
We also have to deal with the unwanted terms. Here we remark that,
since we have chosen $\mathcal{F}^{\alpha}$ to be an eigenvector
of the 6-vertex S-matrix, this also affects the unwanted terms.
One explicitly finds that the unwanted terms are proportional to
\begin{eqnarray}
\left\{\Lambda^{(6v)}(u_\lambda |u_\lambda )A_{\alpha}^{\alpha}(\lambda)-B(\lambda)\right\}|0\rangle_P.
\end{eqnarray}
Cancelling these unwanted terms thus leads us to the following
auxiliary Bethe equations:
\begin{eqnarray}
\prod_{i=1}^{K^{\rm{I}}}\frac{x^+(\lambda)-x^-_i}{x^+(\lambda)-x^+_i}\sqrt{\frac{x^+_i}{x^-_i}}=\Lambda^{(6v)}(u_\lambda |u_\lambda ).
\end{eqnarray}
In order to make contact with the bound state Bethe equations
(\ref{eqn;FullBAE}), let us define $y\equiv x^+(\lambda)$ and
rescale $w \to \frac{g}{2i}w$. We find that $|1\rangle$ is an
eigenstate, provided the auxiliary Bethe equations
hold\footnote{We remark that, for $K^{\rm{III}}=0$, the solution
of (\ref{bete}) correspond to the highest weight state of the
auxiliary six-vertex model, while, for $K^{\rm{III}}=1$, one
formally obtains a solution only if some of the auxiliary roots
are equal to infinity. This corresponds to a descendent of the
highest weight state under the $\alg{su}(2)$ symmetry.}:
\begin{eqnarray}
\label{bete}
\prod_{i=1}^{K^{\rm{I}}}\frac{y-x^-_i}{y-x^+_i}\sqrt{\frac{x^+_i}{x^-_i}}&=&
\prod_{i=1}^{K^{\rm{III}}}\frac{w_i-y-\frac{1}{y}-\frac{i}{g}}{w_i-y-\frac{1}{y}+\frac{i}{g}},
\nonumber\\
\frac{w_i-y-\frac{1}{y}+\frac{i}{g}}{w_i-y-\frac{1}{y}-\frac{i}{g}}
&=& \prod_{j=1,j\neq
i}^{K^{\rm{III}}}\frac{w_i-w_j+\frac{2i}{g}}{w_i-w_j-\frac{2i}{g}}.
\end{eqnarray}
This exactly matches with the auxiliary bound state Bethe ansatz
equations (\ref{eqn;FullBAE}) derived in the previous chapter. The
corresponding eigenvalue is
\begin{eqnarray}\label{eqn;Lambda1exc}
&&\Lambda(q|\vec{p})={\textstyle{\frac{y-x^-_0}{y-x^+_0}\sqrt{\frac{x^+_0}{x^-_0}}
+}}\\
&&
{\textstyle{+}}{\textstyle{\frac{y-x^-_0}{y-x^+_0}\sqrt{\frac{x^+_0}{x^-_0}}
\left[\frac{x^+_0+\frac{1}{x^+_0}-y-\frac{1}{y}}{x^+_0+\frac{1}{x^+_0}-y-\frac{1}{y}-\frac{2i\ell_0}{g}}\right]}}
\prod_{i=1}^{K^{\rm{I}}}
{\textstyle{\left[\frac{(x^-_0-x^-_i)(1-x^-_0
x^+_i)}{(x^-_0-x^+_i)(1-x^+_0
x^+_i)}\sqrt{\frac{x^+_0x^+_i}{x^-_0x^-_i}}\mathscr{X}^{\ell_0,0}_{\ell_0}\right]}}\nonumber\\
&&{\textstyle{+}}
\sum_{k=1}^{\ell_0-1}{\textstyle{\frac{y-x^-_0}{y-x^+_0}\sqrt{\frac{x^+_0}{x^-_0}}
\left[\frac{x^+_0+\frac{1}{x^+_0}-y-\frac{1}{y}}{x^+_0+\frac{1}{x^+_0}-y-\frac{1}{y}-\frac{2ik}{g}}\right]}}
\left\{\prod_{i=1}^{K^{\rm{I}}}{\textstyle{\lambda_+(q,p_i,k)+}}\right.\left.\prod_{i=1}^{K^{\rm{I}}}{\textstyle{\lambda_-(q,p_i,k)}}\right\}\nonumber\\
&&\quad
{\textstyle{-}}\sum_{k=0}^{\ell_0-1}{\textstyle{\frac{y-x^-_0}{y-x^+_0}\sqrt{\frac{x^+_0}{x^-_0}}
\left[\frac{x^+_0+\frac{1}{x^+_0}-y-\frac{1}{y}}{x^+_0+\frac{1}{x^+_0}-y-\frac{1}{y}-\frac{2ik}{g}}\right]}}
\prod_{i=1}^{K^{\rm{I}}}{\textstyle{\frac{x^+_0-x^+_i}{x^-_0-x^+_i}\sqrt{\frac{x^-_0}{x^+_0}}\left[1-\frac{k}{u_0-u_i+\frac{\ell_0-\ell_i}{2}
}\right]}}\times\nonumber\\
&&\quad\times
{\textstyle{\mathscr{X}^{k,0}_k}}\left\{\prod_{i=1}^{K^{\rm{III}}}{\textstyle{\frac{w_i-x^+_0-\frac{1}{x^+_0}+\frac{i(2k-1)}{g}}{w_i-x^+_0-\frac{1}{x^+_0}+\frac{i(2k+1)}{g}}+
}}{\textstyle{\frac{y+\frac{1}{y}-x^+_0-\frac{1}{x^+_0}+\frac{2ik}{g}}{y+\frac{1}{y}-x^+_0-\frac{1}{x^+_0}+\frac{2i(k+1)}{g}}}}\prod_{i=1}^{K^{\rm{III}}}{\textstyle{\frac{w_i-x^+_0-\frac{1}{x^+_0}+\frac{i(2k+3)}{g}}{w_i-x^+_0-\frac{1}{x^+_0}+\frac{i(2k+1)}{g}}}}\right\}.\nonumber
\end{eqnarray}
We stress once again that the above eigenvalue is for the
canonically normalized S-matrix, i.e. it is normalized such that
$\S w_1^{\ell_1}w_2^{\ell_2} = w_1^{\ell_1}w_2^{\ell_2}$. The
dependence of $\Lambda$ on the bound state numbers of the physical
particles is hidden in the parameters $x_i^{\pm}$ and in the
S-matrix element $\mathscr{X}$. Notice that, when projected in the
fundamental representation, the formula above reproduces the
result of \cite{Martins:2007hb}.

\subsection{General result and Bethe equations}

As was stressed before, by comparing our commutation relations
against (34)-(36) from \cite{martins-1997,Ramos:1996us}, one
immediately notices several similarities. It turns out that one
can closely follow the derivation presented in those papers, and
from the diagonal terms read off the general eigenvalue.
Furthermore, cancelling the first few unwanted terms reveals
itself as sufficient to derive the complete set of auxiliary Bethe
equations.

More specifically, the results of appendix B and the
previously known results for the case when all physical legs are
in the fundamental representation indicate the generalization of
the formula for the transfer-matrix eigenvalues to multiple
excitations. In terms of S-matrix elements, this generalization is
given by
\begin{eqnarray}\label{eqn;GeneralEigenv}
\Lambda(q|\vec{p}) &=&
\prod_{m=1}^{K^{\rm{II}}}\frac{\mathscr{X}^{0,0}_0(q,\lambda_m)}{\mathscr{Y}^{0,0;1}_{0;1}(q,\lambda_m)}
+ \prod_{i=1}^{K^{\rm{I}}} \mathscr{Z}^{\ell_0,0;1}_{\ell_0;1}(q,p_i)\prod_{m=1}^{K^{\rm{II}}}\frac{\mathscr{X}^{\ell_0,0}_{\ell_0}(q,\lambda_m)}{\mathscr{Y}^{\ell_0,0;1}_{\ell_0;1}(q,\lambda_m)}+\nonumber\\
&&\sum_{k=1}^{\ell_0-1}
\prod_{m=1}^{K^{\rm{II}}}\frac{\mathscr{X}^{k,0}_k(q,\lambda_m)}{\mathscr{Y}^{k,0;1}_{k;1}(q,\lambda_m)}\left\{\prod_{i=1}^{K^{\rm{I}}}\lambda_+(q,p_i)
+\prod_{i=1}^{K^{\rm{I}}}\lambda_-(q,p_i)\right\}+\nonumber\\
&&-\sum_{k=0}^{\ell_0-1}
\prod_{m=1}^{K^{\rm{II}}}\frac{\mathscr{X}^{k,0}_k(q,\lambda_m)}{\mathscr{Y}^{k,0;1}_{k;1}(q,\lambda_m)}\prod_{i=1}^{K^{\rm{I}}}\mathscr{Y}^{k,0;1}_{k;1}(q,p_i)\Lambda^{(6v)}(u_0+{\textstyle{\frac{\ell_0-1}{2}}}-k,\vec{u}_\lambda
),
\end{eqnarray}
where again $\Lambda^{(6v)}$ is the eigenvalue of the auxiliary
6-vertex model, and we have introduced $\vec{u}_\lambda  =
(u_{\lambda_1 }, \cdots, u_{\lambda_{K^{\rm II}}} )$. The
auxiliary roots satisfy the following equations
\begin{eqnarray}
\Lambda^{(6v)}(u_{\lambda_j } ,\vec{u}_\lambda )\prod_{i=1}^{K^{\rm{I}}}\mathscr{Y}^{0,0;1}_{0;1}(\lambda_j,p_i)&=&1,\\
\prod_{i=1}^{K^{\rm{II}}}b(w_j,u_{\lambda_i} )\prod_{i=1,i\neq
j}^{K^{\rm{III}}} \frac{b(w_i,w_j)}{b(w_j,w_i)}&=&1.
\end{eqnarray}
In appendix B we give a complete derivation of the
eigenvalues $\Lambda(q|\vec{p})$ and of the auxiliary equations
for the case $K^{\rm{III}}=0$. Let us stress that the expression
for $\Lambda(q|\vec{p})$ encodes many eigenvalues that are
labelled by the integer quantum numbers. We would also like to
mention that the form of the eigenvalues appears in the form of
factorized products of single-excitation terms - a somewhat
expected feature, which makes us more confident about the
generalization procedure.


We point out that the dependence of the auxiliary parameters
$\lambda_m$ only appears in the form $x^+(\lambda_m)$. In order to
compare with the known Bethe equations we relabel this to be
$x^+(\lambda_m) \equiv y_m$. We also rescale $w_i\rightarrow
\frac{2i}{g}w_i$. In terms of these parameters, the eigenvalues
(\ref{eqn;GeneralEigenv}) become
\begin{eqnarray}\label{eqn;FullEignvalue}
&&\Lambda(q|\vec{p})=\prod_{i=1}^{K^{\rm{II}}}{\textstyle{\frac{y_i-x^-_0}{y_i-x^+_0}\sqrt{\frac{x^+_0}{x^-_0}}
+}}\\
&&
{\textstyle{+}}\prod_{i=1}^{K^{\rm{II}}}{\textstyle{\frac{y_i-x^-_0}{y_i-x^+_0}\sqrt{\frac{x^+_0}{x^-_0}}
\left[\frac{x^+_0+\frac{1}{x^+_0}-y_i-\frac{1}{y_i}}{x^+_0+\frac{1}{x^+_0}-y_i-\frac{1}{y_i}-\frac{2i\ell_0}{g}}\right]}}
\prod_{i=1}^{K^{\rm{I}}}
{\textstyle{\left[\frac{(x^-_0-x^-_i)(1-x^-_0 x^+_i)}{(x^-_0-x^+_i)(1-x^+_0 x^+_i)}
\sqrt{\frac{x^+_0x^+_i}{x^-_0x^-_i}}\mathscr{X}^{\ell_0,0}_{\ell_0}\right]}}
\nonumber\\
&&{\textstyle{+}}
\sum_{k=1}^{\ell_0-1}\prod_{i=1}^{K^{\rm{II}}}{\textstyle{\frac{y_i-x^-_0}{y_i-x^+_0}\sqrt{\frac{x^+_0}{x^-_0}}
\left[\frac{x^+_0+\frac{1}{x^+_0}-y_i-\frac{1}{y_i}}{x^+_0+\frac{1}{x^+_0}-y_i-\frac{1}{y_i}-\frac{2ik}{g}}\right]}}
\left\{\prod_{i=1}^{K^{\rm{I}}}{\textstyle{\lambda_+(q,p_i,k)+}}\right.\left.\prod_{i=1}^{K^{\rm{I}}}{\textstyle{\lambda_-(q,p_i,k)}}\right\}\nonumber\\
&&\quad
{\textstyle{-}}\sum_{k=0}^{\ell_0-1}\prod_{i=1}^{K^{\rm{II}}}{\textstyle{\frac{y_i-x^-_0}{y_i-x^+_0}\sqrt{\frac{x^+_0}{x^-_0}}
\left[\frac{x^+_0+\frac{1}{x^+_0}-y_i-\frac{1}{y_i}}{x^+_0+\frac{1}{x^+_0}-y_i-\frac{1}{y_i}-\frac{2ik}{g}}\right]}}
\prod_{i=1}^{K^{\rm{I}}}{\textstyle{\frac{x^+_0-x^+_i}{x^-_0-x^+_i}\sqrt{\frac{x^-_0}{x^+_0}}\left[1-\frac{k}{u_0-u_i+\frac{\ell_0-\ell_i}{2}
}\right]}}\times\nonumber\\
&&\quad\times
{\textstyle{\mathscr{X}^{k,0}_k}}\left\{\prod_{i=1}^{K^{\rm{III}}}{\textstyle{\frac{w_i-x^+_0-\frac{1}{x^+_0}+\frac{i(2k-1)}{g}}{w_i-x^+_0-\frac{1}{x^+_0}+\frac{i(2k+1)}{g}}+
}}\prod_{i=1}^{K^{\rm{II}}}{\textstyle{\frac{y_i+\frac{1}{y_i}-x^+_0-\frac{1}{x^+_0}+\frac{2ik}{g}}{y_i+\frac{1}{y_i}-x^+_0-\frac{1}{x^+_0}+\frac{2i(k+1)}{g}}}}\prod_{i=1}^{K^{\rm{III}}}{\textstyle{\frac{w_i-x^+_0-\frac{1}{x^+_0}+\frac{i(2k+3)}{g}}{w_i-x^+_0-\frac{1}{x^+_0}+\frac{i(2k+1)}{g}}}}\right\}.\nonumber
\end{eqnarray}
and the above auxiliary Bethe equations transform into the
well-known ones (\ref{eqn;FullBAE}):
\begin{eqnarray}
\label{bennote}
\prod_{i=1}^{K^{\rm{I}}}\frac{y_k-x^-_i}{y_k-x^+_i}\sqrt{\frac{x^+_i}{x^-_i}}&=&
\prod_{i=1}^{K^{\rm{III}}}\frac{w_i-y_k-\frac{1}{y_k}-\frac{i}{g}}{w_i-y_k-\frac{1}{y_k}+\frac{i}{g}},\\
\prod_{i=1}^{K^{\rm{II}}}\frac{w_k-y_i-\frac{1}{y_i}+\frac{i}{g}}{w_k-y_i-\frac{1}{y_i}-\frac{i}{g}}
&=& \prod_{i=1,i\neq
k}^{K^{\rm{III}}}\frac{w_k-w_i+\frac{2i}{g}}{w_k-w_i-\frac{2i}{g}}.\nonumber
\end{eqnarray}
Once again, we find that for {\it all} particles in the
fundamental representation (including the auxiliary space) this
agrees with what obtained in \cite{Martins:2007hb}. Analogous to
formula (41) from the same paper, one can derive the complete set
of Bethe equations from the transfer matrix. One finds that the
one-particle momenta should satisfy
\begin{eqnarray}
e^{ip_j L} =  \Lambda(p_j|\vec{p}).
\end{eqnarray}
One then notices that, if $q=p_j$ and $\ell_0=\ell_j$, then
$\mathscr{X}^{k,0}_k=0$ if $k>0$. This means that the only
surviving terms is found to be the first one. This gives the
following Bethe equations (after explicitly including the
appropriate scalar factor $S_0$, which was omitted in the
derivation):
\begin{eqnarray}
e^{ip_j L} =  \prod_{i=1,i\neq j}^{K^{\rm{I}}} S_0(p_j,p_i)
\prod_{m=1}^{K^{\rm{II}}}
\frac{y_m-x^-_j}{y_m-x^+_j}\sqrt{\frac{x^+_j}{x^-_j}}.
\end{eqnarray}
Together with the above set of auxiliary Bethe equations, this
indeed reproduces the full set of Bethe equations
(\ref{eqn;FullBAE}).

\section{Different vacua and fusion}

In the previous sections we deduced the spectrum of the transfer
matrix. We found all of its eigenstates and eigenvalues, characterized by the
quantum numbers $K^{\rm{I,II,III}}$. The eigenstates were obtained by
starting with a vacuum with quantum numbers $K^{\rm{II}}=K^{\rm{III}}=0$,
which proved to be an eigenstate, and then applying creation
operators that generate eigenstates with different quantum
numbers. Of course, our choice of vacuum is not unique. We can
build up our algebraic Bethe ansatz starting from a different
vacuum. One trivial example of this would be to start with $w_2$
instead of $w_1$. A more interesting case arises when all physical
particles are fermions.

\subsection{Fermionic vacuum}

Consider a fermionic vacuum with all the physical particles in the
fundamental representation:
\begin{eqnarray}
|0\rangle_P^{\prime}= \theta_{3}\otimes\ldots\otimes\theta_{3}.
\end{eqnarray}
This vacuum has quantum numbers $K^{\rm{II}}=K^{\rm{I}}$ and
$K^{\rm{III}}=0$. One can easily check that this vacuum is also an
eigenstate. The action of the diagonal elements of fermionic type
of the transfer matrix (\ref{eqn;Transfer}) is given by:
\begin{eqnarray}
\begin{aligned}
\mathcal{T}_{3;k}^{3;k}|0\rangle_P^{\prime} &=& \prod_{i=1}^{K^{\rm{I}}}\mathscr{X}^{k,0}_k(q,p_i)|0\rangle_P^{\prime},\nonumber\\
\mathcal{T}_{4;k}^{4;k}|0\rangle_P^{\prime} &=&
\prod_{i=1}^{K^{\rm{I}}}\mathscr{Z}^{k,0;6}_{k;6}(q,p_i)|0\rangle_P^{\prime}.
\end{aligned}
\end{eqnarray}
The explicit values for these scattering elements is given in
section \ref{sect;Reduction}, and one obtains
\begin{eqnarray}
\label{frmi}
\begin{aligned}
\mathcal{T}_{3;k}^{3;k}|0\rangle_P^{\prime} &=
\prod_{i=1}^{K^{\rm{I}}}\frac{x_0^--x_i^+}{x^+_0-x^-_i}\sqrt{\frac{x^+_0x^-_i}{x^-_0x^+_i}}|0\rangle_P^{\prime},
\\
\mathcal{T}_{4;k}^{4;k}|0\rangle_P^{\prime} &=
\prod_{i=1}^{K^{\rm{I}}}\frac{x_0^--x_i^-}{x^+_0-x^-_i}\frac{x_i^--
\frac{1}{x_0^+}}{x_i^+-\frac{1}{x_0^+}}\sqrt{\frac{x^+_0x^+_i}{x^-_0x^-_i}}|0\rangle_P^{\prime}.
\end{aligned}
\end{eqnarray}
Notice that these elements are \emph{independent} of $k$. This
means that, when summing over $k$, this will only give a factor of
$\ell_0$.

The next step is to consider the bosonic elements
$\mathcal{T}^k_k,\mathcal{T}^{34,k}_{34,k}$. Let us again split
off the contributions from $k=0$ and $k=\ell_0$. The corresponding
elements $\mathcal{T}^0_0,\mathcal{T}^{\ell_0}_{\ell_0}$ act on
this new vacuum as
\begin{eqnarray}
\mathcal{T}^0_0|0\rangle_P^{\prime}=\mathcal{T}^{\ell_0}_{\ell_0}|0\rangle_P^{\prime}&=&\prod_{i=1}^{K^{\rm{I}}}\mathscr{Y}^{k,0;2}_{k;2}|0\rangle_P^{\prime},\nonumber\\
&=&\prod_{i=1}^{K^{\rm{I}}}\frac{x_0^--x^-_i}{x_0^+-x^-_i}\sqrt{\frac{x^+_0}{x^-_0}}|0\rangle_P^{\prime}.
\end{eqnarray}
For the remaining elements one finds again, as in the case of the
vacuum (\ref{eqn;BosonicVacuum}) we have been using in the
previous section, an additional matrix structure. More precisely,
this time one needs to compute the eigenvalues of the matrix
\begin{eqnarray}
\begin{pmatrix}
\mathscr{Y}^{k,0;2}_{k;2} & \mathscr{Y}^{k,0;4}_{k;2}\\
\mathscr{Y}^{k,0;2}_{k;4} & \mathscr{Y}^{k,0;4}_{k;4}
\end{pmatrix}.
\end{eqnarray}
Because $\mathscr{Y}^{k,0;4}_{k;2}=\mathscr{Y}^{k,0;2}_{k;4}=0$,
one remarkably finds that this matrix diagonal. Hence, the
eigenvalues are easily read off, and one finds
\begin{eqnarray}
\mathcal{T}^k_k|0\rangle_P^{\prime}&=&\prod_{i=1}^{K^{\rm{I}}}\mathscr{Y}^{k,0;2}_{k;2}(q,p_i) |0\rangle_P^{\prime}
=\prod_{i=1}^{K^{\rm{I}}}\frac{x_0^--x^-_i}{x_0^+-x^-_i}\sqrt{\frac{x^+_0}{x^-_0}}|0\rangle_P^{\prime}
\end{eqnarray}
and
\begin{eqnarray}
\mathcal{T}^{34,k}_{34,k}|0\rangle_P^{\prime}=\prod_{i=1}^{K^{\rm{I}}}
\mathscr{Y}^{k,0;4}_{k;4}(q,p_i) |0\rangle_P^{\prime}
=\prod_{i=1}^{K^{\rm{I}}}\frac{x_0^--x^+_i}{x_0^+-x^-_i}\frac{x_i^--\frac{1}{x_0^+}}{x_i^+-\frac{1}{x_0^+}}\sqrt{\frac{x^+_0}{x^-_0}}|0\rangle_P^{\prime}.
\end{eqnarray}
Similarly to the fermionic contributions (\ref{frmi}), and once
again in contrast to the bosonic vacuum, where we find a very
non-trivial $k$-dependence through $\lambda_{\pm}$
(\ref{eqn;lambda-pm}), one finds that these terms are
\emph{independent} of $k$. Summing everything up finally gives
that $|0\rangle_P^{\prime}$ is an eigenvalue of the transfer
matrix with eigenvalue
\begin{align}
\label{anti} \Lambda(q|\vec{p})=&\
(\ell_0+1)\prod_{i=1}^{K^{\rm{I}}}\frac{x_0^--x^-_i}{x_0^+-x^-_i}
\sqrt{\frac{x^+_0}{x^-_0}}-
\ell_0\prod_{i=1}^{K^{\rm{I}}}\frac{x_0^--x_i^+}{x^+_0-x^-_i}
\sqrt{\frac{x^+_0x^-_i}{x^-_0x^+_i}}\, -\\
\nonumber
&-\ell_0
\prod_{i=1}^{K^{\rm{I}}}\frac{x_0^--x_i^-}{x^+_0-x^-_i}\frac{x_i^--\frac{1}{x_0^+}}
{x_i^+-\frac{1}{x_0^+}}\sqrt{\frac{x^+_0x^+_i}{x^-_0x^-_i}}
+(\ell_0-1)\prod_{i=1}^{K^{\rm{I}}}\frac{x_0^--x^+_i}{x_0^+-x^-_i}
\frac{x_i^--\frac{1}{x_0^+}}{x_i^+-\frac{1}{x_0^+}}\sqrt{\frac{x^+_0}{x^-_0}}.
\end{align}
This precisely agrees with the result of \cite{Beisert:2006qh} for
antisymmetric representations.


Let us remark that the spectrum is clearly independent of the
choice of vacuum. Hence, one should find the same eigenvalues when
starting from the vacuum $|0\rangle_P$ or from the vacuum
$|0\rangle_{P^\prime}$, provided one excites the appropriate set
of auxiliary roots. In particular, if we were to reproduce
(\ref{anti}) starting from the bosonic vacuum and exciting enough
fermions, we would have to  first solve the $K^{\rm{II}}$
auxiliary BAE, and then use these solutions to find the
corresponding eigenvalue, which should therefore agree with
(\ref{anti}). In fact, conversion of one eigenvalue into the other
can be obtained by means of duality transformations
\cite{Beisert:2005di}. We would also like to notice that the
result obtained in this section for fundamental representations in
the physical space happens to have nice fusion properties, and one
can think of combining several of such elementary transfer
matrices to obtain more general ones. This approach has been
followed for instance in \cite{Hatsuda:2008gd}.


\subsection{Bosonic vacuum}

Let us now come back to the bosonic vacuum
(\ref{eqn;BosonicVacuum}) we have been using in the first part of
this chapter. In \cite{Beisert:2006qh}, a prescription for
computing the transfer matrix eigenvalues, for all physical
particles in the fundamental representation, was also given. The
formula was expressed in terms of an expansion of the inverse of a
quantum characteristic function. We have found that this
prescription indeed produces the same eigenvalues as obtained from
our general formula (\ref{eqn;FullEignvalue}), when restricting
the latter to fundamental particles in the physical space. To
demonstrate this fact, we explicitly work out here below the above
mentioned expansion following \cite{Beisert:2006qh}, adapting the
calculation to the notations we use here. We will then
compare the final formula with the suitable restriction of our
result (\ref{eqn;FullEignvalue}), finding perfect agreement.
Indeed, we will be able to relax the condition of physical legs in
the fundamental representation, by making the conjectured
expression for the quantum characteristic function slightly more
general. We will then find agreement with such a formula in the
general case where we are dealing with generic bound state
representations $\ell_i \neq 1$ as well.

Following \cite{Beisert:2006qh}, we define the shift operator $U$
by
\begin{equation}
U \ssp f (u) \, U^{-1} = f \( u + \frac{1}{2} \),
\label{def:shift ops}
\end{equation}
and introduce the notation
\begin{equation}
f^{[\ell]} (u) \equiv U^\ell \ssp f (u) \, U^{-\ell} = f \( u + \frac{\ell}{2} \).
\end{equation}
The spectral parameters of an elementary particle, defined in
\eqref{uvar}, satisfy the relation
\begin{equation}
x^{[1]} + \frac{1}{x^{[1]}} - x^{[-1]} - \frac{1}{x^{[-1]}} = \frac{2i}{g} \,.
\label{def:torus}
\end{equation}
By successive applications of the shift operator to
\eqref{def:torus}, one finds that the pair of variables $\{
x^{[\ell]}, x^{[\ell-2k]} \}$ defines another rapidity torus
\begin{equation}
x^{[\ell]} + \frac{1}{x^{[\ell]}} - x^{[\ell-2k]} - \frac{1}{x^{[\ell-2k]}} = \frac{2ik}{g} \,.
\end{equation}
There are two choices of branch for $x_a^{[\ell-2k]}$ for a given
$x^{[\ell]}$, as can be seen by
\begin{equation}
x^{[\ell-2k]} = \frac12 \( x^{[\ell]} + \frac{1}{x^{[\ell]}} - \frac{2 i k}{g} + \sqrt{ \( x^{[\ell]} + \frac{1}{x^{[\ell]}} - \frac{2 i k}{g} \)^2 - 4} \; \).
\label{x intermediate branches}
\end{equation}
We also use $y_i + 1/y_i \ad{= i v_i}$ in what
follows.\footnote{\ad{Interestingly, the final result
\eqref{symmetric transfer 4} is almost invariant under the map
$y_i \mapsto 1/y_i$, except for an overall factor.}}

Let $\vev{\ell_0-1,0}$ be the $\ell_0$\,-th symmetric
representation of $\alg{su}(2|2)$. The conjecture states that the
transfer matrix for such a representation $T_{\vev{\ell_0-1,0}}
(u_0|\{\vec u, \vec v, \vec w\})$ is generated by $T_{\vev{0,0}}
(u_0|\{\vec u, \vec v, \vec w\})$, where the generating function
is equal to the inverse of the quantum characteristic function:
\begin{alignat}{3}
D_0^{-1} &\defeq \( 1 - U_0 T_4 U_0 \)^{-1} \( 1 - U_0 T_3 U_0 \) \( 1 - U_0 T_2 U_0 \) \( 1 - U_0 T_1 U_0 \)^{-1} , \\[1mm]
&= \( 1 + \sum_{h=1}^\infty (U_0 T_4 U_0)^h \) \( 1 - U_0 T_3 U_0 \) \( 1 - U_0 T_2 U_0 \) \( 1 + \sum_{k=1}^\infty (U_0 T_1 U_0)^k \),
\notag \\[1mm]
&\equiv \sum_{\ell_0=0}^\infty U_0^{\ell_0} \, T_{\vev{\ell_0-1,0}} (u_0|\{\vec u, \vec v, \vec w\}) \, U_0^{\ell_0} \,.
\label{symmetric transfer 1}
\end{alignat}
where the $T_i$s are parts of the fundamental transfer matrix,
which we will specify later. Here $U_0$ is the shift operator for
the variable $u_0$\,. The first few terms can be found as follows:
\begin{alignat}{3}
D_0^{-1} &= 1 + U_0 \( T_4 - T_3 - T_2 + T_1 \) U_0
\label{qch expansion 1} \\
&\hspace{-5mm} + U_0^2 \Bigl\{ T_4^{[-1]} T_4^{[1]} + T_4^{[-1]} T_1^{[1]} + T_1^{[-1]} T_1^{[1]} + T_3^{[-1]} T_2^{[1]} \notag \\
&\hspace{40mm} - T_4^{[-1]} (T_3^{[1]} + T_2^{[1]}) - (T_3^{[-1]}
+ T_2^{[-1]}) T_1^{[1]} \Bigr\} U_0^2 + \ \cdots\nonumber,
\end{alignat}
and, in general,
\begin{equation}
T_{\vev{\ell_0-1,0}} (u_0|\{\vec u, \vec v, \vec w\}) = \tau_{\ell_0,0} - \tau_{\ell_0,1} \[ T_3 \] - \tau_{\ell_0,1} \[ T_2 \] + \tau_{\ell_0,2} \[ T_3 \,, T_2 \],
\label{symmetric transfer 2}
\end{equation}
where
\begin{alignat}{5}
\tau_{\ell_0,0} &= \sum_{k=0}^{\ell_0} T_4^{[-\ell_0+1]} T_4^{[-\ell_0+3]} \cdots T_4^{[\ell_0-2k-3]} \, T_1^{[\ell_0-2k-1]} \cdots T_1^{[\ell_0-1]} \,,
\label{tau m0} \\[1mm]
\tau_{\ell_0,1} \left[ X \right] &= \sum_{k=0}^{\ell_0-1}
T_4^{[-\ell_0+1]} T_4^{[-\ell_0+3]} \cdots T_4^{[-\ell_0+2k-1]}
X^{[\ell_0-2k-1]} \, T_1^{[\ell_0-2k+1]} \cdots T_1^{[\ell_0-1]}
\,,
\label{tau m1} \\[1mm]
\tau_{\ell_0,2} \left[ X, Y \right] &= \sum_{k=0}^{\ell_0-2}
T_4^{[-\ell_0+1]} T_4^{[-\ell_0+3]} \cdots T_4^{[-\ell_0+2k-1]}
X^{[\ell_0-2k-3]} \, Y^{[\ell_0-2k-1]} \, T_1^{[\ell_0-2k+1]}
\cdots T_1^{[\ell_0-1]}. \label{tau m2}
\end{alignat}
The first line of \eqref{qch expansion 1} gives the transfer
matrix for the fundamental representation as
\begin{equation}
T_{\vev{0,0}} (u_0|\{\vec u, \vec v, \vec w\}) = T_1 - T_2 - T_3 + T_4.
\label{T00 expand}
\end{equation}
We recall that the left hand side of this equation is given
explicitly by \eqref{eqn;FullEignvalue} at $\ell_0=1$, which reads
\begin{align}
\Lambda(q|\vec{p}) &= \prod_{i=1}^{K^{\rm{II}}}{\textstyle{\frac{y_i-x^-_0}{y_i-x^+_0}\sqrt{\frac{x^+_0}{x^-_0}}
+}}\\
&{\textstyle{+}}
\prod_{i=1}^{K^{\rm{II}}}{\textstyle{\frac{y_i-x^-_0}{y_i-x^+_0}\sqrt{\frac{x^+_0}{x^-_0}}\left[
\frac{x^+_0+\frac{1}{x^+_0}-y_i-\frac{1}{y_i}}{x^+_0+\frac{1}{x^+_0}-y_i-\frac{1}{y_i}-\frac{2i}{g}} \right]}} \prod_{i=1}^{K^{\rm{I}}}
{\textstyle{\left[\frac{(x^-_0-x^-_i)(1-x^-_0 x^+_i)}{(x^-_0-x^+_i)(1-x^+_0 x^+_i)}
\sqrt{\frac{x^+_0 x^+_i}{x^-_0 x^-_i}}\mathscr{X}^{1,0}_{1}\right]}}
\nonumber\\
&\qquad
- \prod_{i=1}^{K^{\rm{II}}}{\textstyle{\frac{y_i-x^-_0}{y_i-x^+_0}\sqrt{\frac{x^+_0}{x^-_0}} }}\prod_{i=1}^{K^{\rm{I}}}{\textstyle{\frac{x^+_0-x^+_i}{x^-_0-x^+_i}\sqrt{\frac{x^-_0}{x^+_0}} }}\times\nonumber\\
&\qquad \times
{\textstyle{\mathscr{X}^{0,0}_0}}\left\{\prod_{i=1}^{K^{\rm{III}}}{\textstyle{\frac{w_i-x^+_0-\frac{1}{x^+_0}-\frac{i}{g}}{w_i-x^+_0-\frac{1}{x^+_0}+\frac{i}{g}}+
}}\prod_{i=1}^{K^{\rm{II}}}{\textstyle{\frac{y_i+\frac{1}{y_i}-x^+_0-\frac{1}{x^+_0}}{y_i+\frac{1}{y_i}-x^+_0-\frac{1}{x^+_0}+\frac{2i}{g}}}}\prod_{i=1}^{K^{\rm{III}}}{\textstyle{\frac{w_i-x^+_0-\frac{1}{x^+_0}+\frac{3i}{g}}{w_i-x^+_0-\frac{1}{x^+_0}+\frac{i}{g}}}}\right\}.\nonumber
\end{align}
Therefore, $\Lambda(q|\vec{p})$ may be equated with the right hand
side of \eqref{T00 expand} term by term. We simplify the above
expression of $\Lambda(q|\vec{p})$ by introducing variables ${\sf
w}_i$ and ${\sf v}_i$ as follows\footnote{Our notation is $x_0^\pm
= x^{[\pm \ell_0]}$, and $\ell_0=1$ is used when discussing the
fundamental transfer matrix. Note that the shift operator does not
act on $x_i^\pm$.}:
\begin{alignat}{3}
w_i - x_0^\pm - \frac{1}{x_0^\pm} + \frac{in}{g} &\equiv \( {\sf w}_i - u_0 + \frac{\mp \ell_0 + n}{2} \) \frac{2i}{g} \,,
\\[1mm]
y_i + \frac{1}{y_i} - x_0^\pm - \frac{1}{x_0^\pm} + \frac{in}{g} &\equiv \( {\sf v}_i - u_0 + \frac{\mp \ell_0 + n}{2} \) \frac{2i}{g} \,.
\end{alignat}
\ad{With the help of \eqref{xlolook} it produces}
\begin{eqnarray}
&&\Lambda(q|\vec{p}) = \prod_{i=1}^{K^{\rm{II}}} \frac{y_i-x^-_0}{y_i-x^+_0}\sqrt{\frac{x^+_0}{x^-_0}}
\label{Full fundamental}
\\[1mm]
&&
+ \prod_{i=1}^{K^{\rm{II}}} \frac{y_i-x^-_0}{y_i-x^+_0}\sqrt{\frac{x^+_0}{x^-_0}} \left[
\frac{{\sf v}_i - u_0 - \frac12}{{\sf v}_i - u_0 + \frac12} \right]
\prod_{i=1}^{K^{\rm{I}}}
\frac{(x^+_0 - x^+_i) \(1 - \frac{1}{x_0^- x_i^+}\)}{(x^+_0 - x^-_i) \(1 - \frac{1}{x_0^- x_i^-}\)}
\nonumber\\[1mm]
&&\quad
- \prod_{i=1}^{K^{\rm{II}}} \frac{y_i-x^-_0}{y_i-x^+_0}\sqrt{\frac{x^+_0}{x^-_0}}
\prod_{i=1}^{K^{\rm{I}}} \frac{x^+_0-x^+_i}{x_0^+ - x_i^-} \sqrt{\frac{x_i^-}{x_i^+}} \times
\nonumber\\[1mm]
&&\hspace{50mm} \times
\left\{\prod_{i=1}^{K^{\rm{III}}} \frac{{\sf w}_i - u_0 -1}{{\sf w}_i - u_0} +
\prod_{i=1}^{K^{\rm{II}}} \frac{{\sf v}_i - u_0 - \frac12}{{\sf v}_i - u_0 + \frac12}
\prod_{i=1}^{K^{\rm{III}}} \frac{{\sf w}_i - u_0 + 1}{{\sf w}_i - u_0} \right\}.
\nonumber
\end{eqnarray}
It is useful to separate a common factor in the following fashion:
\begin{equation}
T_i = S_{\vev{0,0}} \, \tilde T_i \,,\qquad
S_{\vev{0,0}} \equiv \prod_{i=1}^{K^{\rm{II}}} \frac{y_i-x^-_0}{y_i-x^+_0}\sqrt{\frac{x^+_0}{x^-_0}} \,,\qquad (i=1, \ldots , 4).
\label{def:tilde Ti}
\end{equation}
Then, the tilded functions can be written as
\begin{alignat}{3}
\tilde T_1 &= \prod_{i=1}^{K^{\rm{II}}}
\frac{{\sf v}_i - u_0 - \frac12}{{\sf v}_i - u_0 + \frac12} \,
\prod_{i=1}^{K^{\rm{I}}}
\frac{\(1 - \frac{1}{x_0^- x_i^+}\) (x^+_0 - x^+_i)}{\(1 - \frac{1}{x_0^- x_i^-}\) (x^+_0 - x^-_i)} \,,
\label{T1 Full} \\[1mm]
\tilde T_2 &= \prod_{i=1}^{K^{\rm{III}}} \frac{{\sf w}_i - u_0 + 1}{{\sf w}_i - u_0} \,
\prod_{i=1}^{K^{\rm{II}}} \frac{{\sf v}_i - u_0 - \frac12}{{\sf v}_i - u_0 + \frac12} \,
\prod_{i=1}^{K^{\rm{I}}} \frac{x^+_0-x^+_i}{x_0^+ - x_i^-} \sqrt{\frac{x_i^-}{x_i^+}} \,,
\label{T2 Full} \\[2mm]
\tilde T_3 &= \prod_{i=1}^{K^{\rm{III}}} \frac{{\sf w}_i - u_0 -1}{{\sf w}_i - u_0} \,
\prod_{i=1}^{K^{\rm{I}}} \frac{x^+_0-x^+_i}{x_0^+ - x_i^-} \sqrt{\frac{x_i^-}{x_i^+}} \,,
\label{T3 Full} \\[1mm]
\tilde T_4 &= 1.
\label{T4 Full}
\end{alignat}
Note that different identification of $\tilde T_i$'s would produce
the transfer matrix for different representations
\cite{Beisert:2005di}.

One can now explicitly evaluate the the function $\tau$'s
appearing in the conjectured transfer matrix for the $\ell_0$\,-th
symmetric representation \eqref{symmetric transfer 2}. We will
only state the final result here and refer for details to
\cite{Arutyunov:2009iq}. The transfer matrix derived from the
conjecture on the quantum characteristic function:
\begin{alignat}{3}
&T_{\vev{\ell_0-1,0}} (u_0|\{\vec u,\vec v,\vec w\}) =
\prod_{i=1}^{K^{\rm{II}}} {\textstyle
\frac{y_i-x^{[-\ell_0]}_0}{y_i -
x^{[\ell_0]}_0}\sqrt{\frac{x^{[\ell_0]}_0}{x^{[-\ell_0]}_0}} } \
\times
\label{symmetric transfer 4} \\[2mm]
&\quad \Biggl( 1 + \prod_{i=1}^{K^{\rm{II}}} {\textstyle
\frac{{\sf v}_i - u_0 - \frac{\ell_0}{2}}{{\sf v}_i - u_0 +
\frac{\ell_0}{2}} } \, \prod_{i=1}^{K^{\rm{I}}} {\textstyle \left[
\frac{x^{[-\ell_0]}_0 - x^-_i}{x^{[\ell_0]}_0 - x^-_i} \, \frac{1
- \frac{1}{x^{[-\ell_0]}_0 x^+_i}}{1 - \frac{1}{x^{[\ell_0]}_0
x^+_i}} \, \frac{\mathscr{X}^{\ell_0,0}_{\ell_0}}{\mathcal{D}}
\right] }
\notag \\[1mm]
&\quad + \sum_{k=1}^{\ell_0-1} \pare{ \prod_{i=1}^{K^{\rm{II}}}
{\textstyle \frac{{\sf v}_i - u_0 - \frac{\ell_0}{2}}{{\sf v}_i -
u_0 - \frac{\ell_0-2k}{2}} } \, \prod_{i=1}^{K^{\rm{I}}}
{\textstyle \[ \frac{x^{[\ell_0-2k]}_0 - x^-_i}{x^{[\ell_0-2k]}_0
- x^+_i} \, \frac{u_0-u_i+\frac{\ell_0-\ell_i -
2k}{2}}{u_0-u_i+\frac{\ell_0-\ell_i}{2}} \,
\frac{x^{[\ell_0]}_0-x^+_i}{x^{[\ell_0]}_0-x^-_i} \,
\frac{\mathscr{X}^{k,0}_k}{\mathcal{D}} \] }}
\notag \\[1mm]
&\quad + \sum_{k=1}^{\ell_0-1} \pare{ \prod_{i=1}^{K^{\rm{II}}}
{\textstyle \frac{{\sf v}_i - u_0 - \frac{\ell_0}{2}}{{\sf v}_i -
u_0 - \frac{\ell_0-2k}{2}} }\, \prod_{i=1}^{K^{\rm{I}}}
{\textstyle \[ \frac{x^-_i}{x^+_i} \, \frac{1 -
\frac{1}{x^{[\ell_0-2k]}_0 x^-_i}}{1 - \frac{1}{x^{[\ell_0-2k]}_0
x^+_i}} \, \frac{u_0-u_i+\frac{\ell_0-\ell_i -
2k}{2}}{u_0-u_i+\frac{\ell_0-\ell_i}{2}} \,
\frac{x^{[\ell_0]}_0-x^+_i}{x^{[\ell_0]}_0-x^-_i} \,
\frac{\mathscr{X}^{k,0}_k}{\mathcal{D}} \] } }
\notag \\[1mm]
&\quad - \sum_{k=0}^{\ell_0-1} \pare{ \prod_{i=1}^{K^{\rm{III}}}
{\textstyle \frac{{\sf w}_i - u_0 - \frac{\ell_0-2k+1}{2}}{{\sf
w}_i - u_0 - \frac{\ell_0-2k-1}{2}} } \, \prod_{i=1}^{K^{\rm{II}}}
{\textstyle \frac{{\sf v}_i - u_0 - \frac{\ell_0}{2}}{{\sf v}_i -
u_0 - \frac{\ell_0-2k}{2}} } \, \prod_{i=1}^{K^{\rm{I}}}
{\textstyle \[ \sqrt{\frac{x^-_i}{x^+_i}} \,
\frac{u_0-u_i+\frac{\ell_0-\ell_i -
2k}{2}}{u_0-u_i+\frac{\ell_0-\ell_i}{2}}  \,
\frac{x^{[\ell_0]}_0-x^+_i}{x^{[\ell_0]}_0-x^-_i}\frac{\mathscr{X}^{k,0}_k}{\mathcal{D}}
\] }}
\notag \\[1mm]
&\quad - \sum_{k=0}^{\ell_0-1} \pare{ \prod_{i=1}^{K^{\rm{III}}}
{\textstyle \frac{{\sf w}_i - u_0 - \frac{\ell_0-2k-3}{2}}{{\sf
w}_i - u_0 - \frac{\ell_0-2k-1}{2}} } \, \prod_{i=1}^{K^{\rm{II}}}
{\textstyle \frac{{\sf v}_i - u_0 - \frac{\ell_0}{2}}{{\sf v}_i -
u_0 - \frac{\ell_0-2k-2}{2}} } \, \prod_{i=1}^{K^{\rm{I}}}
{\textstyle
\[ \sqrt{\frac{x^-_i}{x^+_i}} \,
\frac{u_0-u_i+\frac{\ell_0-\ell_i -
2k}{2}}{u_0-u_i+\frac{\ell_0-\ell_i}{2}}  \,
\frac{x^{[\ell_0]}_0-x^+_i}{x^{[\ell_0]}_0-x^-_i}\frac{\mathscr{X}^{k,0}_k}{\mathcal{D}}
\] }} \Biggr) . \notag
\end{alignat}
Finally by substituting $x^{[\ell_0-2k]}$ of \eqref{x intermediate
branches} into the definition of $\lambda_\pm (q,p_i,k)$ in
\eqref{eqn;lambda-pm}, we find
\begin{equation}
\lambda_\pm (q,p_i,k) = \begin{cases}
\ds \ \frac{x_0^{[\ell_0-2k]} - x_i^-}{x_0^{[\ell_0-2k]} - x_i^+} \, \frac{u_0 - u_i + \frac{\ell_0 - \ell_i - 2k}{2}}{u_0 - u_i + \frac{\ell_0 - \ell_i}{2}} \,
\frac{x_0^{[\ell_0]} - x_i^+}{x_0^{[\ell_0]} - x_i^-} \,
\frac{\mathscr{X}^{k,0}_k}{\mathcal{D}} \,,
\\[6mm]
\ds \ \frac{x_i^-}{x_i^+} \, \frac{1 - \frac{1}{x_0^{[\ell_0-2k]} x_i^-}}{1 - \frac{1}{x_0^{[\ell_0-2k]} x_i^+}} \, \frac{u_0 - u_i + \frac{\ell_0 - \ell_i - 2k}{2}}{u_0 - u_i + \frac{\ell_0 - \ell_i}{2}} \,
\frac{x_0^{[\ell_0]} - x_i^+}{x_0^{[\ell_0]} - x_i^-} \,
\frac{\mathscr{X}^{k,0}_k}{\mathcal{D}} \,.
\end{cases}
\end{equation}
From this one can compare the above result term by term with the
previously derived result and find agreement.

How the agreement works can be understood in the following way.
From the expression (\ref{symmetric transfer 4}) we see that,
apparently, a spurious dependence on the parameters $x_0^{[\ell_0
-2k]}$ is left among the different blocks of the quantum
characteristic function. However, one can make use of (\ref{x
intermediate branches}) to re-express each of these variables only
in terms of the bound state variable $x_0^{[\ell_0]}$, provided
one chooses a branch of the quadratic map. The remarkable
observation is that, after this replacement, one can recast the
above expression in a form that precisely agrees with our result
(\ref{eqn;FullEignvalue}). This happens for both choices of
branch, consistent with the fact that the formula we have obtained
{\it via} the alternative route of the ABA does not bear any
dependence on such a choice.


\section*{Acknowledgements}

First and foremost I would like to thank G. Arutyunov for many
valuable discussions and for sharing insights on this subject. I
am grateful to A. Torrielli, S. Frolov and R. Suzuki for fruitful
discussions and collaborations. I would also like to thank Z.
Bajnok, N. Beisert, B. de Wit, S. Frolov and  M. Staudacher for
giving useful comments on the manuscript.

\addcontentsline{toc}{section}{Appendix A: Algebraic Bethe ansatz
for the 6-vertex model}

\section*{Appendix A: Algebraic Bethe ansatz for the 6-vertex
model}\label{sect;6Vertex}

In this chapter we used the algebraic Bethe ansatz approach to
diagonalize the $\ads$ superstring transfer matrix for bound
states. We closely followed the discussion for the Hubbard model
\cite{martins-1997}. In this model, just as in our case, the
6-vertex model plays an important role. In this section we will
discuss the algebraic Bethe ansatz for this model, for
completeness and to fix notations.

The algebraic Bethe ansatz for the 6-vertex model is a standard
chapter of the theory of integrable systems, and it is treated for
example in \cite{Faddeev:1996iy,Korepin}. The scattering matrix of
the model is given by
\begin{eqnarray}\label{eqn;6VertexSmat}
r_{12}(u_1,u_2) = \begin{pmatrix}
  1 & 0 & 0 & 0 \\
  0 & b(u_1,u_2) & a(u_1,u_2) & 0 \\
  0 & a(u_1,u_2) & b(u_1,u_2) & 0 \\
  0 & 0 & 0 & 1
\end{pmatrix},
\end{eqnarray}
where
\begin{eqnarray}
a = \frac{U}{u_1-u_2+U}, \qquad b = \frac{u_1-u_2}{u_1-u_2+U}.
\end{eqnarray}
It is convenient to write it as
\begin{eqnarray}\nonumber
r_{12}(u_1,u_2)
&=& r_{\alpha\beta}^{\gamma\delta}(u_1,u_2)E^{\alpha}_{\gamma}\otimes E^{\beta}_{\delta}\\
&=& \frac{u_1-u_2}{u_1-u_2+U}\left[E^\alpha_\alpha\otimes
E^\beta_\beta +\frac{U}{u_1-u_2} E^\alpha_\beta\otimes
E^\beta_\alpha\right]\label{eqn;6vertexComponents},
\end{eqnarray}
with $E^\alpha_\beta$ the standard matrix unities (the matrices
where all entries are zero except for a $1$ at position
$(\beta,\alpha)$). Let us consider $K$ particles, with rapidities
$u_i$. One can construct the monodromy matrix
\begin{eqnarray}
\mathcal{T}(u_0|\vec{u})= \prod_{i=1}^{K} r_{0i}(u_0|u_i).
\end{eqnarray}
Let us write it as a matrix in the auxiliary space
\begin{eqnarray}
\mathcal{T}^{(1)}(u_0|\vec{u})= \begin{pmatrix}
  A(u_0|\vec{u}) & B(u_0|\vec{u}) \\
  C(u_0|\vec{u}) & D(u_0|\vec{u})
\end{pmatrix}.
\end{eqnarray}
In the algebraic Bethe Ansatz, one constructs the eigenvalues of
the transfer matrix by first specifying a ground state
$|0\rangle$. The ground state, in this case, is defined as
\begin{eqnarray}
|0\rangle = \bigotimes_{i=1}^{K}{1\choose0}.
\end{eqnarray}
It is easily checked that it is an eigenstate of the transfer
matrix. More precisely, the action of the different
elements of the monodromy matrix on $|0\rangle$ is given by
\begin{eqnarray}
 A(u_0|\vec{u})|0\rangle &=& |0\rangle,\nonumber\\
 C(u_0|\vec{u})|0\rangle &=& 0,\\
 D(u_0|\vec{u})|0\rangle &=& \prod_{i=1}^{K}b(u_0,u_i)|0\rangle\nonumber.
\end{eqnarray}
Thus, $|0\rangle$ is an eigenstate of the transfer matrix with the
following eigenvalue
\begin{eqnarray}
1+\prod_{i=1}^{K}b(u_0,u_i).
\end{eqnarray}
The operator $B$ from the monodromy matrix will be considered as a
creation operator. It will create all the other eigenstates out of
the vacuum. We introduce additional parameters $w_i$ and consider
the state
\begin{eqnarray}
|M\rangle := \phi_M(w_1,\ldots,w_M)|0\rangle,\qquad
\phi_M(w_1,\ldots,w_M):=\prod_{i=1}^{M} B(w_i|\vec{u}).
\end{eqnarray}
In the context of the Heisenberg spin chain the vacuum corresponds
to all spins down and the state $|M\rangle$ corresponds to the
eigenstate of the transfer matrix that has $M$ spins turned up.

In order to evaluate the action of the transfer matrix
$\mathscr{T}(u_0|\vec{u}) = A(u_0|\vec{u})+D(u_0|\vec{u})$ on the
state $|M\rangle$, one needs the commutation relations between the
fields $A,B,D$. From (\ref{eqn;YBE-operators}) applied to this
S-matrix (\ref{eqn;6VertexSmat}) one reads
\begin{eqnarray}
A(u_0|\vec{u})B(w|\vec{u})&=&\frac{1}{b(w,u_0)}B(w|\vec{u})A(u_0|\vec{u})
-\frac{a(w,u_0)}{b(w,u_0)}B(u_0|\vec{u})A(w|\vec{u})\nonumber\\
B(w_1|\vec{u})B(w_2|\vec{u})&=&B(w_2|\vec{u})B(w_1|\vec{u})\\
D(u_0|\vec{u})B(w|\vec{u})&=&\frac{1}{b(u_0,w)}B(w|\vec{u})D(u_0|\vec{u})
-\frac{a(u_0,w)}{b(u_0,w)}B(u_0|\vec{u})D(w|\vec{u}).\nonumber
\end{eqnarray}
From this, one can determine exactly when $|M\rangle$ is an
eigenstate of the transfer matrix. By definition we have that
\begin{eqnarray}\label{eqn;XXXinduction}
|M\rangle = B(w_M|\vec{u})|M-1\rangle,
\end{eqnarray}
and this allows us to use induction. By using the identity
\begin{eqnarray}
\frac{1}{b(w_M,u_0)}\frac{a(w_i,u_0)}{b(w_i,u_0)}-\frac{a(w_M,u_0)}{b(w_Mu_0)}\frac{a(w_i,w_M)}{b(w_i,w_M)}
= \frac{a(w_i,u_0)}{b(w_i,u_0)}\frac{1}{b(w_M,w_i)}
\end{eqnarray}
in (\ref{eqn;XXXinduction}) one can prove
\begin{eqnarray}
A(u_0|\vec{u})\phi_M(w_1,\ldots,w_M) &=& \prod_{i=1}^{M}\frac{1}{b(w_i,u_0)}\phi_M(w_1,\ldots,w_M)A(u_0|\vec{u})\\
&&-
\sum_{i=1}^{M}\left[\frac{a(w_i,u_0)}{b(w_i,u_0)}\prod_{j=1,j\neq
i}^{M}\frac{1}{b(w_j,w_i)}\hat{\phi}_M
A(w_i|\vec{u})\right],\nonumber
\end{eqnarray}
where $\hat{\phi}_M $ stands for
$\phi_M(\ldots,w_{i-1},u_0,w_{i+1},\ldots)$. One can find a
similar relation for the commutator between $D$ and $B$. Using
these relations gives
\begin{eqnarray}
\mathscr{T}(u_0|\vec{u})|M\rangle &=&
\left\{A(u_0|\vec{u})+D(u_0|\vec{u})\right\}|M\rangle\\
&=&\phi_M(w_1,\ldots,w_M)\left\{A(u_0|\vec{u})\prod_{i=1}^{M}\frac{1}{b(w_i,u_0)}
+D(u_0|\vec{u})\prod_{i=1}^{M}\frac{1}{b(u_0,w_i)}\right\}|0\rangle \nonumber\\
&&-\sum_{i=1}^{M}\left[\frac{a(w_i,u_0)}{b(w_i,u_0)}\, \hat{\phi}_M \left\{\prod_{j\neq
i}\frac{1}{b(w_j,w_i)} A(w_i|\vec{u})-\prod_{j\neq
i}\frac{1}{b(w_i,w_j)}
D(w_i|\vec{u})\right\}\right]|0\rangle.\nonumber
\end{eqnarray}
From this we find that $|M\rangle$ is an eigenstate of the
transfer matrix with eigenvalue
\begin{eqnarray}\label{eqn;eigenvalue6V}
\Lambda^{(6v)}(u_0|\vec{u}) = \prod_{i=1}^M
\frac{1}{b(w_i,u_0)}+\prod_{i=1}^M
\frac{1}{b(u_0,w_i)}\prod_{i=1}^K b(u_0,u_i)
\end{eqnarray}
provided that the auxiliary parameters $w_i$ satisfy the following
equations
\begin{eqnarray}\label{eqn;AuxEqns6V}
\prod_{i=1}^K b(w_j,u_i)=\prod_{i=1,i\neq j}^M
\frac{b(w_j,w_i)}{b(w_i,w_j)}.
\end{eqnarray}
This now completely determines the spectrum of the 6-vertex model.

To conclude, we briefly explain how these eigenvalues are used to
generate an infinite tower of conserved charges. From
(\ref{eqn;YBE-operators}) one finds that
\begin{eqnarray}
\mathscr{T}(u_0|\vec{u})\mathscr{T}(\mu|\vec{u})=\mathscr{T}(\mu|\vec{u})\mathscr{T}(u_0|\vec{u}).
\end{eqnarray}
This means that if one writes $\mathscr{T}(u_0|\vec{u})$ as a
series the auxiliary parameter $u_0$, the coefficients of this
series will depend on $\vec{u}$ and they mutually commute. It
actually turns out that the 6-vertex model Hamiltonian can be
written in terms of these coefficients.


\addcontentsline{toc}{section}{Appendix B: Excited states,
$K^{{\rm III}}=0$}

\section*{Appendix B: Excited states, $K^{{\rm III}}=0$}\label{sec;TasBA}

In this section we will discuss the class of higher excited states
with $K^{{\rm III}}=0$. We will present for these states a full
derivation of transfer matrix eigenvalues and auxiliary Bethe
equations. From the general construction it is easily seen that a
more general eigenvector of the transfer matrix is given by
\begin{eqnarray}
|a\rangle = \Phi(\lambda_1,\ldots,\lambda_a)|0\rangle_P, \qquad
\Phi(\lambda_1,\ldots,\lambda_a) = B_3(\lambda_1)\ldots
B_3(\lambda_a).
\end{eqnarray}
These states have quantum number $K^{\rm{III}}=0$. This allows for
a similar inductive procedure as applied to the 6-vertex model in
appendix A. Furthermore, because of the properties
of the creation operators (\ref{eqn;CommRelCreation}), we find
that
\begin{eqnarray}\label{eqn;symmetryTas}
\Phi(\lambda_1,\ldots,\lambda_{j-1},\lambda_{j},\ldots\lambda_a) =
-\mathscr{X}^{0,0}_0
r^{33}_{33}(\lambda_{j-1},\lambda_{j})\Phi(\lambda_1,\ldots,\lambda_{j},\lambda_{j-1},\ldots\lambda_a).
\end{eqnarray}
This means that all permutations of the momenta $\lambda_i$ are
related to each other by a simple multiplication by a scalar
prefactor. We will exploit this property later on. Let us first
derive some useful identities. One uses induction to show that
\begin{eqnarray}
A^{4}_3|a\rangle = C^*_3|a\rangle = C_4|a\rangle = C|a\rangle=0,
\end{eqnarray}
for any $a$. This vastly simplifies the computations, since we can
discard any term proportional to the above operators from the
commutation relations. Let us first turn to (\ref{eqn;CommRel}).
This now becomes, after discarding the term proportional to
$C^*_3$,
\begin{eqnarray}
\mathcal{T}_{3,k}^{3,k}(q)B_{3}(\lambda)
=\frac{\mathscr{X}^{k,0}_k}{\mathscr{Y}^{k,0;1}_{k;1}}B_{3}(\lambda)\mathcal{T}_{3,k}^{3,k}(q)
+\frac{\mathscr{Y}^{k,1;1}_{k;2}}{\mathscr{Y}^{k,0;1}_{k;1}}\mathcal{T}_{3,k}^{k}(q)A_{3}^{3}(\lambda)
+\frac{\mathscr{Y}^{k,1;1}_{k;4}}{\mathscr{Y}^{k,0;1}_{k;1}}\mathcal{T}_{3,k}^{34,k-1}(q)A_{3}^{3}(\lambda).\nonumber
\end{eqnarray}
Applying this to $\Phi(\lambda_1,\ldots,\lambda_a) =
B_3(\lambda_1)\Phi(\lambda_2,\ldots,\lambda_a)$ we find
\begin{eqnarray}\label{eqn;TasComm1}
\mathcal{T}_{3,k}^{3,k}(q)\Phi(\lambda_1,\ldots,\lambda_a)&=&
\frac{\mathscr{X}^{k,0}_k}{\mathscr{Y}^{k,0;1}_{k;1}}B_{3}(\lambda_1)\mathcal{T}_{3,k}^{3,k}(q)\Phi(\lambda_2,\ldots,\lambda_a)\nonumber\\
&&+\frac{\mathscr{Y}^{k,1;1}_{k;2}}{\mathscr{Y}^{k,0;1}_{k;1}}\mathcal{T}_{3,k}^{k}(q)A_{3}^{3}(\lambda_1)\Phi(\lambda_2,\ldots,\lambda_a)\\
&&+\frac{\mathscr{Y}^{k,1;1}_{k;4}}{\mathscr{Y}^{k,0;1}_{k;1}}\mathcal{T}_{3,k}^{34,k-1}(q)A_{3}^{3}(\lambda_1)\Phi(\lambda_2,\ldots,\lambda_a).\nonumber
\end{eqnarray}
Obviously, by applying this relation recursively one finds
\begin{eqnarray}
\mathcal{T}_{3,k}^{3,k}(q)\Phi(\lambda_1,\ldots,\lambda_a)&=&\prod_{i=1}^{a}\frac{\mathscr{X}^{k,0}_k(q,\lambda_i)}{\mathscr{Y}^{k,0;1}_{k;1}(q,\lambda_i)}\Phi(\lambda_1,\ldots,\lambda_a)\mathcal{T}_{3,k}^{3,k}(q)\nonumber\\
&&+\sum_{i=1}^{a}c_i\Phi_{k;i}(q,\lambda)A_{3}^{3}(\lambda_i)+\sum_{i=1}^{a}d_i\Psi_{k;i}(q,\lambda)A_{3}^{3}(\lambda_i),\qquad
\end{eqnarray}
where $c_i$ are some numerical coefficients and
$\Phi_{k;i}(q,\lambda) = \mathcal{T}_{3,k}^{k}(q) \prod_{j\neq i}
B_3(\lambda_i)$, \\
\noindent $\Psi_{k;i}(q,\lambda) = \mathcal{T}_{3,k}^{34,k-1}(q)
\prod_{j\neq i} B_3(\lambda_i)$. It is easily seen from
(\ref{eqn;TasComm1}) that the numerical coefficients in front of
$\Phi_{k;1}(q,\lambda),\Psi_{k;1}(q,\lambda)$ are given by
\begin{eqnarray}
c_1 =
\frac{\mathscr{Y}^{k,1;1}_{k;2}(q,\lambda_1)}{\mathscr{Y}^{k,0;1}_{k;1}(q,\lambda_1)}
\prod_{i=2}^{a}\frac{\mathscr{X}^{k,0}_k(q,\lambda_i)}{\mathscr{Y}^{k,0;1}_{k;1}(q,\lambda_i)},
\qquad d_1 =
\frac{\mathscr{Y}^{k,1;1}_{k;4}(q,\lambda_1)}{\mathscr{Y}^{k,0;1}_{k;1}(q,\lambda_1)}
\prod_{i=2}^{a}\frac{\mathscr{X}^{k,0}_k(q,\lambda_i)}{\mathscr{Y}^{k,0;1}_{k;1}(q,\lambda_i)}.
\end{eqnarray}
Here we can exploit the symmetry property (\ref{eqn;symmetryTas})
to relate all the other coefficients to this one. Let us denote
these proportionality coefficients by $\mathcal{P}_{1i}$. We
find
\begin{eqnarray}
\mathcal{T}_{3,k}^{3,k}(q)\Phi(\lambda_1,\ldots,\lambda_a)&=&\prod_{i=1}^{a}\frac{\mathscr{X}^{k,0}_k(q,\lambda_i)}{\mathscr{Y}^{k,0;1}_{k;1}(q,\lambda_i)}\Phi(\lambda_1,\ldots,\lambda_a)\mathcal{T}_{3,k}^{3,k}(q)\nonumber\\
&&+\sum_{i=1}^{a}c_iP_{1i}\Phi_{k;i}(q,\lambda)A_{3}^{3}(\lambda_i)+\sum_{i=1}^{a}d_iP_{1i}\Psi_{k;i}(q,\lambda)A_{3}^{3}(\lambda_i),\qquad~
\end{eqnarray}
where
\begin{eqnarray}
c_j =
\frac{\mathscr{Y}^{k,1;1}_{k;2}(q,\lambda_j)}{\mathscr{Y}^{k,0;1}_{k;1}(q,\lambda_j)}
\prod_{i=1,i\neq j
}^{a}\frac{\mathscr{X}^{k,0}_k(q,\lambda_i)}{\mathscr{Y}^{k,0;1}_{k;1}(q,\lambda_i)},\qquad
d_j =
\frac{\mathscr{Y}^{k,1;1}_{k;4}(q,\lambda_j)}{\mathscr{Y}^{k,0;1}_{k;1}(q,\lambda_j)}
\prod_{i=1,i\neq j
}^{a}\frac{\mathscr{X}^{k,0}_k(q,\lambda_i)}{\mathscr{Y}^{k,0;1}_{k;1}(q,\lambda_i)}.~~
\end{eqnarray}
Next, we consider the commutator with $\mathcal{T}^k_k +
\mathcal{T}^{34,k-1}_{34,k-1}$. Upon dismissing terms that vanish
because they have annihilation operators acting on the vacuum, we
find
\begin{eqnarray}
\left[\mathcal{T}^k_k +
\mathcal{T}^{34,k-1}_{34,k-1}\right]B_3(\lambda) &=&
\frac{\mathscr{X}^{k,0}_k}{\mathscr{Y}^{k,0;1}_{k;1}}B_3(\lambda)\left[\mathcal{T}^k_k
+ \mathcal{T}^{34,k-1}_{34,k-1}\right] +\\
&&\frac{\mathscr{Y}^{k,1;1}_{k;2}}{\mathscr{Y}^{k,0;1}_{k;1}}\left\{\mathcal{T}^{k}_{3,k}B-\mathcal{T}^{4,k-1}_{34,k-1}A^3_3\right\}
+\frac{\mathscr{Y}^{k,1;1}_{k;4}}{\mathscr{Y}^{k,0;1}_{k;1}}\left\{\mathcal{T}^{34,k-1}_{3,k}B+\mathcal{T}^{4,k-1}_{k}A^3_3\right\}.\nonumber
\end{eqnarray}
If we now define $\hat{\Phi}_{k;i}(q,\lambda) =
\mathcal{T}^{4,k-1}_{k}(q) \prod_{j\neq i}
\ad{B_3(\lambda_j)},\hat{\Psi}_{k;i}(q,\lambda) =
\mathcal{T}^{4,k-1}_{34,k-1}(q) \prod_{j\neq i} \ad{B_3(\lambda_j)}$,
then we can repeat the above steps to find
\begin{eqnarray}
\left[\mathcal{T}^k_k +
\mathcal{T}^{34,k-1}_{34,k-1}\right]\Phi(\lambda_1,\ldots,\lambda_a)
&=&
\prod_{i=1}^{a}\frac{\mathscr{X}^{k,0}_k(q,\lambda_i)}{\mathscr{Y}^{k,0;1}_{k;1}(q,\lambda_i)}\Phi(\lambda_1,\ldots,\lambda_a)\left[\mathcal{T}^k_k
+ \mathcal{T}^{34,k-1}_{34,k-1}\right] +\nonumber\\
&&\sum_{i=1}^{a}
c_iP_{1i}\left\{\Phi_{k;i}(q,\lambda)B(\lambda_i)-\hat{\Psi}_{k;i}(q,\lambda)A^3_3(\lambda_i)\right\}
+\\
&&\sum_{i=1}^{a}
d_iP_{1i}\left\{\Psi_{k;i}(q,\lambda)B(\lambda_i)+\hat{\Phi}_{k;i}(q,\lambda)A^3_3(\lambda_i)\right\}.\qquad\nonumber
\end{eqnarray}
The last commutation relation finally gives
\begin{eqnarray}
\mathcal{T}_{4,k}^{4,k}(q)\Phi(\lambda_1,\ldots,\lambda_a)&=&
\frac{\mathscr{X}^{k+1,0}_{k+1}}{\mathscr{Y}^{k,0;1}_{k;1}}\frac{u_q-u_{\lambda_1}+\frac{\ell_0-1}{2}-k}{u_q-u_{\lambda_1}+\frac{\ell_0-3}{2}-k}B_{3}(\lambda_1)\mathcal{T}_{4,k}^{4,k}(q)\Phi(\lambda_2,\ldots,\lambda_a)\nonumber\\
&&-\frac{\mathscr{Y}^{k+1,1;1}_{k+1;2}}{\mathscr{Y}^{k+1,0;1}_{k+1;1}}\mathcal{T}_{34,k}^{4,k}(q)B(\lambda_1)\Phi(\lambda_2,\ldots,\lambda_a)\\
&&+\frac{\mathscr{Y}^{k+1,1;1}_{k+1;4}}{\mathscr{Y}^{k+1,0;1}_{k+1;1}}\mathcal{T}_{k+1}^{4,k}(q)B(\lambda_1)\Phi(\lambda_2,\ldots,\lambda_a).\nonumber
\end{eqnarray}
By summing all the terms, we find that $|a\rangle$ is indeed an
eigenstate of the transfer matrix, provided that the parameters
$\lambda_i$ satisfy
\begin{eqnarray}
B(\lambda_i)|0\rangle_P = A^3_3(\lambda_i)|0\rangle_P.
\end{eqnarray}
When working this out, we only find a dependence on
$x^+(\lambda_i)$, which we denote as $y_i\equiv x^+(\lambda_i)$.
The explicit formula is given by
\begin{eqnarray}
\prod_{j=1}^{K^{\rm{I}}}
\frac{y_i-x^+_j}{y_i-x^-_j}\sqrt{\frac{x_j^-}{x^+_j}}= 1,
\end{eqnarray}
which agrees with the known auxiliary BAE for $K^{\rm{III}}=0$
from (\ref{eqn;FullEignvalue}). The explicit eigenvalue of
$|a\rangle$ is given by
\begin{eqnarray}
\Lambda(q|\vec{p}) &=&
\prod_{m=1}^{K^{\rm{II}}}\frac{\mathscr{X}^{0,0}_0(q,\lambda_m)}{\mathscr{Y}^{0,0;1}_{0;1}(q,\lambda_m)}
+ \prod_{i=1}^{K^{\rm{I}}} \mathscr{Z}^{\ell_0,0;1}_{\ell_0;1}(q,p_i)\prod_{m=1}^{K^{\rm{II}}}\frac{\mathscr{X}^{\ell_0,0}_{\ell_0}(q,\lambda_m)}{\mathscr{Y}^{\ell_0,0;1}_{\ell_0;1}(q,\lambda_m)}+\nonumber\\
&&\sum_{k=1}^{\ell_0-1}
\prod_{m=1}^{K^{\rm{II}}}\frac{\mathscr{X}^{k,0}_k(q,\lambda_m)}{\mathscr{Y}^{k,0;1}_{k;1}(q,\lambda_m)}\left\{\prod_{i=1}^{K^{\rm{I}}}\lambda_+(q,p_i)
+\prod_{i=1}^{K^{\rm{I}}}\lambda_-(q,p_i)\right\}+\\
&&-\sum_{k=0}^{\ell_0-1}
\prod_{m=1}^{K^{\rm{II}}}\frac{\mathscr{X}^{k,0}_k(q,\lambda_m)}{\mathscr{Y}^{k,0;1}_{k;1}(q,\lambda_m)}
\left[1+\frac{u_q-u_{\lambda_m}+\frac{\ell_0}{2}-k}{u_q-u_{\lambda_m}+\frac{\ell_0-2}{2}-k}\right]\prod_{i=1}^{K^{\rm{I}}}\mathscr{Y}^{k,0;1}_{k;1}(q,p_i).\nonumber
\end{eqnarray}
This is indeed the case $K^{\rm{III}}=0$ of
(\ref{eqn;FullEignvalue}).